\documentclass[onecolumn]{aastex6}
\usepackage{graphicx}
\usepackage{natbib}
\usepackage{amsmath,amsthm,amssymb}
\usepackage{color}
\usepackage{ulem}
\usepackage{epstopdf}
\newcommand{\gps}{\ensuremath{g_{\rm P1}}}
\newcommand{\rps}{\ensuremath{r_{\rm P1}}}
\newcommand{\ips}{\ensuremath{i_{\rm P1}}}
\newcommand{\zps}{\ensuremath{z_{\rm P1}}}
\newcommand{\yps}{\ensuremath{y_{\rm P1}}}

\newcommand{\grizy}{\gps\rps\ips\zps\yps}
\newcommand{\jtwo}{\ensuremath{J_{\rm 2MASS}}}
\newcommand{\htwo}{\ensuremath{H_{\rm 2MASS}}}
\newcommand{\ktwo}{\ensuremath{K_{S,{\rm 2MASS}}}}
\newcommand{\PS}{\protect \hbox {Pan-STARRS1}}
\newcommand{\WISE}{{\it WISE}}
\newcommand{\um}{$\mu$m}
\newcommand{\rchi}{\ensuremath{\chi^2_\nu}}
\newcommand{\mua}{\ensuremath{\mu_\alpha{\rm cos}\,\delta}}     % pm_ra * cos(dec)
\newcommand{\mud}{\ensuremath{\mu_\delta}}                                % pm_dec
\newcommand{\my}{\protect \hbox {mas yr$^{-1}$}}
\newcommand{\kms}{\protect \hbox {km s$^{-1}$}}

\newcommand{\vtan}{\ensuremath{v_{\rm tan}}}
\newcommand{\stan}{\ensuremath{\sigma_{\rm tan}}}
\newcommand{\gr}{$g_{\rm P1}-r_{\rm P1}$}
\newcommand{\gi}{$g_{\rm P1}-i_{\rm P1}$}
\newcommand{\gy}{$g_{\rm P1}-y_{\rm P1}$}
\newcommand{\ri}{$r_{\rm P1}-i_{\rm P1}$}
\newcommand{\rz}{$r_{\rm P1}-z_{\rm P1}$}
\newcommand{\ry}{$r_{\rm P1}-y_{\rm P1}$}
\newcommand{\iz}{$i_{\rm P1}-z_{\rm P1}$}
\newcommand{\iy}{$i_{\rm P1}-y_{\rm P1}$}
\newcommand{\ijt}{$i_{\rm P1}-J_{\rm 2MASS}$}
\newcommand{\zy}{$z_{\rm P1}-y_{\rm P1}$}
\newcommand{\zjt}{$z_{\rm P1}-J_{\rm 2MASS}$}
\newcommand{\yjt}{$y_{\rm P1}-J_{\rm 2MASS}$}

\newcommand{\jht}{$(J-H)_{\rm 2MASS}$}
\newcommand{\jkt}{$(J-K_S)_{\rm 2MASS}$}
\newcommand{\hkt}{$(H-K_S)_{\rm 2MASS}$}
\newcommand{\ywa}{$y_{\rm P1}-W1$}
\newcommand{\wawb}{$W1-W2$}
\newcommand{\wbwc}{$W2-W3$}
\newcommand{\fldg}{\mbox{\textsc{fld-g}}}
\newcommand{\intg}{\mbox{\textsc{int-g}}}

\newcommand{\varnobjtot}{9888}
\newcommand{\varnmdwarf}{8271}
\newcommand{\varnmdwarfwest}{7808}
\newcommand{\varnmdwarfmega}{463}
\newcommand{\varnldwarf}{1265}
\newcommand{\varntdwarf}{352}
\newcommand{\varnltdwarf}{1617}
\newcommand{\varnlitmltdwarf}{2080}
\newcommand{\varnbinary}{81}
\newcommand{\varpblend}{4}
\newcommand{\varnrbandldwarf}{494}
\newcommand{\varnrbandldwarfnorm}{433}
\newcommand{\varngaia}{7772}
\newcommand{\varngaialdwarf}{284}
\newcommand{\varnplxjmag}{234}
\newcommand{\varnphotcols}{51}

\newcommand{\varnpmtot}{9770}
\newcommand{\varnpmerr}{2.9}
\newcommand{\varnpmucool}{2405}
\newcommand{\varnpml}{1242}
\newcommand{\varnpmt}{260}

\newcommand{\varnnewpmucool}{406}
\newcommand{\varnbetterpmucool}{1176}

\shorttitle{M, L, and T Dwarfs from Pan-STARRS1}
\shortauthors{Best, W. M. J. et al}

\slugcomment{\large Accepted by {\it The Astrophysical Journal Supplement}, 2017 October 4}

\begin{document}

\title{Photometry and Proper Motions of M, L, and T Dwarfs from the \PS\
  3$\pi$ Survey}
\author{William M. J. Best\altaffilmark{1}, 
  Eugene A. Magnier\altaffilmark{1}, 
  Michael C. Liu\altaffilmark{1},
  Kimberly M. Aller\altaffilmark{1}, 
  Zhoujian Zhang\altaffilmark{1}, 
  W. S. Burgett\altaffilmark{2}, 
  K. C. Chambers\altaffilmark{1}, 
  P. Draper\altaffilmark{3}, 
  H. Flewelling\altaffilmark{1}, 
  N. Kaiser\altaffilmark{1}, 
  R.-P. Kudritzki\altaffilmark{1}, 
  N. Metcalfe\altaffilmark{3}, 
  J. L. Tonry\altaffilmark{1}, 
  R. J. Wainscoat\altaffilmark{1},
  C. Waters\altaffilmark{1}
}

% The ordering here should be sequential, matching the sequence in the list of authors:
\altaffiltext{1}{Institute for Astronomy, University of Hawaii, 2680 Woodlawn Drive, Honolulu, HI 96822, USA; wbest@ifa.hawaii.edu}
\altaffiltext{2}{GMTO Corporation, 465 N. Halstead St., Suite 250, Pasadena, CA 91107, USA}
\altaffiltext{3}{Department of Physics, Durham University, South Road, Durham DH1 3LE, UK}

\begin{abstract}
  We present a catalog of \varnobjtot\ M, L and T dwarfs detected in the
  \PS~3$\pi$ Survey (PS1), covering three-quarters of the sky.
  Our catalog contains nearly all known objects of spectral types L0--T2 in the
  PS1 field, with objects as early as M0 and as late as T9, and includes PS1,
  2MASS, AllWISE, and \textit{Gaia}~DR1 photometry.
  We analyze the different types of photometry reported by PS1, and use two
  types in our catalog to maximize both depth and accuracy.
  Using parallaxes from the literature, we construct empirical SEDs for field
  ultracool dwarfs spanning $0.5-12$~\um.
  We determine typical colors of M0--T9 dwarfs, and we highlight the distinctive
  colors of subdwarfs and young objects.
  Our catalog includes \varnrbandldwarf~L~dwarfs detected in \rps, the largest
  sample of L~dwarfs detected at such blue wavelengths.
  We combine astrometry from PS1, 2MASS, and \textit{Gaia}~DR1 to calculate new
  proper motions for our catalog.
  We achieve a median precision of \varnpmerr~\my, a factor of $\approx$3$-$10
  improvement over previous large catalogs.
  Our catalog contains proper motions for \varnpmucool~M6--T9 dwarfs and
  includes the largest set of homogeneous proper motions for L and T dwarfs
  published to date, \varnnewpmucool~objects for which there were no previous
  measurements, and \varnbetterpmucool~objects for which we improve upon
  previous literature values.
  We analyze the kinematics of ultracool dwarfs in our catalog and find evidence
  that bluer but otherwise generic late-M and L~field dwarfs (i.e., not
  subdwarfs) tend to have higher tangential velocities compared to typical field
  objects.
  With the public release of the PS1 data, this survey will continue to be an
  essential tool for characterizing the ultracool dwarf population.
\end{abstract}


\section{Introduction}
\label{intro}
Ultracool dwarfs (spectral types M6 and later) are the lowest-mass members of
the stellar population, encompassing the coolest stars, brown dwarfs, and
planetary-mass objects.  The discovery of brown dwarfs over 20 years ago
launched an understanding of the complex properties and evolution of ultracool
atmospheres \citep[e.g.,][]{Burrows:2006ia}, and allowed us to constrain the
low-mass end of the stellar mass and luminosity functions in the solar
neighborhood \citep[and references therein]{Marocco:2015iz}.  In addition, the
youngest ($\approx$10--100~Myr) ultracool dwarfs in the field appear to be our
best analogs to directly-imaged giant planets \citep[e.g.,][]{Liu:2013gy}, and
they are far easier to observe without the drowning glare of host stars.  The
major drivers for ultracool discoveries, which now include $\approx$2,000 L and
T dwarfs and many thousands of late-M dwarfs, have been wide-field imaging
surveys such as the Deep Near Infrared Survey of the Southern Sky
\citep[DENIS,][]{Epchtein:1999tz}, the Sloan Digital Sky Survey
\citep[SDSS;][]{York:2000gn}, the Two Micron All Sky Survey
\citep[2MASS;][]{Skrutskie:2006hl}, the UKIRT Infrared Deep Sky Survey
\citep[UKIDSS;][]{Lawrence:2007hu}, and the Wide-Field Infrared Survey Explorer
\citep[\WISE;][]{Wright:2010in}.

Large photometric samples obtained from these imaging surveys have provided much
of our fundamental knowledge about ultracool dwarfs.  Samples of L~dwarfs have
revealed a surprising diversity of near-IR colors
\citep[e.g.][]{Leggett:2002cd,Knapp:2004ji,Gizis:2012kv} which are believed to
be caused by variations in surface gravity and/or dusty clouds
\citep[e.g.][]{Kirkpatrick:2008ec,Allers:2013hk} or thermo-chemical
instabilities \citep{Tremblin:2016hi}.  Objects transitioning from L to T
spectral types undergo a dramatic shift to bluer near-IR colors thought to be
driven by the clearing of clouds and the formation of methane
\citep[e.g.,][]{Burgasser:2002fy,Chiu:2006jd,Saumon:2008im}.  UKIDSS and \WISE\
have illustrated the diversity of late-T and Y dwarf near- and mid-IR colors
\citep[e.g.,][]{Burningham:2010dh,Kirkpatrick:2011ey,Mace:2013jh}, and \WISE\
has enabled the discovery of the coolest known substellar objects
\citep[e.g.,][]{Cushing:2011dk,Kirkpatrick:2012ha,Luhman:2014jd}.  Large samples
have revealed the mass and luminosity functions of the local ultracool
population \citep[e.g.,][]{Allen:2005jf,Cruz:2007kb,Burningham:2010dh}.
Measurements of the space density of brown dwarfs
\citep[e.g.][]{Reid:2008fz,Metchev:2008gx} have identified a relative paucity of
L/T transition dwarfs, indicating that this evolutionary phase is short-lived
and constraining the birth history of substellar objects
\citep[e.g.,][]{DayJones:2013hm,Marocco:2015iz}. The surveys have also enhanced
brown dwarf searches in star-forming regions
\citep[e.g.,][]{Lodieu:2009ep,Martin:2010cx}, important for determination of the
substellar initial mass function.  Photometric samples encompassing more than
one survey have enabled us to determine ultracool colors across a broad range of
wavelengths \citep[e.g.,][]{Schmidt:2015hv,Skrzypek:2015kp} and to measure
bolometric luminosities that yield effective temperatures and constraints on
atmospheric and evolutionary models
\citep[e.g.][]{Leggett:2002cd,Golimowski:2004en}.

Similarly, large samples of proper motions have contributed significantly to our
discovery and understanding of the ultracool population.  Proper motions have
enabled searches to distinguish ultracool dwarfs from distant luminous red
objects such as giants and galaxies
\citep[e.g.,][]{Kirkpatrick:2000gi,Lepine:2005jx,Theissen:2016gn,Theissen:2017df}
and to detemine whether individual discoveries are members of star-forming
regions \citep[e.g.,][]{Lodieu:2007if,Lodieu:2012jw}.  Proper motions have
helped to find objects in crowded areas of the sky such as the Galactic plane
\citep[e.g.,][]{Luhman:2013bp,Smith:2014df} and to identify ultracool dwarfs
with atypical colors that were missed by color cuts used in photometry-only
searches \citep[e.g.,][]{Kirkpatrick:2010dc}.  Several studies have found clear
evidence for dynamically cold (slow-moving) and hot (fast-moving) populations of
ultracool dwarfs that are consistent with thin disk and thick disk/halo
populations \citep[e.g.,][]{Faherty:2009kg,Schmidt:2010ex,Dupuy:2012bp},
implying that ultracool dwarfs form in the same manner as hotter stars.
Searches for high-proper motion objects, often using surveys with shorter time
baselines, have identified rare fast-moving objects that are typically members
of the older, low-metallicity populations
\citep[e.g.,][]{Jameson:2008ha,Smith:2014gw,Kirkpatrick:2014kv} or very nearby,
previously overlooked objects
\citep[e.g.,][]{Luhman:2014hj,Luhman:2014jd,Schneider:2016bs,Kirkpatrick:2016jt}.
Proper motions measured from the large surveys have enabled us to identify the
substellar members of nearby young moving groups
\citep[e.g.,][]{Gagne:2015ij,Gagne:2015dc,Faherty:2016fx,Liu:2016co}, a
population crucial to our understanding of brown dwarf evolution over their
first few hundred million years.  Proper motions from large catalogs have also
identified wide comoving companions to higher-mass stars whose ages and
metallicities can more easily be determined
\citep[e.g.,][]{Zhang:2013kq,Luhman:2012ir,Burningham:2013gt,Smith:2014df},
making the ultracool companions important benchmarks for constraining
atmospheric and evolutionary models.

The Panoramic Survey Telescope And Rapid Response System (\PS) is a large
multi-epoch, multi-wavelength, optical imaging survey using a 1.8-m wide-field
telescope on Haleakala, Maui \citep{Kaiser:2010gr}.  \PS\ uses a 1.4~gigapixel
camera (GPC1) with a $0.\!\!''258$ pixel scale.  The \PS\ 3$\pi$~Survey
\citep[PS1;][]{Chambers:2017vk} observed the entire sky north of
$\delta=-30^\circ$ (three-quarters of the sky) in five filters (\grizy) over
four years (2010 May -- 2014 March), imaging $\approx$12 times in each filter
and achieving a median angular resolution of $\approx$$1.\!\!''1$ with a floor
of $\approx$~$0.\!\!''7$ \citep{Magnier:2017vq}.  PS1 images are $\sim$1~mag
deeper in $z$-band than the most comparable optical survey to date (SDSS), and
the novel \yps\ filter (0.918$–-$1.001~\um) extends further toward the
near-infrared than previous optical surveys.  This long-wavelength sensitivity
allows PS1 to better detect and characterize red objects such as ultracool
dwarfs.  In addition, the multi-epoch astrometry of PS1 enables precise
measurement of proper motions and parallaxes that help to distinguish faint,
nearby ultracool dwarfs from reddened background stars and galaxies.

Significant ultracool discoveries from PS1 include many wide ultracool
companions to main sequence stars
\citep{Deacon:2012eg,Deacon:2012gf,Deacon:2014ey} and young stars
\citep{Aller:2013bc}, L/T transition dwarfs that are difficult to identify with
near-IR photometry alone \citep{Deacon:2011gz,Best:2013bp,Best:2015em}, new
low-mass members of the Hyades \citep{Goldman:2013jl} and Praesepe
\citep{Wang:2014bj}, and new brown dwarf members of nearby young moving groups
\citep{Liu:2013gy,Aller:2016kg}.  PS1 has also enabled studies with large
samples of more massive stars, including fiducial sequences of Galactic star
clusters \citep{Bernard:2014cj}, proper motions and wide binaries in the
\textit{Kepler} field \citep[who also present SEDs for spectral types B9V
through M9V in the PS1 photometric system]{Deacon:2016bh}, and photometric
distances and reddening for all stars detected by PS1
\citep{Green:2014hv,Schlafly:2014hh}.  PS1 can detect ultracool dwarfs at larger
distances than SDSS and 2MASS, so its optical photometry helps to create a rich
multi-color catalog that will enable even bigger searches based solely on
photometry, a precursor to science with the Large Synoptic Survey Telescope
\citep[LSST;][]{Ivezic:2008ub}.  In addition, the proper motions and parallaxes
in PS1 should be fertile ground for identifying more ultracool dwarfs that have
eluded detection due to their locations in crowded areas of the sky
\citep[e.g.,][]{Liu:2011hc}, or are too red and faint to be measured by
\textit{Gaia} \citep{GaiaCollaboration:2016cu}.

In this paper we present a comprehensive catalog of ultracool dwarfs observed by
PS1, including photometry, proper motions, spectral types, gravity
classifications, and multiplicity.  Section~\ref{catalog} describes the contents
and assembly of our catalog.  The PS1 photometry and proper motions are
discussed in detail in Sections \ref{photometry} and~\ref{pm}, respectively.  We
briefly describe a binary M7 dwarf newly identified by PS1 in
Section~\ref{binary}.  We summarize our catalog and its features in
Section~\ref{summary}.

\section{Catalog}
\label{catalog}
Our catalog of ultracool dwarfs in \PS\ contains photometry and proper motions
from PS1 for \varnobjtot\ M, L, and T dwarfs, along with photometry from 2MASS,
AllWISE, and \textit{Gaia}~DR1 whenever available.  The catalog includes all
published L and T dwarfs as of 2015 December with photometry in at least one of
the five PS1 bands (\grizy).  The catalog does not contain all known M dwarfs,
but does include a large sample in order to accurately represent the colors and
kinematics of M dwarfs in PS1.

We describe the construction of our catalog in
Section~\ref{catalog.construction}.  In Sections \ref{catalog.lt}
and~\ref{catalog.m}, we provide more details about our selection of L+T and M
dwarfs, respectively.  In Section~\ref{catalog.spt}, we discuss the spectral
types used in our catalog.  We describe our identification of young objects in
Section~\ref{catalog.young} and our treatment of binaries in
Section~\ref{catalog.binaries}.  In Section~\ref{catalog.completeness} we assess
the completeness of our catalog.

\subsection{Construction}
\label{catalog.construction}
To create our catalog, we compiled a list of late-M, L and T dwarfs from
DwarfArchives\footnote{Hosted at \url{http://DwarfArchives.org}.  Last updated
  2013-05-29.}, M dwarfs from \citet{West:2008eq}, and numerous literature
sources from 2012--2017.  We included positions, proper motions, spectral types,
and photometry from 2MASS \citep{Cutri:2003vr}, AllWISE \citep{Cutri:2014wx},
and \textit{Gaia}~DR1 \citep{Lindegren:2016gr} when available.  We also tracked
objects identified as binaries and those with spectroscopic or other indications
of youth.  The catalog includes new discoveries through 2015 December and a
handful of updates to photometry, astrometry, and spectral types from 2016 and
2017.

In order to ensure that every object in our catalog is a bona fide M, L, or T
dwarf, we included only published objects with spectroscopic classification.  We
have therefore excluded objects with only photometric spectral types (e.g.,
based on optical or near-infrared colors or methane imaging).  Our catalog also
does not include close substellar companions to main sequence stars detected by
high-angular resolution imaging and/or radial velocity because these objects are
not resolved by PS1.

We cross-matched our list with the full PS1 Processing Version 3.3 database
(PV3.3, 2017 March) by position using a 3''~matching radius, retaining the
closest object matched in PS1.  PV3.3 includes an update to the 2016 December
public data release (PS1~DR1) that reduced the astrometric errors but did not
affect photometry \citep{Magnier:2017vq}.  In order to maximize the number of
accurate matches, we used PS1 positions published in the literature (from
earlier processing versions) or AllWISE positions (nearly contemporaneous with
PS1) whenever possible for the objects in our list.  If neither of those were
available, we used the most recent positions reported in the literature;
frequently these came from 2MASS, SDSS, or UKIDSS.  When these objects had
reported proper motions, we used the proper motions to project expected PS1
coordinates and adopted those for our cross-match.

To ensure that our catalog contains only secure PS1 measurements of real
astrophysical objects, we applied photometric quality cuts described in detail
in Section~\ref{phot.ps1.chipwarp}.  Briefly, we required our PS1 matches to
have photometric errors less than 0.2~mag in at least one PS1 band, with
detections at two or more epochs in that band, and we excluded objects likely to
be saturated in all bands.  In addition, we excluded any sources flagged as
having poor PSF fits ({$\tt psf\_qf<0.85$}), and we verified that none of our
PS1 matches were marked as quasars, transients, periodic variables, or solar
system objects.  Any object without a PS1 match within 3'' of the expected
coordinates was removed from our catalog.

To check for incorrect matches, we calculated colors using the 2MASS, AllWISE,
and PS1 photometry for our matches (Section~\ref{phot.colors}).  We sorted our
list into bins of one spectral sub-type, and inspected every object with a \gy,
\ri, \rz, \ry, \iz, \iy, \zy, \yjt, or \ywa\ color differing from the mean for
its spectral type by more than 3~times the rms color for that spectral type bin.
We also inspected every T dwarf with a reported \gps, \rps, or \ips\ detection.
In addition, since the cool temperatures of M, L, and T dwarfs necessitate
reddish PS1 (optical) colors, we inspected all objects having a secure detection
in a bluer band but not in a redder band (e.g., some nearby M~dwarfs were
saturated in \ips, \zps, and \yps, but not in \gps\ or \rps).  To inspect an
object, we examined stacked images from PS1, 2MASS, and AllWISE, and searched in
all three surveys within a 60'' radius around the PS1 position for other
possible matches.  We discarded PS1 matches for which an image artifact, a
nearby brighter star, or a blue or extended background object had clearly
contaminated the detection (i.e., the source of contamination lay within the
visible PSF of the object from our list).  In cases where contamination affected
some but not all of the PS1 bands, we retained the object in our catalog and
rejected photometry only from the contaminated bands. (All \gps\ and \rps\
detections of T~dwarfs were discarded in this manner.)  We corrected a match
when the images and colors clearly pointed to a different PS1 source, but we did
not make corrections in ambiguous cases in order to minimize rejection of
objects with naturally-occurring unusual photometry.

For most outliers we found nothing to indicate the object was anything other
than an object with unusual colors.  Many red outliers were young objects in
star-forming regions and/or with low gravity spectral classifications, both
associated with redder-than-typical colors for L dwarfs
\citep[e.g.,][]{Faherty:2013bc}.  We discovered a few cases in which the 2MASS
or AllWISE photometry was for a different nearby object, often a brighter source
with which the ultracool object was blended.  In the case of blends we rejected
the 2MASS or AllWISE photometry; otherwise we adopted the photometry of the
correct ultracool object.

\subsection{L and T Dwarfs}
\label{catalog.lt}
Although ultracool dwarfs are normally brightest in the near-infrared, the depth
and red-optical sensitivity of \PS\ have allowed PS1 to detect \varnltdwarf~L
and T dwarfs, including spectral types as late as T9.  Barring unintentional
omissions, our catalog contains all spectroscopically confirmed L and T dwarfs
published through 2015 December and meeting our detection standards in PS1.  The
L and T dwarfs in our catalog are primarily drawn from DwarfArchives and
\citet{Mace:2014wq}, supplemented by other literature sources.  The final
catalog includes \varnldwarf~L~dwarfs and \varntdwarf~T~dwarfs.

\subsection{M Dwarfs}
\label{catalog.m}
M dwarfs comprise the majority of the stars in our galaxy, so a clear
understanding of M dwarf properties is essential for characterizing the local
stellar population and constraining models of star formation and evolution.  In
addition, M dwarfs provide context for the photometric and kinematic properties
of L and T dwarfs, and more massive brown dwarfs younger than $\sim$200~Myr will
have late-M spectral types.  Compiling a complete list of known M dwarfs would
require an effort far beyond what is needed to accurately characterize the PS1
photometry and proper motions of the nearby field population.  Instead, we built
a representative sample of the field population from two sub-samples.

The first sub-sample comprises objects with well-studied and/or potentially
distinctive photometry and kinematics from the recent literature, and contains
\varnmdwarfmega~M6--M9~dwarfs.  These objects were included in order to sample
the diversity of colors and kinematics in late-M dwarfs.  We included M dwarfs
from the proper motion and parallax compilations of
\citet{Faherty:2009kg,Faherty:2012cy} and \citet{Dupuy:2012bp}, the young object
list from \citet{Allers:2013hk}, the young moving group members and non-members
from \citet{Gagne:2015dc}, the catalog of SpeX spectra from
\citet{BardalezGagliuffi:2014fl}, and wide ultracool companions to main sequence
stars from \citet{Deacon:2014ey}.

The second sub-sample is a large set of M dwarfs with high-quality photometry,
representative of the generic field population.  We cross-matched all M0--M9
dwarfs listed in the catalog of \citet{West:2008eq} with PS1, 2MASS, and AllWISE
using a matching radius of $5''$.  We required sources to have photometric
errors~$<0.05$~mag in at least five of the eight total PS1 and 2MASS bands
($grizyJHK$), and at least two detections in individual exposures in each PS1
band.  To avoid saturated objects, we rejected any sources brighter than the
limits listed in Section~\ref{phot.ps1.catalog} or flagged by PS1 for poor PSF
fits ({$\tt psf\_qf<0.85$}).  We then removed objects with non-zero confusion,
saturation, extendedness, or de-blending flags in either 2MASS or AllWISE.
These cuts are more stringent than for the late-M, L and T dwarfs in our catalog
in order to ensure a very clean field M dwarf sample with high-quality
photometry.  This sub-sample contains \varnmdwarfwest~M~dwarfs from
\citet{West:2008eq}, bringing the total for M dwarfs in our catalog to
\varnmdwarf.

\subsection{Spectral Types}
\label{catalog.spt}
For the objects in our catalog, we use spectral types from the literature.
These spectral types were determined by a variety of methods, based on visual or
numerical analysis of red-optical ($\approx$$0.65-1$~\um)
or near-infrared ($\approx$$1-2.5$~\um)
spectra.  In cases where an object has both an optical and a near-IR spectral
type, we adopt the optical type for M and L dwarfs and the near-IR type for T
dwarfs.  The spectral types for the M dwarfs drawn from \citet{West:2008eq} were
all derived from optical spectra.

There are seven objects in our catalog with an optical L type and a near-IR
T~type.  For these we use the T spectral type (all T0--T1).  All seven objects
show clear methane absorption at 1.6~\um\ and/or 2.2~\um\ in their near-IR
spectra, a hallmark of T dwarfs.  We note that these objects are all confirmed
binaries (by high-resolution imaging) or candidate binaries (based on peculiar
spectra) with components spanning the L/T transition.  The spectral types are
therefore based on unresolved spectral blends, explaining the disagreement
between the optical and near-IR types.

\begin{figure*}
  \begin{center}
    \includegraphics[width=7.2 in]{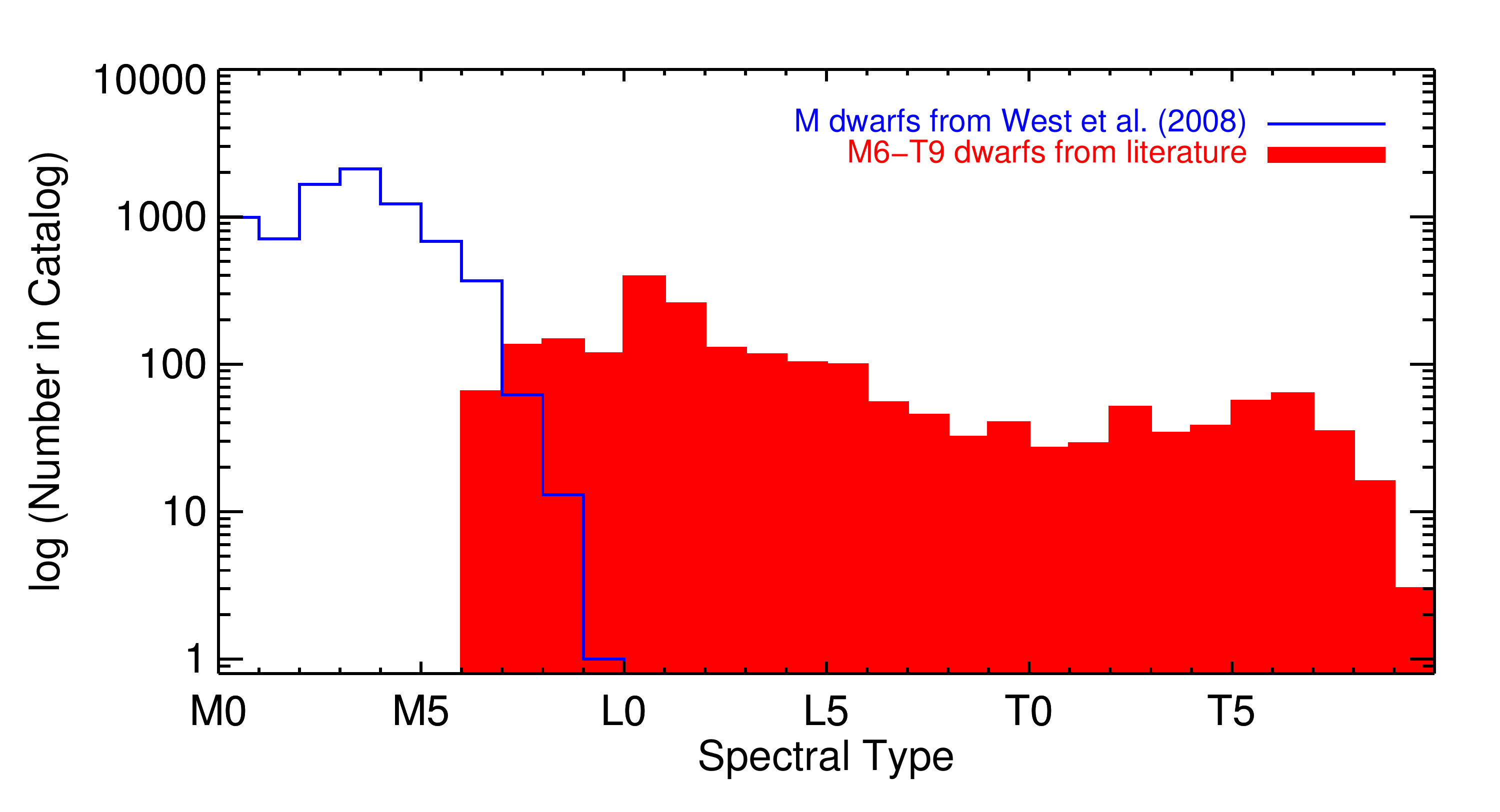}
    \caption{The distribution of spectral types in our catalog.  The late-M, L
      and T dwarfs compiled from the literature are shown in solid red, while
      the M dwarfs from \citet{West:2008eq} are shown with a blue outline.  The
      catalog robustly samples the temperature range of all but the coolest
      brown dwarfs, and includes objects with spectral types as late as T9.}
    \label{hist.spt.all}
  \end{center}
\end{figure*}

\begin{figure}
  \begin{center}
    \includegraphics[width=1.00\columnwidth]{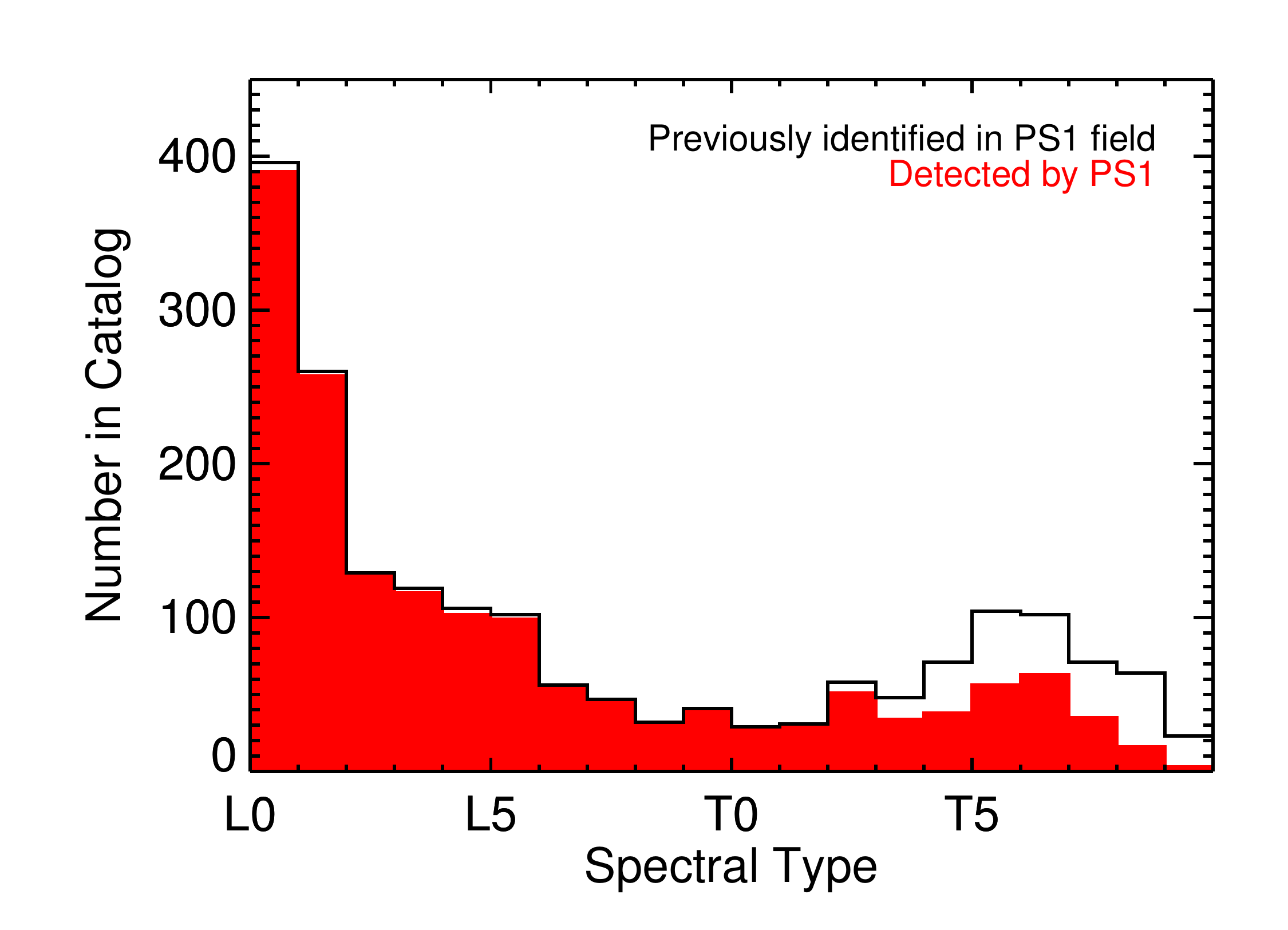}
    \caption{The distribution of L and T spectral types in our catalog, compared
      to previously identified L and T dwarfs in the PS1 field.  PS1 has
      detected nearly all previously-known L dwarfs; the handful of
      non-detections are mostly faint objects in star-forming regions or
      unresolved companions to higher-mass stars.  The T dwarfs (mostly
      later-type) not detected by PS1 have primarily been discovered by deeper
      near-IR searches over narrower fields or in the mid-IR using the \WISE\
      survey.}
    \label{hist.spt.lt}
  \end{center}
\end{figure}

We show the distribution of all spectral types in our catalog in
Figure~\ref{hist.spt.all}.  The earliest type in our catalog is M0 (by
construction), and the latest spectral type detected by PS1 is T9.  Our catalog
contains more than 20 objects of each spectral sub-type through T7, robustly
sampling the ultracool dwarf population for all but the coolest objects.  We
compare the distribution of L and T dwarfs in our catalog to all L and T dwarfs
in the PS1 field (north of $\delta=-30^\circ$) in Figure~\ref{hist.spt.lt}.  The
known objects not present in our catalog are mostly later-T dwarfs too faint to
be detected by PS1.  These have chiefly been discovered by deeper near-IR
searches over narrower fields \citep[e.g.,][]{Albert:2011bc,Burningham:2013gt}
or by searches for late-T and Y dwarfs using \WISE\
\citep[e.g.,][]{Kirkpatrick:2011ey}.  The $\approx$30~L~dwarfs not detected by
PS1 are mostly unresolved companions to higher-mass stars or discoveries from
deep imaging of star-forming regions.

\subsection{Young Objects}
\label{catalog.young}
Our catalog includes many young objects (ages $\lesssim200$~Myr), which are
known to have distinctive colors and kinematics
\citep[e.g.,][]{Kirkpatrick:2008ec,Faherty:2009kg}.  We identify young objects
primarily by low-gravity classifications reported in the literature: $\beta$,
$\gamma$, and $\delta$ classes based on optical
\citep{Kirkpatrick:2005cv,Cruz:2009gs} or near-IR \citep{Gagne:2015dc} spectra,
and \textsc{int-g} and \textsc{vl-g} based on near-IR spectra
\citep{Allers:2013hk}.  We also identify any object in a star-forming region as
young.  In addition, we include objects in our young sample that lack formal
low-gravity classifications but have other evidence for youth: NLTT~13728,
LP~423-31, and 2MASS~J19303829$-$1335083 \citep{Shkolnik:2009dx},
LSPM~J1314+1320 \citep{Schlieder:2014cm}, 2MASS~J17081563+2557474
\citep{Kellogg:2015cf}, and 2MASS~J22344161+4041387 \citep{Allers:2009ib} show
spectroscopic signs of low gravity; SDSSp~J111010.01+011613.1
\citep{Gagne:2015kf} and WISEA~J114724.10$-$204021.3 \citep{Schneider:2016iq}
are members of young moving groups; and LP~261-75B \citep{Reid:2006eg} and
Gl~417BC \citep{Kirkpatrick:2001bi} are wide companions to young stars.

\subsection{Binaries}
\label{catalog.binaries}
Our catalog naturally includes ultracool binaries with separations wide enough
to be resolved in PS1, as well as many that are unresolved.  In our catalog, we
assigned the term ''binary'' only to pairs that are unresolved in PS1 but
confirmed by high-resolution imaging or radial velocity measurements.  We treat
these as single objects, reporting their blended photometry.  We note that
peculiar spectral features have been used to identify candidate unresolved
binaries \citep{Burgasser:2010df,BardalezGagliuffi:2014fl}, but this technique
has not been demonstrated to robustly distinguish actual blends from single
objects with unusual atmospheric properties.  Given our conservative approach,
we expect our catalog to contain some unidentified binaries.  Our catalog
identifies a total of \varnbinary\ unresolved binaries and two unresolved triple
systems among the \varnlitmltdwarf\ late-M, L, and T~dwarfs from the literature.
This binary fraction of only \varpblend\% is less than the $\approx20\%$
estimated by population studies \citep[e.g.,][]{Marocco:2015iz}, implying that
our catalog indeed contains unrecognized binaries.  This is not surprising given
that many ultracool dwarfs have not yet been targeted with high-resolution
imaging.

The $\approx$$1.\!\!''1$ angular resolution of PS1 allows it to resolve
binaries that were not resolved in either 2MASS
\citep[$\approx$$2.\!\!''5$][]{Skrutskie:2006hl} or AllWISE
\citep[$\approx$$6''$ for the W1 and W2 bands][]{Wright:2010in}.  However, the
literature contains fewer than twenty binaries with separations wider than $1''$
for which both components are ultracool dwarfs.  Most of these binaries do not
appear in our catalog because they are too far south for PS1
($\delta<-30^{\circ}$) or because both components are M~dwarfs (for which our
catalog is not complete).  Our catalog contains a single instance of a binary
resolved in PS1 and 2MASS, but not in AllWISE: UScoCTIO~108 and UScoCTIO~108b
\citep{Bejar:2008ev}.  For this pair, we report the AllWISE photometry (blended)
only for the primary (treating it as an unresolved binary), and no AllWISE
photometry for the secondary.

Our catalog also contains three ultracool binaries that are resolved in PS1,
2MASS, and AllWISE.  We report the photometry and proper motion for each
component individually.  One pair, LP~704-48 and SDSS~J000649.16$-$085246.3
(itself an unresolved binary), is well-separated at $27.\!\!''4$
\citep{Burgasser:2012ga}.  Another pair, the blue L6 dwarf
SDSS~J141624.08+134826.7 and the T7.5 dwarf ULAS~J141623.94+134836.3, has a
separation of $9''$ \citep{Burningham:2010du} and is well resolved in PS1.
ULAS~J141623.94+134836.3 was not detected in 2MASS despite lying well outside
the PSF of the brighter L6 primary; we include synthetic 2MASS photometry from
\citet{Dupuy:2012bp} in our catalog.  ULAS~J141623.94+134836.3 appears barely
resolved in AllWISE images, and we include pipeline-deblended AllWISE photometry
for each component in our catalog.  Finally, VHS~J125601.92$-$125723.9AB
\citep[also an unresolved binary;][]{Stone:2016fz} and its companion
VHS~J125601.92-125723.9~b \citep[separation $8.\!\!''1$;][]{Gauza:2015fw} are
well-resolved in both PS1 and 2MASS.  They appear partially resolved in AllWISE
images. We include in our catalog the deblended photometry from AllWISE for the
primary and the decontaminated photometry for the wide companion (removing a
diffraction spike) from \citet{Gauza:2015fw}.

\bibpunct[ ]{(}{)}{;}{a}{}{,}
We also note two previously known binaries in our catalog with separations
$\approx$$1''$ that appear resolved in PS1 images, but are each represented by
only a single object in the PS1 database: DENIS-P~J220002.05$-$303832.9 and
2MASS~J17072343$-$0558249.  These cases result from the algorithm by which
the PS1 database assembles multiple detections over the four years of PS1
observations into individual objects \citep[see Section~\ref{pm.method.recalc}
and][for more details]{Magnier:2017vq}.  Close binaries may be combined into a
single object in the database, especially if the binary's proper motion over the
PS1 survey period is comparable in amplitude and direction to the binary's
separation.  We have not attempted to de-blend these objects in our catalog, and
we mark both as binaries (unresolved) in our catalog.  More $\approx$1''
binaries such as these are sure to appear in the PS1 database as single objects;
we describe our discovery of one such binary in Section~\ref{binary}.
\bibpunct[, ]{(}{)}{;}{a}{}{,}

\subsection{Completeness}
\label{catalog.completeness}
Our catalog is a combination of discoveries from many searches for M, L, and
T~dwarfs, conducted using a variety of methods and therefore containing a
variety of biases.  Our selection of a representative sample of M~dwarfs means
that our catalog will be far from complete for this spectral type, especially
for types M0--M5 for which our photometric quality cuts include only bright
objects.  While we have included all previously identified L and T~dwarfs
observed by PS1, there are some L and T dwarfs beyond the detection limit or
angular resolution of PS1 (Figure~\ref{hist.spt.lt}), and there are sure to be
undiscovered objects remaining in the PS1 field.

\begin{figure*}
\begin{center}
  \begin{minipage}[t]{0.49\textwidth}
    \includegraphics[width=1.00\columnwidth, trim = 0 0 10mm 0]{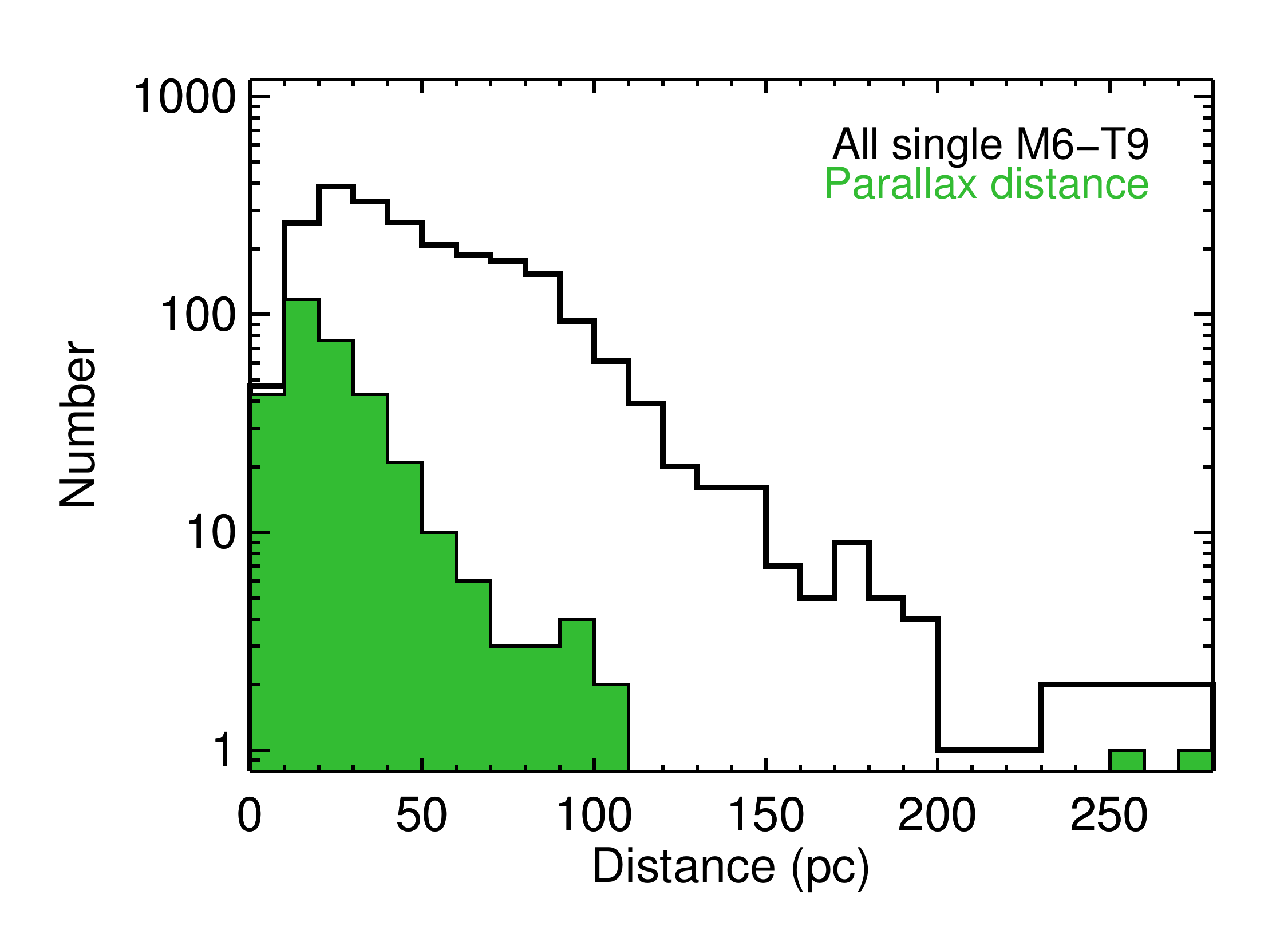}
  \end{minipage}
  \hfill
  \begin{minipage}[t]{0.49\textwidth}
    \includegraphics[width=1.00\columnwidth, trim = 10mm 0 0 0]{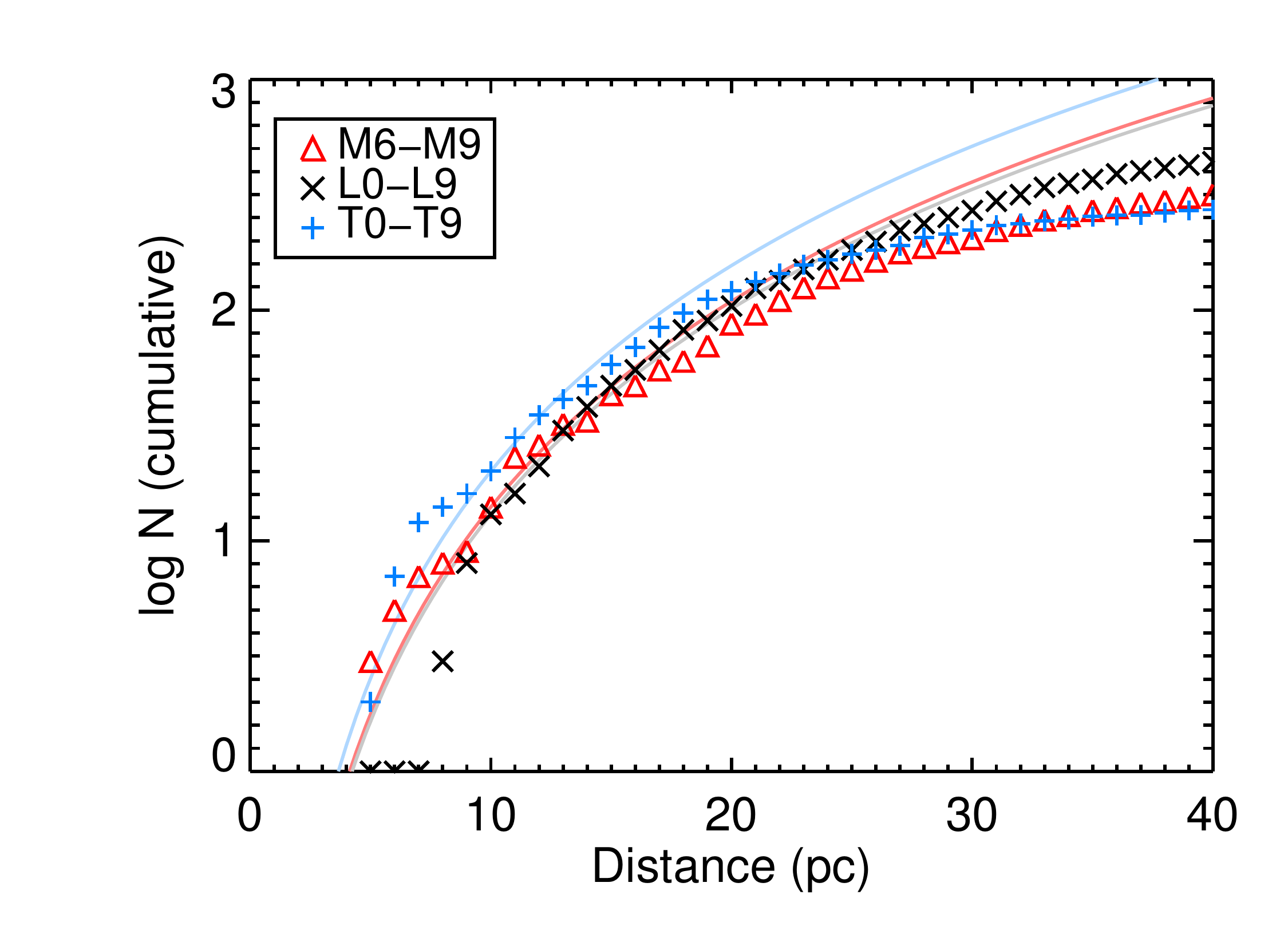}
  \end{minipage}
  \caption{\textit{Left}: Distribution of the distances of single M6--T9 dwarfs
    in our catalog (black outline).  Where available, we use a parallax distance
    from the literature (solid green).  For other objects we use $W2$-based
    photometric distances.  \textit{Right}: Cumulative distribution of these
    distances for M (red triangles), L (black $\times$), and T (blue $+$)
    dwarfs, using a format similar to Figure~5 in \citet{Faherty:2009kg}.  The
    curves indicate density distributions from a simple Galactic thin disk model
    with a scale height of 300~pc, normalized at 10~pc.  We use light red, grey,
    and blue curves for M, L, and T~dwarfs, respectively.  Our catalog is not
    consistent with the expected density distribution beyond 10~pc for late-M
    and T~dwarfs and 20~pc for L~dwarfs, implying that our catalog is incomplete
    beyond these distances.}
  \label{fig.pm.dist}
\end{center}
\end{figure*}

We assess the completeness of our catalog for spectral types M6--T9 by examining
the number of objects as a function of distance, shown in
Figure~\ref{fig.pm.dist}.  We use trigonometric parallax distances when
available from the literature.  For the remaining objects we use photometric
distances calculated from $W2$~magnitudes and the spectral type-absolute
magnitude polynomial from \citet{Dupuy:2012bp}.  Photometric distances for
unresolved binaries will be systematically too small, so we exclude known
binaries from our assessment.  Figure~\ref{fig.pm.dist} also compares the
cumulative distributions of late-M, L, and T~dwarf distances to distributions
from a simple Galactic thin disk model for space density
\hbox{$\rho=\rho_0\exp{[-Z/H_0]}$}, where $\rho_0$ is the space density at the
Galactic plane, $Z$ is the distance from the plane and $H_0=300$~pc is the scale
height \citep{Bochanski:2010kq}.  We integrate this model over the PS1 survey
area to account for varying lines of sight relative to the galactic plane.  We
normalize this model distribution with the cumulative numbers of known late-M,
L, and T~dwarfs at 10~pc.  The numbers of objects begin to deviate from our
model distributions at $\approx$10~pc for late-M and T~dwarfs and $\approx$20~pc
for L~dwarfs, implying that our catalog is not volume-complete beyond these
distances.

\section{Photometry}
\label{photometry}
We present the PS1, 2MASS, and AllWISE photometry for our catalog in
Table~\ref{tbl.phot}.  PS1 photometry is on the AB magnitude scale
\citep{Tonry:2012gq}, calibrated using the procedures outlined in
\citet{Schlafly:2012da} and \citet{Magnier:2013hk}.  2MASS, AllWISE, and
\textit{Gaia} photometry are calibrated on the Vega magnitude scale
\citep[respectively]{Cohen:2003gg,Wright:2010in,Carrasco:2016el}.

The full table contains \varnphotcols\ columns, and is available for download in
electronic form from the online journal.  Table~\ref{tbl.phot} is arranged in two
parts: (1)~the late-M, L, and T dwarfs compiled from the literature, followed by
(2)~the M dwarfs from \citet{West:2008eq}.  For reference, Table~\ref{tbl.phot}
includes spectral types (with notation for subdwarfs), and indicates whether an
object has been classified as a low-gravity object based on its optical or
near-IR spectrum, identified as a young object (due to low gravity or other
reasons), or confirmed as a binary.  Table~\ref{tbl.phot.sample} shows a sample
of the rows and columns of Table~\ref{tbl.phot} for guidance regarding format
and content.

\begin{deluxetable*}{lll}
\centering
\tablecaption{Photometry of M, L, and T Dwarfs in the \PS\ 3$\pi$ Survey \label{tbl.phot}}
\tablewidth{0pt}
\tabletypesize{\scriptsize}
\tablehead{   % column headings
  \colhead{Column} &
  \colhead{Label} &
  \colhead{Description}
}
\startdata
 1 & Name & Name used in the object's discovery or spectral confirmation paper \\
 2 & Spectral Type: Opt & Optical spectral type\tablenotemark{a,b} \\
 3 & Spectral Type: NIR & Near-infrared spectral type\tablenotemark{a,b} \\
 4 & Spectral Type: Adopted & Adopted spectral type\tablenotemark{a,b} \\
 5 & Gravity: Opt & Low-gravity classification from an optical spectrum\tablenotemark{b} \\
 6 & Gravity: NIR & Low-gravity classification from a near-infrared spectrum\tablenotemark{b} \\
 7 & Binary & ''Y'' or ''triple'' for known binary or triple systems not resolved in PS1 \\
 8 & Young & ''Y'' for known young objects\tablenotemark{c} \\
 9 & \PS\ Name & PS1 Designation\tablenotemark{d},  rrr.rrrr+dd.dddd (J2000) \\
10 & \gps & PS1 $g$ magnitude \\
11 & err$_{\gps}$ & Error in PS1 $g$ magnitude \\
12 & $N_g$ & Number of measurements used in the \gps\ photometry \\
13 & S$_g$ & Source of the \gps\ photometry: chip (C), recalculated chip\tablenotemark{e} (R), or forced warp (W) \\
14 & \rps & PS1 $r$ magnitude \\
15 & err$_{\rps}$ & Error in PS1 $r$ magnitude \\
16 & $N_r$ & Number of measurements used in the \rps\ photometry \\
17 & S$_r$ & Source of the \rps\ photometry: chip (C), recalculated chip\tablenotemark{e} (R), or forced warp (W) \\
18 & \ips & PS1 $i$ magnitude \\
19 & err$_{\ips}$ & Error in PS1 $i$ magnitude \\
20 & $N_i$ & Number of measurements used in the \ips\ photometry \\
21 & S$_i$ & Source of the \ips\ photometry: chip (C), recalculated chip\tablenotemark{e} (R), or forced warp (W) \\
22 & \zps & PS1 $z$ magnitude \\
23 & err$_{\zps}$ & Error in PS1 $z$ magnitude \\
24 & $N_z$ & Number of measurements used in the \zps\ photometry \\
25 & S$_z$ & Source of the \zps\ photometry: chip (C), recalculated chip\tablenotemark{e} (R), or forced warp (W) \\
26 & \yps & PS1 $y$ magnitude \\
27 & err$_{\yps}$ & Error in PS1 $y$ magnitude \\
28 & $N_y$ & Number of measurements used in the \yps\ photometry \\
29 & S$_y$ & Source of the \yps\ photometry: chip (C), recalculated chip\tablenotemark{e} (R), or forced warp (W) \\
30 & 2MASS Name & 2MASS catalog designation \\
31 & $J$ & $J$ magnitude or upper limit (2MASS) \\
32 & err$_J$ & Error in $J$ magnitude \\
33 & $H$ & $H$ magnitude or upper limit (2MASS) \\
34 & err$_H$ & Error in $H$ magnitude \\
35 & $K_S$ & $K_S$ magnitude or upper limit (2MASS) \\
36 & err$_{K_S}$ & Error in $K_S$ magnitude \\
37 & 2MASS Clfg & 2MASS contamination and confusion flags: three-character string corresponding to JHK bands. \\
38 & AllWISE Name & AllWISE catalog designation \\
39 & $W1$ & $W1$ magnitude or upper limit (AllWISE) \\
40 & err$_{W1}$ & Error in $W1$ magnitude \\
41 & $W2$ & $W2$ magnitude or upper limit (AllWISE) \\
42 & err$_{W2}$ & Error in $W2$ magnitude \\
43 & $W3$ & $W3$ magnitude or upper limit (AllWISE) \\
44 & err$_{W3}$ & Error in $W3$ magnitude \\
45 & $W4$ & $W4$ magnitude or upper limit (AllWISE) \\
46 & err$_{W4}$ & Error in $W4$ magnitude \\
47 & AllWISE cc\_flags & AllWISE contamination and confusion flags: four-character string corresponding to W1W2W3W4 bands. \\
48 & AllWISE neighbor & Number of other AllWISE objects detected within $8''$ of the AllWISE position. \\
49 & $G$ & \textit{Gaia}~DR1 $G$ magnitude \\
50 & err$_G$ & Error in \textit{Gaia}~DR1 $G$ magnitude \\
51 & References & References: Discovery, Spectral Type, Gravity, Binarity, 2MASS photometry, AllWISE photometry \\
\enddata
\tablenotetext{a}{Spectral types taken from the literature
  (Section~\ref{catalog.spt}).  When both optical and near-IR types are
  available, we adopt the optical type for M and L~dwarfs and the near-IR type
  for T~dwarfs.  Most spectral types have an uncertainty of $\pm0.5$~subtypes;
  ``:'' = $\pm1$~subtype; ``::'' = $\pm2$ or more~subtypes. ``sd'' = subdwarf;
  ``esd'' = extreme subdwarf \citep{Gizis:1997cj}.}
\tablenotetext{b}{$\beta$, $\gamma$, and $\delta$ indicate classes of
  increasingly low gravity based on optical
  \citep{Kirkpatrick:2005cv,Cruz:2009gs} or near-infrared \citep{Gagne:2015dc}
  spectra.  \textsc{fld-g} indicates near-infrared spectral signatures of
  field-age gravity, \textsc{int-g} indicates intermediate gravity, and
  \textsc{vl-g} indicates very low gravity \citep{Allers:2013hk}.}
\tablenotetext{c}{Young objects identified by low-gravity classifications or
  other spectroscopic evidence for youth, membership in star-forming regions or
  young moving groups, or companionship to a young star
  (Section~\ref{catalog.young}).}
\tablenotetext{d}{Pan-STARRS names are from the $3\pi$ Survey, Processing
  Version~3.3 (PV3.3). Photometry listed here is from PV3.3 and supersedes
  values in previous publications.}
\tablenotetext{e}{Chip photometry recalculated by combining the measurements for
  an object that is split into two or more ``partial objects'' in PS1 (Sections
  \ref{phot.fast} and~\ref{pm.method.recalc}).}
\tablenotetext{f}{Although classified as \fldg, the spectrum shows hints of
  intermediate gravity \citep[as described in][]{Aller:2016kg}.}
\tablenotetext{g}{Photometry rejected for this band after visual inspection of
  stack images found no detection at the PS1 coordinates.}
\tablenotetext{h}{Photometry rejected for this band after visual inspection of
  stack images found obvious contamination by a background object.}
\tablenotetext{i}{Photometry rejected for this band after visual inspection of
  stack images found an image processing artifact at the PS1 coordinates.}
\tablenotetext{j}{Photometry rejected for this band after visual inspection of
  stack images found obvious contamination from a nearby bright star.}
\tablenotetext{k}{UScoCTIO 108 and UScoCTIO 108b \citep{Bejar:2008ev} are not
  resolved in AllWISE.  For this binary, we report AllWISE photometry (blended)
  only for the primary, and no AllWISE photometry for the secondary.}
\tablecomments{This table is available in its entirety in machine-readable form
  in the online journal.  A sample of the rows and columns is shown in
  Table~\ref{tbl.phot.sample}.}
\end{deluxetable*}

\floattable
\begin{deluxetable*}{lccchhcchhhhhhhhhCCccCCccCCcchhhhhhhhhhhhhhhhhhhhhc}
\centering
\tablecaption{Sample of columns in Table~\ref{tbl.phot} \label{tbl.phot.sample}}
\tablewidth{0pt}
\rotate
\tabletypesize{\tiny}
\tablehead{   % column headings
  \colhead{} &
  \multicolumn{3}{c}{Spectral Type\tablenotemark{a,b}} &
  \nocolhead{} &
  \nocolhead{} &
  \nocolhead{} &
  \nocolhead{} &
  \multicolumn{21}{c}{\PS} &
  \colhead{} \\
  \cline{2-4}
  \cline{5-6}
  \cline{9-29}
  \colhead{Discovery Name} &
  \colhead{Opt} &
  \colhead{NIR} &
  \colhead{Adopted} &
  \nocolhead{Opt} &
  \nocolhead{NIR} &
  \colhead{Binary} &
  \colhead{Young\tablenotemark{c}} &
  \nocolhead{\PS\ Name\tablenotemark{d}} &
  \nocolhead{\gps} &
  \nocolhead{err$_{\gps}$} &
  \nocolhead{$N_g$} &
  \nocolhead{S$_g$} &
  \nocolhead{\rps} &
  \nocolhead{err$_{\rps}$} &
  \nocolhead{$N_r$} &
  \nocolhead{S$_r$} &
  \colhead{\ips} &
  \colhead{err$_{\ips}$} &
  \colhead{$N_i$} &
  \colhead{S$_i$} &
  \colhead{\zps} &
  \colhead{err$_{\zps}$} &
  \colhead{$N_z$} &
  \colhead{S$_z$} &
  \colhead{\yps} &
  \colhead{err$_{\yps}$} &
  \colhead{$N_y$} &
  \colhead{S$_y$} &
  \nocolhead{2MASS Name} &
  \nocolhead{$J$} &
  \nocolhead{err$_J$} &
  \nocolhead{$H$} &
  \nocolhead{err$_H$} &
  \nocolhead{$K_S$} &
  \nocolhead{err$_{K_S}$} &
  \nocolhead{2MASS Cflg} &
  \nocolhead{AllWISE Name} &
  \nocolhead{$W1$} &
  \nocolhead{err$_{W1}$} &
  \nocolhead{$W2$} &
  \nocolhead{err$_{W2}$} &
  \nocolhead{$W3$} &
  \nocolhead{err$_{W3}$} &
  \nocolhead{$W4$} &
  \nocolhead{err$_{W4}$} &
  \nocolhead{AllWISE cc\_flg} &
  \nocolhead{AllWISE neighbor} &
  \nocolhead{$G$} &
  \nocolhead{err$_{G}$} &
  \colhead{References} \\
  \multicolumn{8}{c}{} &
  \nocolhead{} &
  \nocolhead{(mag)} &
  \nocolhead{(mag)} &
  \nocolhead{} &
  \nocolhead{} &
  \nocolhead{(mag)} &
  \nocolhead{(mag)} &
  \nocolhead{} &
  \nocolhead{} &
  \colhead{(mag)} &
  \colhead{(mag)} &
  \multicolumn{2}{c}{} &
  \colhead{(mag)} &
  \colhead{(mag)} &
  \multicolumn{2}{c}{} &
  \colhead{(mag)} &
  \colhead{(mag)} &
  \multicolumn{2}{c}{} &
  \nocolhead{} &
  \nocolhead{(mag)} &
  \nocolhead{(mag)} &
  \nocolhead{(mag)} &
  \nocolhead{(mag)} &
  \nocolhead{(mag)} &
  \nocolhead{(mag)} &
  \nocolhead{} &
  \nocolhead{} &
  \nocolhead{(mag)} &
  \nocolhead{(mag)} &
  \nocolhead{(mag)} &
  \nocolhead{(mag)} &
  \nocolhead{(mag)} &
  \nocolhead{(mag)} &
  \nocolhead{(mag)} &
  \nocolhead{(mag)} &
  \nocolhead{} &
  \nocolhead{} &
  \nocolhead{(mag)} &
  \nocolhead{(mag)} &
  \colhead{(Disc; SpT; Grav; Bin;)} \\
  \multicolumn{50}{c}{} &
  \colhead{(2MASS; AllWISE)}
}
\startdata
SDSS J000013.54+255418.6 & T5 & T4.5 & T4.5 & \nodata & \nodata & \nodata & \nodata & PSO J000.0563+25.9054 & \nodata & \nodata & \nodata & \nodata & \nodata & \nodata & \nodata & \nodata & \nodata & \nodata & \nodata & \nodata & 19.17 & 0.01 & 10 & C & 17.42 & 0.01 & 11 & C & 2MASS J00001354+2554180 & 15.06 & 0.04 & 14.73 & 0.12 & 14.84 & 0.12 & ccc & \nodata & \nodata & \nodata & \nodata & \nodata & \nodata & \nodata & \nodata & \nodata & \nodata & \nodata & \nodata & \nodata & 175; 247,51; --; --; 88; -- \\
SDSS J000112.18+153535.5 & \nodata & L3.7 \intg & L3.7 \intg & \nodata & \intg & \nodata & Y & PSO J000.3012+15.5925 & \nodata & \nodata & \nodata & \nodata & \nodata & \nodata & \nodata & \nodata & 20.37 & 0.01 & 13 & C & 18.85 & 0.02 & 7 & C & 17.81 & 0.01 & 10 & C & 2MASS J00011217+1535355 & 15.52 & 0.06 & 14.51 & 0.04 & 13.71 & 0.04 & 000 & WISEA J000112.27+153533.6 & 12.97 & 0.02 & 12.54 & 0.02 & 11.67 & 0.24 & >\!8.93 & \nodata & 0000 & 0 & \nodata & \nodata & 175; 124; 124; --; 88; 89 \\
WISEA J000131.93$-$084126.9 & \nodata & L1 pec (blue) & L1 pec (blue) & \nodata & \nodata & \nodata & \nodata & PSO J000.3831$-$08.6909 & \nodata & \nodata & \nodata & \nodata & \nodata & \nodata & \nodata & \nodata & 20.21 & 0.03 & 7 & C & 18.57 & 0.01 & 12 & C & 17.57 & 0.01 & 10 & C & 2MASS J00013166$-$0841234 & 15.71 & 0.05 & 15.03 & 0.09 & 14.70 & 0.09 & c00 & WISEA J000131.93$-$084127.2 & 14.34 & 0.03 & 13.99 & 0.05 & >\!12.36 & \nodata & >\!8.45 & \nodata & 0000 & 0 & 20.39 & 0.02 & 218; 218; --; --; 88; 89 \\
SDSS J000250.98+245413.8 & \nodata & L5.5 & L5.5 & \nodata & \nodata & \nodata & \nodata & PSO J000.7124+24.9037 & \nodata & \nodata & \nodata & \nodata & \nodata & \nodata & \nodata & \nodata & 22.30 & 0.08 & 11 & W & 20.30 & 0.04 & 9 & W & 19.31 & 0.03 & 16 & W & 2MASS J00025097+2454141 & 17.17 & 0.22 & 16.06 & 0.22 & 15.66 & 0.22 & 000 & WISEA J000250.98+245413.6 & 14.82 & 0.03 & 14.57 & 0.06 & >\!11.73 & \nodata & >\!8.10 & \nodata & 0000 & 0 & \nodata & \nodata & 74; 74; --; --; 88; 89 \\
2MASSI J0003422$-$282241 & M7.5 & M7: \fldg & M7.5 & \nodata & \fldg & \nodata & \nodata & PSO J000.9273$-$28.3785 & 20.92 & 0.03 & 6 & C & 19.41 & 0.01 & 8 & C & 16.77 & 0.01 & 7 & C & 15.43 & 0.01 & 6 & C & 14.67 & 0.01 & 6 & C & 2MASS J00034227$-$2822410 & 13.07 & 0.02 & 12.38 & 0.03 & 11.97 & 0.03 & 000 & WISEA J000342.53$-$282242.7 & 11.71 & 0.02 & 11.52 & 0.02 & 11.01 & 0.12 & >\!8.64 & \nodata & dd00 & 0 & 17.19 & 0.01 & 83; 83,11; 11; --; 88; 89 \\
2MASS J00044144$-$2058298 & M8 & \nodata & M8 & \nodata & \nodata & \nodata & \nodata & PSO J001.1753$-$20.9747 & 20.85 & 0.06 & 4 & C & 19.15 & 0.01 & 5 & C & 16.41 & 0.01 & 8 & C & 14.94 & 0.01 & 5 & C & 14.06 & 0.01 & 7 & C & 2MASS J00044144$-$2058298 & 12.40 & 0.02 & 11.83 & 0.02 & 11.40 & 0.02 & 000 & WISEA J000442.04$-$205828.9 & 11.06 & 0.02 & 10.75 & 0.02 & 10.27 & 0.07 & >\!8.93 & \nodata & 0000 & 0 & \nodata & \nodata & 160; 161; --; --; 88; 89 \\
2MASS J00054844$-$2157196 & M9 & \nodata & M9 & \nodata & \nodata & \nodata & \nodata & PSO J001.4543$-$21.9558 & 21.34 & 0.06 & 5 & C & 19.95 & 0.01 & 9 & C & 17.11 & 0.01 & 3 & C & 15.69 & 0.01 & 10 & C & 14.84 & 0.01 & 8 & C & 2MASS J00054844$-$2157196 & 13.27 & 0.03 & 12.62 & 0.02 & 12.20 & 0.03 & 000 & WISEA J000549.00$-$215721.0 & 11.89 & 0.02 & 11.62 & 0.02 & 11.31 & 0.17 & >\!8.83 & \nodata & 0000 & 0 & \nodata & \nodata & 270; 268; --; --; 88; 89 \\
ULAS J000613.24+154020.7 & \nodata & L9 & L9 & \nodata & \nodata & \nodata & \nodata & PSO J001.5553+15.6723 & \nodata & \nodata & \nodata & \nodata & \nodata & \nodata & \nodata & \nodata & \nodata & \nodata & \nodata & \nodata & 21.05 & 0.08 & 5 & C & 19.87 & 0.06 & 9 & C & \nodata & \nodata & \nodata & \nodata & \nodata & \nodata & \nodata & \nodata & WISEA J000613.26+154020.7 & 15.34 & 0.04 & 14.70 & 0.07 & >\!12.03 & \nodata & >\!8.41 & \nodata & 0000 & 0 & \nodata & \nodata & 94; 94; --; --; --; 89 \\
SDSS J000614.06+160454.5 & L0 & \nodata & L0 & \nodata & \nodata & \nodata & \nodata & PSO J001.5586+16.0816 & \nodata & \nodata & \nodata & \nodata & \nodata & \nodata & \nodata & \nodata & 20.75 & 0.02 & 18 & W & 19.23 & 0.01 & 9 & C & 18.30 & 0.01 & 9 & C & 2MASS J00061406+1604546 & 16.59 & 0.13 & 15.84 & 0.16 & 15.09 & 0.16 & 000 & WISEA J000614.06+160454.1 & 15.07 & 0.04 & 14.90 & 0.07 & >\!12.26 & \nodata & >\!8.72 & \nodata & 0000 & 0 & \nodata & \nodata & 332; 332; --; --; 88; 89 \\
\enddata
\tablerefs{(1) This work, (2) \citet{Aberasturi:2014cc}, (3) \citet{Aganze:2016fi}, (4) \citet{Albert:2011bc}, (5) \citet{Allen:2007hd}, (6) \citet{Allen:2012dn}, (7) \citet{Aller:2013bc}, (8) \citet{Aller:2016kg}, (9) \citet{Allers:2009ib}, (10) \citet{Allers:2010cg}, (11) \citet{Allers:2013hk}, (12) \citet{Allers:2013vv}, (13) \citet{AlvesdeOliveira:2013dx}, (14) \citet{Artigau:2006bh}, (15) \citet{Artigau:2011hr}, (16) \citet{BardalezGagliuffi:2014fl}, (17) \citet{BardalezGagliuffi:2015fd}, (18) \citet{Baron:2015fn}, (19) \citet{BarradoYNavascues:2002ee}, (20) \citet{Basri:2000dg}, (21) \citet{Beamin:2013fx}, (22) \citet{Becklin:1988kz}, (23) \citet{Bejar:2008ev}, (24) \citet{Bessell:1991bg}, (25) \citet{Best:2013bp}, (26) \citet{Best:2015em}, (27) \citet{Best:2017br}, (28) \citet{Best:2017ih}, (29) \citet{Bihain:2010gm}, (30) \citet{Bihain:2013gw}, (31) \citet{Boeshaar:1976wm}, (32) \citet{Boudreault:2013bb}, (33) \citet{Bouvier:2008kf}, (34) \citet{Bouy:2003eg}, (35) \citet{Bowler:2010gd}, (36) \citet{Bryja:1992kh}, (37) \citet{Bryja:1994ir}, (38) \citet{Burgasser:1999fp}, (39) \citet{Burgasser:2000bm}, (40) \citet{Burgasser:2000dp}, (41) \citet{Burgasser:2002fy}, (42) \citet{Burgasser:2003dh}, (43) \citet{Burgasser:2003dx}, (44) \citet{Burgasser:2003ew}, (45) \citet{Burgasser:2003ij}, (46) \citet{Burgasser:2003jf}, (47) \citet{Burgasser:2004fh}, (48) \citet{Burgasser:2004hg}, (49) \citet{Burgasser:2005gj}, (50) \citet{Burgasser:2005jp}, (51) \citet{Burgasser:2006cf}, (52) \citet{Burgasser:2006db}, (53) \citet{Burgasser:2006hd}, (54) \citet{Burgasser:2006jj}, (55) \citet{Burgasser:2007jw}, (56) \citet{Burgasser:2007ki}, (57) \citet{Burgasser:2008cj}, (58) \citet{Burgasser:2008ei}, (59) \citet{Burgasser:2009db}, (60) \citet{Burgasser:2009gs}, (61) \citet{Burgasser:2010bj}, (62) \citet{Burgasser:2010df}, (63) \citet{Burgasser:2011jx}, (64) \citet{Burgasser:2012ga}, (65) \citet{Burgasser:2015fd}, (66) \citet{Burgasser:2016cx}, (67) \citet{Burningham:2010dh}, (68) \citet{Burningham:2010du}, (69) \citet{Burningham:2011kh}, (70) \citet{Burningham:2013gt}, (71) \citet{Castro:2012dj}, (72) \citet{Castro:2013bb}, (73) \citet{Castro:2016dr}, (74) \citet{Chiu:2006jd}, (75) \citet{Chiu:2008dt}, (76) \citet{Close:2002fu}, (77) \citet{Close:2002jp}, (78) \citet{Close:2003ie}, (79) \citet{Crifo:2005fz}, (80) \citet{Cruz:2002dw}, (81) \citet{Cruz:2003fi}, (82) \citet{Cruz:2004gz}, (83) \citet{Cruz:2007kb}, (84) \citet{Cruz:2009gs}, (85) \citet{Cushing:2006bu}, (86) \citet{Cushing:2011dk}, (87) \citet{Cushing:2014be}, (88) \citet{Cutri:2003vr}, (89) \citet{Cutri:2014wx}, (90) \citet{Dahn:1986dt}, (91) \citet{Dahn:2002fu}, (92) \citet{Dahn:2008bo}, (93) \citet{Dawson:2014hl}, (94) \citet{DayJones:2013hm}, (95) \citet{Deacon:2005jf}, (96) \citet{Deacon:2007jl}, (97) \citet{Deacon:2009bb}, (98) \citet{Deacon:2011gz}, (99) \citet{Deacon:2012eg}, (100) \citet{Deacon:2012gf}, (101) \citet{Deacon:2014ey}, (102) \citet{Deacon:2017ja}, (103) \citet{Deacon:2017kd}, (104) \citet{Delfosse:1997uj}, (105) \citet{Delfosse:1999bx}, (106) \citet{Delorme:2008dl}, (107) \citet{Dobbie:2002dr}, (108) \citet{Dupuy:2009jb}, (109) \citet{Dupuy:2010ch}, (110) \citet{Dupuy:2012bp}, (111) \citet{Dupuy:2015gl}, (112) \citet{Dupuy:2016bg}, (113) \citet{Dupuy:2017ke}, (114) \citet{Faherty:2009kg}, (115) \citet{Faherty:2010gt}, (116) \citet{Faherty:2012cy}, (117) \citet{Faherty:2013bc}, (118) \citet{Faherty:2016fx}, (119) \citet{Fan:2000iu}, (120) \citet{Folkes:2012ep}, (121) \citet{Forveille:2005ee}, (122) \citet{Freed:2003js}, (123) \citet{Gagne:2014dt}, (124) \citet{Gagne:2015dc}, (125) \citet{Gagne:2017gy}, (126) \citet{Gauza:2012en}, (127) \citet{Gauza:2015fw}, (128) \citet{Geballe:2002kw}, (129) \citet{Geissler:2011gg}, (130) \citet{Gelino:2011cw}, (131) \citet{Gelino:2014ie}, (132) \citet{Giampapa:1986er}, (133) \citet{Giclas:1967ui}, (134) \citet{Gillon:2016hl}, (135) \citet{Gilmore:1985wb}, (136) \citet{Gizis:1997cj}, (137) \citet{Gizis:1997gq}, (138) \citet{Gizis:2000hq}, (139) \citet{Gizis:2000kz}, (140) \citet{Gizis:2001jp}, (141) \citet{Gizis:2002je}, (142) \citet{Gizis:2003fj}, (143) \citet{Gizis:2011fq}, (144) \citet{Gizis:2011jv}, (145) \citet{Gizis:2012kv}, (146) \citet{Gizis:2013ik}, (147) \citet{Gizis:2015ey}, (148) \citet{Goldman:2010ct}, (149) \citet{Gomes:2013eg}, (150) \citet{Hall:2002jk}, (151) \citet{Hawley:2002jc}, (152) \citet{Henry:1990ct}, (153) \citet{Henry:2004jj}, (154) \citet{Henry:2006jp}, (155) \citet{Huelamo:2015fz}, (156) \citet{Irwin:1991wf}, (157) \citet{Kellogg:2015cf}, (158) \citet{Kendall:2003bd}, (159) \citet{Kendall:2004kb}, (160) \citet{Kendall:2007fd}, (161) \citet{Kendall:2007hi}, (162) \citet{Kirkpatrick:1991da}, (163) \citet{Kirkpatrick:1993fs}, (164) \citet{Kirkpatrick:1994bl}, (165) \citet{Kirkpatrick:1995cl}, (166) \citet{Kirkpatrick:1997kt}, (167) \citet{Kirkpatrick:1997va}, (168) \citet{Kirkpatrick:1999ev}, (169) \citet{Kirkpatrick:2000gi}, (170) \citet{Kirkpatrick:2001cy}, (171) \citet{Kirkpatrick:2008ec}, (172) \citet{Kirkpatrick:2010dc}, (173) \citet{Kirkpatrick:2011ey}, (174) \citet{Kirkpatrick:2014kv}, (175) \citet{Knapp:2004ji}, (176) \citet{Koerner:1999fb}, (177) \citet{Kraus:2009hc}, (178) \citet{Lachapelle:2015cx}, (179) \citet{Law:2006dm}, (180) \citet{Leggett:1992dk}, (181) \citet{Leggett:1996eb}, (182) \citet{Leggett:2000ja}, (183) \citet{Leinert:1994ti}, (184) \citet{Lepine:2002gc}, (185) \citet{Lepine:2002kp}, (186) \citet{Lepine:2003fn}, (187) \citet{Lepine:2003hu}, (188) \citet{Lepine:2003ku}, (189) \citet{Lepine:2005jx}, (190) \citet{Lepine:2009hi}, (191) \citet{Liebert:1979io}, (192) \citet{Liebert:2003bx}, (193) \citet{Liebert:2006kp}, (194) \citet{Liu:2002kj}, (195) \citet{Liu:2005cu}, (196) \citet{Liu:2006ce}, (197) \citet{Liu:2010cw}, (198) \citet{Liu:2011hc}, (199) \citet{Liu:2013gy}, (200) \citet{Liu:2016co}, (201) \citet{Lodieu:2002bp}, (202) \citet{Lodieu:2005kd}, (203) \citet{Lodieu:2007fr}, (204) \citet{Lodieu:2008hm}, (205) \citet{Lodieu:2010iu}, (206) \citet{Lodieu:2012go}, (207) \citet{Lodieu:2012il}, (208) \citet{Lodieu:2013eo}, (209) \citet{Lodieu:2014jo}, (210) \citet{Looper:2007ee}, (211) \citet{Looper:2008hs}, (212) \citet{Looper:2011ur}, (213) \citet{Loutrel:2011ft}, (214) \citet{Lucas:2010iq}, (215) \citet{Luhman:2007fu}, (216) \citet{Luhman:2009cn}, (217) \citet{Luhman:2012ir}, (218) \citet{Luhman:2014hj}, (219) \citet{Luyten:1979vs}, (220) \citet{Mace:2013jh}, (221) \citet{Manjavacas:2013cg}, (222) \citet{Marocco:2013kv}, (223) \citet{Marocco:2015iz}, (224) \citet{Marshall:2008il}, (225) \citet{Martin:1994it}, (226) \citet{Martin:1998hr}, (227) \citet{Martin:1999er}, (228) \citet{Martin:1999fc}, (229) \citet{Martin:2000jp}, (230) \citet{Martin:2010cx}, (231) \citet{Matsuoka:2011kp}, (232) \citet{McCarthy:1964tp}, (233) \citet{McCaughrean:2002ip}, (234) \citet{McElwain:2006ca}, (235) \citet{McGovern:2004cc}, (236) \citet{Metchev:2008gx}, (237) \citet{Mohanty:2003fg}, (238) \citet{Montagnier:2006dt}, (239) \citet{Mugrauer:2006iy}, (240) \citet{Murphy:2015kv}, (241) \citet{Murray:2011ey}, (242) \citet{Muzic:2012hc}, (243) \citet{PhanBao:2001ij}, (244) \citet{PhanBao:2006bg}, (245) \citet{PhanBao:2006ev}, (246) \citet{PhanBao:2008kz}, (247) \citet{Pineda:2016ku}, (248) \citet{Pinfield:2008jx}, (249) \citet{Pokorny:2004fd}, (250) \citet{Probst:1983fy}, (251) \citet{Radigan:2008jd}, (252) \citet{Radigan:2013cv}, (253) \citet{Rebolo:1998el}, (254) \citet{Reid:1981hd}, (255) \citet{Reid:1995kw}, (256) \citet{Reid:2000iw}, (257) \citet{Reid:2001ch}, (258) \citet{Reid:2002bt}, (259) \citet{Reid:2002ep}, (260) \citet{Reid:2003cz}, (261) \citet{Reid:2003hw}, (262) \citet{Reid:2003kx}, (263) \citet{Reid:2004be}, (264) \citet{Reid:2005bm}, (265) \citet{Reid:2006da}, (266) \citet{Reid:2006dy}, (267) \citet{Reid:2007gl}, (268) \citet{Reid:2008fz}, (269) \citet{Reiners:2006ey}, (270) \citet{Reyle:2004hn}, (271) \citet{Rice:2010ke}, (272) \citet{Rodono:1980fc}, (273) \citet{Rodriguez:2013fv}, (274) \citet{Ruiz:1997cw}, (275) \citet{Ruiz:2001hv}, (276) \citet{Sahlmann:2015ku}, (277) \citet{Salim:2003eg}, (278) \citet{Sarro:2014ci}, (279) \citet{Schmidt:2007ig}, (280) \citet{Schmidt:2010ex}, (281) \citet{Schmidt:2015hv}, (282) \citet{Schneider:1991eb}, (283) \citet{Schneider:2002jp}, (284) \citet{Schneider:2011eq}, (285) \citet{Schneider:2014jd}, (286) \citet{Schneider:2016bs}, (287) \citet{Schneider:2016iq}, (288) \citet{Scholz:2001bf}, (289) \citet{Scholz:2002by}, (290) \citet{Scholz:2004bc}, (291) \citet{Scholz:2004fr}, (292) \citet{Scholz:2009fp}, (293) \citet{Scholz:2010cy}, (294) \citet{Scholz:2010gm}, (295) \citet{Scholz:2011gs}, (296) \citet{Scholz:2012ek}, (297) \citet{Scholz:2014hg}, (298) \citet{Scholz:2014hk}, (299) \citet{Schweitzer:1999wb}, (300) \citet{Seifahrt:2010jd}, (301) \citet{Sheppard:2009ed}, (302) \citet{Shkolnik:2009dx}, (303) \citet{Siegler:2003ki}, (304) \citet{Siegler:2005is}, (305) \citet{Siegler:2007ki}, (306) \citet{Silvestri:2007bx}, (307) \citet{Sivarani:2009bu}, (308) \citet{Skrutskie:2006hl}, (309) \citet{Stern:2007ge}, (310) \citet{Stone:2016fz}, (311) \citet{Strauss:1999iw}, (312) \citet{Stumpf:2009gx}, (313) \citet{Stumpf:2010de}, (314) \citet{Stumpf:2011ie}, (315) \citet{Thompson:2013kv}, (316) \citet{Thorstensen:2003hn}, (317) \citet{Tinney:1993dx}, (318) \citet{Tinney:1993hq}, (319) \citet{Tinney:1993jm}, (320) \citet{Tinney:1996vd}, (321) \citet{Tinney:1998tz}, (322) \citet{Tinney:2005hz}, (323) \citet{Tsvetanov:2000cg}, (324) \citet{vanBiesbroeck:1961cw}, (325) \citet{West:2008eq}, (326) \citet{Wilson:2001db}, (327) \citet{Wilson:2003tk}, (328) \citet{Wright:2013bo}, (329) \citet{ZapateroOsorio:1999iu}, (330) \citet{ZapateroOsorio:2000ds}, (331) \citet{Zhang:2009kw}, (332) \citet{Zhang:2010bq}, (333) \citet{Zhang:2013kq}.}
\tablecomments{Table \ref{tbl.phot} is published in its entirety in machine
  readable format in the online journal.  A portion is shown here for guidance
  regarding its form and content. The full table contains \varnphotcols~columns
  and \varnobjtot~rows.}
\end{deluxetable*}

\subsection{PS1 Photometry}
\label{phot.ps1}

\subsubsection{Chip and Forced Warp Photometry}
\label{phot.ps1.chipwarp}
Our catalog uses two types of PSF photometry from the PS1 database, known as
``chip'' and ``forced warp.''  These types are described in detail in
\citet{Magnier:2017vq}, and we explain them briefly here.

During PS1 data processing, each raw image was individually detrended and
calibrated to create a ``chip'' image, and each detected object on a chip was
fitted with a PSF model to determine its photometry and astrometry.  The chip
pixels were geometrically transformed onto a grid with uniform
$0.25''\!$~pixel$^{-1}$ scale representing pre-defined sky coordinates (R.A. and
Decl.), creating ``warp'' images.  The warps for each filter matching the same
portions of the sky were then summed together, forming ``stack'' images.
Detections in the warps and stacks were again fit with PSFs to measure
photometry and astrometry.

Chip photometry is the mean measurement from all chips in which an object was
detected, and is likely to be the most accurate photometry for a well-detected
object due to the individual calibration of each chip.  Stack photometry is
measured from the single fit to a stack detection.  Stack photometry will
generally be less accurate because individual images forming a stack were taken
in varying conditions and at different locations on the \PS\ detector, creating
poorly-defined PSFs.  However, the stacks can identify objects too faint to be
detected in individual images, as long as the objects do not move significantly
over the 4-year time baseline of the survey.  To take advantage of the greater
depth of the stacks without sacrificing too much of the calibration of the chip
images, the PS1 data pipeline fit a model PSF on every warp image at the
location of each object detected in a stack.  The warp photometry reported by
PS1 is the mean of the fluxes from the forced PSF fits at a given location,
excluding cases where the warp pixels were excessively masked.  Warp photometry
will not have the full accuracy of the chip measurements, but achieves the depth
of the stack photometry with more accuracy than the stack image alone.  Warp
photometry is therefore most useful, at least in theory, for slow-moving objects
with magnitudes comparable to or fainter than the chip detection limit.

\begin{deluxetable}{cC}
\tablewidth{100pt}
\tablecaption{Flux overestimation bias thresholds for PS1 chip photometry \label{tbl.faint.bias}}
\tablehead{   % column headings
  \colhead{Band} &
  \colhead{Threshold} \\
  \colhead{} &
  \colhead{(mag)} 
}
\startdata
\phm{xxxxxx}\gps\phm{xxxxxx} & \phm{xxxxxx}21.0\phm{xxxxxx} \\
\rps\ & 21.0 \\
\ips\ & 20.5 \\
\zps\ & 20.0 \\
\yps\ & 19.0 
\enddata
\tablecomments{PS1 chip magnitudes fainter than these thresholds may be
  significantly affected by flux overestimation bias.  For slow-moving objects
  ($<$100~\my), warp photometry is more likely to be accurate.}
\end{deluxetable}

\begin{figure}
  \begin{center}
  \begin{minipage}[t]{0.49\textwidth}
    \includegraphics[width=1.00\columnwidth]{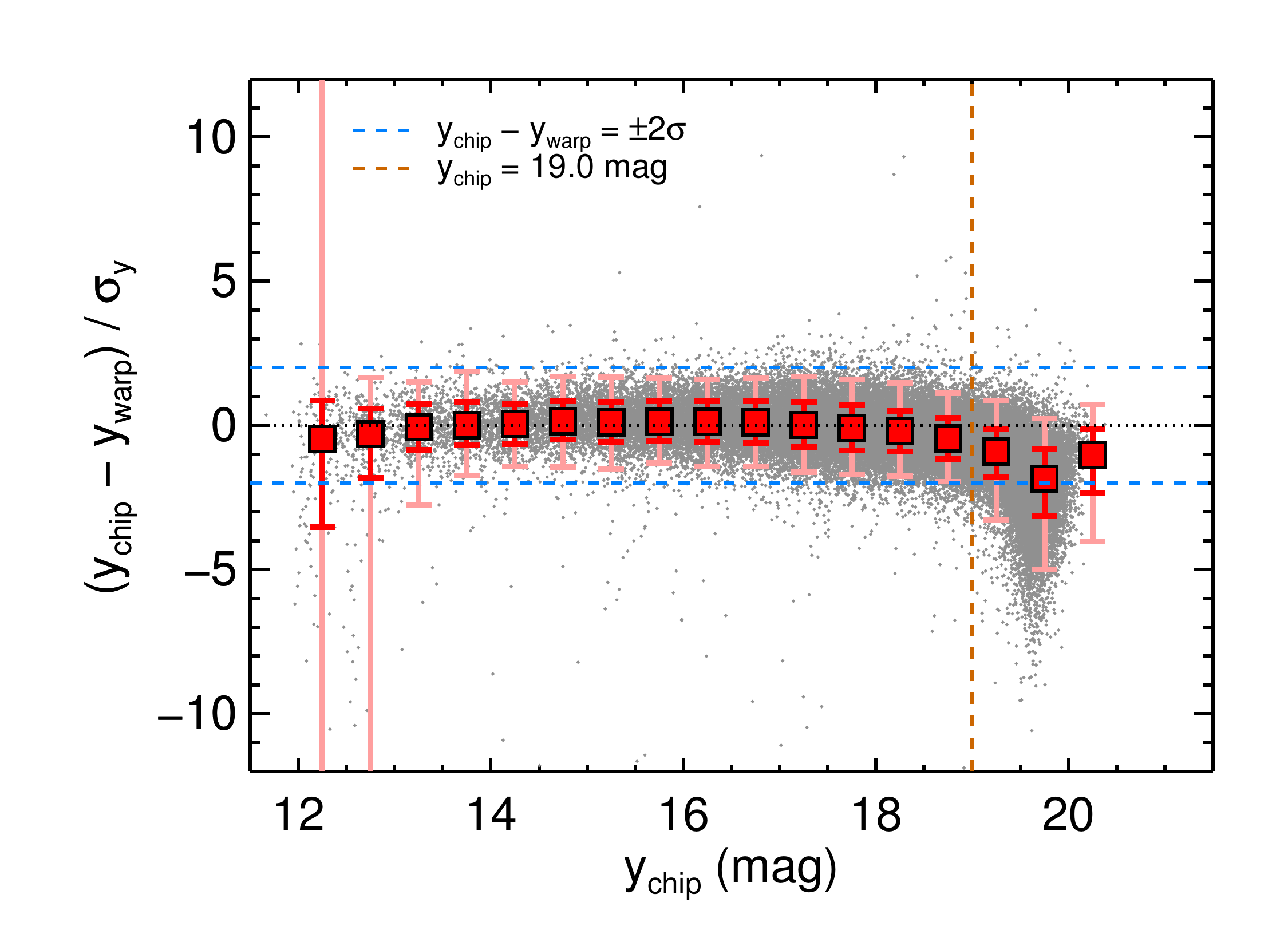}
  \end{minipage}
  \hfill
  \begin{minipage}[t]{0.49\textwidth}
    \includegraphics[width=1.00\columnwidth]{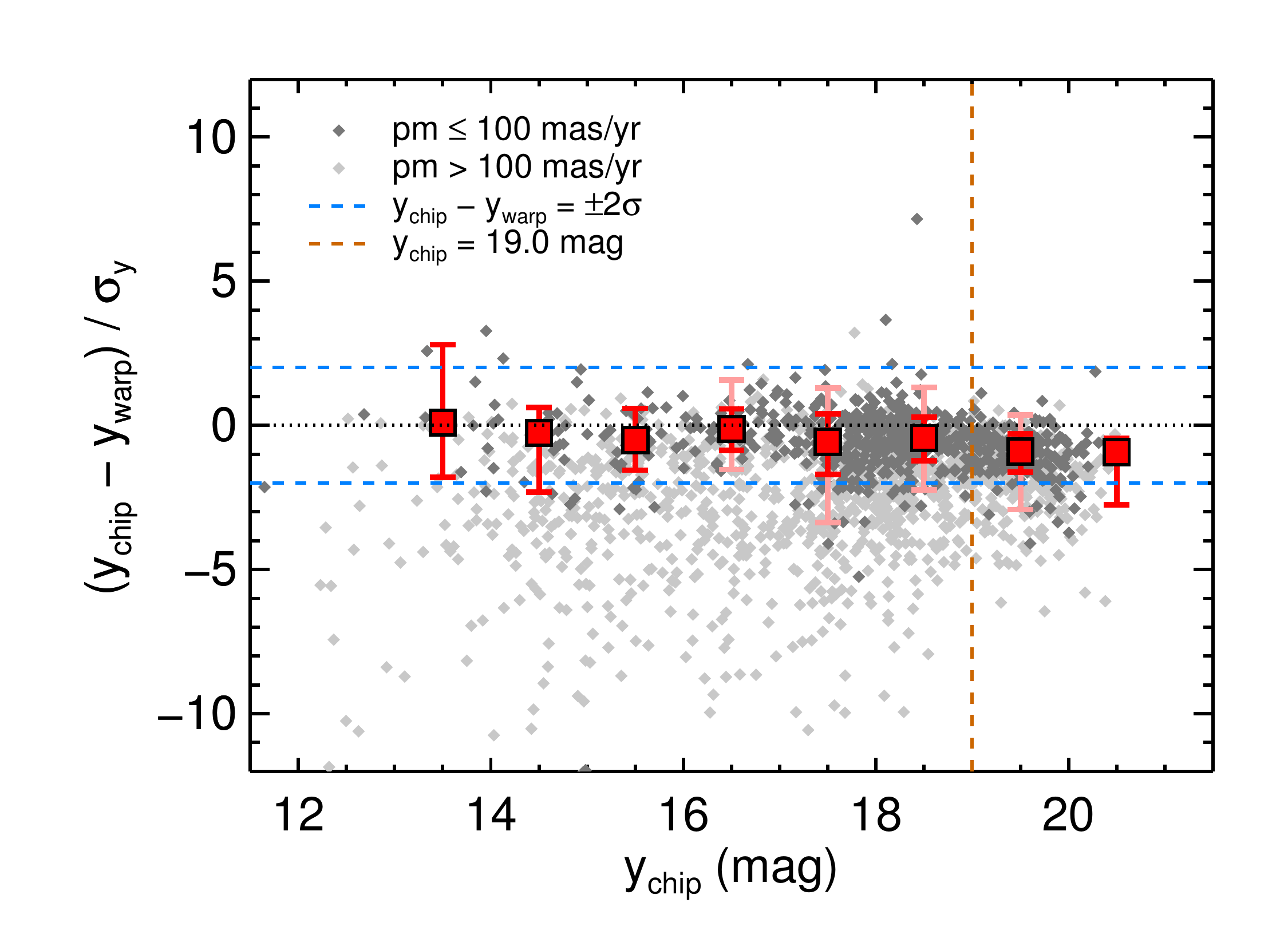}
  \end{minipage}
  \caption{{\it Left}: Differences in \yps\ chip and warp mean magnitudes (gray
    dots) for objects in an arbitrary 4~deg$^2$ patch of sky, normalized by the
    uncertainties.  $>$99\% of these objects have proper motions less than
    100~\my. The red boxes show the median differences for objects in bins of
    0.5~mag; the dark and light red error bars indicate 68\% and 95\% confidence
    limits, respectively.  For reference, the blue dashed lines mark
    $y_{\rm chip}-y_{\rm warp}$ differences of $\pm2\sigma$, and the dotted
    black line indicates no difference.  For objects of moderate brightness, the
    differences between chip and warp magnitudes are nearly always less than
    $2\sigma$ significance. At the bright end, saturation causes significant
    scatter in magnitudes.  At the faint end, many chip magnitudes become
    significantly brighter due to flux overestimation bias.  The vertical brown
    dashed line at $y=19$~mag marks the onset of this bias. {\it Right}: The
    same plot for objects from our ultracool catalog.  Those having proper
    motions less than 100~\my\ are plotted in dark grey, with faster moving
    objects in light grey.  The red boxes indicate the median
    $y_{\rm chip}-y_{\rm warp}$ differences only for the slower
    ($\mu\le100$~\my) objects.  Warp photometry is significantly fainter than
    chip photometry for most of the faster moving objects because an object is
    not at the same force-fit position in all warps.  We use chip photometry in
    our catalog as our default, but use warp photometry for objects with
    $y_{\rm chip}>19.0$~mag and proper motion $<100$~\my.  Analogous plots for
    the other four PS1 bands show similar results.}
  \label{fig.chipwarp}
  \end{center}
\end{figure}

To quantify where PS1 chip and warp photometry differ significantly, we examined
the photometry of a large sample of well-detected objects in PS1 chip images.
For each of the five PS1 bands, we extracted the chip and warp magnitudes for
all objects having at least three chip detections in a 4~deg$^2$ patch of sky
(centered at $\alpha=80^\circ$, $\delta=5^\circ$) at moderate galactic latitude
($\approx$$-18^\circ$) and away from regions of significant reddening.  This
gave us a sample of more than 60,000~objects in each band, $\gtrsim$99\% having
proper motions less than 100~\my.  Figure~\ref{fig.chipwarp} plots the
differences between \yps\ chip and warp magntudes for each object, normalized by
the quadrature sum of the chip and warp errors, as a function of magnitude.  For
brighter unsaturated objects ($\yps\approx13-19$~mag), the difference between
chip and warp magnitudes is nearly always less than $2\sigma$.  For objects
fainter than $\yps\approx19$~mag, however, the chip photometry becomes
significantly brighter for many objects, due to a flux overestimation bias for
objects near the chip detection threshold.  This well-known bias is discussed in
depth in the context of 2MASS in \citet{Cutri:2006ws} and
\citet{Kellogg:2015cf}, and is important for detections with S/N~$\lesssim10$.
Briefly, since the chips have brighter detection limits than the warps, mean
chip photometry may not include the fainter measurements that are present in the
warp photometry.  For example, an object near the chip detection limit may have
10~forced warp measurements but only 3~chip detections due to noise at the
detection threshold.  Those chip detections will be the 3~brightest measurements
of the object, so the mean chip magnitude will be brighter than the mean warp
magnitude.  Warp photometry is therefore more likely to be accurate for faint
objects near the chip detection limits in each band.  We list the magnitudes at
which the flux overestimation bias becomes significant for chip photometry in
each PS1 band in Table~\ref{tbl.faint.bias}, based on visual inspection of
Figure~\ref{fig.chipwarp} and analogous plots for the other PS1 bands.  We note
that for even fainter objects, flux overestimation bias will also impact the
warp photometry, but for those objects there is no deeper PS1 photometry
available.

Objects that moved significantly over the four years of PS1 observations will be
smeared on the stack images.  This smearing will also impact the warp
photometry, because the warp PSF fits are applied at the same location on each
warp but a moving object will not be centered at that location in every warp.
To assess the impact of proper motion on warp photometry, we examined the
difference between chip and warp magnitudes for objects brighter than the faint
thresholds in Table~\ref{tbl.faint.bias}, sorted into bins of PS1-measured
proper motion.  We determined that nearly all objects for which chip and warp
photometry are $>$2$\sigma$ different have proper motions exceeding 100~\my\
(Figure~\ref{fig.chipwarp}), so we do not use warp photometry for objects moving
faster than 100~\my.

\subsubsection{PS1 Photometry Reported in Our Catalog}
\label{phot.ps1.catalog}
For each object and PS1 band (\grizy) in our catalog, we report a single
magnitude, either chip or warp.  By default, we use the chip photometry for
objects with chip errors~$<0.2$~mag and detected in at least two chip exposures.
We use the warp photometry only in specific cases, when either:
\begin{enumerate}
\item Chip photometry is fainter than the thresholds listed in
  Table~\ref{tbl.faint.bias}, to avoid flux overestimation bias in the chip
  photometry.
\item Chip photometry is either not measured or of insufficient quality (i.e.,
  fewer than two detections or error~$\ge0.2$~mag).
\end{enumerate}
However, since warp photometry degrades for faster-moving objects, we only use
warp photometry when both of the following are true:
\begin{enumerate}
\item A proper motion of $\mu>100$~\my\ with $\frac{\mu}{\sigma_\mu}>3$ has not
  been measured in PS1 or the literature, to avoid fast-moving objects.
\item A proper motion with $\mu-\sigma_\mu>100$~\my\ has not been measured in
  PS1 or the literature, to avoid most fast-moving objects with poorly-measured
  proper motions.
\end{enumerate}
Finally, we only use warp photometry with errors~$<0.2$~mag and calculated from
at least two successful warp fits, the same standards we use for chip
photometry.

On the bright end, we rejected any photometry with $\gps<14.5$~mag,
$\rps<14.5$~mag, $\ips<14.5$~mag, $\zps<13.5$~mag, or $\yps<12.5$~mag to avoid
saturation.

For many objects in our catalog we use chip photometry for some bands and warp
for others, depending on the values and quality of the chip photometry.  If
neither the chip nor the warp photometry meet or quality standards in a given
band, we report no photometry for that band.  Objects with no chip or warp
photometry of sufficient quality in any of the five PS1 bands do not appear in
our catalog.

The photometric errors reported in the PS1 database are formal errors that do
not include systematics.  Given the sensitivity of the \PS\ camera and the
multiple epochs of photometry, these formal errors can be very small, less than
0.0005~mag in some cases.  A full assessment of the systematic errors for PV3.3
has not yet been completed, but a calibration of the first 1.5 years of PS1
photometry performed by \citet{Schlafly:2012da} found per-image zeropoints had
rms scatter $\approx10$~mmag in all five PS1 filters, so we adopt this value
(0.01~mag) as a floor for our catalog.

\subsubsection{False Warp Detections of Faint Objects}
\label{phot.ps1.faintwarp}
While inspecting the PS1 colors of our catalog objects, we discovered a few
dozen instances where very faint objects near the stack detection limit (e.g.,
$\gps\gtrsim23$~mag) had warp photometry with implausibly small errors (as small
as 0.01~mag) despite only two or three warps contributing to the mean
photometry.  In particular, we found T~dwarfs with high-S/N warp photometry
reported for \gps\ and \rps.  T~dwarfs are much too faint in these visual bands
to be detected by PS1.  We confirmed that the \gps\ and \rps\ stack images
showed no objects at the locations of these false detections.

We traced the source of these false high-S/N detections to the method used to
calculate PS1 photometry errors.  This method is described in detail in
\citet{Magnier:2017vq}; here we give a brief summary.  Each individual
photometry measurement (chip, warp, or stack) includes a measurement
uncertainty, which naturally is large for faint objects.  Mean chip and warp
magnitudes are computed using an iterative reweighting process to reject
outliers, and the errors for the mean photometry are calculated by bootstrap
resampling of the non-outlier measurements.  Bootstrapping uses the individual
measurements but not their uncertainties, instead sampling the outlier-cleaned
photometry measurements to determine the error.  For well-detected objects with
multiple measurements, bootstrapping is demonstrated to calculate errors
consistent with standard errors on the mean photometry while effectively
clipping outliers such as transient image artifacts \citep{Magnier:2017vq}.
However, in cases where photometry was measured only a few times for an object,
and those measurements are very similar (a statistical possibility even for
objects at the detection limit), the bootstrapping process will produce a very
small error, even if the individual measurements had large uncertainties.  The
false \gps\ and \rps\ warp detections in our catalog were the result of
$\approx$2--3 low-S/N forced warp measurements of background noise that happened
to find similar values, resulting in errors $<0.2$~mag (from bootstrapping) on
the mean warp photometry.

To identify and remove these false detections from our catalog in a systematic
fashion, we extracted the individual warp flux measurements used to calculate
the mean PS1 warp photometry for each object in our catalog, along with the
formal uncertainties for the measurements.  For each object and band, we
calculated a weighted-mean warp magnitude (using inverse variance weighting) and
the standard error on this weighted mean.  If the reported warp error in PS1
(from bootstrapping) was less than 0.2~mag but our calculated standard error was
greater than 0.2~mag, we discarded the warp photometry for that object and band.
This procedure removed 12\% of the PS1-reported warp measurements for the
late-M, L, and T~dwarfs from the literature that passed our original criteria
for inclusion in the catalog (i.e., slow-moving objects with secure proper
motion measurements, Section~\ref{phot.ps1.catalog}), affecting 255 objects in
at least one band and demonstrating that false detections in the warps can be a
significant source of contamination at the faint end.  Specifically, we
discarded 129 out of 199 measurements in \gps, 123 out of 410 in \rps, 27 out of
601 in \ips, 19 out of 677 in \rps, and 0 out of 698 in \yps, consistent with
the red nature of the objects in our sample (nearly all have strong \yps\
detections).  If the reported warp error in PS1 and the standard error were both
less than 0.2~mag, we retained the PS1-reported photometry and error for our
catalog.  We used our calculated standard errors only to assess the reliability
of small warp photometry errors reported by PS1, and do not include them in our
catalog.

\subsubsection{Fast-Moving Objects}
\label{phot.fast}
As discussed in detail in Section~\ref{pm.method.recalc}, we found that objects
with proper motions $\gtrsim200$~\my\ were often split into two or more distinct
``partial objects'' in the PS1 database.  In these cases we use the photometry
from the partial object with the chip photometry in the most PS1 bands, giving
preference to the redder bands if no partial object had photometry in all bands.
In a few dozen cases where a chosen partial object had no photometry or
photometry of insufficient quality in a PS1 band, we used photometry
recalculated by combining the measurements from the partial objects into a
single object (Section~\ref{pm.method.recalc}) for that band.

\subsection{2MASS and AllWISE Photometry}
\label{phot.2mwise}
Our catalog contains photometry from 2MASS and AllWISE for all objects matched
to PS1 detections (Section~\ref{catalog.construction}).  We include 2MASS and
AllWISE photometry with nonzero contamination and confusion flags for
completeness' sake, and note that some of this potentially contaminated
photometry has been used in previous studies.  We include columns for the 2MASS
and AllWISE contamination and confusion flags in Table~\ref{tbl.phot}, and refer
readers to the Explanatory
Supplements\footnote{\url{http://www.ipac.caltech.edu/2mass/releases/allsky/doc/sec2_2a.html}}\textsuperscript{,}\footnote{\url{http://wise2.ipac.caltech.edu/docs/release/allwise/expsup/sec2_1a.html}}
for these surveys for details.

In addition, since the large $\approx$6$''$ beam of WISE makes blending with
nearby objects a frequent issue, we include a column in Table~\ref{tbl.phot}
indicating whether each object has an AllWISE neighbor within $8''$ of the
AllWISE position.  The AllWISE catalog includes deblended photometry for objects
with overlapping PSFs, but the deblending may not be completely successful when
objects are within $8''$ of each other \citep[see their
Figure~6]{Theissen:2016gn}.

While we include photometry with non-zero contamination and confusion flags and
potential contamination from neighbors in our catalog, we exclude such
photometry from our analysis of colors and SEDs in Section~\ref{phot.colors}.

\subsection{\textit{Gaia} DR1 Photometry}
To obtain \textit{Gaia}~DR1 $G$-band photometry, we cross-matched our catalog
(PS1 coordinates) with \textit{Gaia}~DR1 using a $2''$ matching radius.  We
found matches for \varngaia~objects including \varngaialdwarf~L~dwarfs.  As
\textit{Gaia}~DR1 is preliminary and does not cover the entire PS1 survey area,
we do not evaluate the DR1 $G$-band photometry for M, L, and T~dwarfs here, but
we include it in our catalog for reference.  We rejected \textit{Gaia}
photometry for objects in our catalog for which we had identified contamination
to \gps, \rps, or \ips\ photometry by a bluer object
(Section~\ref{catalog.construction}), but performed no other quality inspection
of the \textit{Gaia}-PS1 matches.  Many $G$ magnitudes have reported errors less
than 0.01~mag, but as \textit{Gaia}'s systematic photometric uncertainties are
not yet fully understood \citep{GaiaCollaboration:2016gd}, we adopt a minimum
error of 0.01~mag.

\subsection{Colors and SEDs}
\label{phot.colors}
We show multiple colors from our catalog in
Figures~\ref{fig.colorspt}--\ref{fig.colorcolor.unusual}, spanning \gps\ through
$W3$.  We show 11 colors using at least one PS1 band, and four more (\jkt,
$J_{\rm 2MASS}-W1$, \wawb, and \wbwc) that have been used in many previous
studies.  To create these figures, we have extracted all objects known to be
young, subdwarfs, or binaries, and used these to form a sample of ``unusual''
objects.  The remaining objects are our ``normal field'' sample.
Figure~\ref{fig.colorspt} shows colors as a function of spectral type for the
normal field sample, and Figure~\ref{fig.colorspt.unusual} shows the same colors
for the unusual sample.  In Figures \ref{fig.colorcolor}
and~\ref{fig.colorcolor.unusual}, we use the same format to show colors
vs. colors for the normal field and unusual samples, respectively.

For the figures and analysis presented in this section, we only use 2MASS and
AllWISE photometry with errors less than 0.2~mag, the same standard we use for
PS1 photometry (Section~\ref{phot.ps1.catalog}).  (Note that
Table~\ref{tbl.phot} includes 2MASS and AllWISE photometry with larger errors.)
In addition, we exclude any 2MASS or AllWISE magnitude with a non-zero
contamination flag, and we exclude all AllWISE magnitudes for each object having
an AllWISE neighbor within $8''$ (Section~\ref{phot.2mwise}).

%%% Color vs. Spectral Type plots
\begin{figure*}
\begin{center}
  \begin{minipage}[t]{0.49\textwidth}
    \includegraphics[width=1.00\columnwidth, trim = 20mm 0 10mm 0]{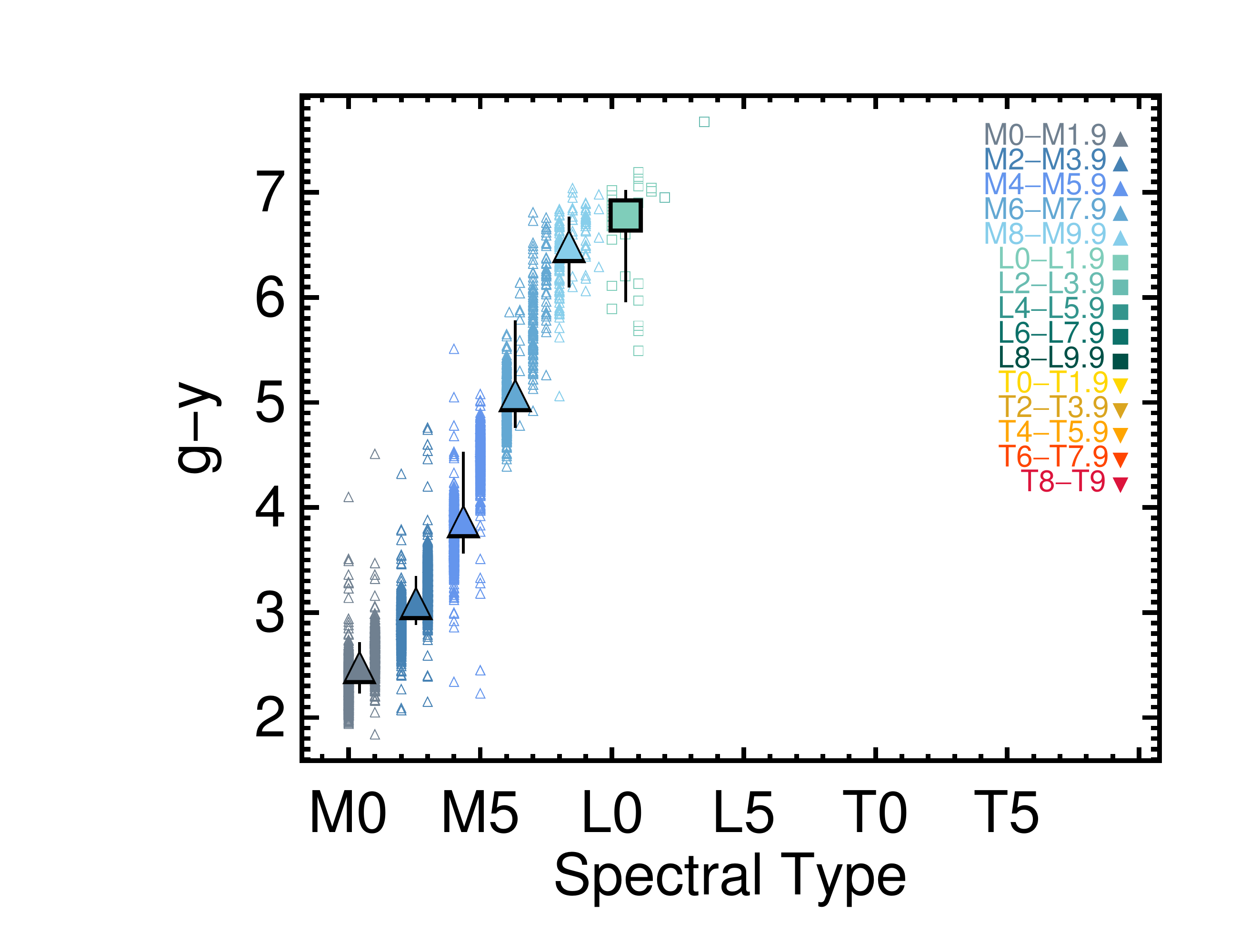}
  \end{minipage}
  \hfill
  \begin{minipage}[t]{0.49\textwidth}
    \includegraphics[width=1.00\columnwidth, trim = 20mm 0 10mm 0]{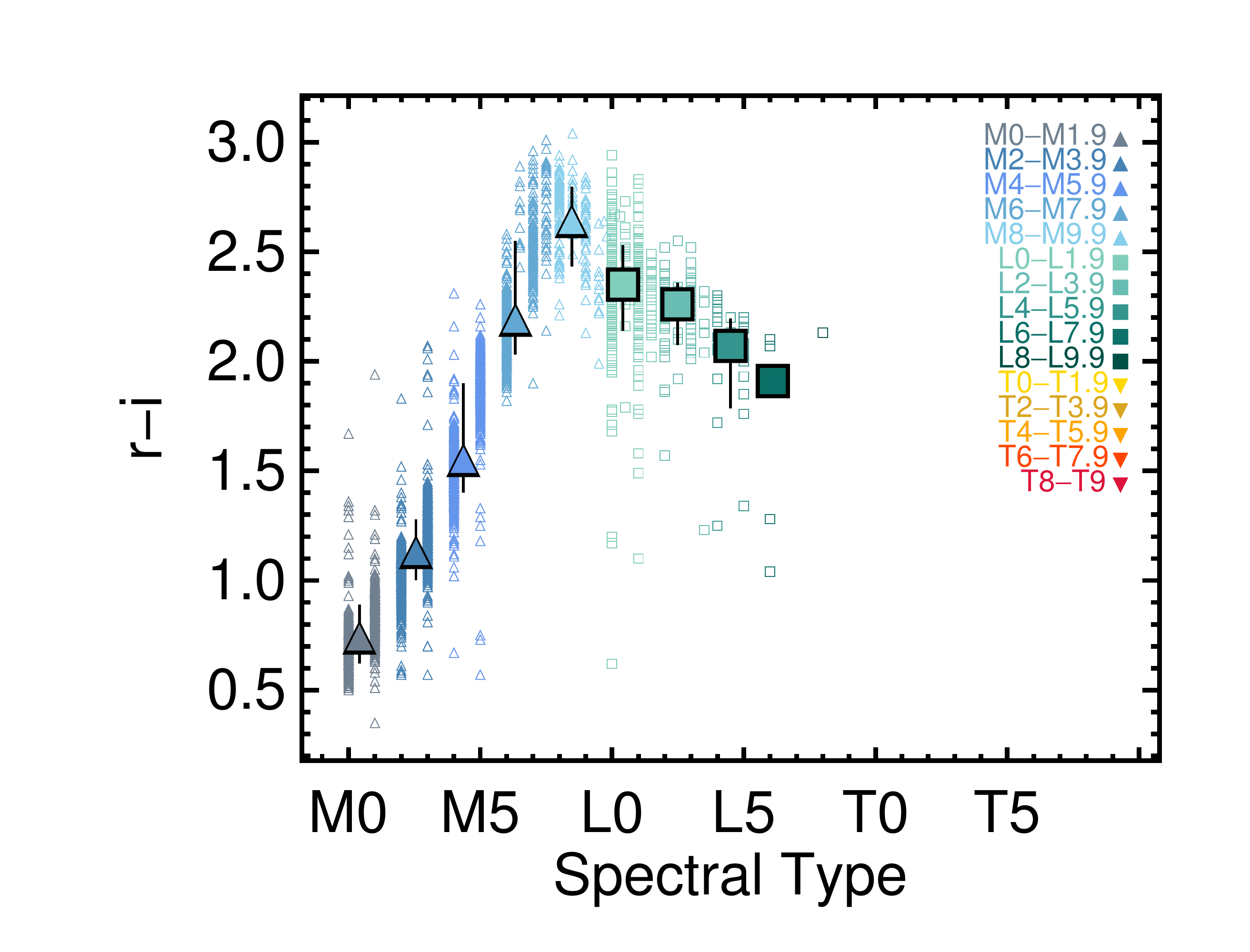}
  \end{minipage}
  \begin{minipage}[t]{0.49\textwidth}
    \includegraphics[width=1.00\columnwidth, trim = 20mm 0 10mm 0]{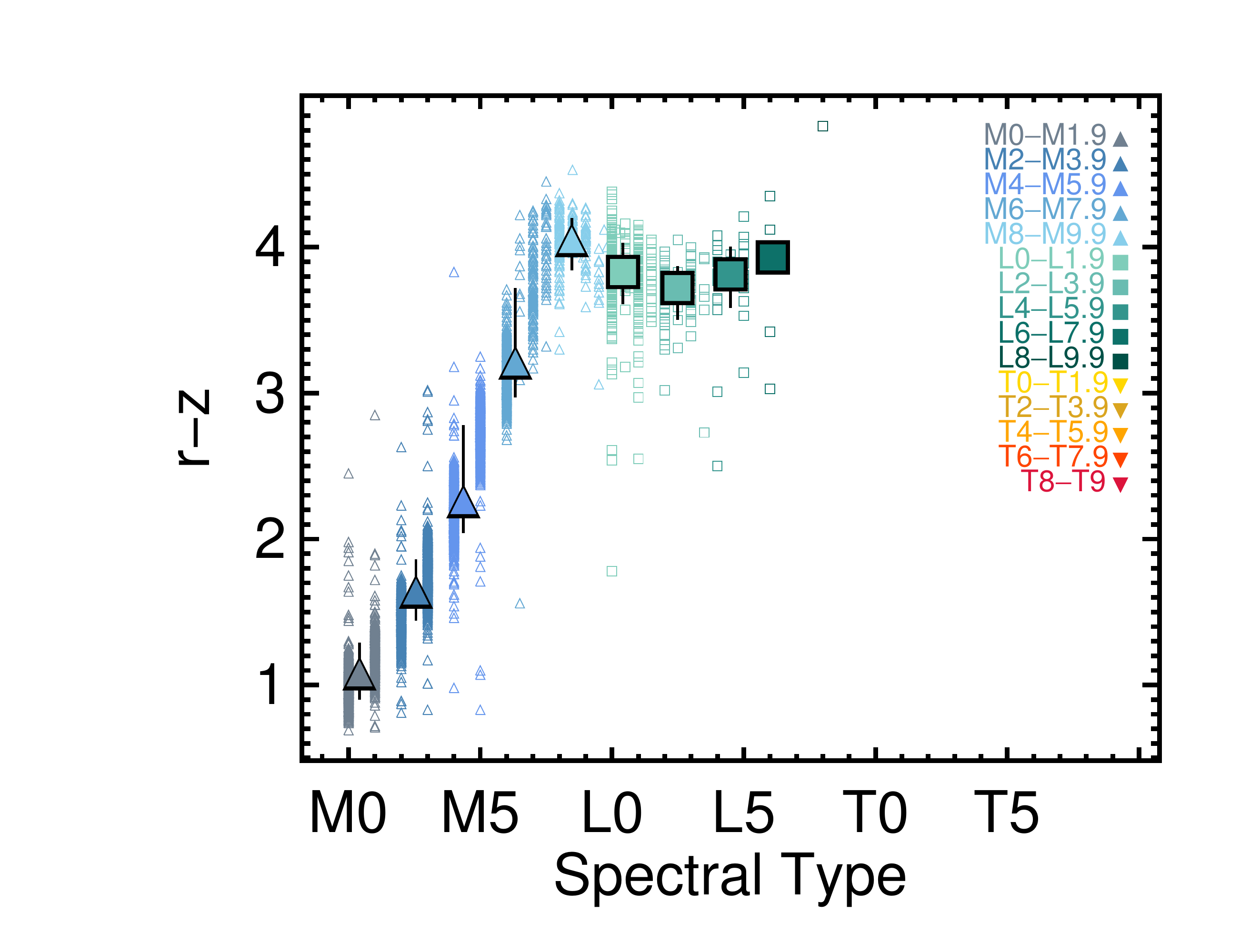}
  \end{minipage}
  \hfill
  \begin{minipage}[t]{0.49\textwidth}
    \includegraphics[width=1.00\columnwidth, trim = 20mm 0 10mm 0]{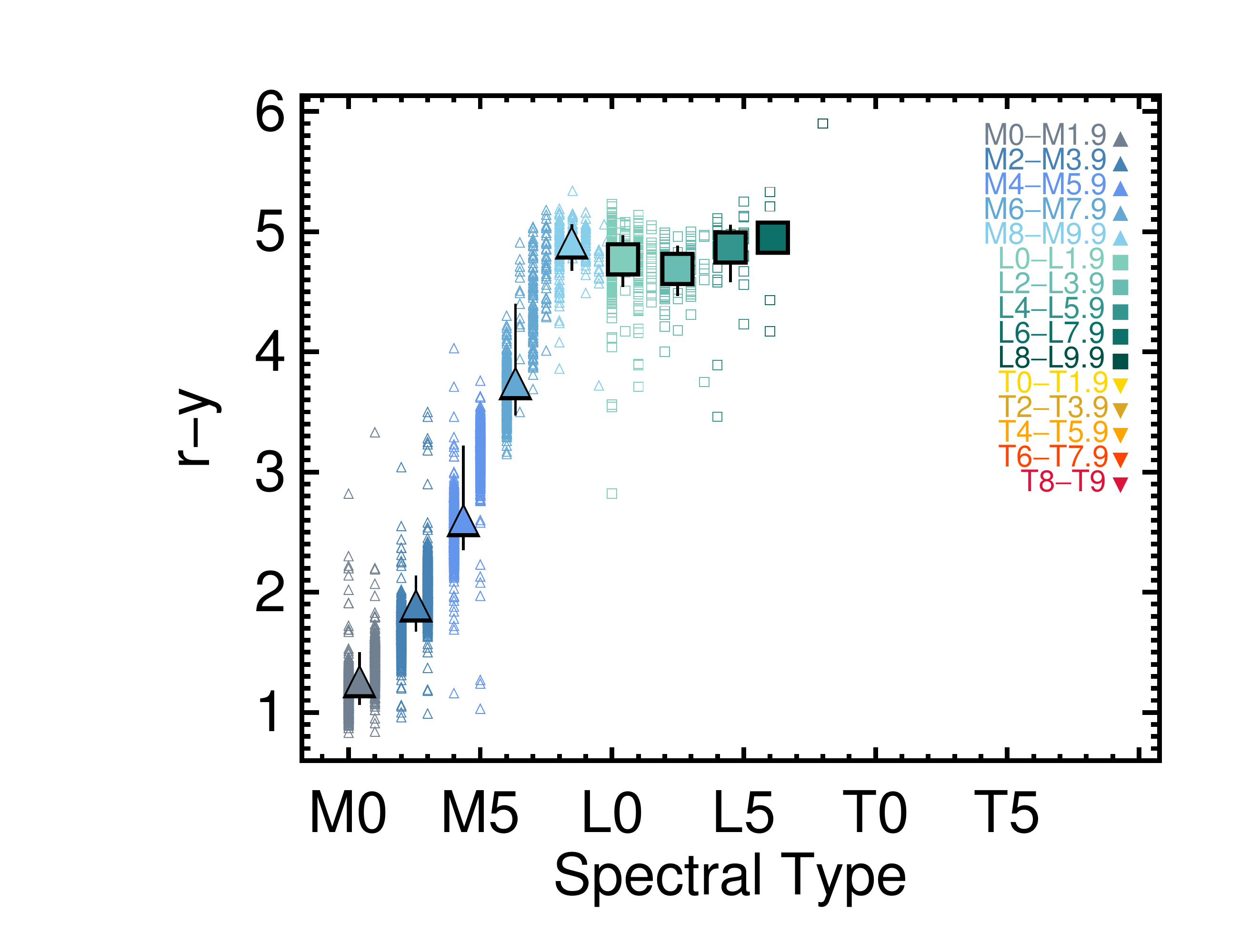}
  \end{minipage}
  \begin{minipage}[t]{0.49\textwidth}
    \includegraphics[width=1.00\columnwidth, trim = 20mm 0 10mm 0]{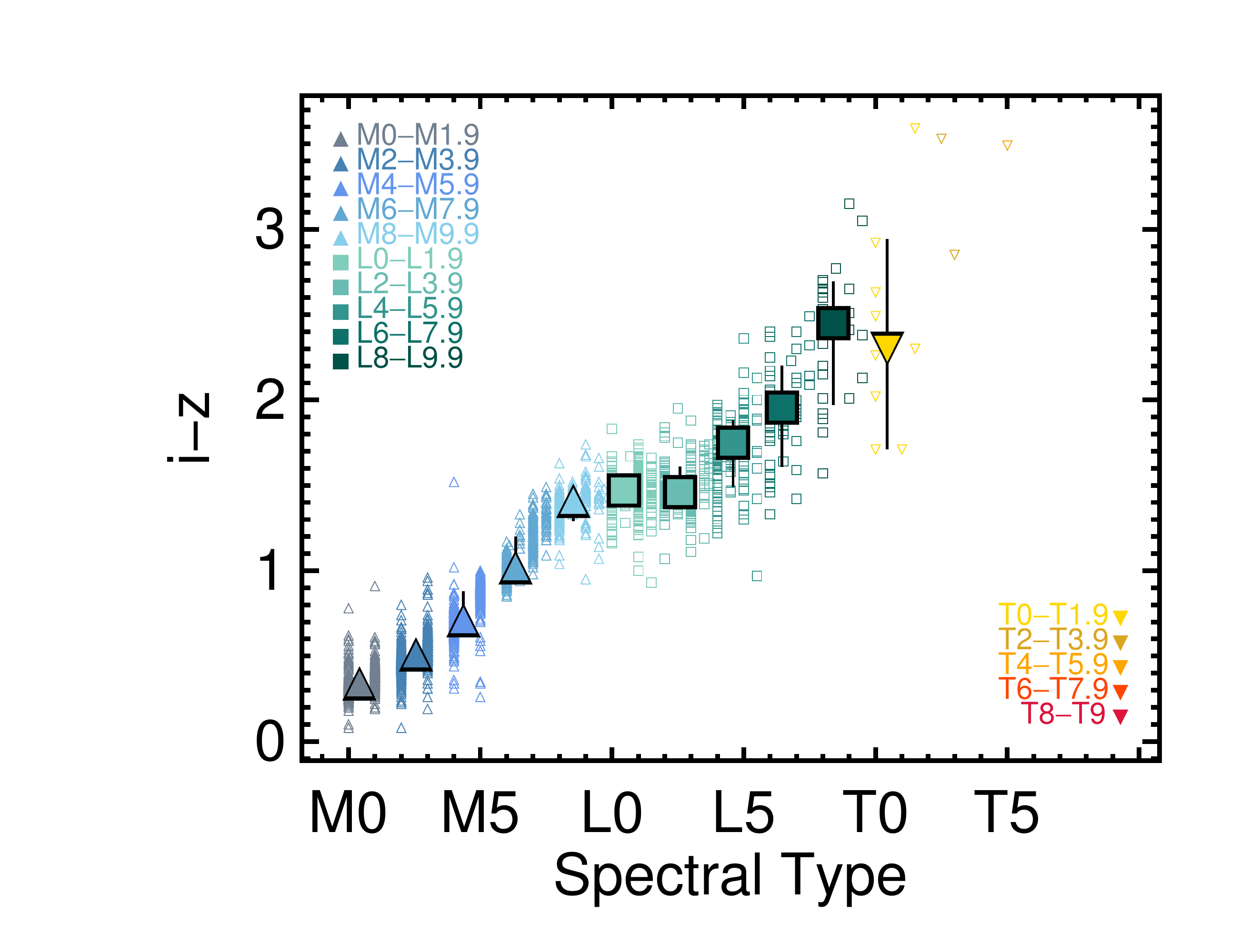}
  \end{minipage}
  \hfill
  \begin{minipage}[t]{0.49\textwidth}
    \includegraphics[width=1.00\columnwidth, trim = 20mm 0 10mm 0]{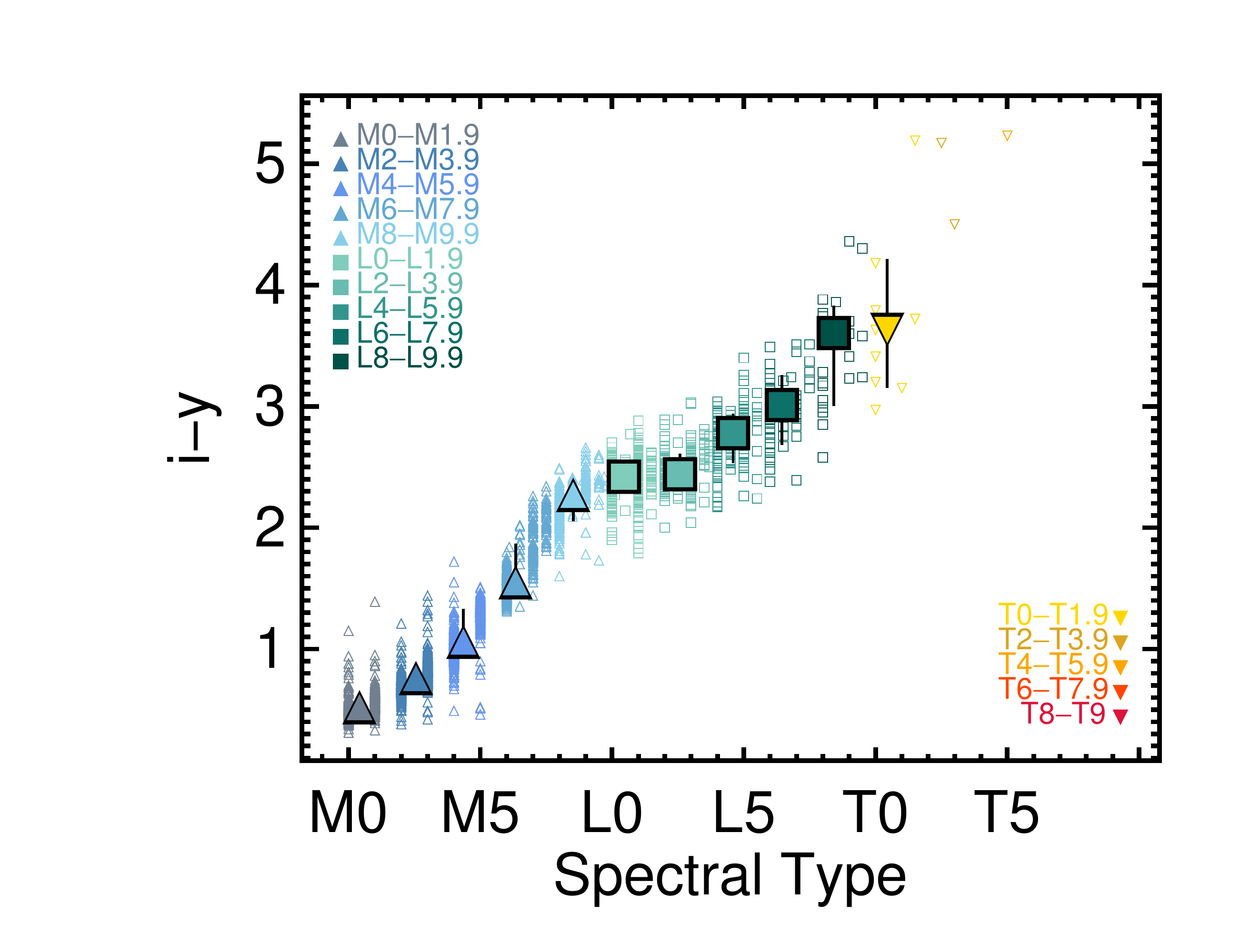}
  \end{minipage}
  \caption{Color vs. spectral type plots for the M, L, and T dwarfs in our
    PS1-detected catalog, excluding objects known to be binaries, subdwarfs, or
    young.  We use only photometry with errors $<0.2$~mag.  For objects having
    both optical and near-IR spectral types, we use the optical type for M and L
    dwarfs and the near-IR type for T~dwarfs. Colors of individual objects are
    shown with small open symbols, while median colors and 68\% confidence
    limits for bins of two spectral subtypes are shown with large filled symbols
    (see legend in each figure).  Median symbols are plotted for bins with at
    least three objects, and confidence limits for bins with at least seven
    objects.  Most PS1 colors plateau through the L dwarfs but become redder for
    T dwarfs (when detected); \ri\ and \ywa\ are notable exceptions.}
  \figurenum{fig.colorspt.1}
\end{center}
\end{figure*}

\begin{figure*}
\begin{center}
  \begin{minipage}[t]{0.49\textwidth}
    \includegraphics[width=1.00\columnwidth, trim = 20mm 0 10mm 0]{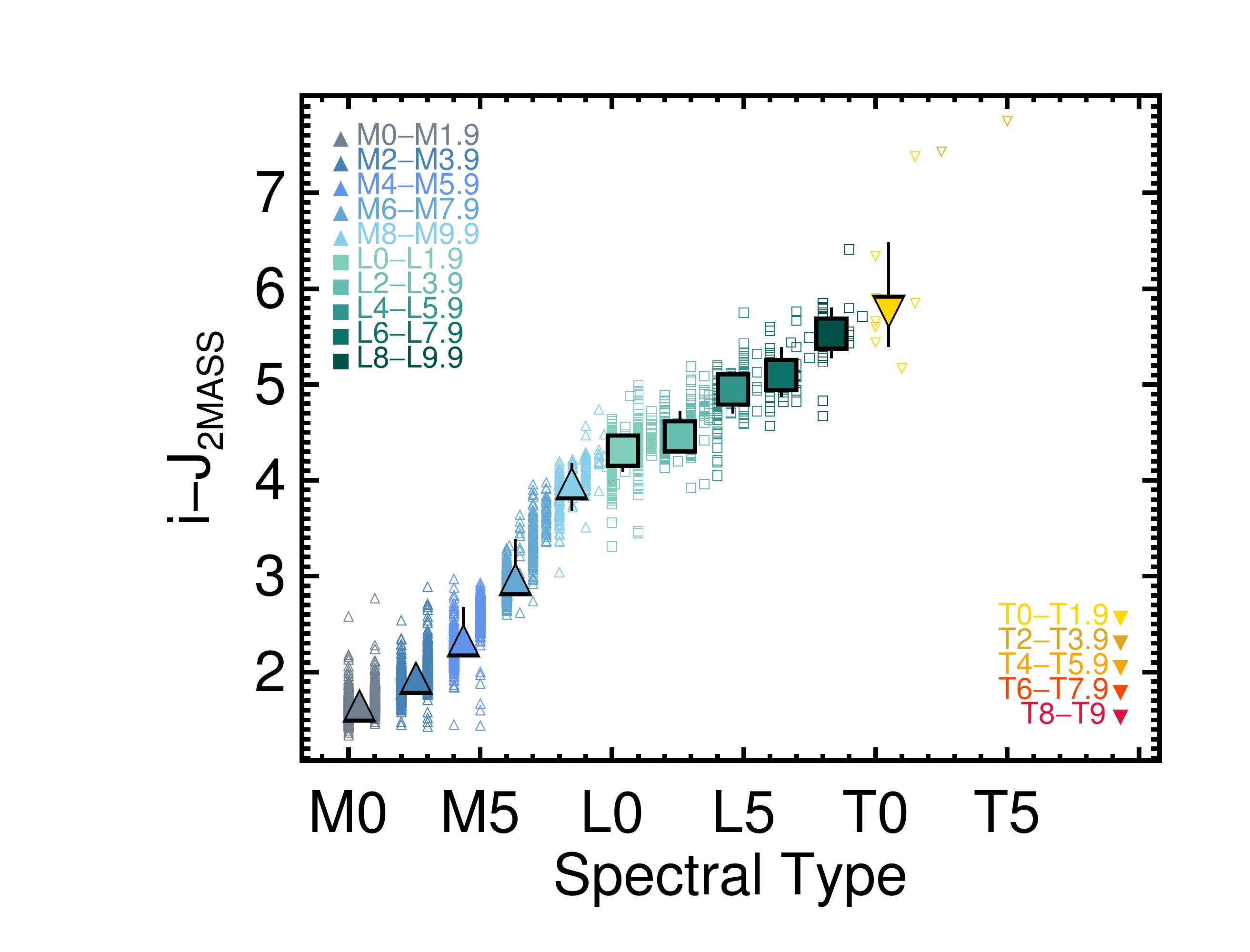}
  \end{minipage}
  \hfill
  \begin{minipage}[t]{0.49\textwidth}
    \includegraphics[width=1.00\columnwidth, trim = 20mm 0 10mm 0]{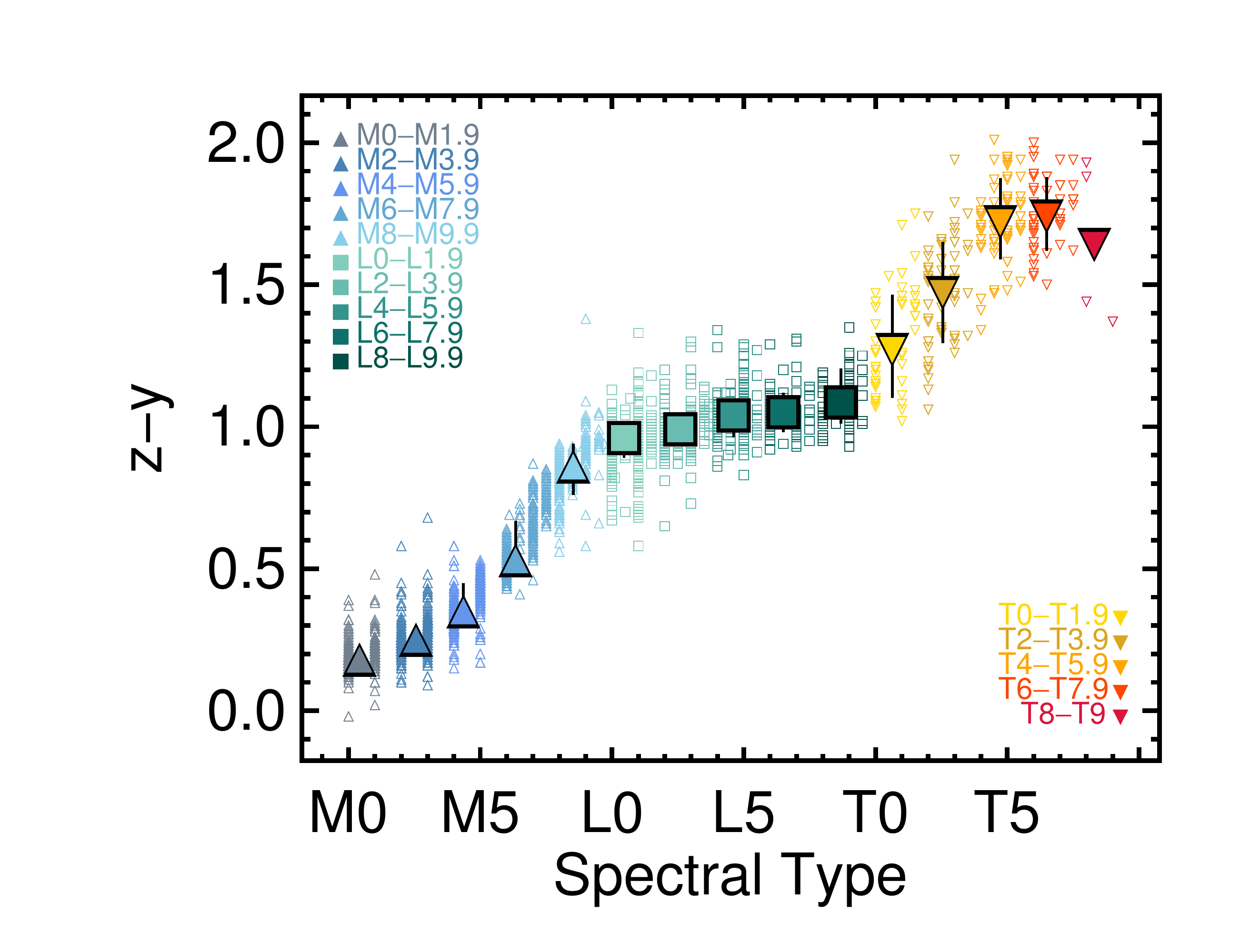}
  \end{minipage}
  \begin{minipage}[t]{0.49\textwidth}
    \includegraphics[width=1.00\columnwidth, trim = 20mm 0 10mm 0]{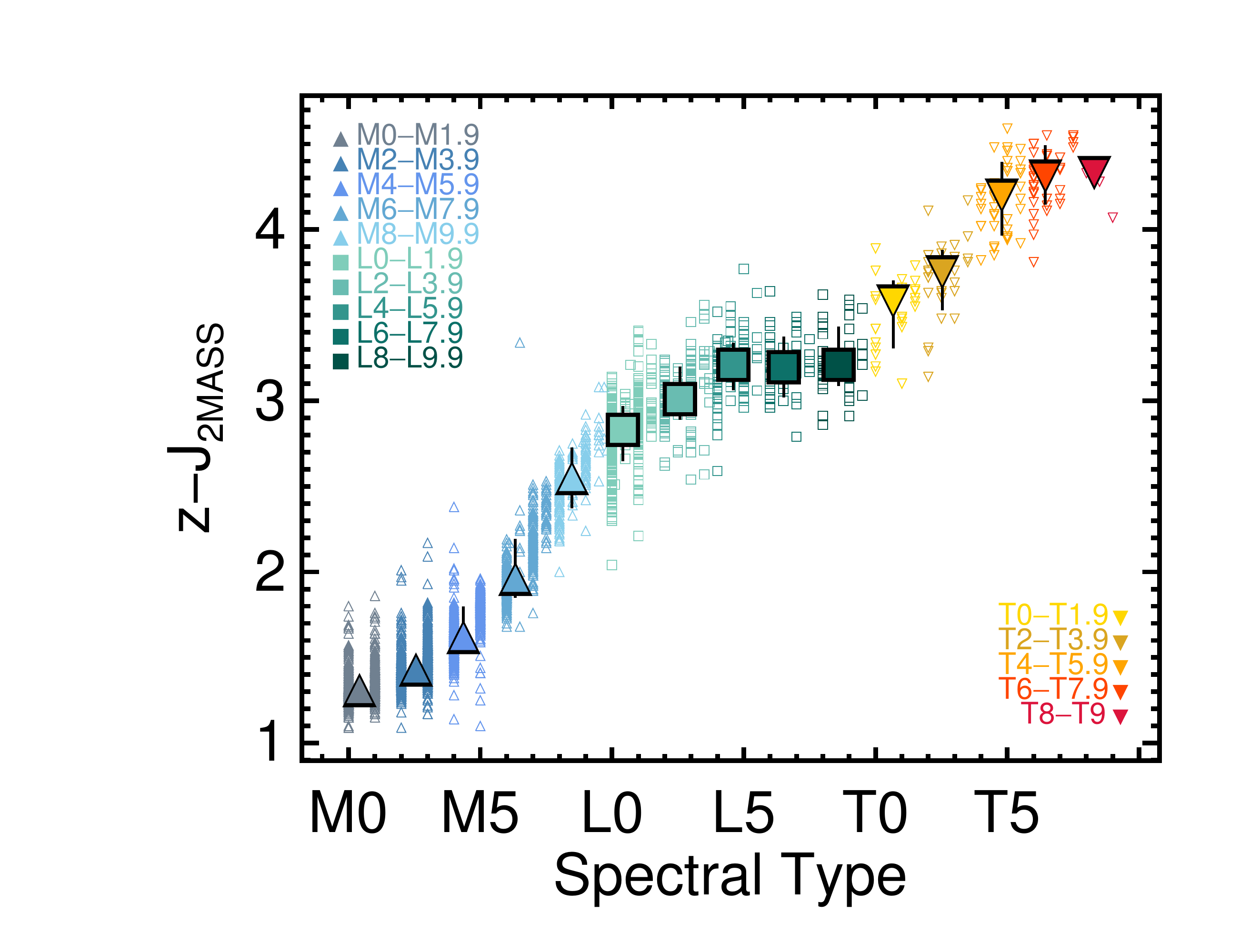}
  \end{minipage}
  \hfill
  \begin{minipage}[t]{0.49\textwidth}
    \includegraphics[width=1.00\columnwidth, trim = 20mm 0 10mm 0]{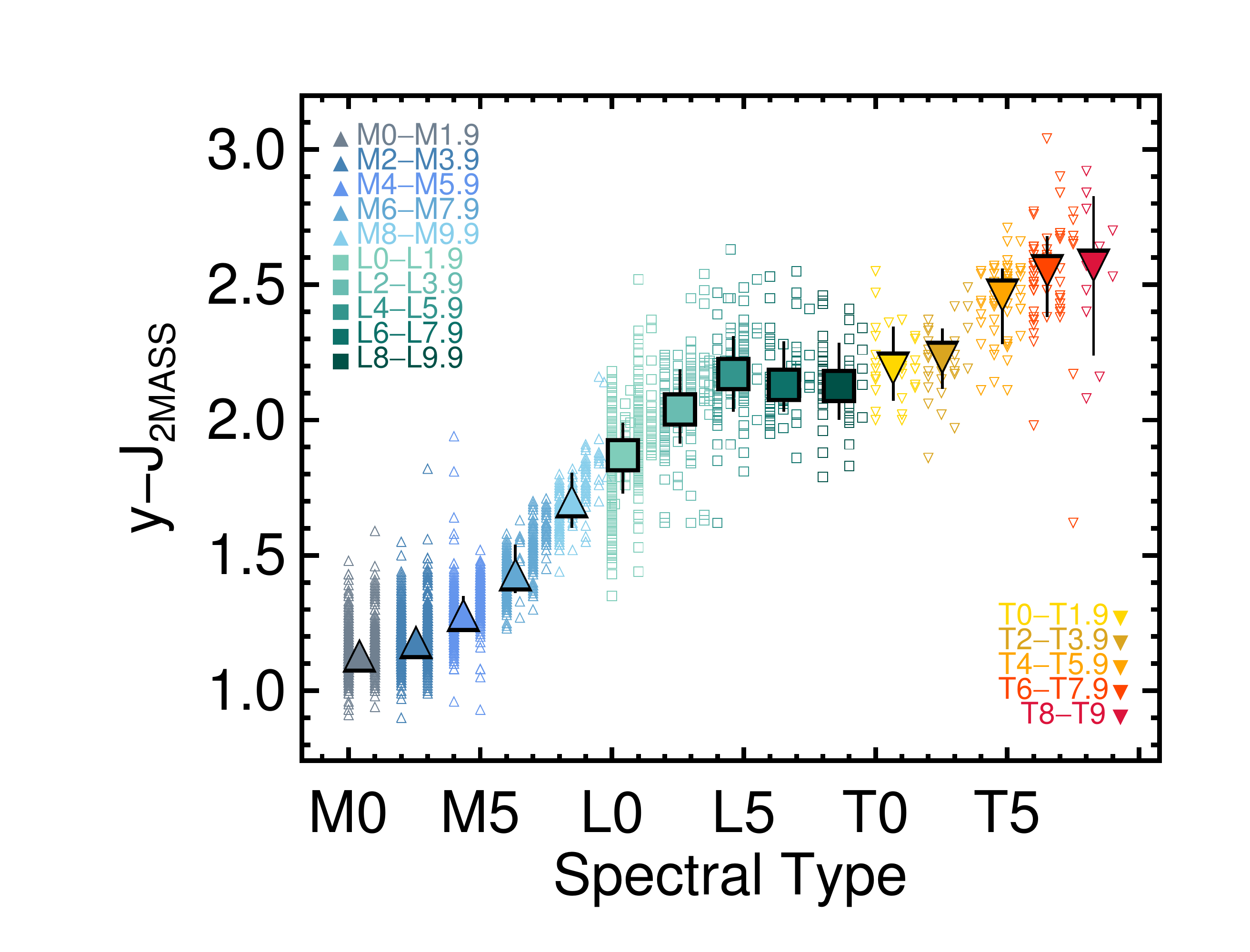}
  \end{minipage}
  \begin{minipage}[t]{0.49\textwidth}
    \includegraphics[width=1.00\columnwidth, trim = 20mm 0 10mm 0]{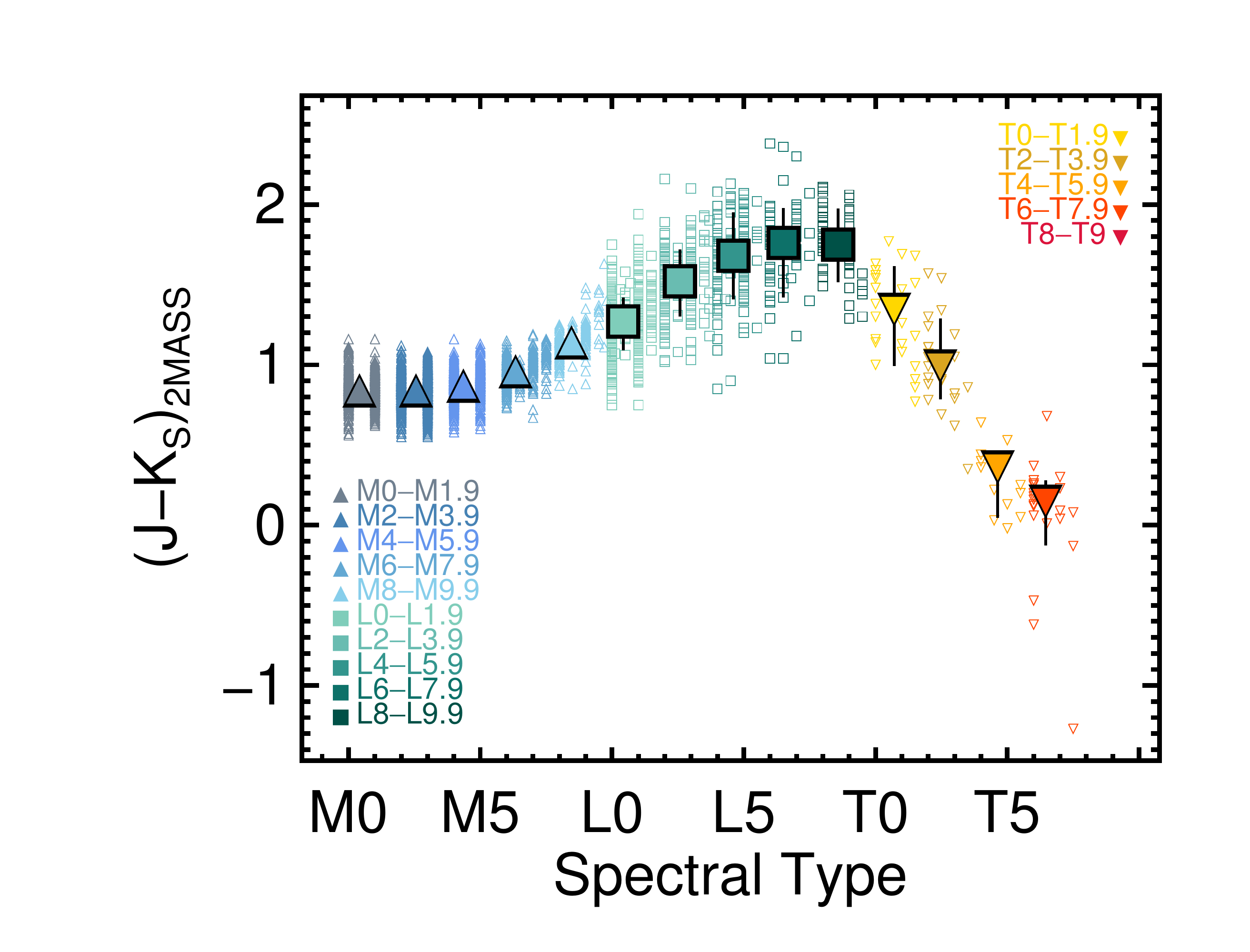}
  \end{minipage}
  \hfill
  \begin{minipage}[t]{0.49\textwidth}
    \includegraphics[width=1.00\columnwidth, trim = 20mm 0 10mm 0]{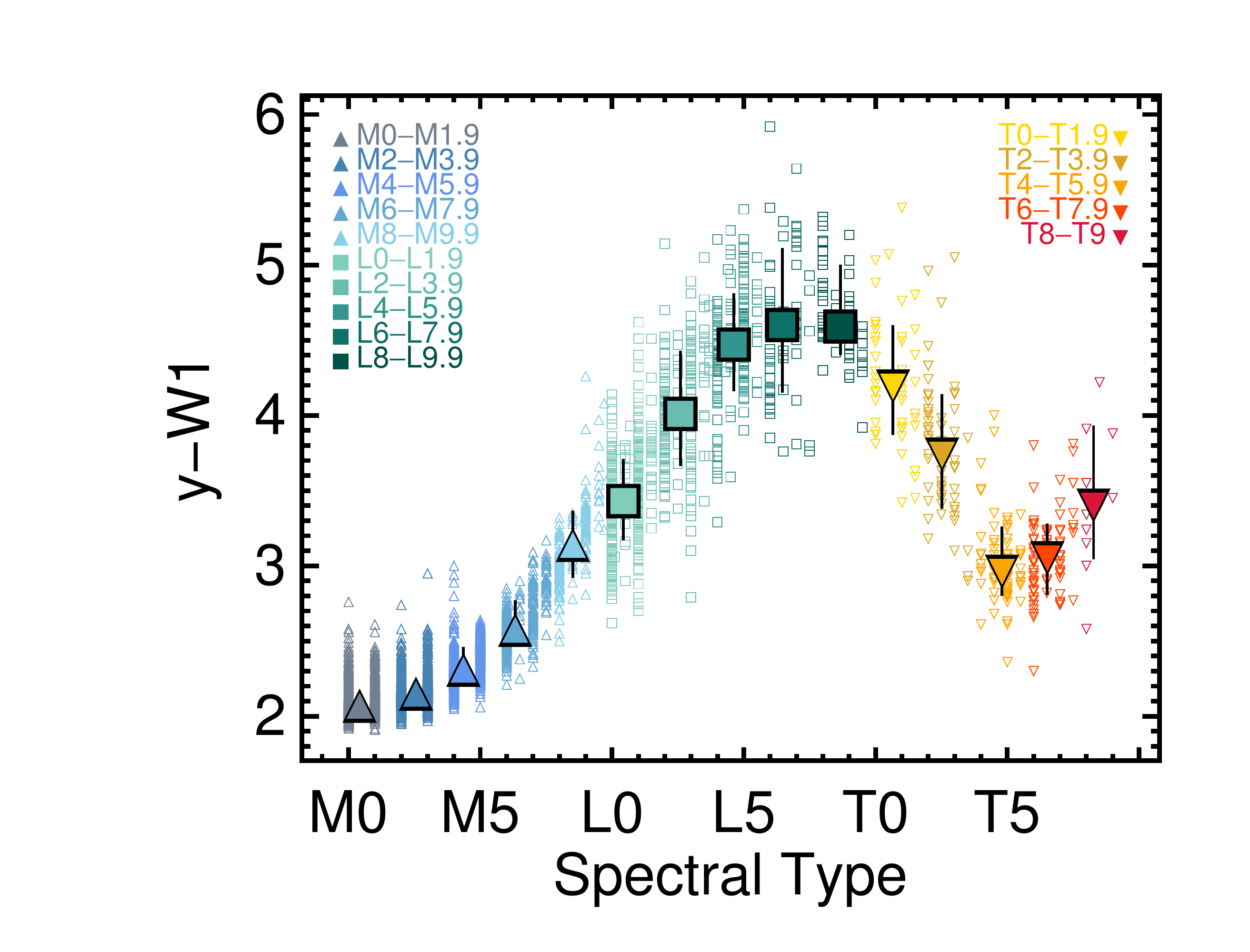}
  \end{minipage}
  \caption{continued.}
  \figurenum{fig.colorspt.2}
\end{center}
\end{figure*}

\begin{figure*}
\begin{center}
  \begin{minipage}[t]{0.49\textwidth}
    \includegraphics[width=1.00\columnwidth, trim = 20mm 0 10mm 0]{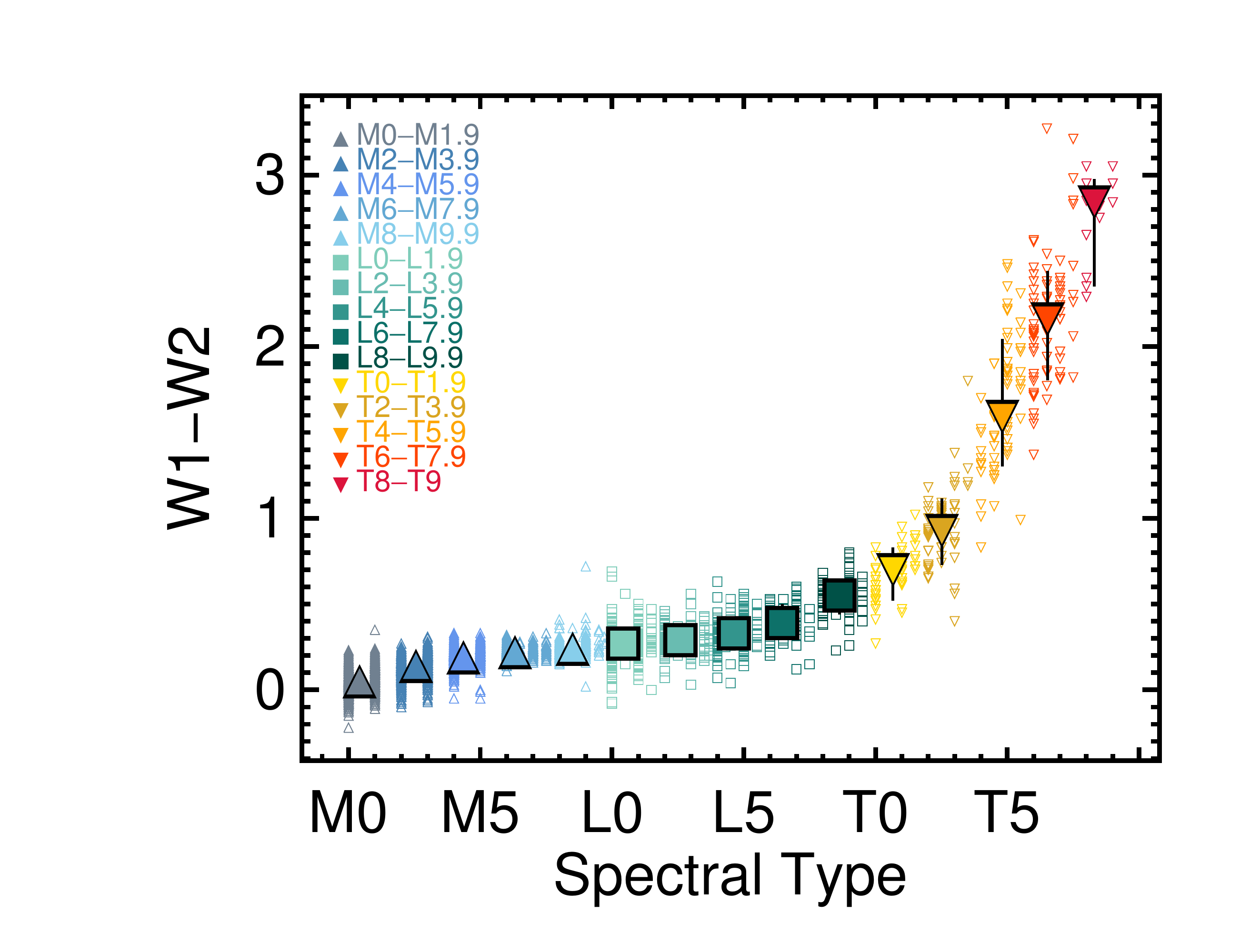}
  \end{minipage}
  \hfill
  \begin{minipage}[t]{0.49\textwidth}
    \includegraphics[width=1.00\columnwidth, trim = 20mm 0 10mm 0]{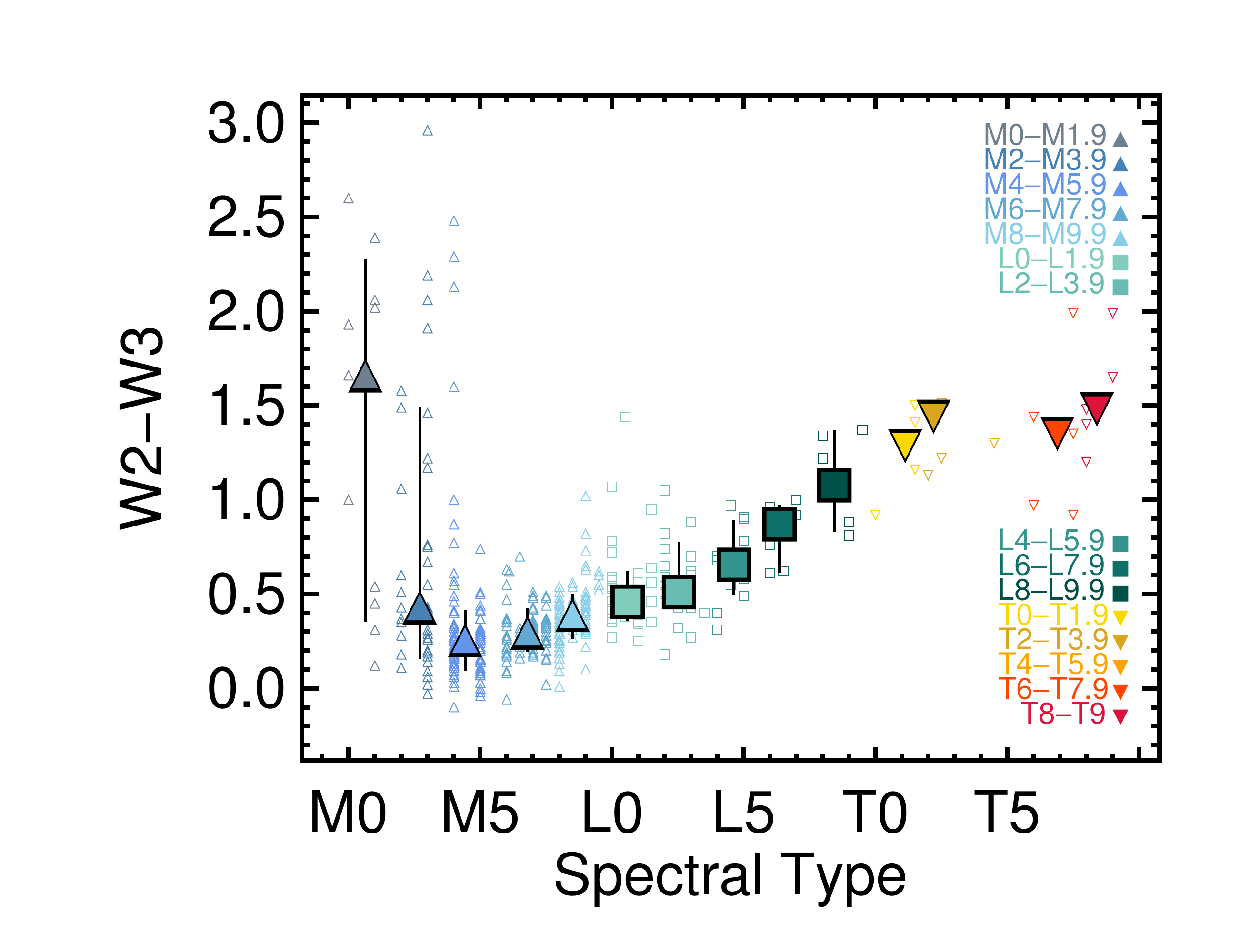}
  \end{minipage}
  \caption{continued.}
  \label{fig.colorspt}
\end{center}
\end{figure*}

%%% Color vs. Spectral Type plots for unusual objects
\begin{figure*}
\begin{center}
  \begin{minipage}[t]{0.49\textwidth}
    \includegraphics[width=1.00\columnwidth, trim = 20mm 0 10mm 0]{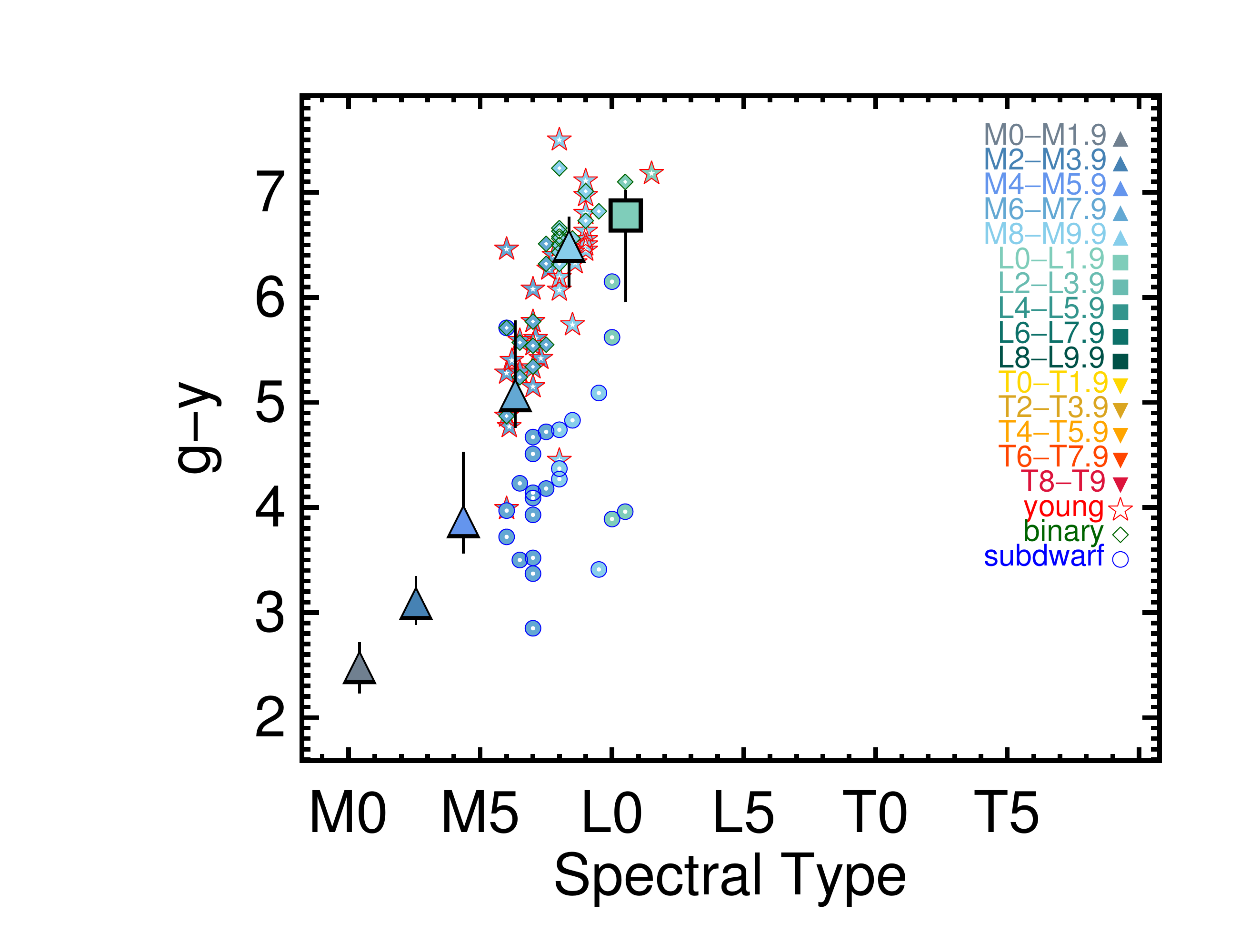}
  \end{minipage}
  \hfill
  \begin{minipage}[t]{0.49\textwidth}
    \includegraphics[width=1.00\columnwidth, trim = 20mm 0 10mm 0]{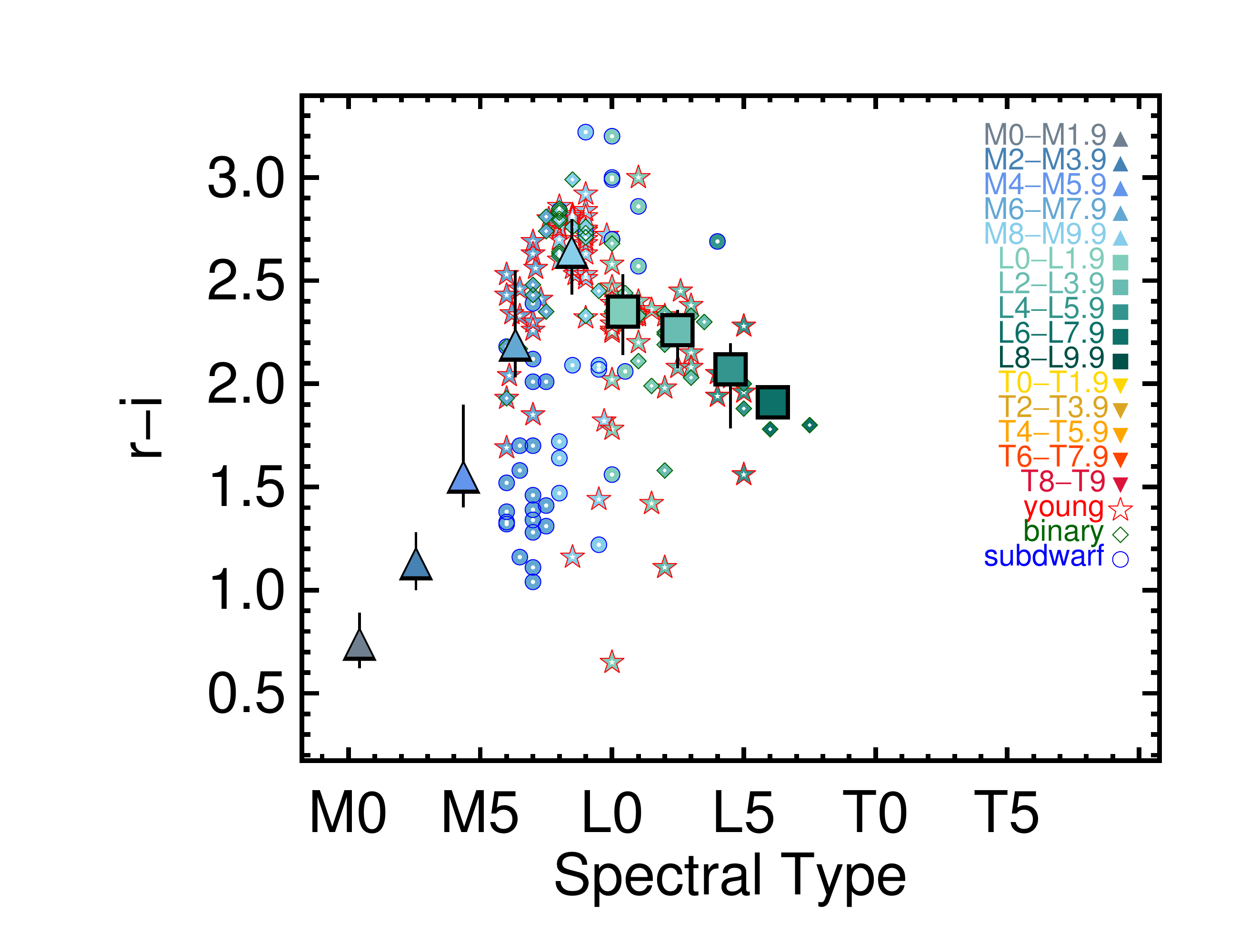}
  \end{minipage}
  \begin{minipage}[t]{0.49\textwidth}
    \includegraphics[width=1.00\columnwidth, trim = 20mm 0 10mm 0]{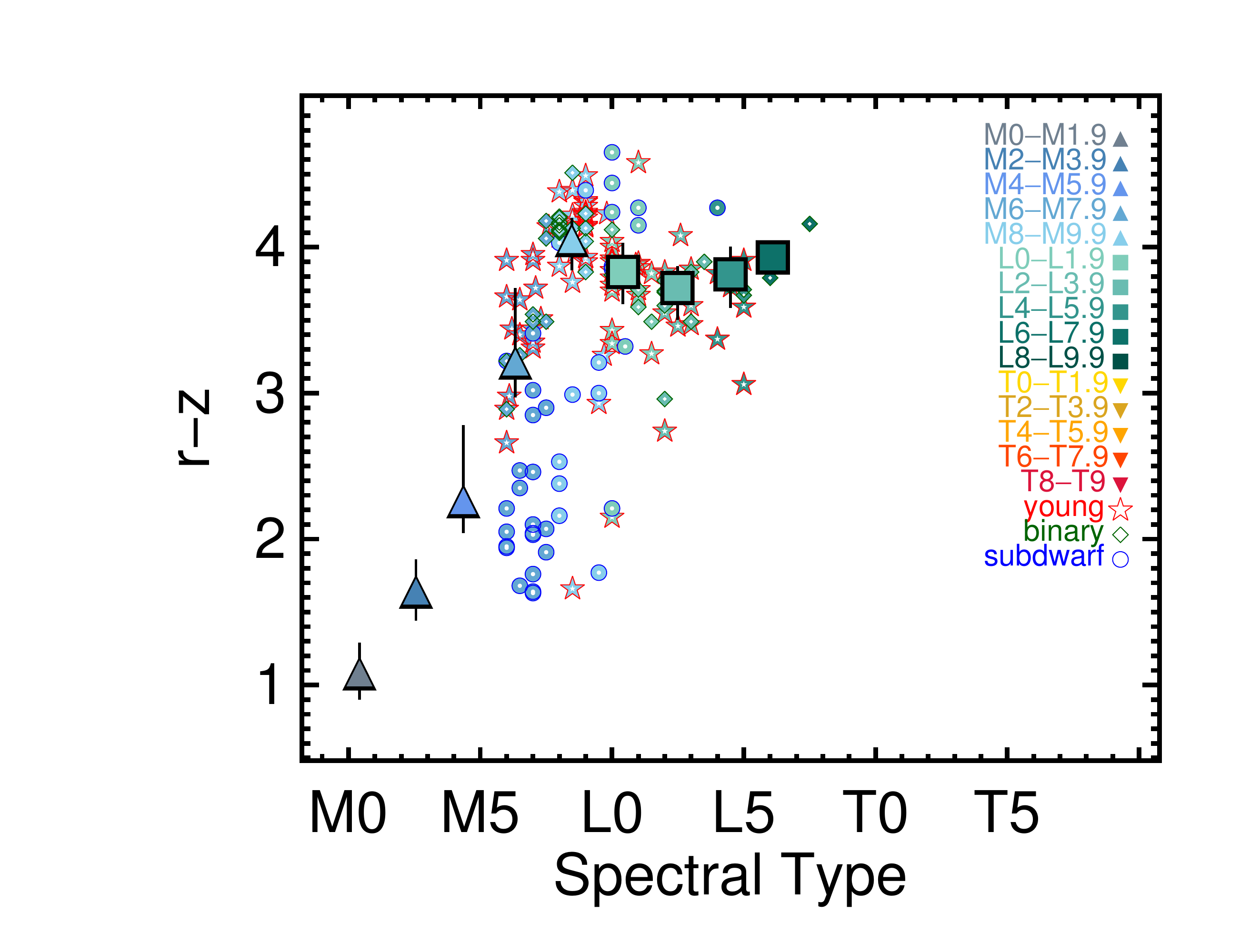}
  \end{minipage}
  \hfill
  \begin{minipage}[t]{0.49\textwidth}
    \includegraphics[width=1.00\columnwidth, trim = 20mm 0 10mm 0]{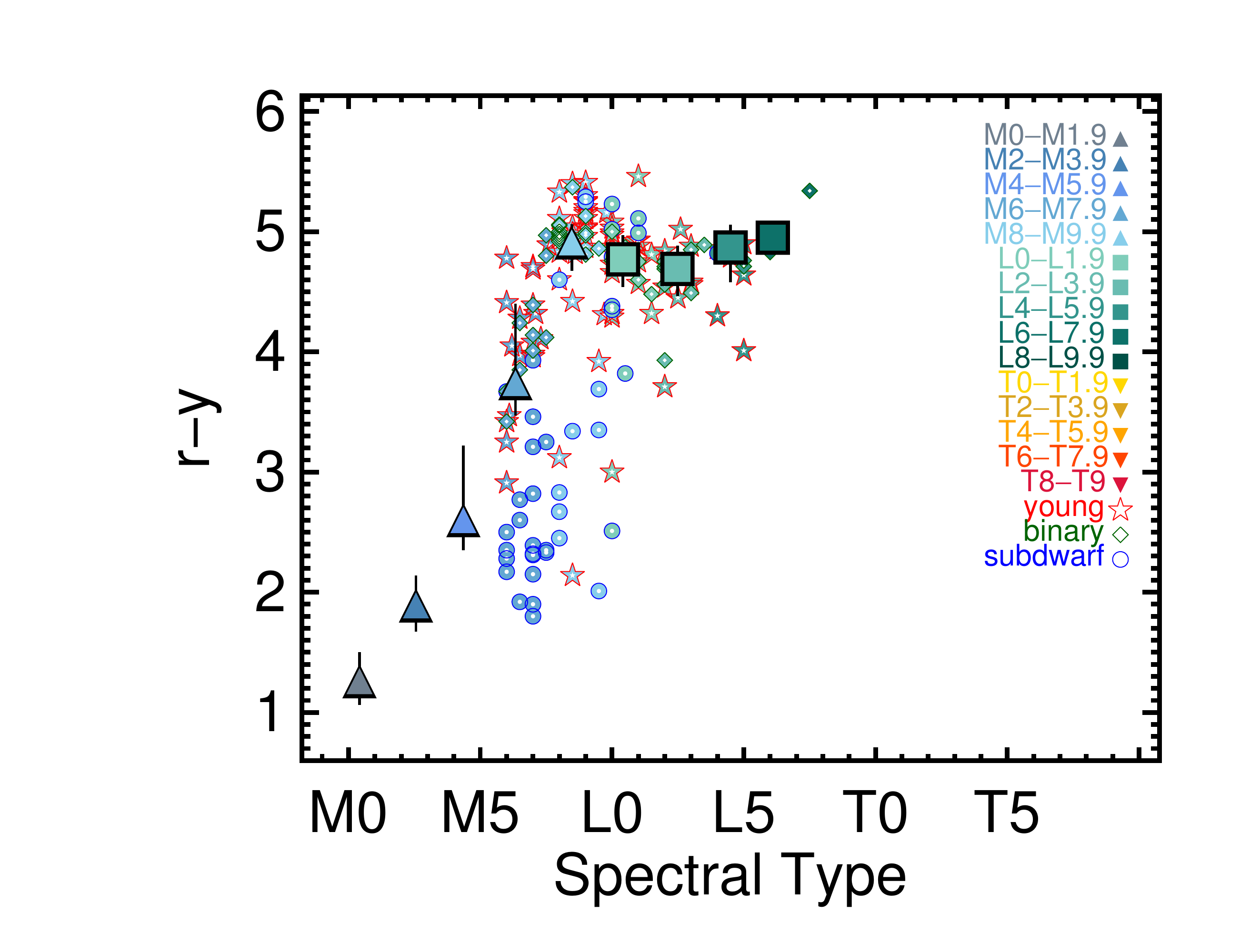}
  \end{minipage}
  \begin{minipage}[t]{0.49\textwidth}
    \includegraphics[width=1.00\columnwidth, trim = 20mm 0 10mm 0]{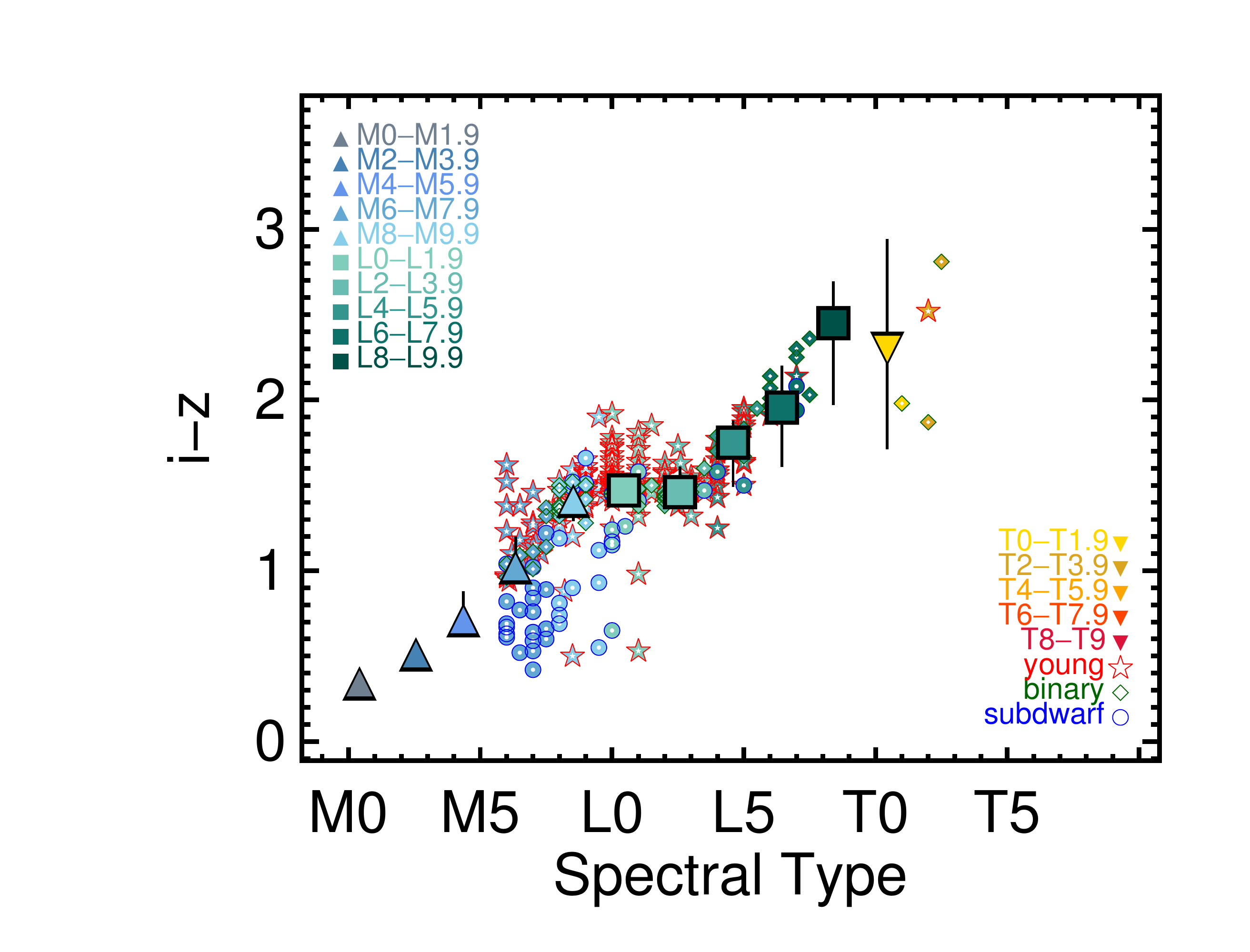}
  \end{minipage}
  \hfill
  \begin{minipage}[t]{0.49\textwidth}
    \includegraphics[width=1.00\columnwidth, trim = 20mm 0 10mm 0]{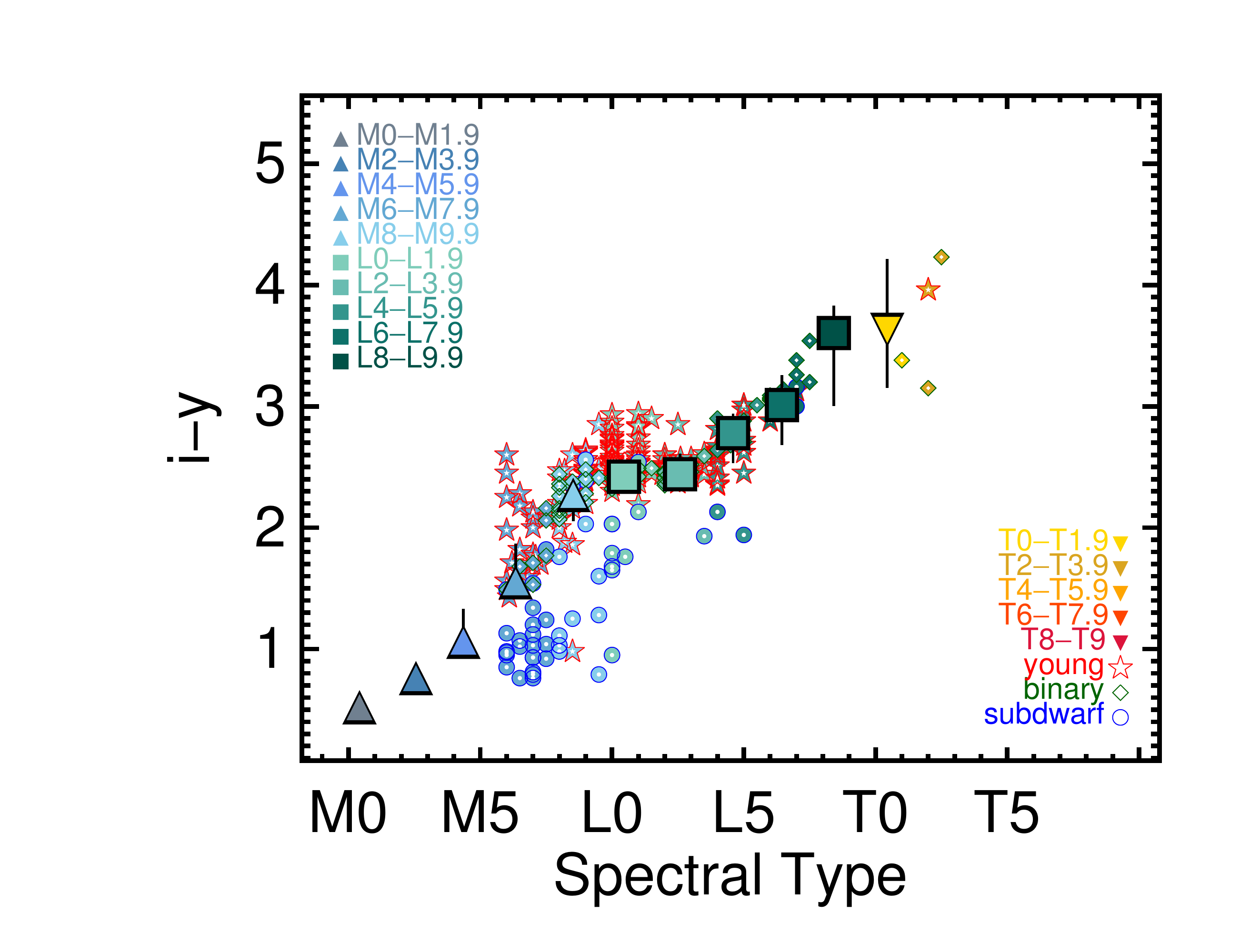}
  \end{minipage}
  \caption{Color vs. spectral type plots for known young objects (outlined with
    red stars), binaries (green diamonds), and subdwarfs (blue circles) in our
    PS1-detected catalog.  Interiors of the symbols use the same color scheme as
    in Figure~\ref{fig.colorspt} (see legends).  Median colors and 68\%
    confidence limits for normal field objects from Figure~\ref{fig.colorspt}
    are overplotted for reference.  Typically the young objects have field-like
    or redder colors while the subdwarfs have bluer-than-field colors, but the
    \ri, \rz, and \ry\ vs. SpT plots show a number of exceptions to both of
    these norms.}
  \figurenum{fig.colorspt.unusual.1}
\end{center}
\end{figure*}

\begin{figure*}
\begin{center}
  \begin{minipage}[t]{0.49\textwidth}
    \includegraphics[width=1.00\columnwidth, trim = 20mm 0 10mm 0]{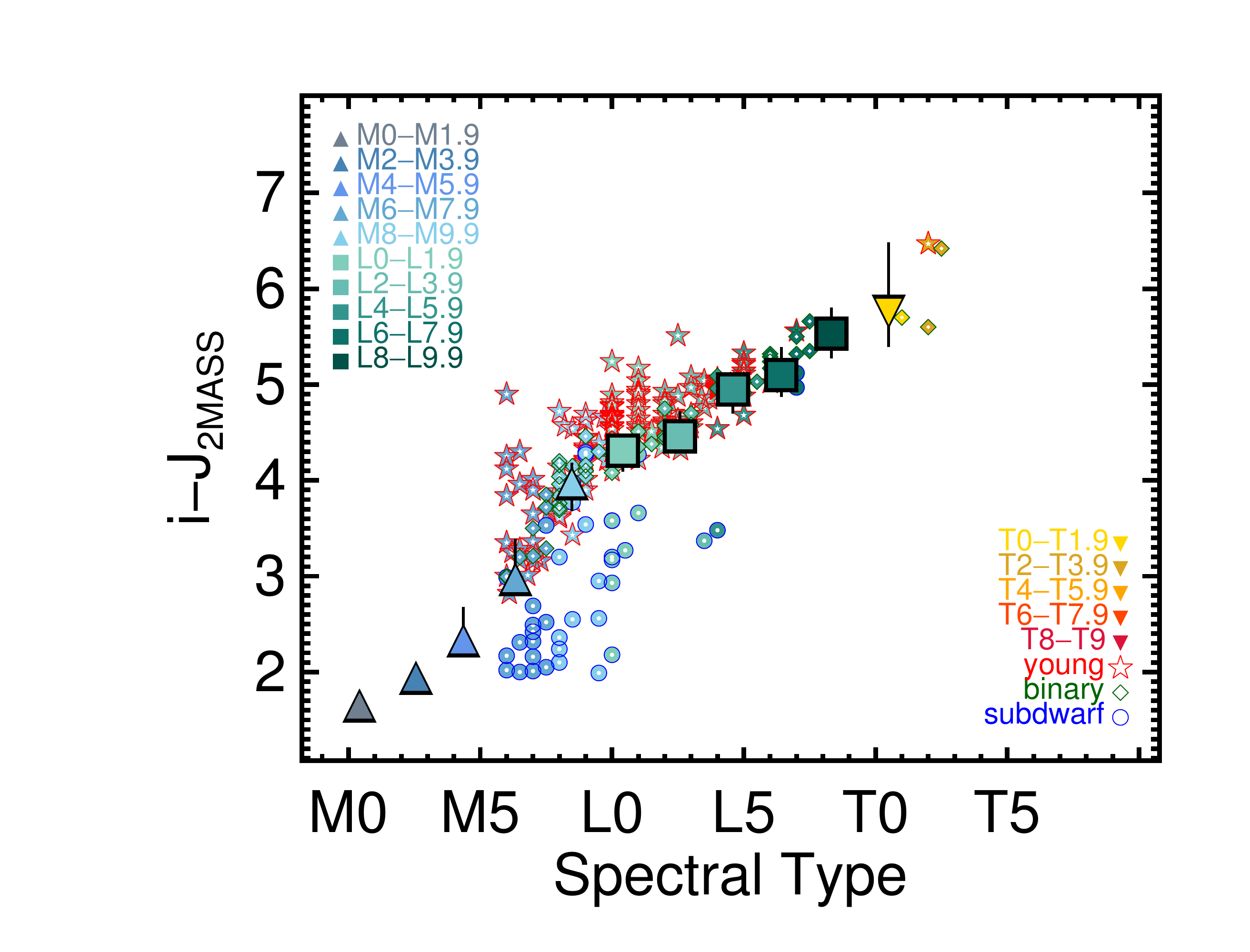}
  \end{minipage}
  \hfill
  \begin{minipage}[t]{0.49\textwidth}
    \includegraphics[width=1.00\columnwidth, trim = 20mm 0 10mm 0]{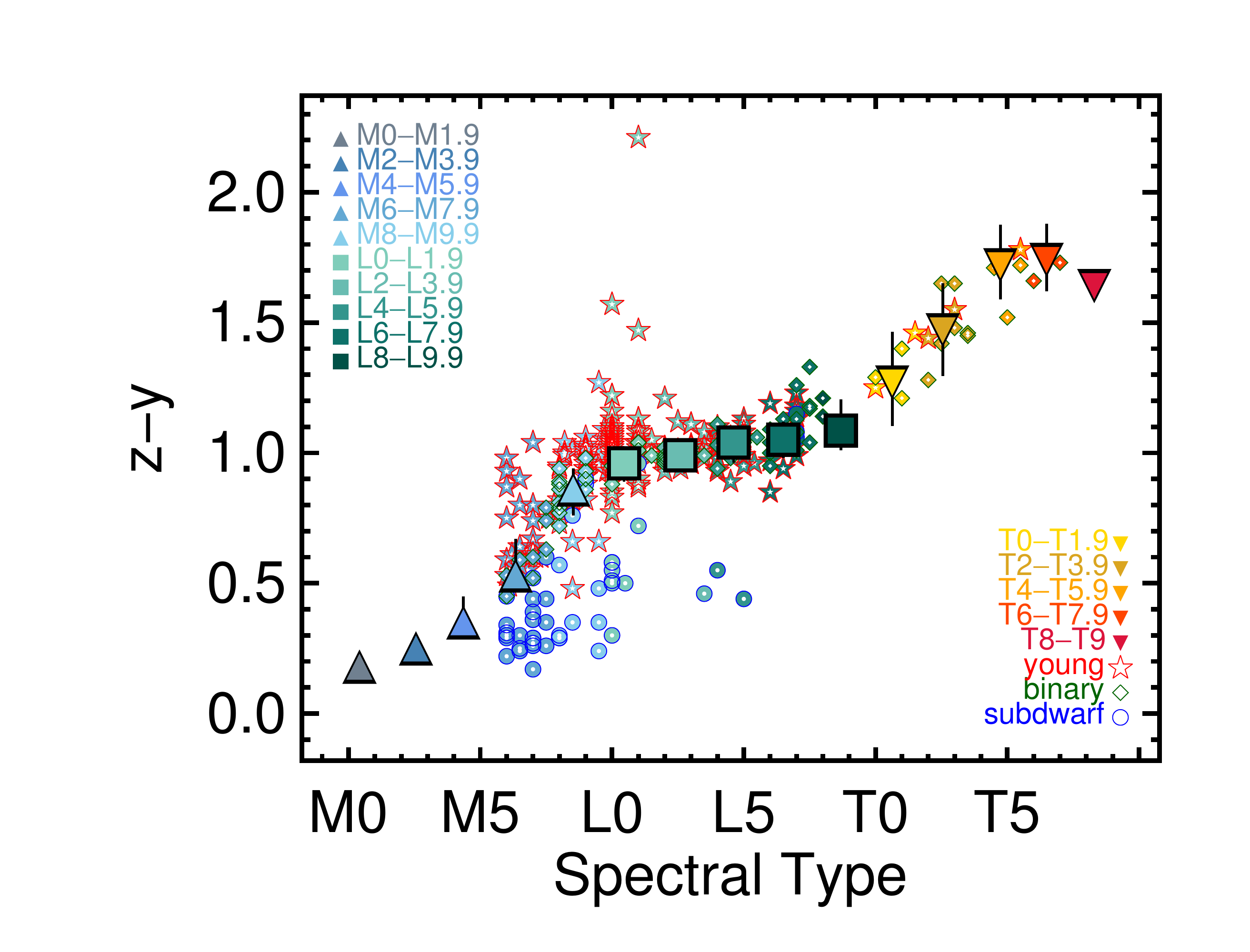}
  \end{minipage}
  \begin{minipage}[t]{0.49\textwidth}
    \includegraphics[width=1.00\columnwidth, trim = 20mm 0 10mm 0]{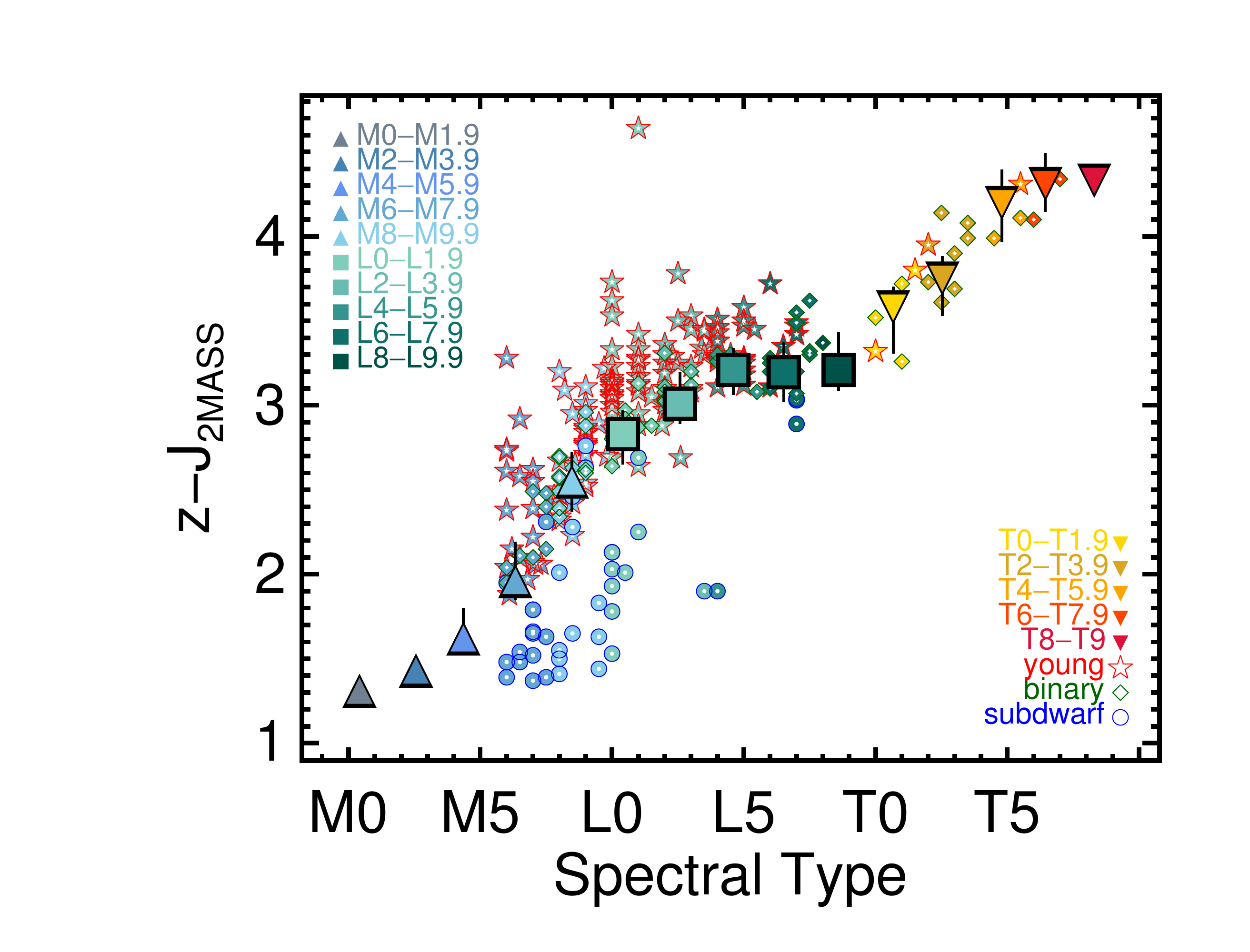}
  \end{minipage}
  \hfill
  \begin{minipage}[t]{0.49\textwidth}
    \includegraphics[width=1.00\columnwidth, trim = 20mm 0 10mm 0]{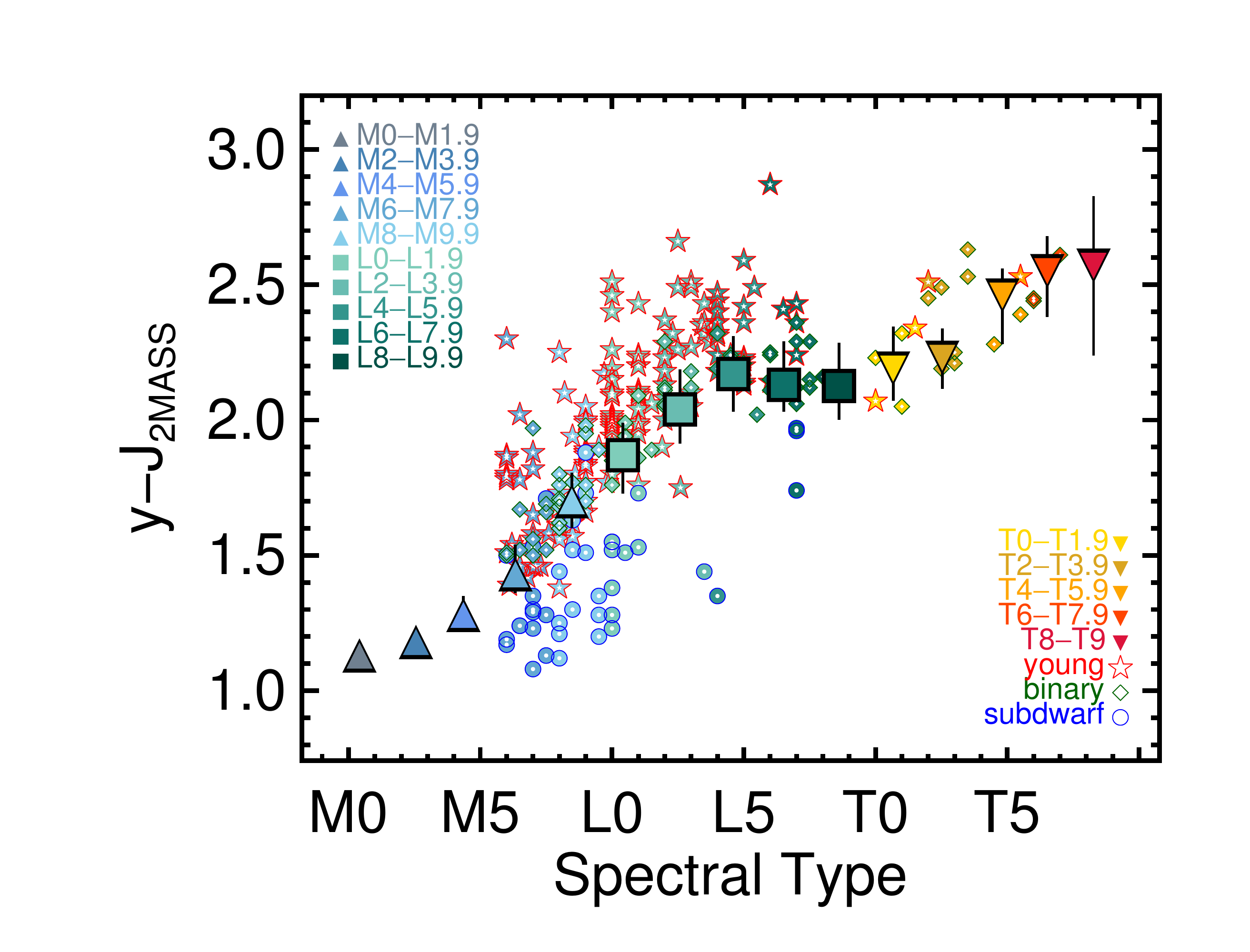}
  \end{minipage}
  \begin{minipage}[t]{0.49\textwidth}
    \includegraphics[width=1.00\columnwidth, trim = 20mm 0 10mm 0]{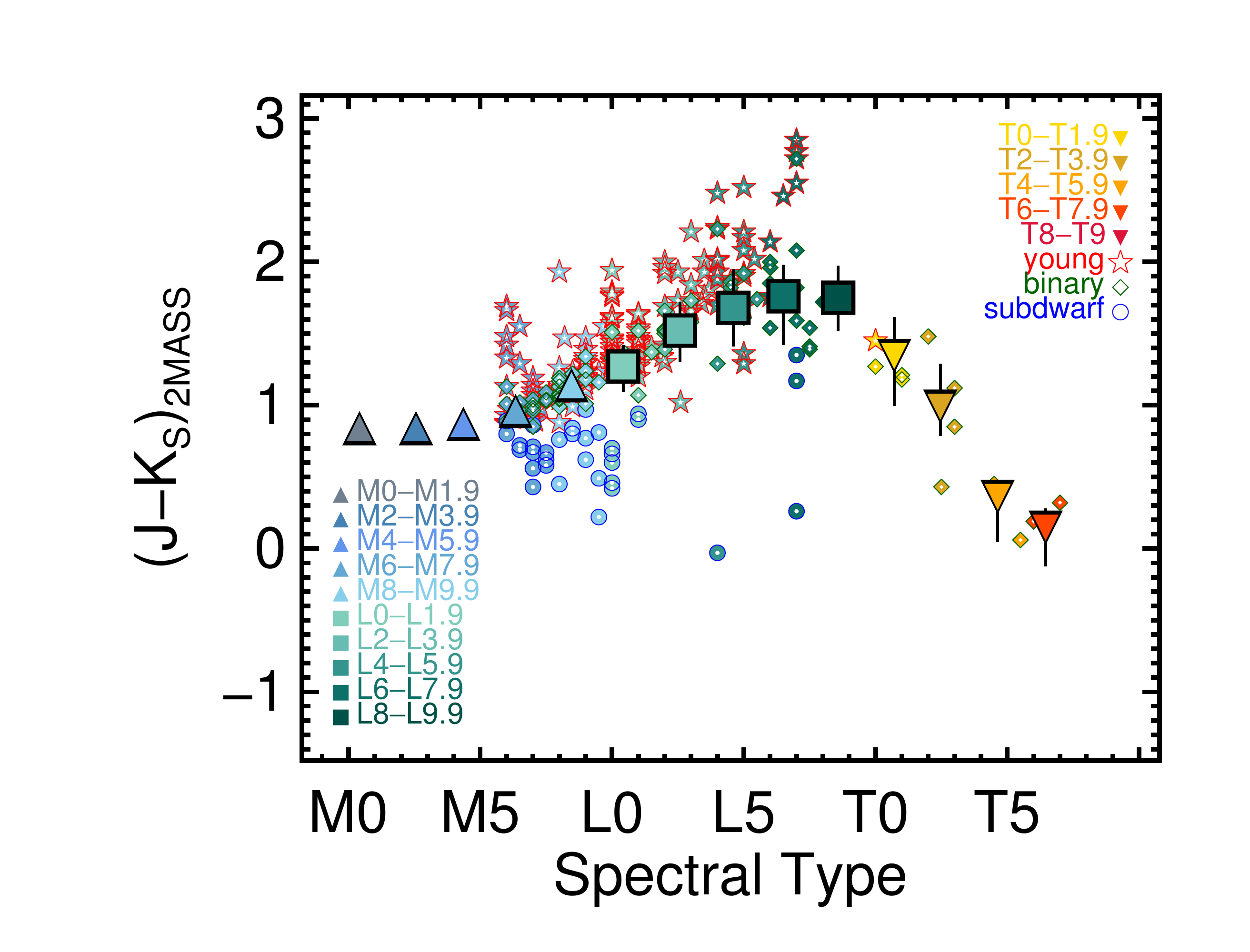}
  \end{minipage}
  \hfill
  \begin{minipage}[t]{0.49\textwidth}
    \includegraphics[width=1.00\columnwidth, trim = 20mm 0 10mm 0]{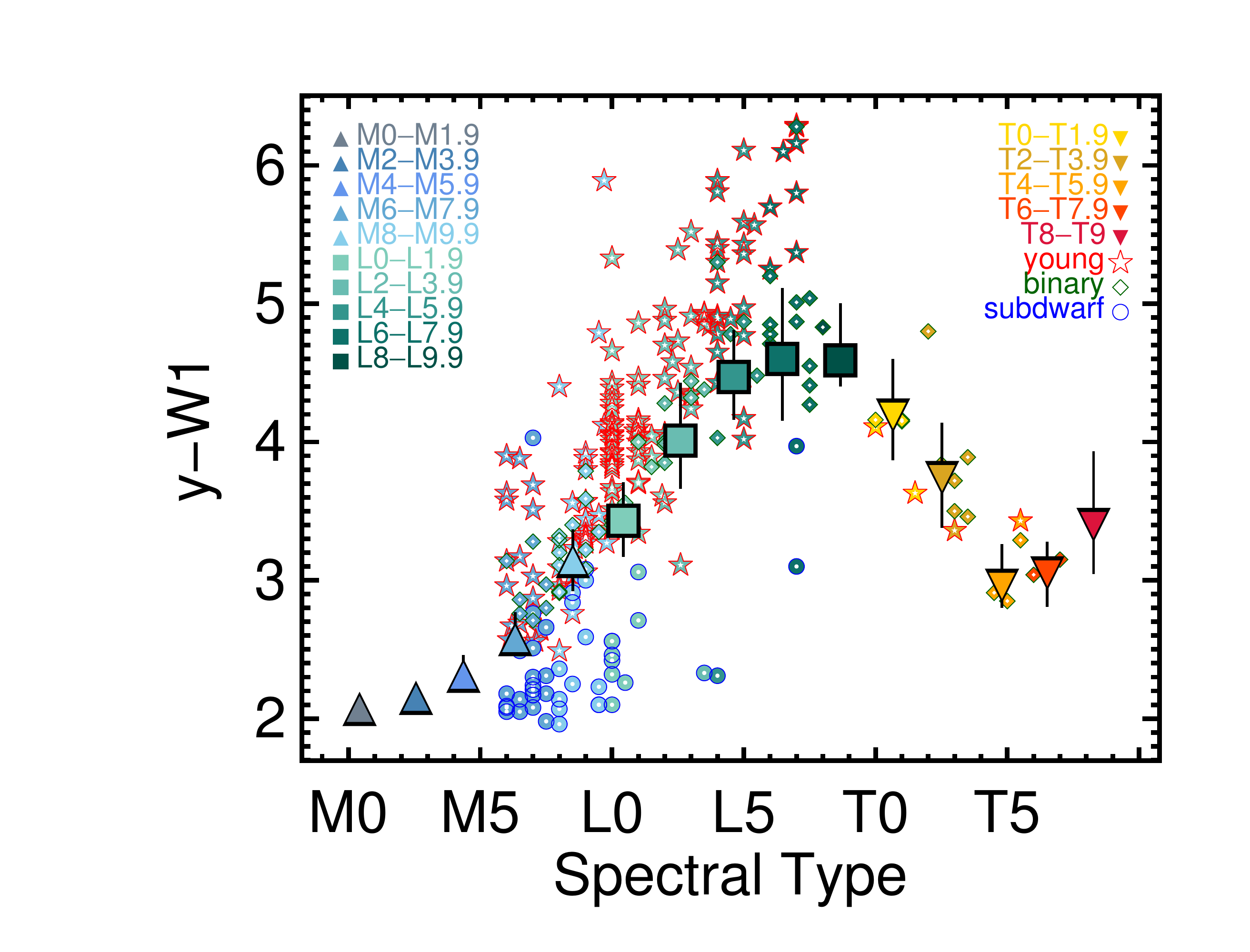}
  \end{minipage}
  \caption{continued.}
  \figurenum{fig.colorspt.unusual.2}
\end{center}
\end{figure*}

\begin{figure*}
\begin{center}
  \begin{minipage}[t]{0.49\textwidth}
    \includegraphics[width=1.00\columnwidth, trim = 20mm 0 10mm 0]{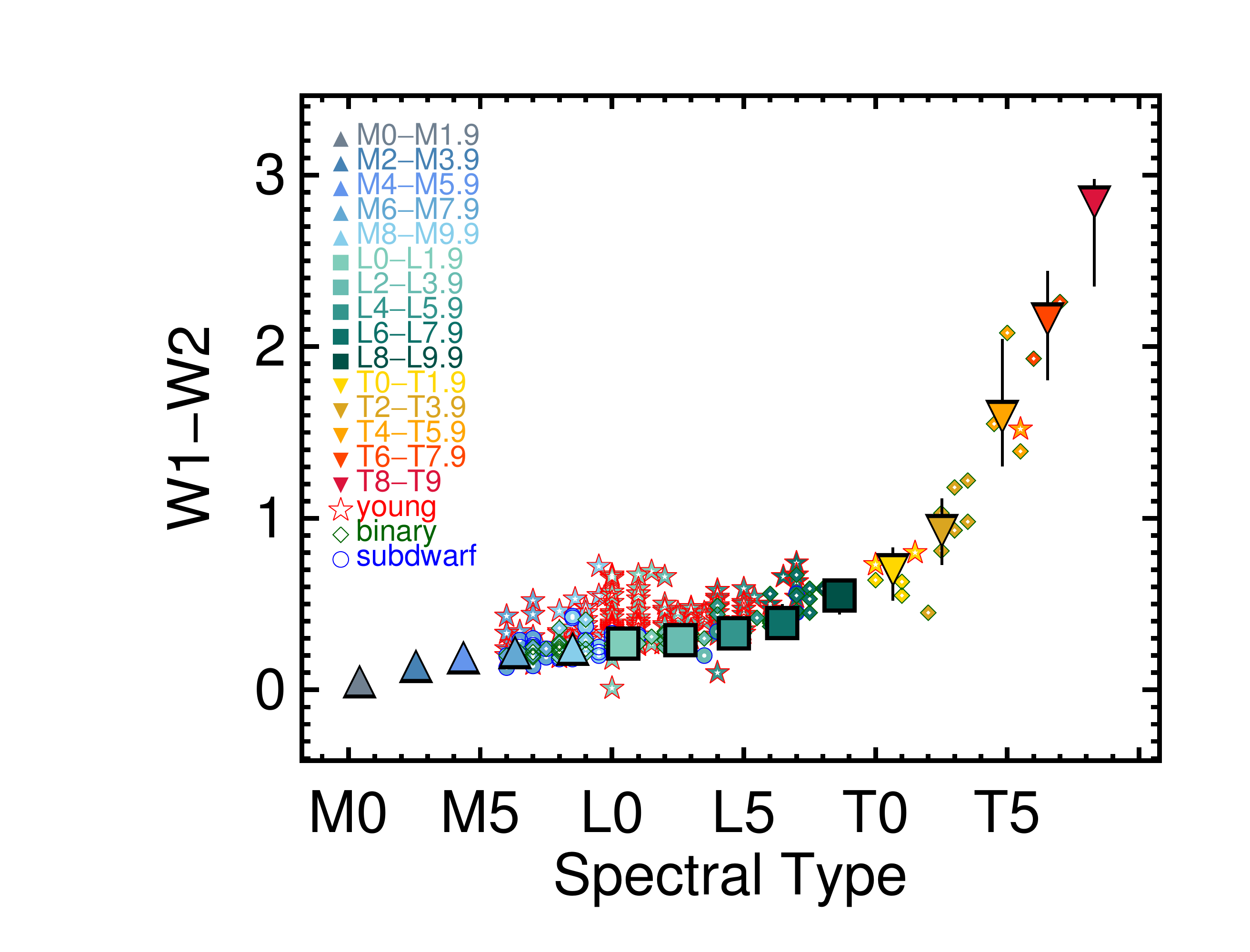}
  \end{minipage}
  \hfill
  \begin{minipage}[t]{0.49\textwidth}
    \includegraphics[width=1.00\columnwidth, trim = 20mm 0 10mm 0]{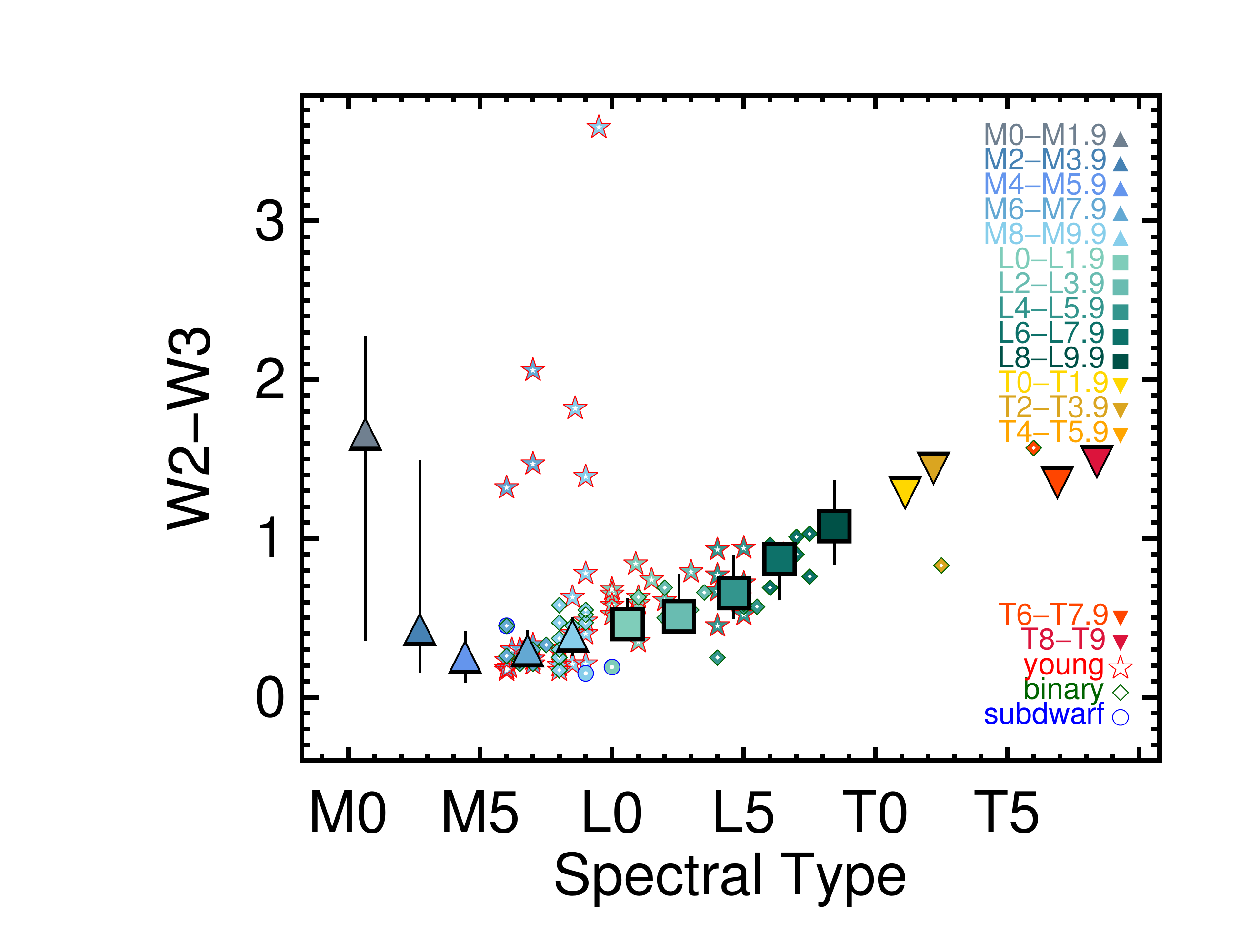}
  \end{minipage}
  \caption{continued.}
  \label{fig.colorspt.unusual}
\end{center}
\end{figure*}

%%% Color vs. Color plots
\begin{figure*}
\begin{center}
  \begin{minipage}[t]{0.49\textwidth}
    \includegraphics[width=1.00\columnwidth, trim = 20mm 0 10mm 0]{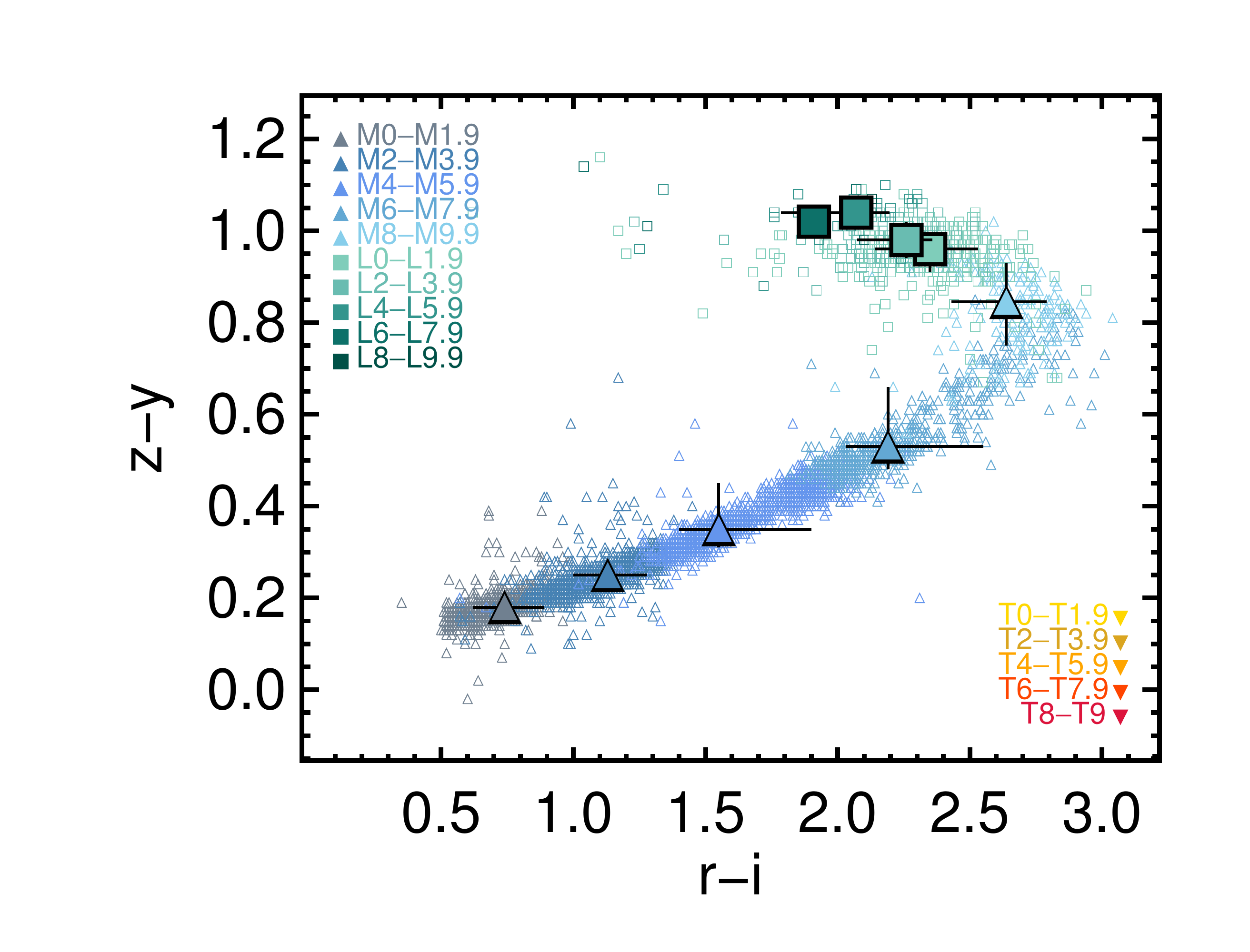}
  \end{minipage}
  \hfill
  \begin{minipage}[t]{0.49\textwidth}
    \includegraphics[width=1.00\columnwidth, trim = 20mm 0 10mm 0]{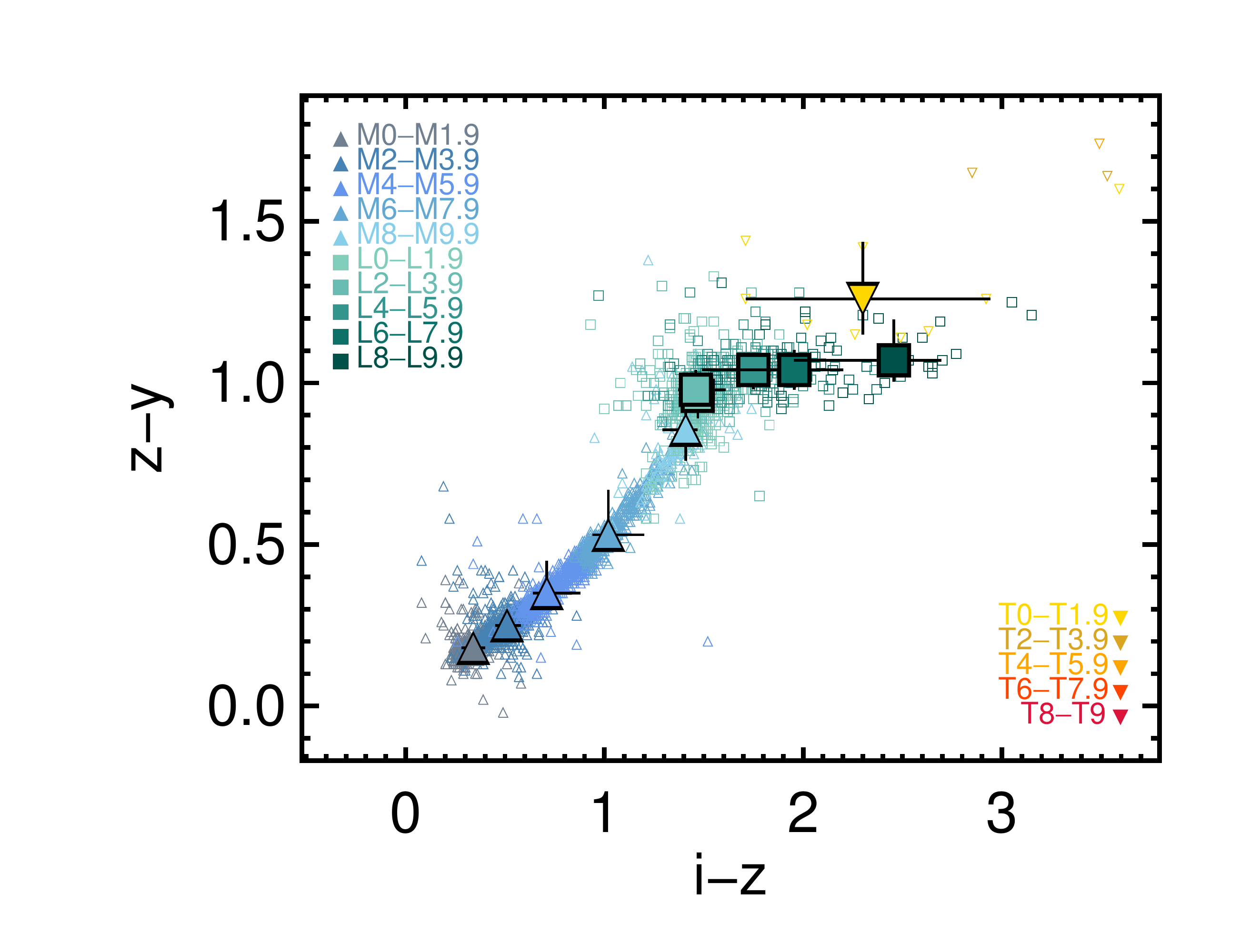}
  \end{minipage}
  \begin{minipage}[t]{0.49\textwidth}
    \includegraphics[width=1.00\columnwidth, trim = 20mm 0 10mm 0]{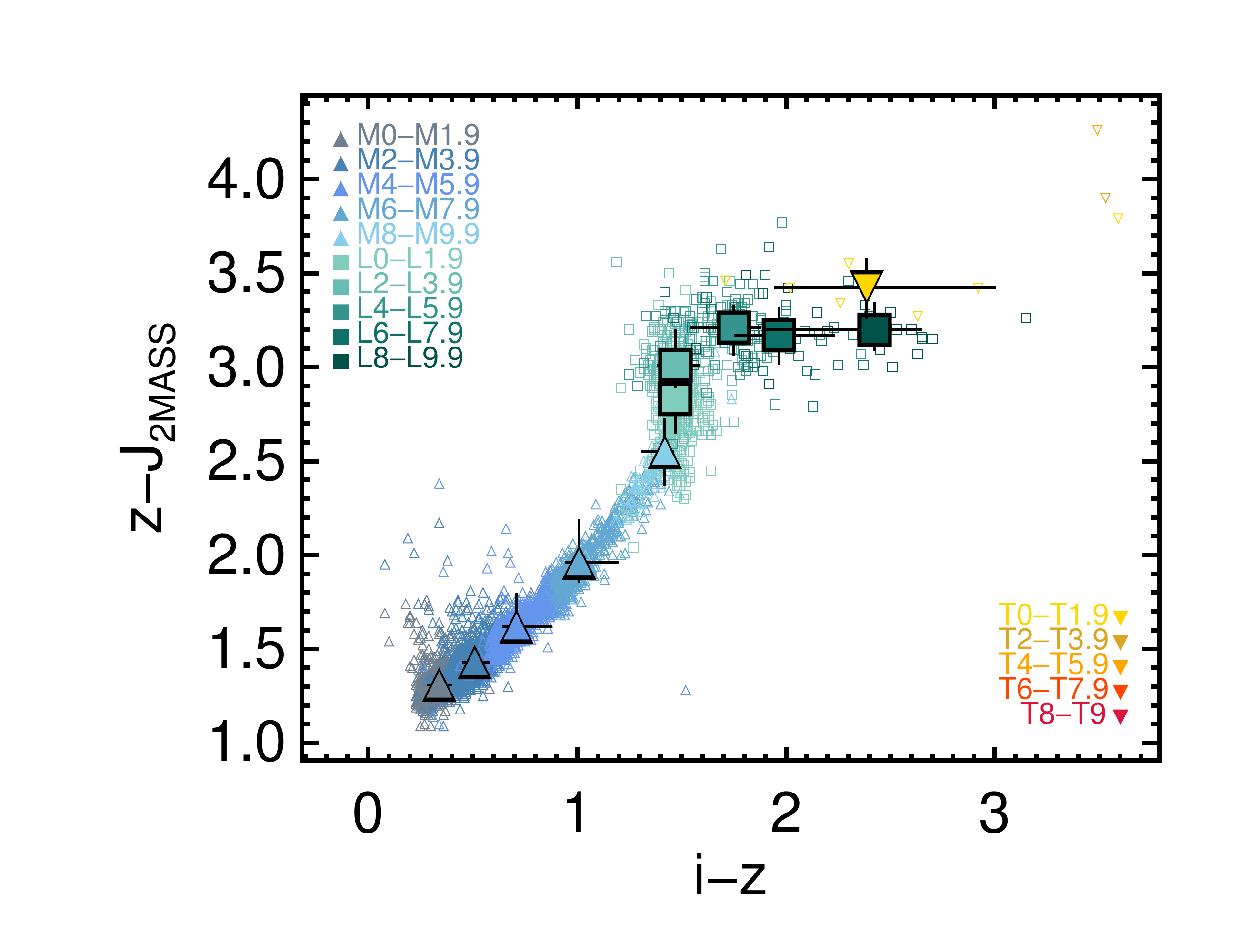}
  \end{minipage}
  \hfill
  \begin{minipage}[t]{0.49\textwidth}
    \includegraphics[width=1.00\columnwidth, trim = 20mm 0 10mm 0]{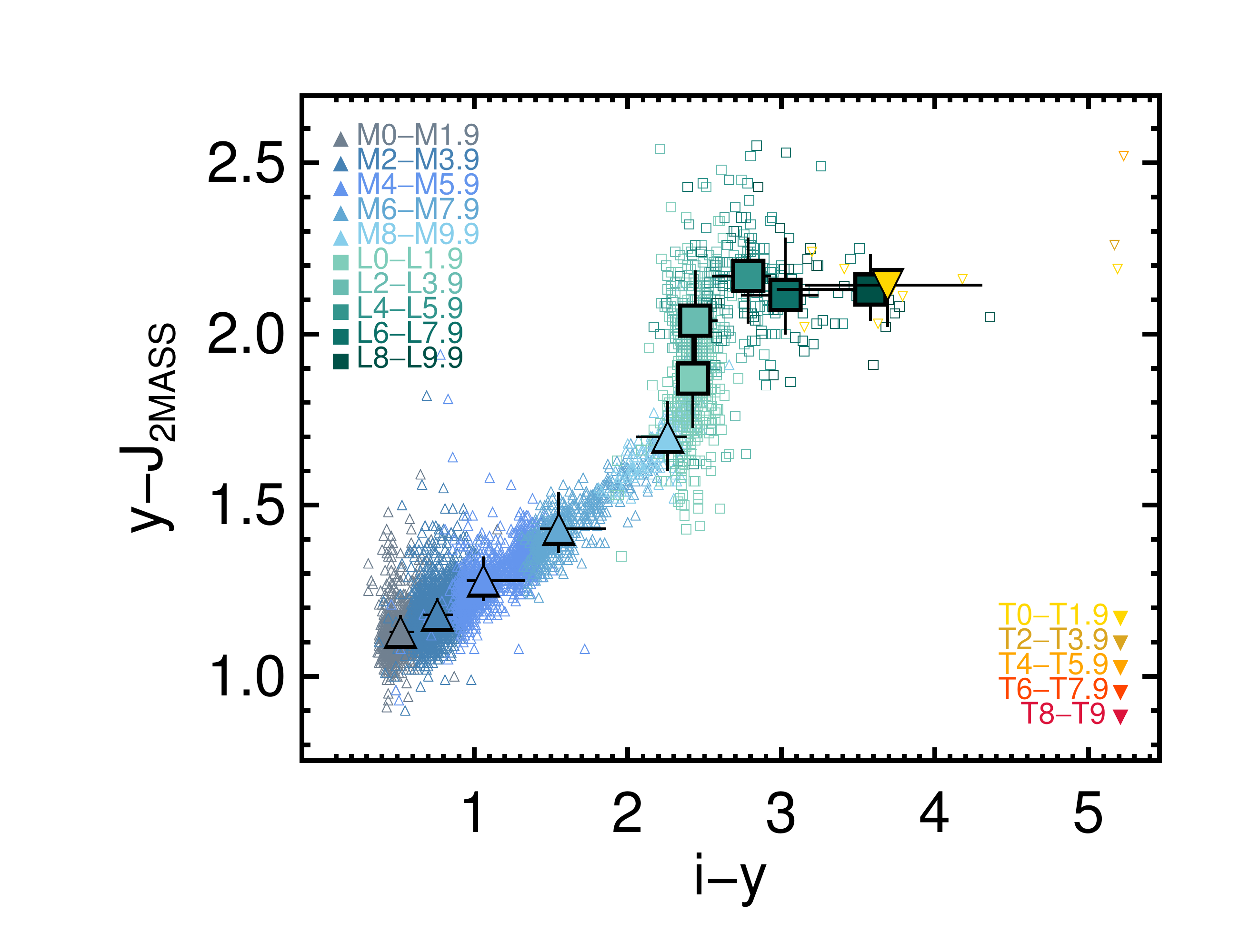}
  \end{minipage}
  \begin{minipage}[t]{0.49\textwidth}
    \includegraphics[width=1.00\columnwidth, trim = 20mm 0 10mm 0]{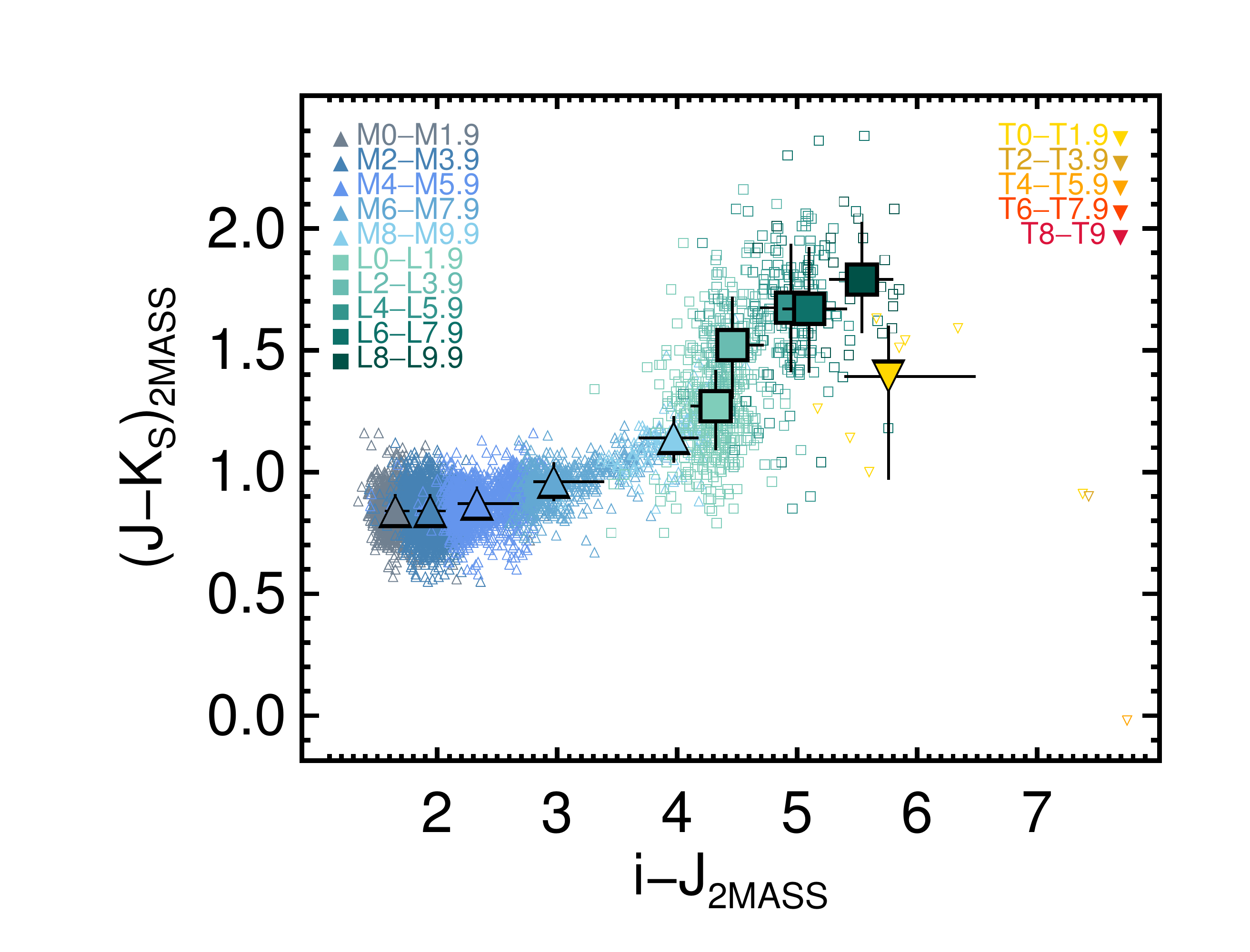}
  \end{minipage}
  \hfill
  \begin{minipage}[t]{0.49\textwidth}
    \includegraphics[width=1.00\columnwidth, trim = 20mm 0 10mm 0]{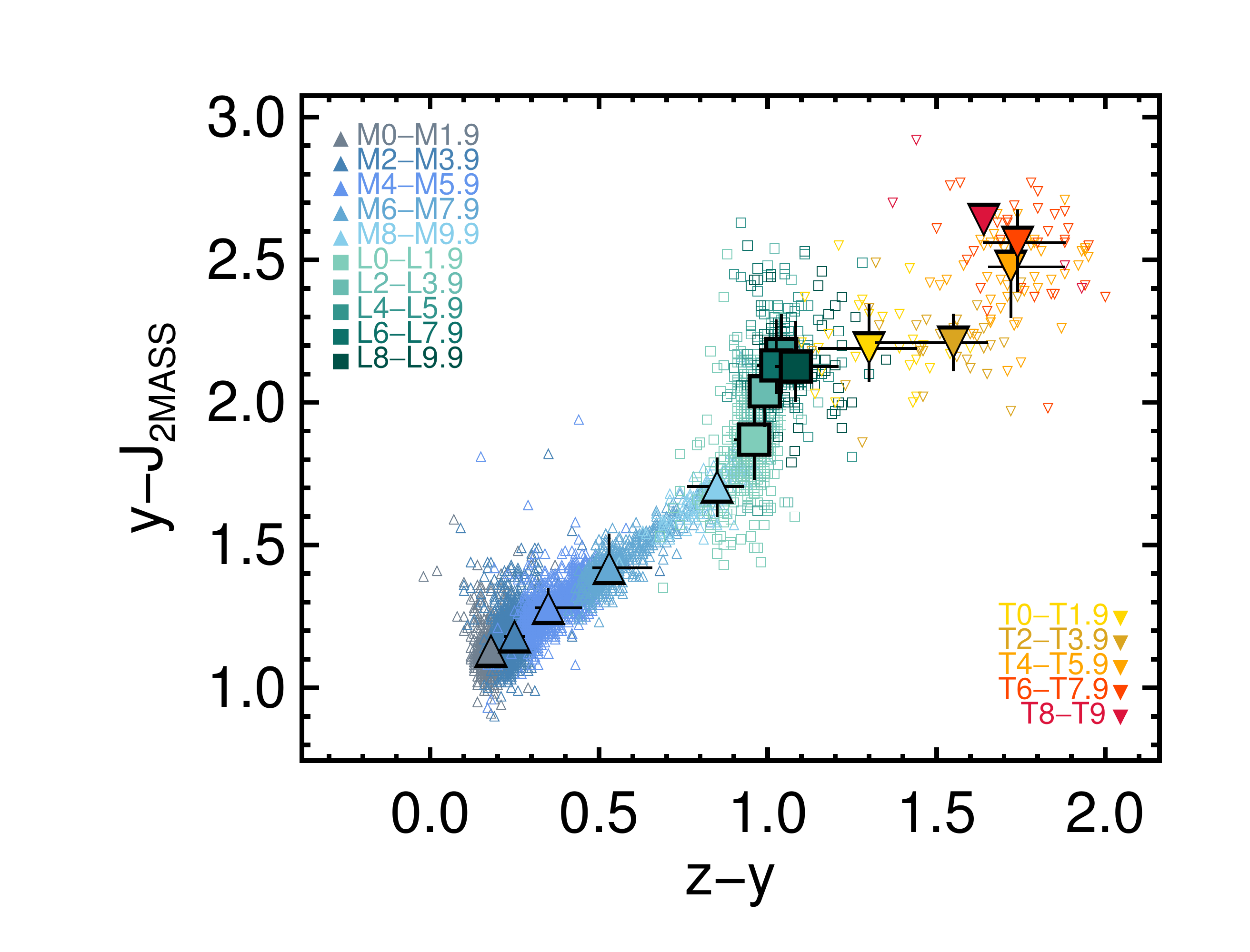}
  \end{minipage}
  \caption{Color-color plots for the M, L, and T dwarfs in our PS1-detected
    catalog, using the same format as in Figure~\ref{fig.colorspt}.  The L dwarf
    color plateau is especially evident for \zy\ and \zjt.}
  \figurenum{fig.colorcolor.1}
\end{center}
\end{figure*}

\begin{figure*}
\begin{center}
  \begin{minipage}[t]{0.49\textwidth}
    \includegraphics[width=1.00\columnwidth, trim = 20mm 0 10mm 0]{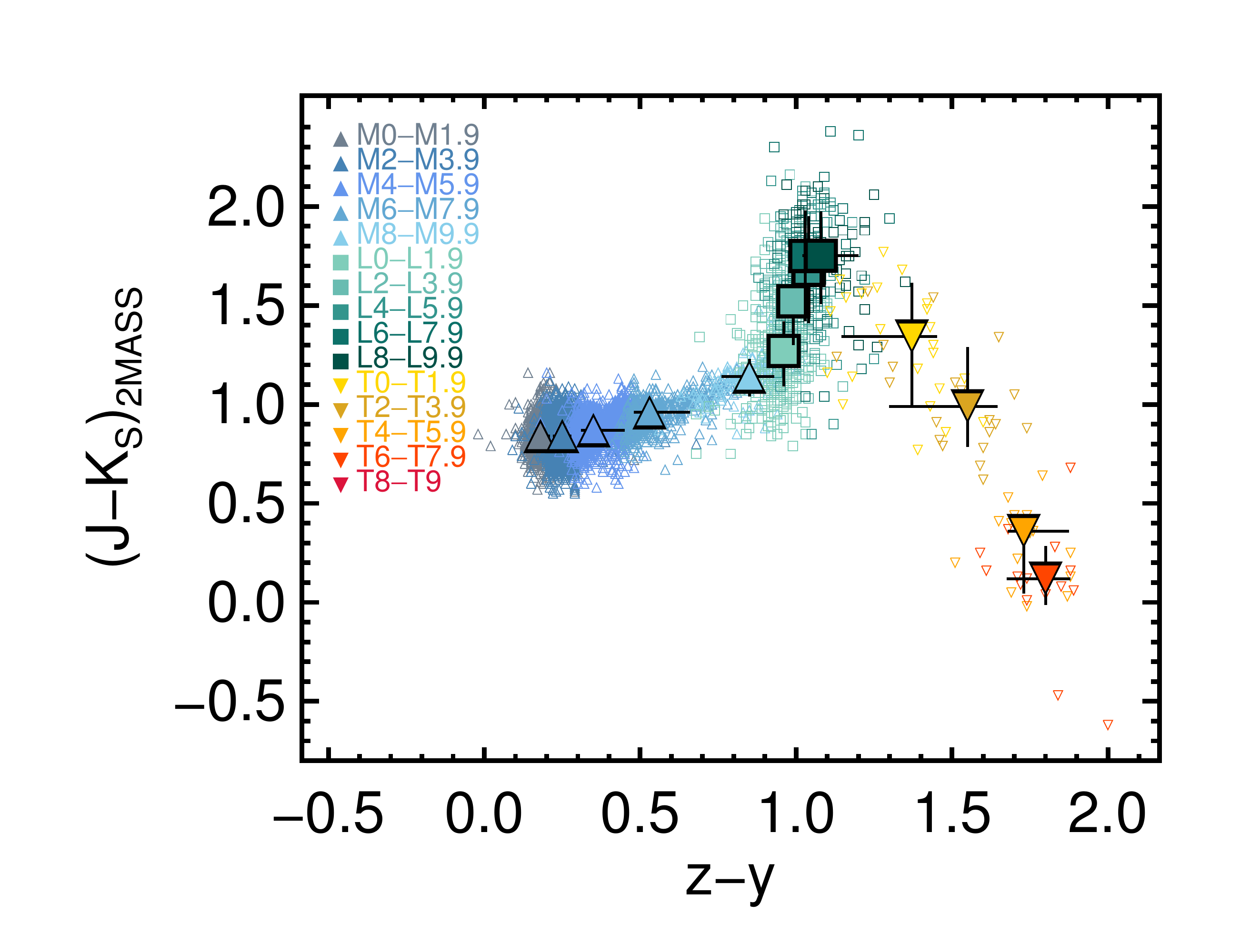}
  \end{minipage}
  \hfill
  \begin{minipage}[t]{0.49\textwidth}
    \includegraphics[width=1.00\columnwidth, trim = 20mm 0 10mm 0]{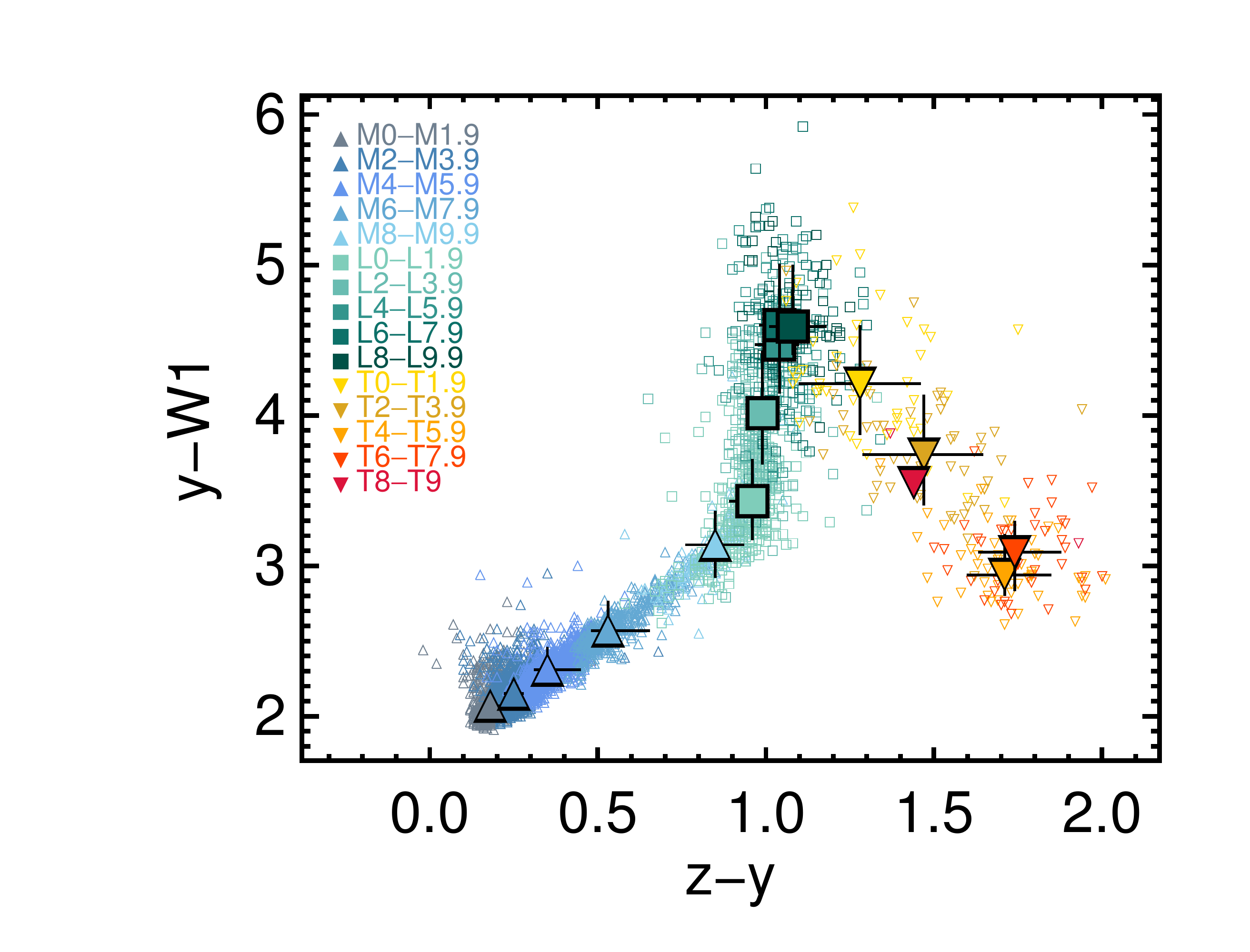}
  \end{minipage}
  \begin{minipage}[t]{0.49\textwidth}
    \includegraphics[width=1.00\columnwidth, trim = 20mm 0 10mm 0]{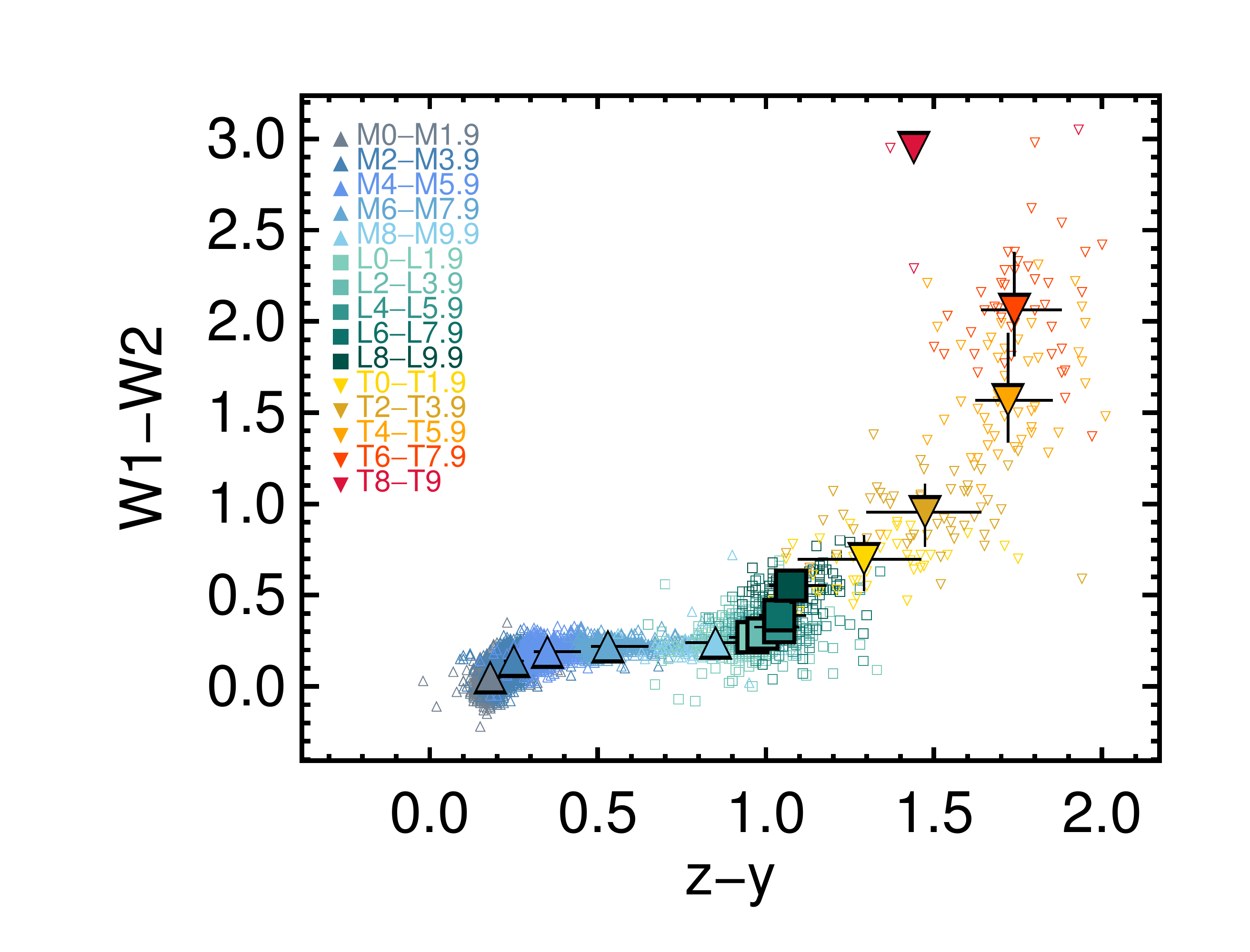}
  \end{minipage}
  \hfill
  \begin{minipage}[t]{0.49\textwidth}
    \includegraphics[width=1.00\columnwidth, trim = 20mm 0 10mm 0]{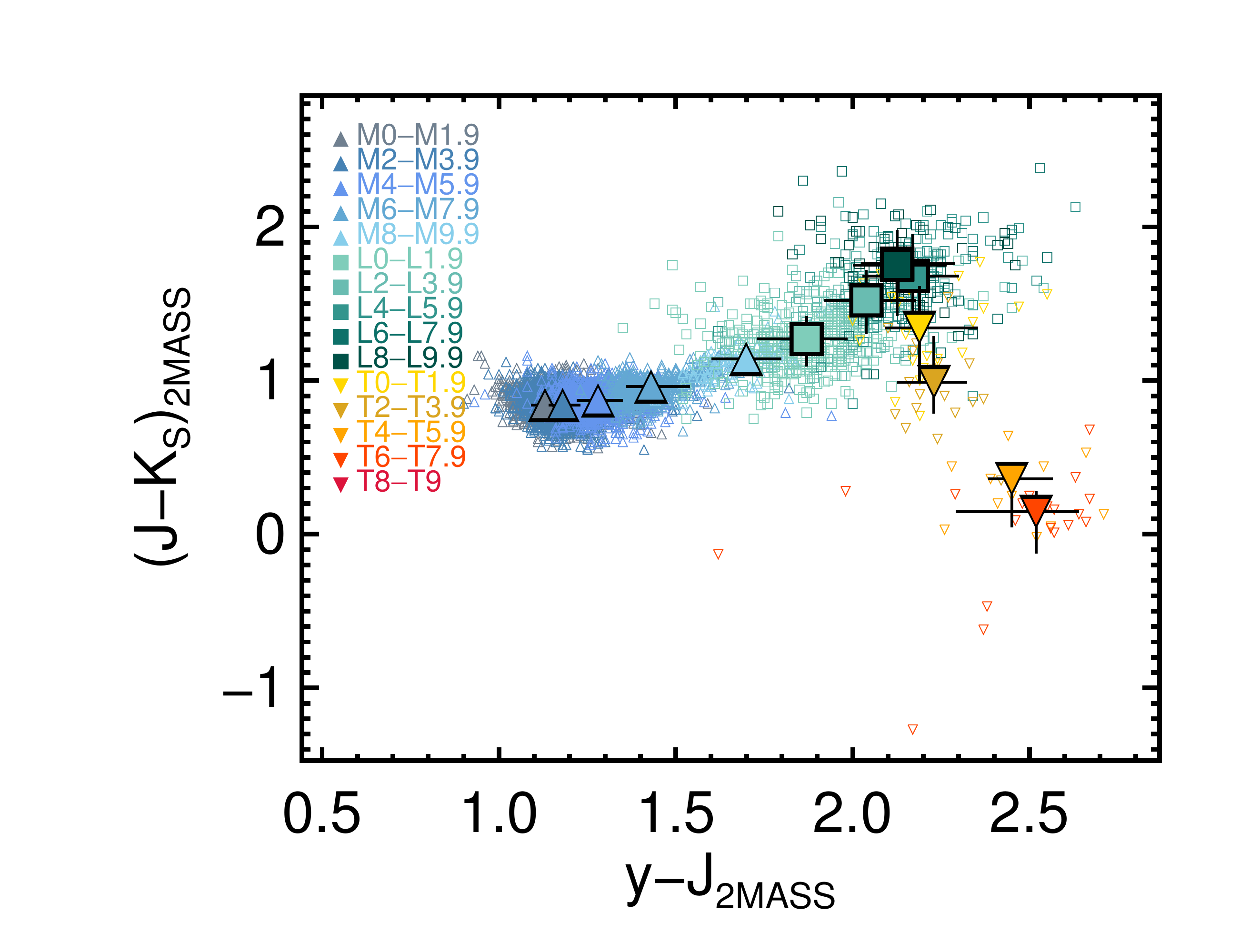}
  \end{minipage}
  \begin{minipage}[t]{0.49\textwidth}
    \includegraphics[width=1.00\columnwidth, trim = 20mm 0 10mm 0]{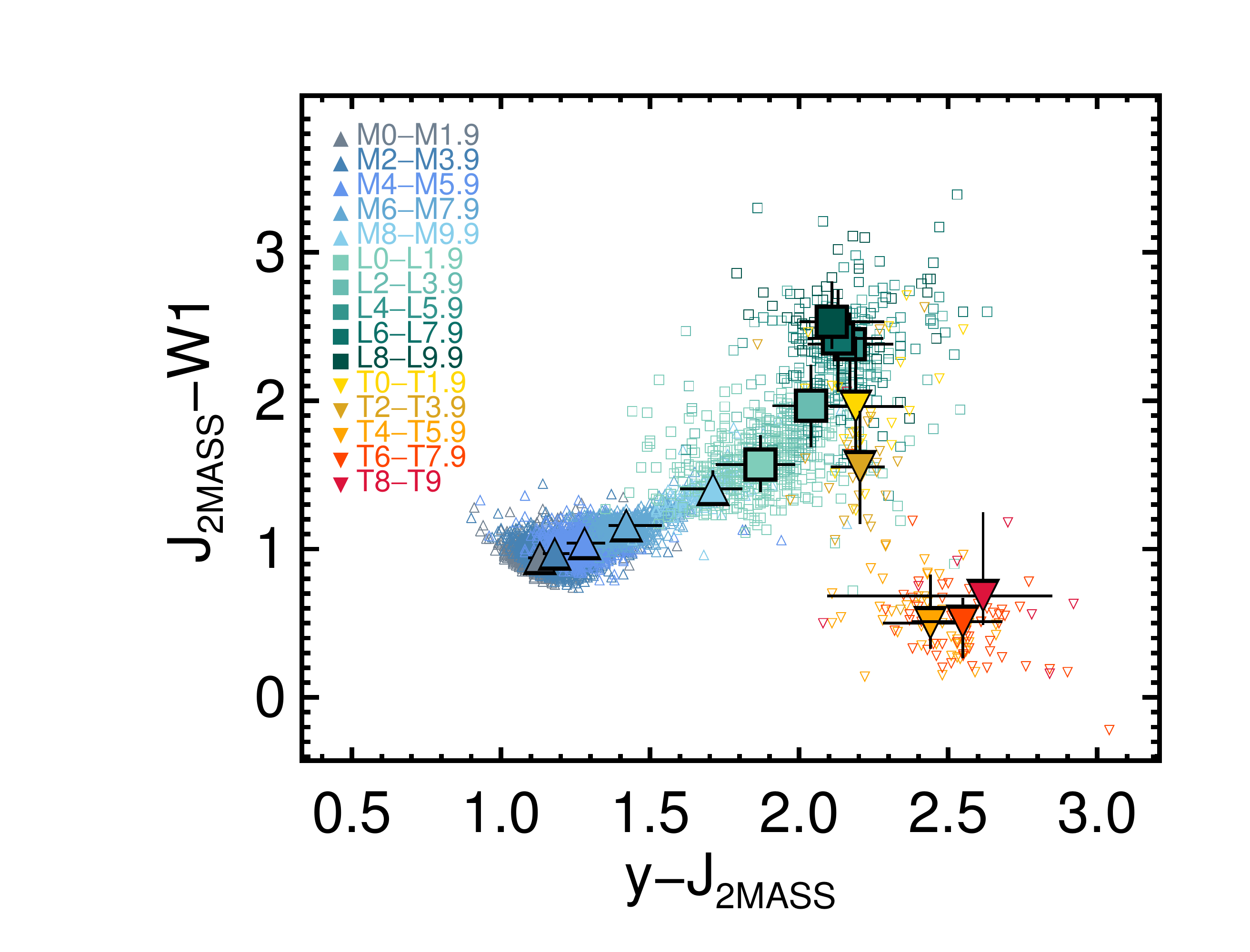}
  \end{minipage}
  \hfill
  \begin{minipage}[t]{0.49\textwidth}
    \includegraphics[width=1.00\columnwidth, trim = 20mm 0 10mm 0]{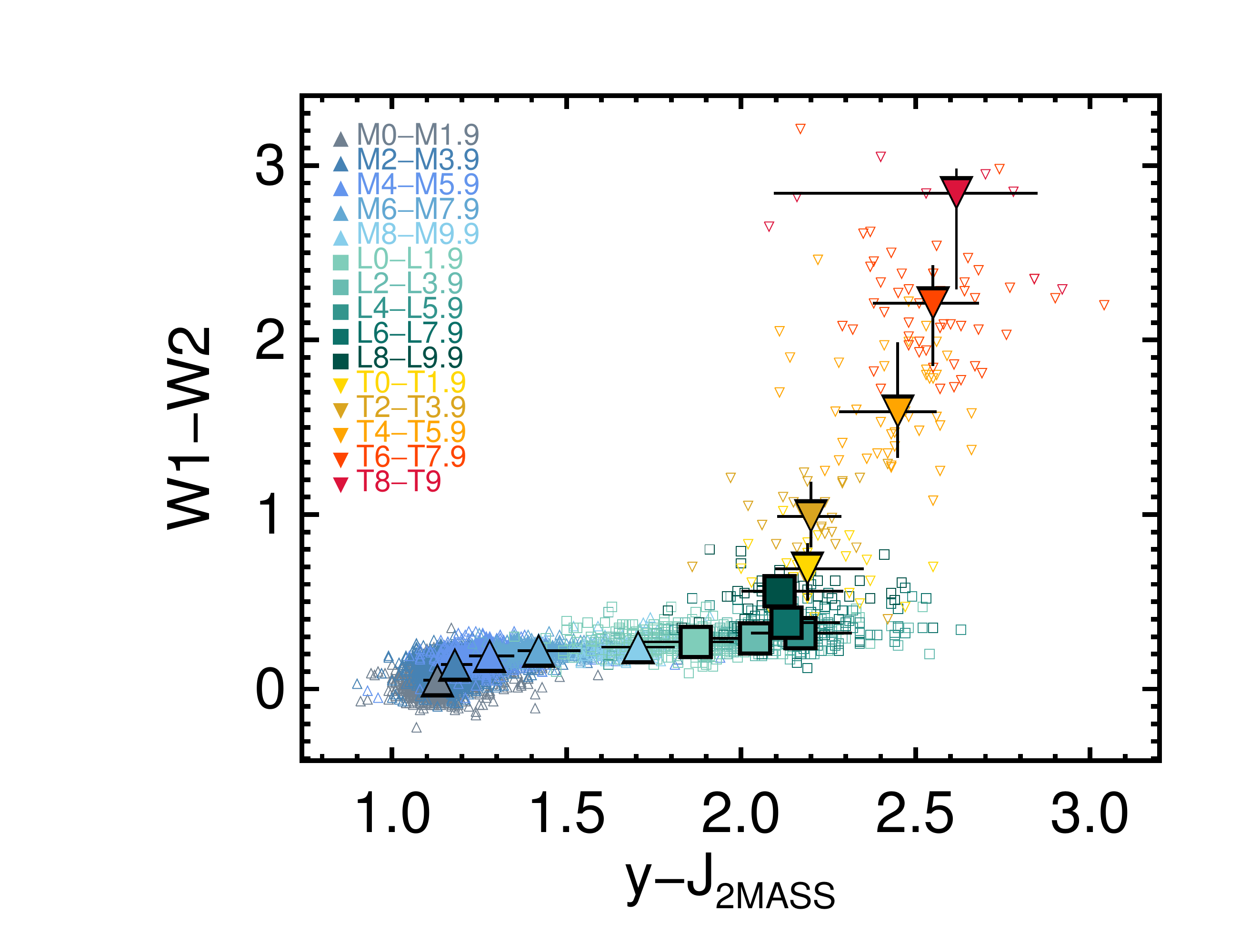}
  \end{minipage}
  \caption{continued.}
  \figurenum{fig.colorcolor.2}
\end{center}
\end{figure*}

\begin{figure*}
\begin{center}
  \begin{minipage}[t]{0.49\textwidth}
    \includegraphics[width=1.00\columnwidth, trim = 20mm 0 10mm 0]{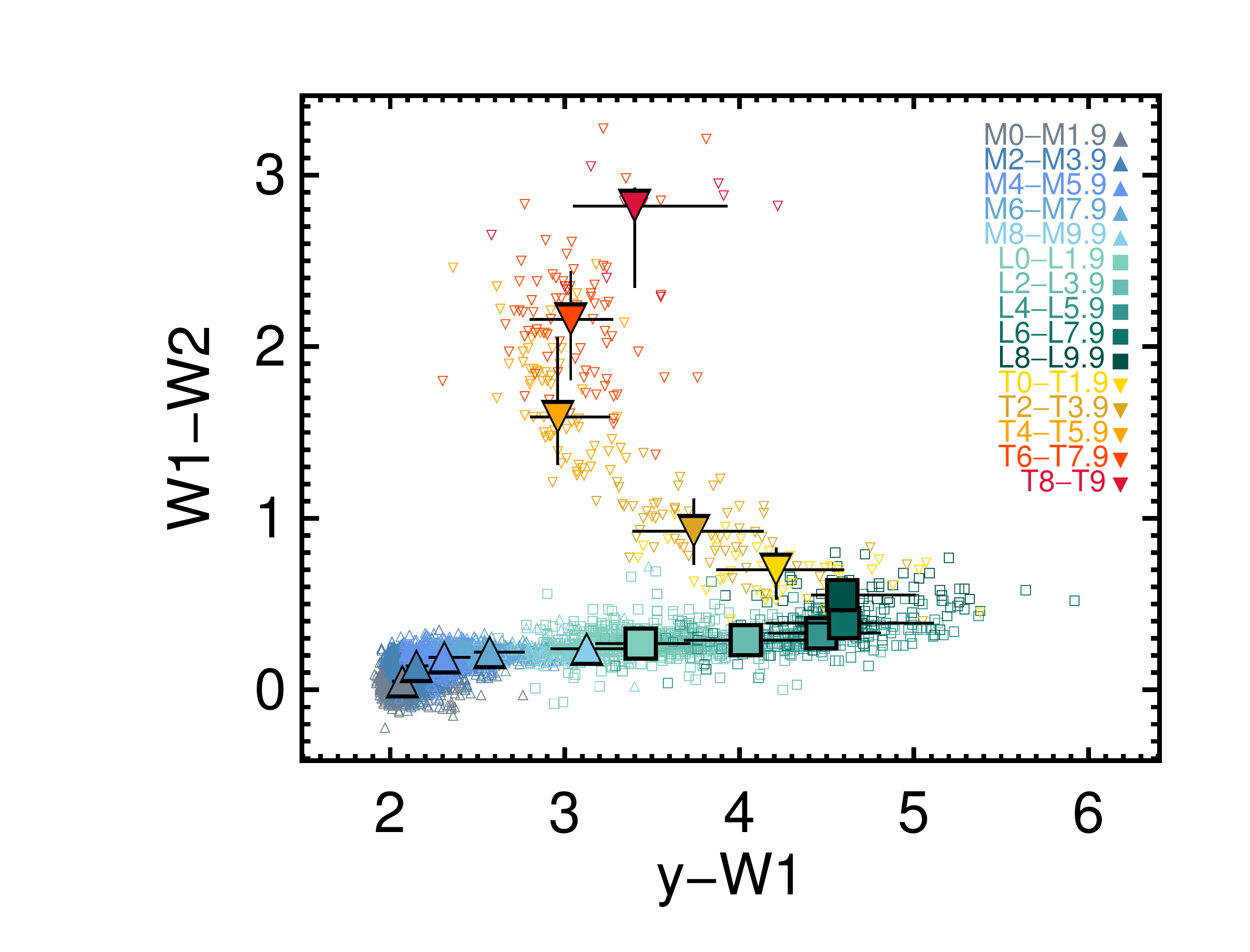}
  \end{minipage}
  \caption{continued.}
  \label{fig.colorcolor}
\end{center}
\end{figure*}

%%% Color vs. Color plots for unusual objects
\begin{figure*}
\begin{center}
  \begin{minipage}[t]{0.49\textwidth}
    \includegraphics[width=1.00\columnwidth, trim = 20mm 0 10mm 0]{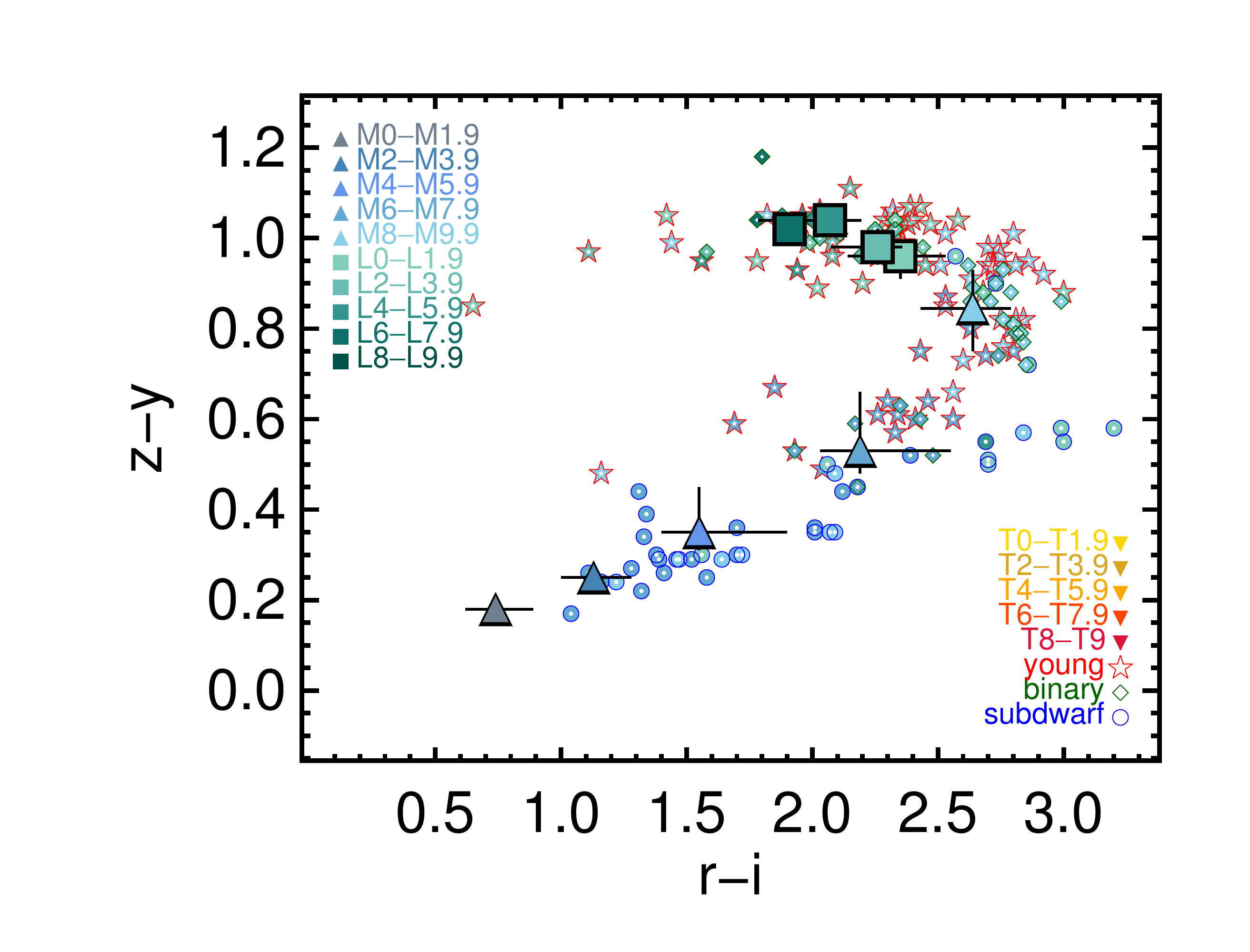}
  \end{minipage}
  \hfill
  \begin{minipage}[t]{0.49\textwidth}
    \includegraphics[width=1.00\columnwidth, trim = 20mm 0 10mm 0]{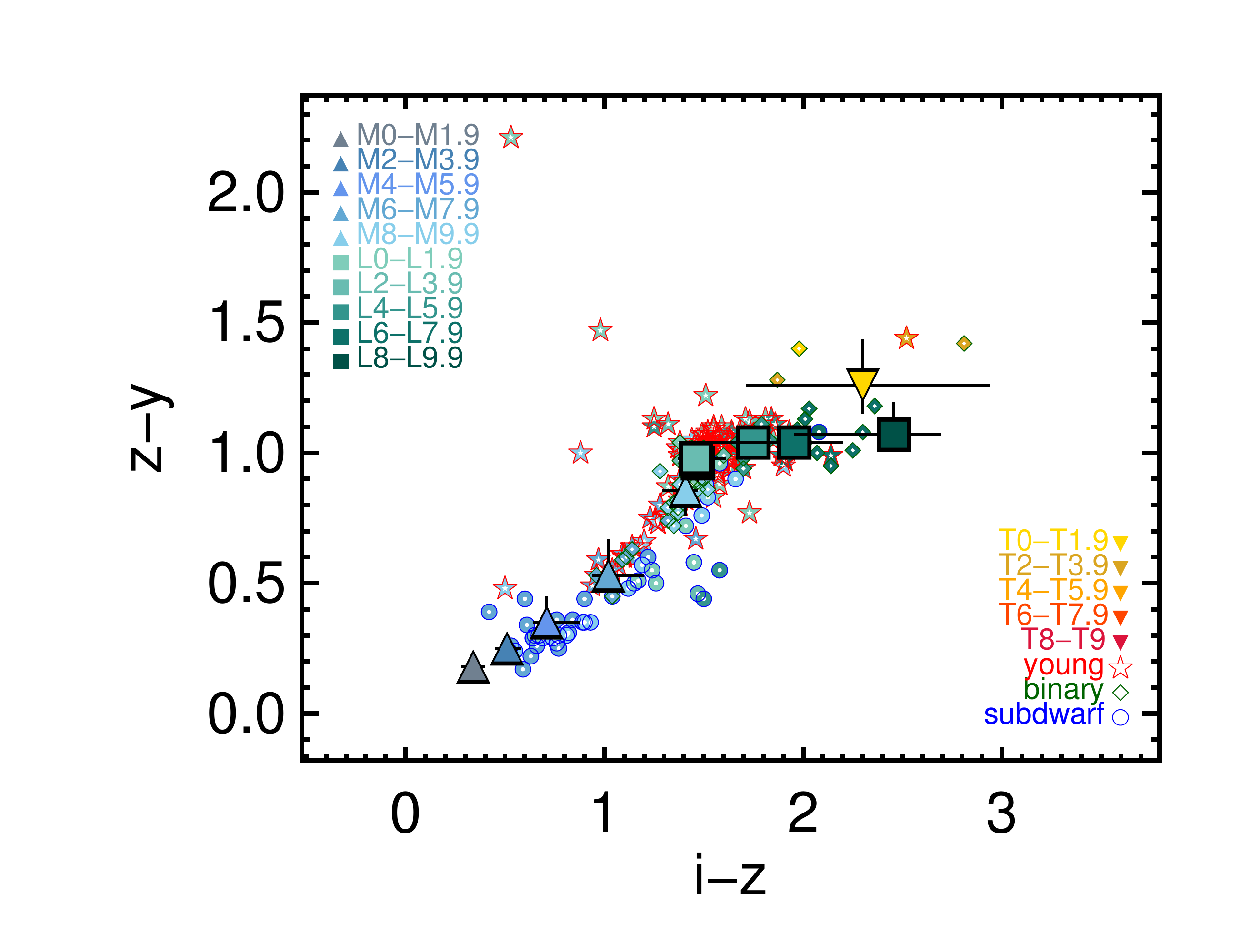}
  \end{minipage}
  \begin{minipage}[t]{0.49\textwidth}
    \includegraphics[width=1.00\columnwidth, trim = 20mm 0 10mm 0]{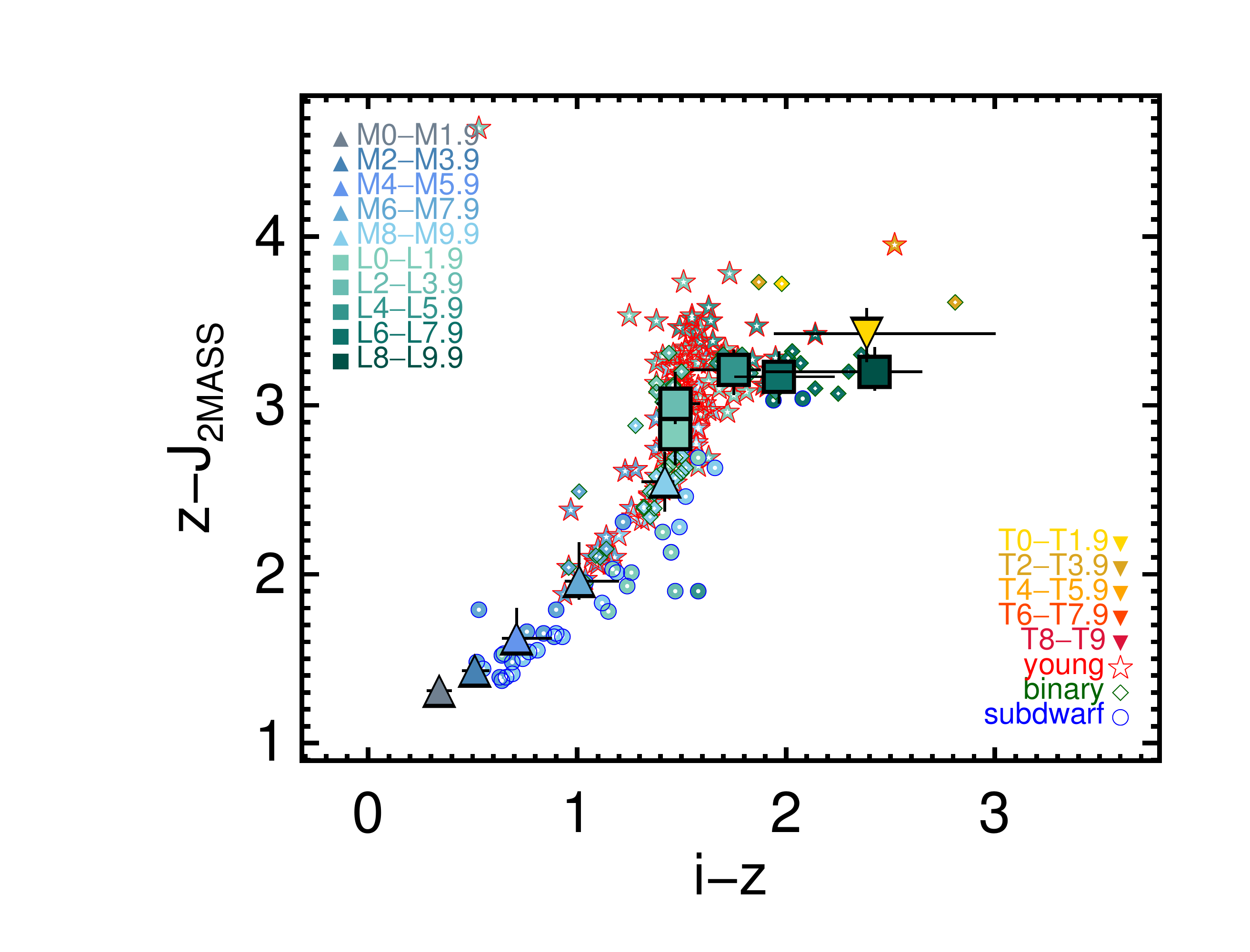}
  \end{minipage}
  \hfill
  \begin{minipage}[t]{0.49\textwidth}
    \includegraphics[width=1.00\columnwidth, trim = 20mm 0 10mm 0]{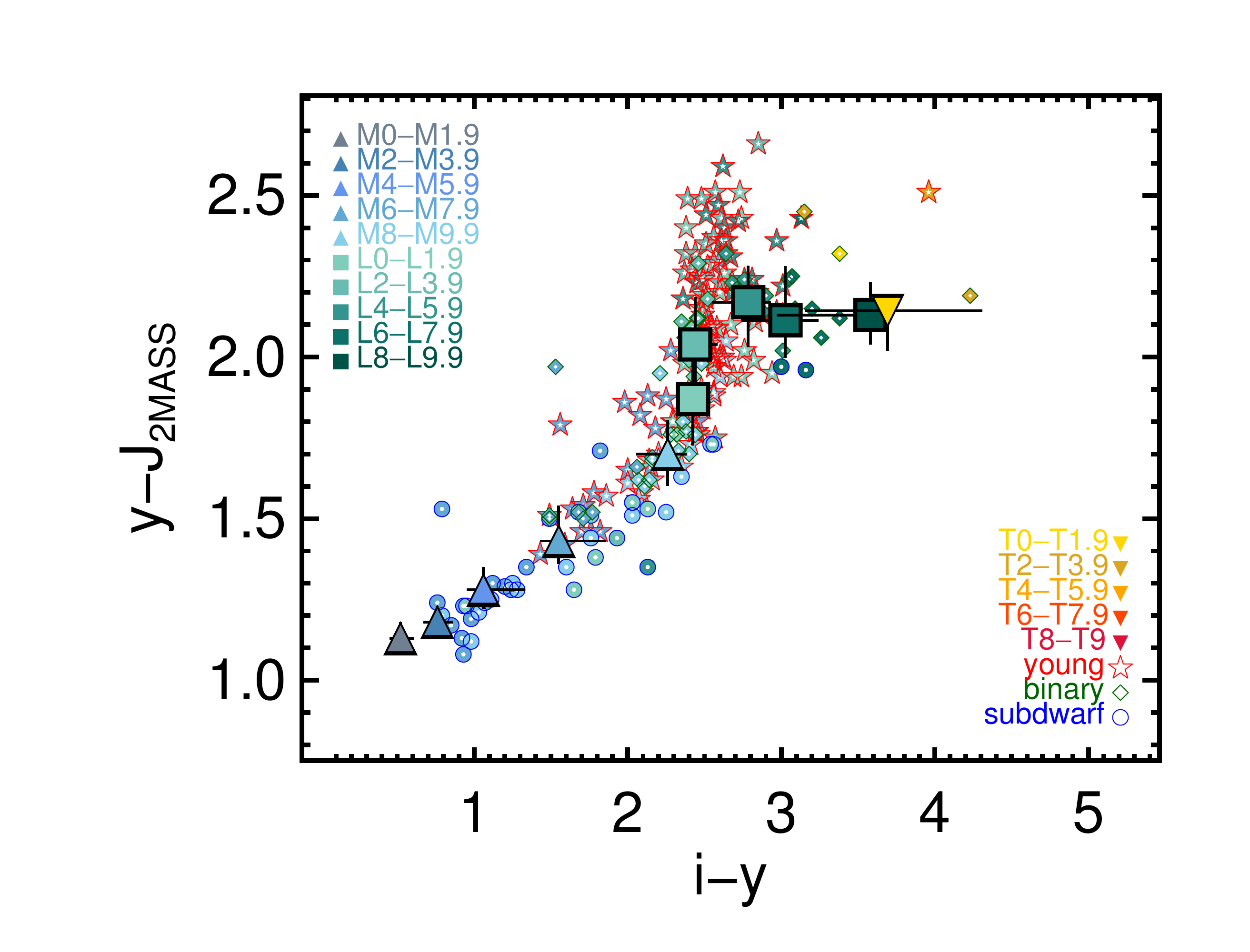}
  \end{minipage}
  \begin{minipage}[t]{0.49\textwidth}
    \includegraphics[width=1.00\columnwidth, trim = 20mm 0 10mm 0]{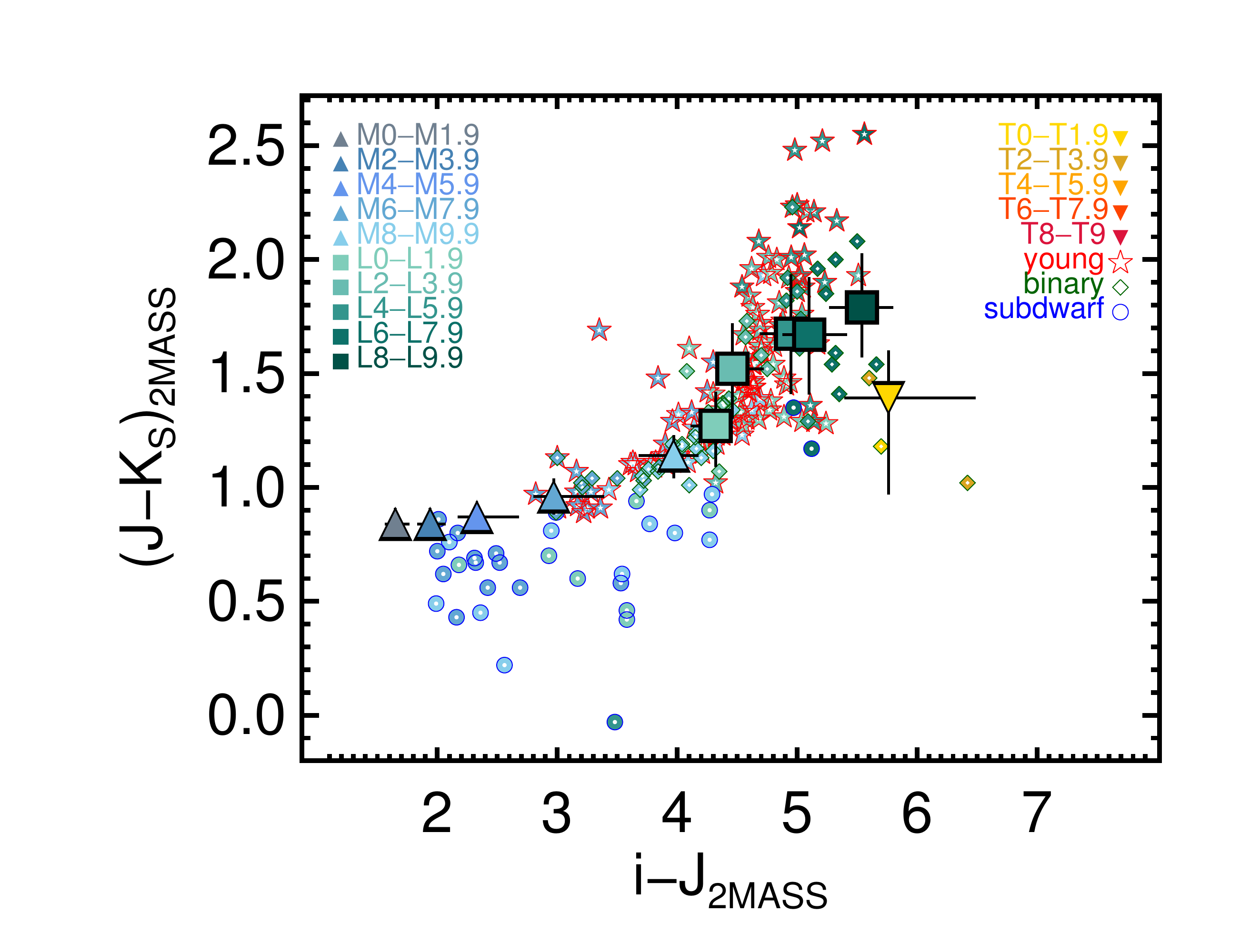}
  \end{minipage}
  \hfill
  \begin{minipage}[t]{0.49\textwidth}
    \includegraphics[width=1.00\columnwidth, trim = 20mm 0 10mm 0]{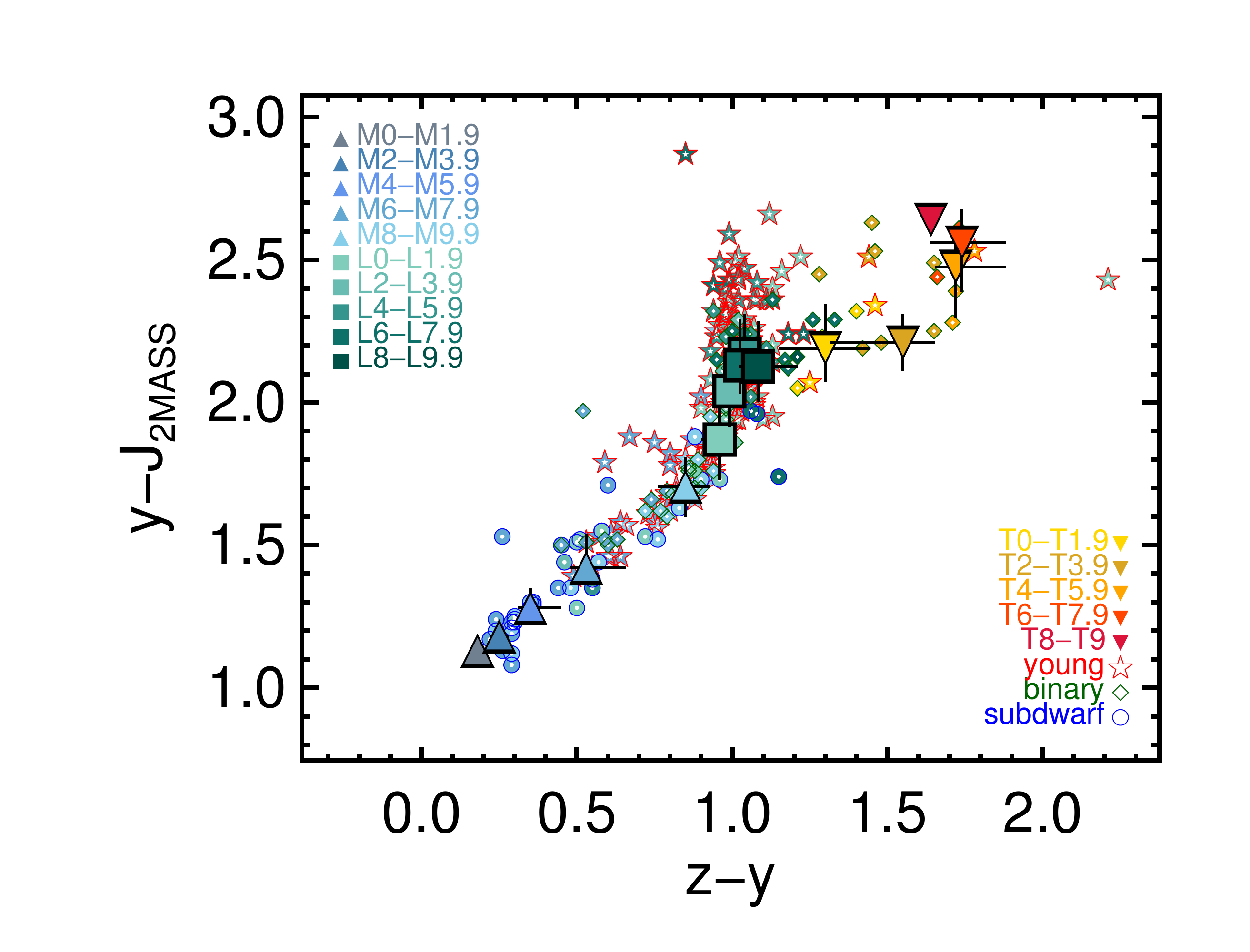}
  \end{minipage}
  \caption{Color-color plots for known young objects, binaries, and subdwarfs in
    our PS1-detected catalog, using the same format as in
    Figure~\ref{fig.colorspt.unusual}.  Median colors and 68\% confidence limits
    for normal field objects from Figure~\ref{fig.colorcolor} are overplotted
    for reference.}
  \figurenum{fig.colorcolor.unusual.1}
\end{center}
\end{figure*}

\begin{figure*}
\begin{center}
  \begin{minipage}[t]{0.49\textwidth}
    \includegraphics[width=1.00\columnwidth, trim = 20mm 0 10mm 0]{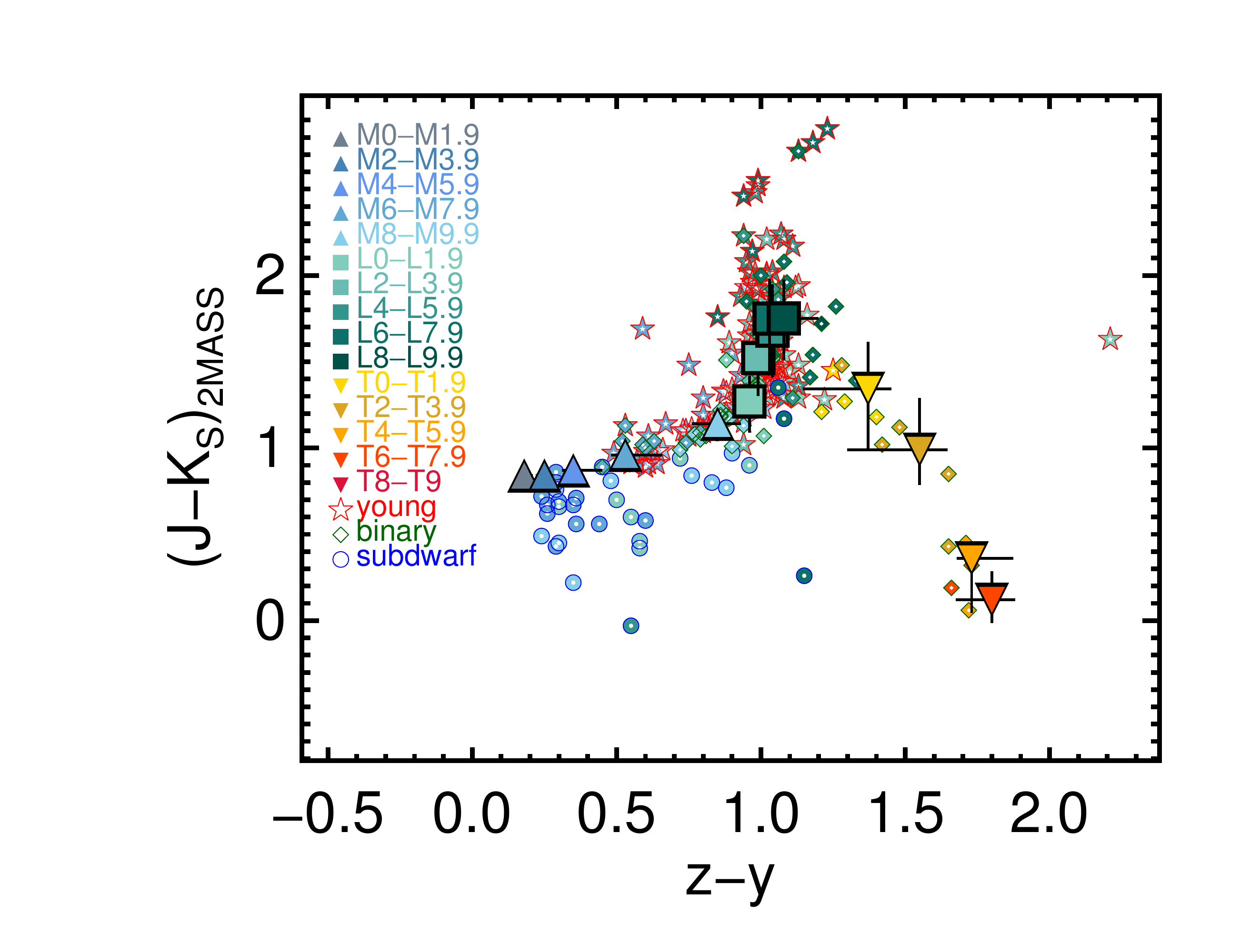}
  \end{minipage}
  \hfill
  \begin{minipage}[t]{0.49\textwidth}
    \includegraphics[width=1.00\columnwidth, trim = 20mm 0 10mm 0]{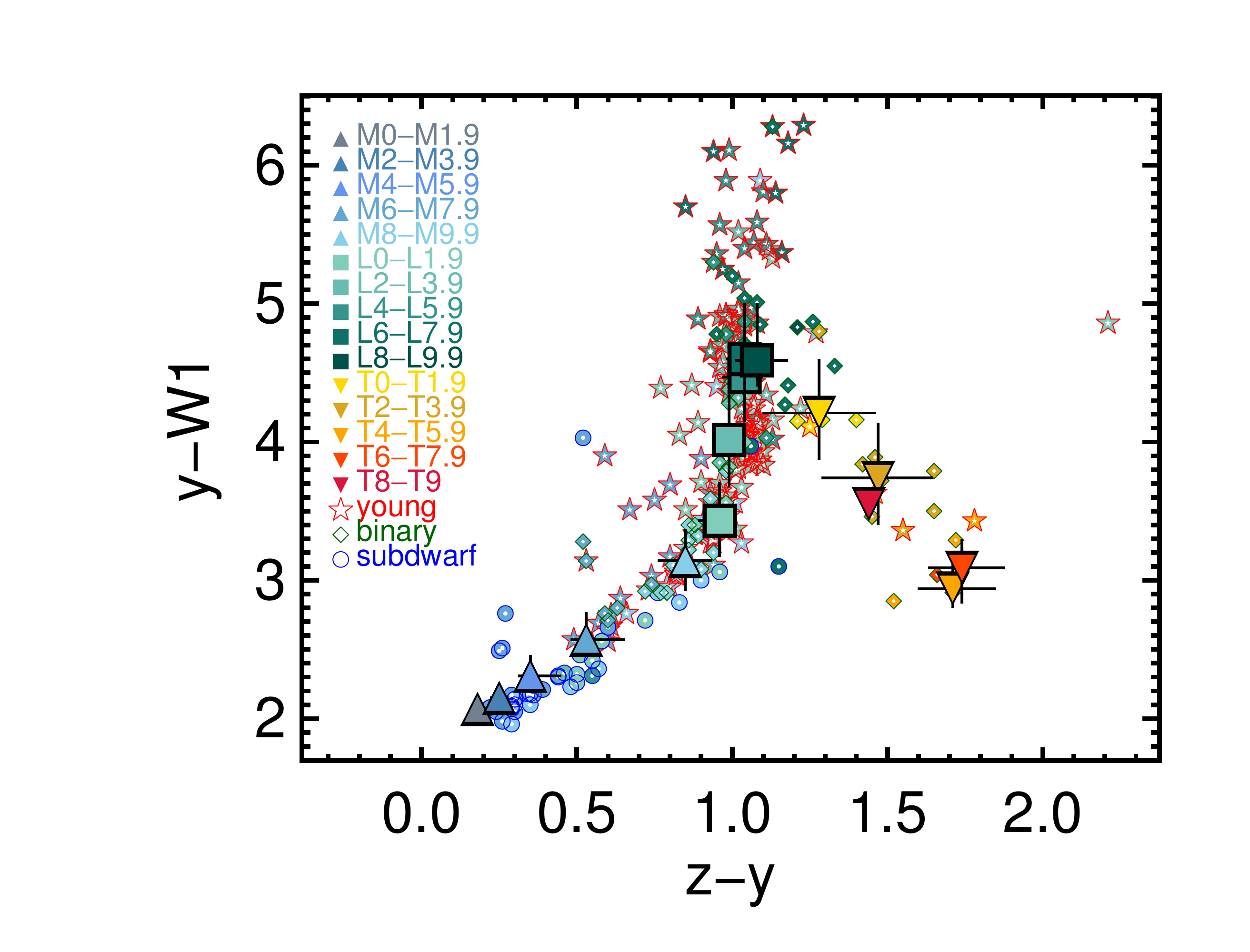}
  \end{minipage}
  \begin{minipage}[t]{0.49\textwidth}
    \includegraphics[width=1.00\columnwidth, trim = 20mm 0 10mm 0]{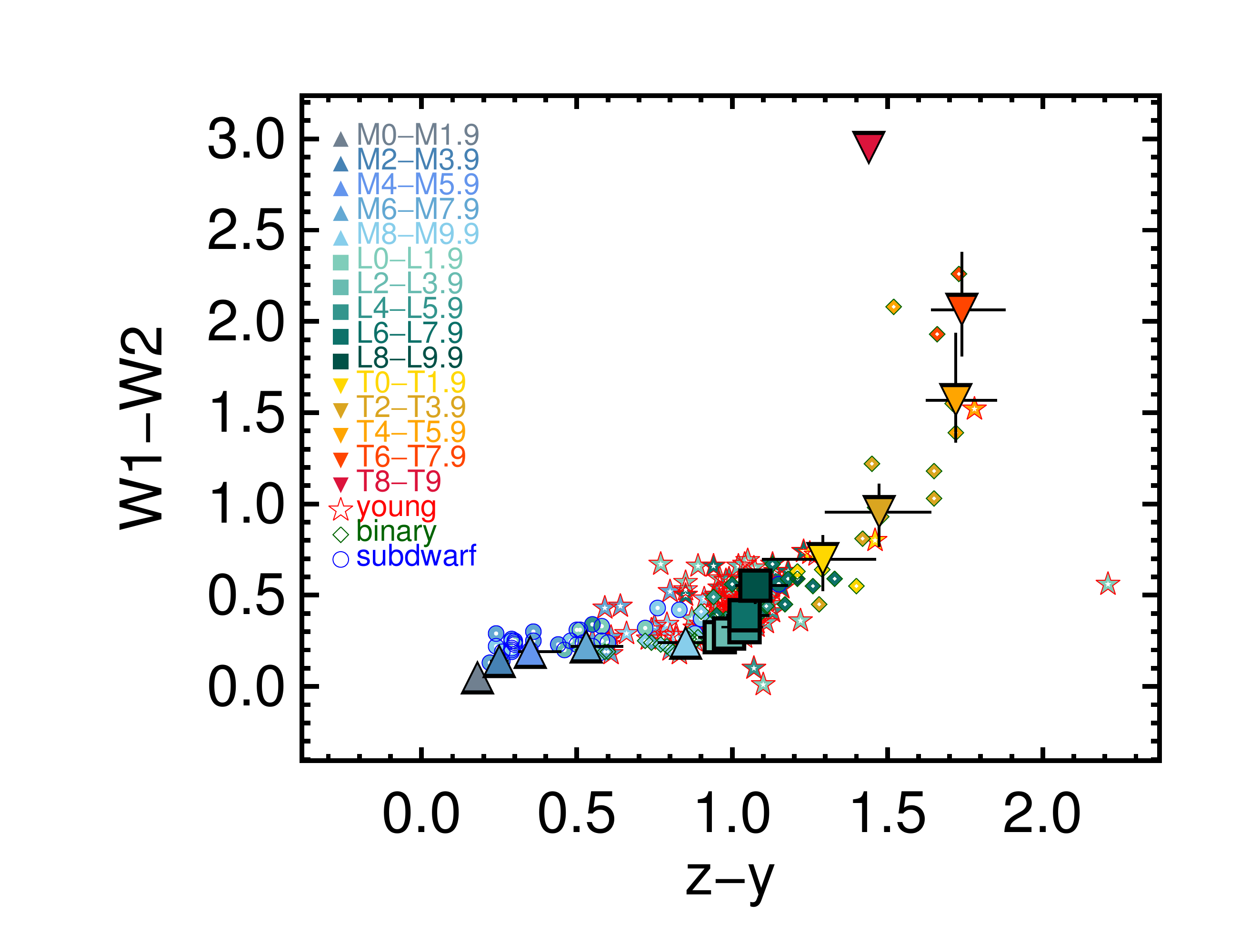}
  \end{minipage}
  \hfill
  \begin{minipage}[t]{0.49\textwidth}
    \includegraphics[width=1.00\columnwidth, trim = 20mm 0 10mm 0]{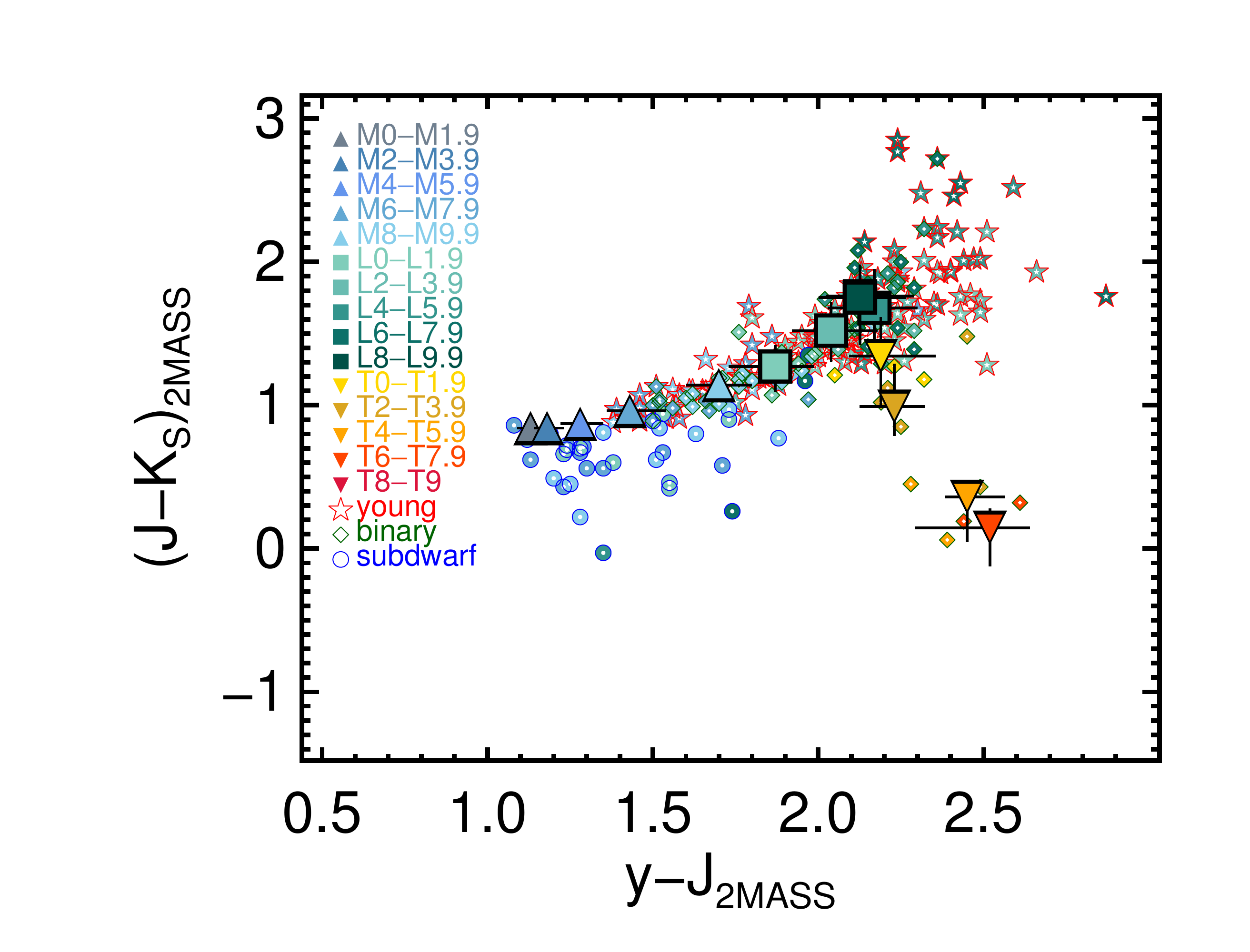}
  \end{minipage}
  \begin{minipage}[t]{0.49\textwidth}
    \includegraphics[width=1.00\columnwidth, trim = 20mm 0 10mm 0]{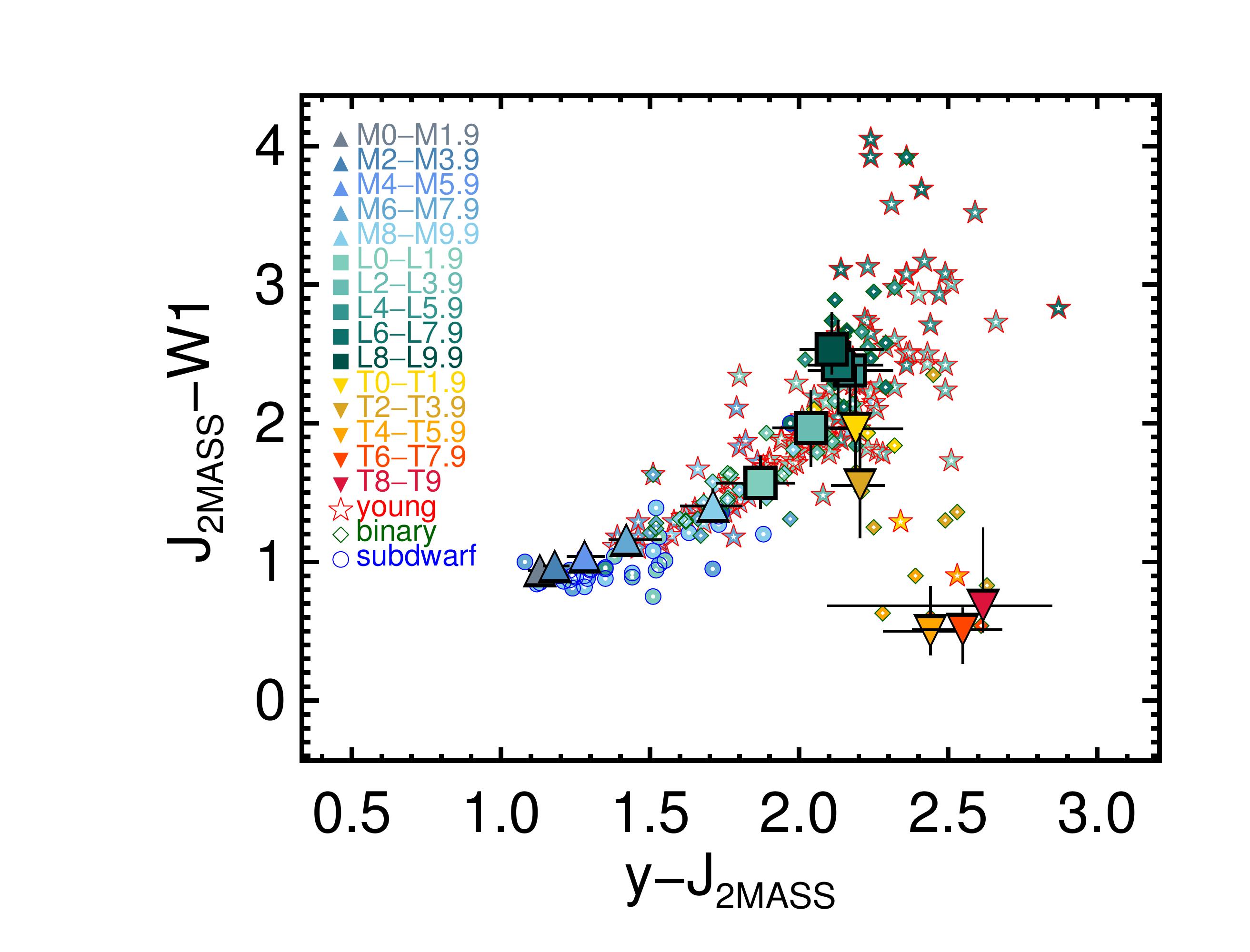}
  \end{minipage}
  \hfill
  \begin{minipage}[t]{0.49\textwidth}
    \includegraphics[width=1.00\columnwidth, trim = 20mm 0 10mm 0]{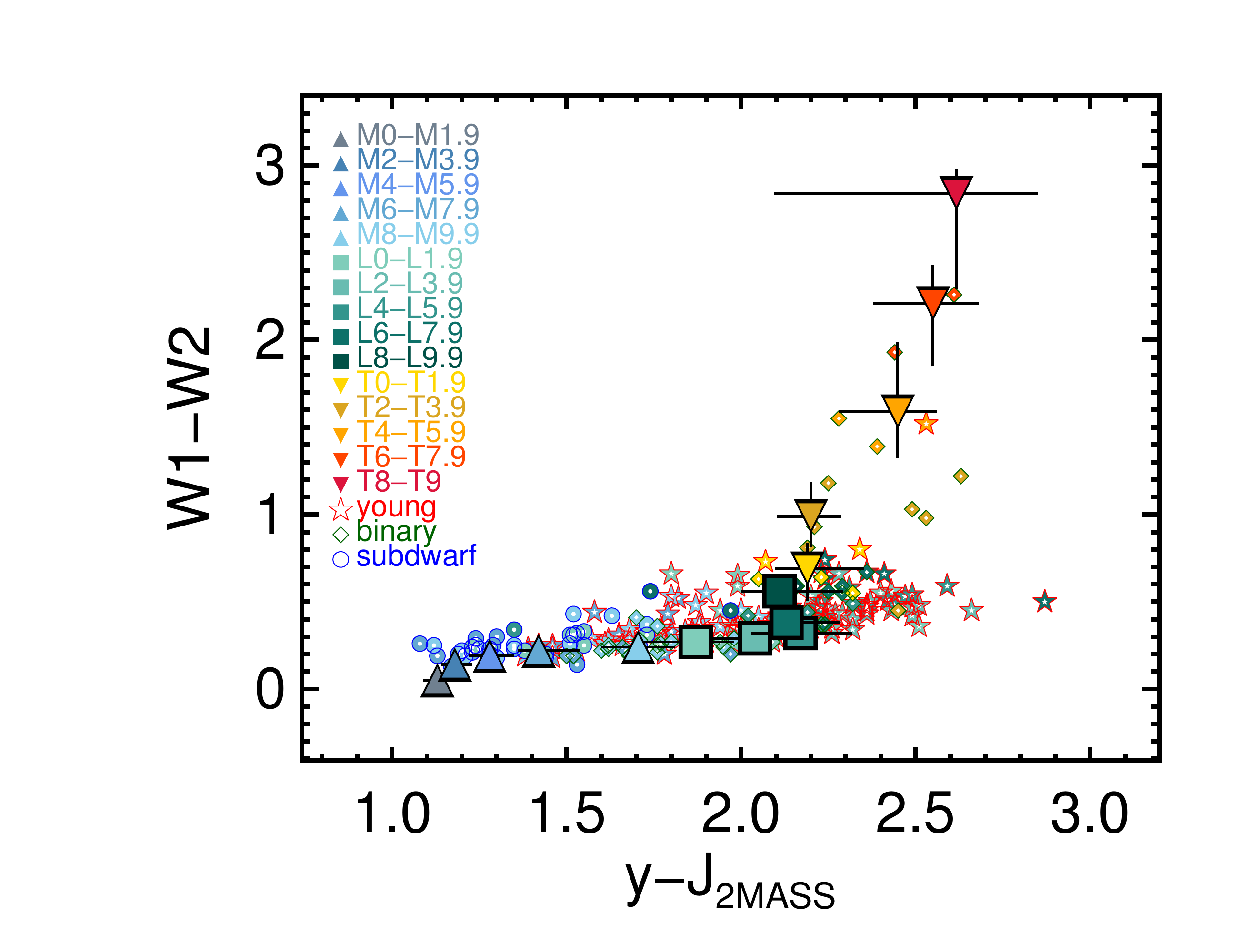}
  \end{minipage}
  \caption{continued.}
  \figurenum{fig.colorcolor.unusual.2}
\end{center}
\end{figure*}

\begin{figure*}
\begin{center}
  \begin{minipage}[t]{0.49\textwidth}
    \includegraphics[width=1.00\columnwidth, trim = 20mm 0 10mm 0]{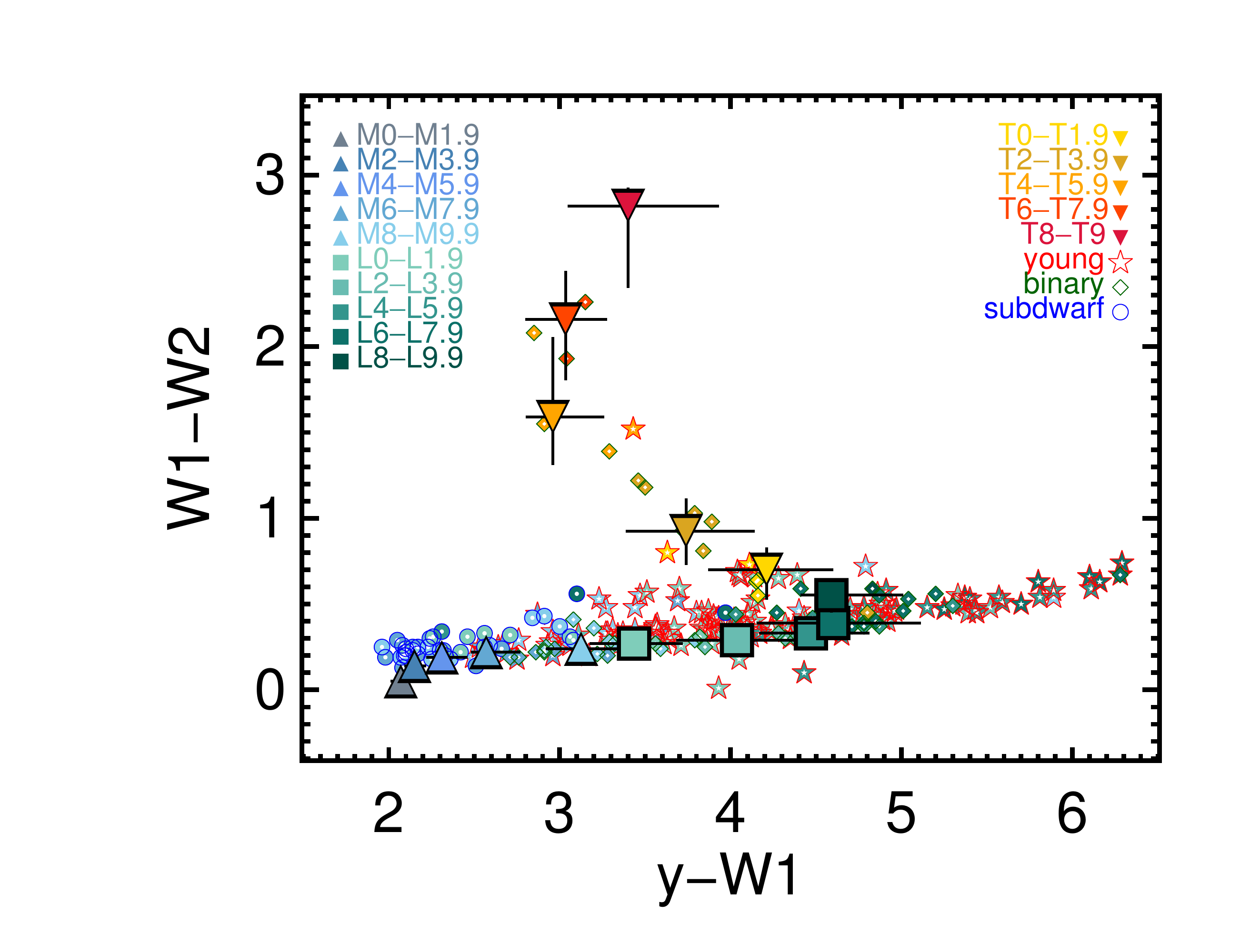}
  \end{minipage}
  \caption{continued.}
  \label{fig.colorcolor.unusual}
\end{center}
\end{figure*}

Our catalog contains \varnrbandldwarf~L dwarfs with \rps\ detections, including
\varnrbandldwarfnorm\ in the normal field sample.  With these we present the
largest set of $r$-band colors of L dwarfs to date, a tenfold increase over the
sample presented by \citet{Liebert:2006kp} and the compilation of
\citet{Koen:2013dz}.  Most PS1 colors become redder through the M dwarfs,
plateau for the L dwarfs and become redder again for T dwarfs (when detected),
but the L dwarfs show a different behavior with \rps.  \ri\ features a blueward
turn at spectral type $\approx$M8, becoming $\approx$0.8~mag bluer by spectral
type L5 where the objects become too faint for \rps\ detection, robustly
confirming previous findings using much smaller samples
\citep{Hawley:2002jc,Liebert:2006kp}.  \rz\ and \ry\ show similar but less
amplified trends.  \citet{Liebert:2006kp} explain these unusual blueward trends
as a consequence of decreasing TiO absorption, which strongly suppresses
$r$-band flux in M7-M8~dwarfs but weakens in later spectral types as Ti-bearing
dust grains form.  The resulting reduction in \rps\ opacity largely cancels out
the drop in flux expected from cooler objects, while the flux in \ips, \zps, and
\yps\ continues to decrease.  In addition, {K\,{\sc i}} absorption doublets
centered in the \ips-band (7665~\AA\ and 7699~\AA) increase in strength through
the L~types \citep{Kirkpatrick:1999ev}, enhancing the trend towards bluer \ri\
colors.  We note that another color, \ywa, also takes a blueward turn, peaking
at spectral type $\approx$L7 and becoming more than 1.5~mag bluer through
spectral type T5.  This trend arises from the appearance of the methane
fundamental band at 3.3~\um\ in late-L dwarfs \citep{Noll:2000ht}, which
broadens to a deep trough spanning $3.1-4.0$~\um\ by mid-T spectral types
\citep{Kirkpatrick:2005cv}.

We also note in Figure~\ref{fig.colorspt} that early-M dwarfs have few
detections and redder colors in \wbwc.  M~dwarfs are relatively faint in $W3$,
and the early-M dwarfs in our catalog are more distant than other spectral
types, so most were not detected in $W3$.  In contrast, all were detected with
errors $\le0.05$~mag in $W2$.  The few early-M dwarfs that have $W3$ detections
are the brightest ones in that band in our catalog, so the observed colors are
redder than the overall population.  The source of the $W3$ emission for the
reddest objects is most likely to be debris disks or contamination from
background objects, and we note that the \wbwc\ colors are consistent with those
of other debris disks \citep[e.g.,][]{Theissen:2014ec}.  We therefore do not
interpret the \wbwc\ colors in our catalog as representative of stellar
photospheres for spectral types earlier than M5.

We use the photometry from our catalog to calculate median colors of field
M0--T9 dwarfs spanning $\approx$0.55 to 12~\um.  As with Figures
\ref{fig.colorspt} and~\ref{fig.colorcolor}, we have excluded all known
binaries, young objects, and subdwarfs from these calculations in order to
produce colors representative of the normal field population.
Table~\ref{tbl.meancols} presents these colors in single steps of adjacent
filter pairs from \gps\ to $W3$ (excluding spectral types M0--M4 for \wbwc\ as
explained in the preceding paragraph).  Table~\ref{tbl.meancols2} presents five
additional colors previously used to study ultracool dwarfs: \iy, \ijt, \zjt,
\jkt, and \ywa.  For both tables, we list the median colors and 68\% confidence
limits for single spectral subtypes, along with the number of objects used to
determine each color.

\floattable
\begin{deluxetable}{lCCCcCCCcCCCcCCCcCCCcCCCcCCCcCCCcCCCcCCC}
\tablecaption{Median Colors of M0--T9 Dwarfs\label{tbl.meancols}}
\rotate
\tabletypesize{\tiny}
\tablewidth{0pt}
\setlength{\tabcolsep}{0.02in}
\tablehead{
  \colhead{} &
  \multicolumn{3}{c}{$\gps-\rps$} &
  \colhead{} &
  \multicolumn{3}{c}{$\rps-\ips$} &
  \colhead{} &
  \multicolumn{3}{c}{$\ips-\zps$} &
  \colhead{} &
  \multicolumn{3}{c}{$\zps-\yps$} &
  \colhead{} &
  \multicolumn{3}{c}{$\yps-\jtwo$} &
  \colhead{} &
  \multicolumn{3}{c}{\jht} &
  \colhead{} &
  \multicolumn{3}{c}{\hkt} &
  \colhead{} &
  \multicolumn{3}{c}{$\ktwo-W1$} &
  \colhead{} &
  \multicolumn{3}{c}{$W1-W2$} &
  \colhead{} &
  \multicolumn{3}{c}{$W2-W3$} \\
  \cline{2-4}
  \cline{6-8}
  \cline{10-12}
  \cline{14-16}
  \cline{18-20}
  \cline{22-24}
  \cline{26-28}
  \cline{30-32}
  \cline{34-36}
  \cline{38-40}
  \colhead{SpT} &
  \colhead{Median} &
  \colhead{68\%} &
  \colhead{$N$} &
  \colhead{} &
  \colhead{Median} &
  \colhead{68\%} &
  \colhead{$N$} &
  \colhead{} &
  \colhead{Median} &
  \colhead{68\%} &
  \colhead{$N$} &
  \colhead{} &
  \colhead{Median} &
  \colhead{68\%} &
  \colhead{$N$} &
  \colhead{} &
  \colhead{Median} &
  \colhead{68\%} &
  \colhead{$N$} &
  \colhead{} &
  \colhead{Median} &
  \colhead{68\%} &
  \colhead{$N$} &
  \colhead{} &
  \colhead{Median} &
  \colhead{68\%} &
  \colhead{$N$} &
  \colhead{} &
  \colhead{Median} &
  \colhead{68\%} &
  \colhead{$N$} &
  \colhead{} &
  \colhead{Median} &
  \colhead{68\%} &
  \colhead{$N$} &
  \colhead{} &
  \colhead{Median} &
  \colhead{68\%} &
  \colhead{$N$} \\
  \colhead{} &
  \colhead{(mag)} &
  \colhead{(mag)} &
  \colhead{} &
  \colhead{} &
  \colhead{(mag)} &
  \colhead{(mag)} &
  \colhead{} &
  \colhead{} &
  \colhead{(mag)} &
  \colhead{(mag)} &
  \colhead{} &
  \colhead{} &
  \colhead{(mag)} &
  \colhead{(mag)} &
  \colhead{} &
  \colhead{} &
  \colhead{(mag)} &
  \colhead{(mag)} &
  \colhead{} &
  \colhead{} &
  \colhead{(mag)} &
  \colhead{(mag)} &
  \colhead{} &
  \colhead{} &
  \colhead{(mag)} &
  \colhead{(mag)} &
  \colhead{} &
  \colhead{} &
  \colhead{(mag)} &
  \colhead{(mag)} &
  \colhead{} &
  \colhead{} &
  \colhead{(mag)} &
  \colhead{(mag)} &
  \colhead{} &
  \colhead{} &
  \colhead{(mag)} &
  \colhead{(mag)} &
  \colhead{}
}
\startdata
M0--M0.9\phn & 1.19 & $^{+0.04}_{-0.05}$ & 991 &  & 0.67 & $^{+0.08}_{-0.08}$ & 991 &  & 0.31 & $^{+0.04}_{-0.04}$ & 991 &  & 0.17 & $^{+0.02}_{-0.02}$ & 991 &  & 1.12 & $^{+0.04}_{-0.04}$ & 991 &  & 0.66 & $^{+0.05}_{-0.06}$ & 991 &  & 0.18 & $^{+0.06}_{-0.06}$ & 991 &  & 0.10 & $^{+0.05}_{-0.06}$ & 969 &  & 0.02 & $^{+0.06}_{-0.05}$ & 969 &  & \nodata & \nodata & 0 \\
M1--M1.9\phn & 1.22 & $^{+0.04}_{-0.04}$ & 707 &  & 0.85 & $^{+0.08}_{-0.08}$ & 707 &  & 0.39 & $^{+0.03}_{-0.04}$ & 707 &  & 0.20 & $^{+0.02}_{-0.02}$ & 707 &  & 1.14 & $^{+0.05}_{-0.04}$ & 707 &  & 0.64 & $^{+0.06}_{-0.05}$ & 707 &  & 0.21 & $^{+0.05}_{-0.07}$ & 707 &  & 0.11 & $^{+0.07}_{-0.06}$ & 700 &  & 0.07 & $^{+0.05}_{-0.06}$ & 700 &  & \nodata & \nodata & 0 \\
M2--M2.9\phn & 1.21 & $^{+0.04}_{-0.04}$ & 1657 &  & 1.02 & $^{+0.07}_{-0.07}$ & 1657 &  & 0.46 & $^{+0.03}_{-0.03}$ & 1657 &  & 0.23 & $^{+0.02}_{-0.02}$ & 1657 &  & 1.16 & $^{+0.05}_{-0.04}$ & 1657 &  & 0.62 & $^{+0.06}_{-0.07}$ & 1657 &  & 0.22 & $^{+0.06}_{-0.06}$ & 1657 &  & 0.12 & $^{+0.06}_{-0.05}$ & 1633 &  & 0.12 & $^{+0.05}_{-0.06}$ & 1633 &  & \nodata & \nodata & 0 \\
M3--M3.9\phn & 1.21 & $^{+0.04}_{-0.05}$ & 2107 &  & 1.22 & $^{+0.09}_{-0.08}$ & 2107 &  & 0.55 & $^{+0.05}_{-0.04}$ & 2107 &  & 0.27 & $^{+0.02}_{-0.03}$ & 2107 &  & 1.20 & $^{+0.04}_{-0.05}$ & 2106 &  & 0.60 & $^{+0.06}_{-0.06}$ & 2106 &  & 0.24 & $^{+0.05}_{-0.06}$ & 2107 &  & 0.14 & $^{+0.05}_{-0.06}$ & 2079 &  & 0.16 & $^{+0.03}_{-0.05}$ & 2079 &  & \nodata & \nodata & 0 \\
M4--M4.9\phn & 1.23 & $^{+0.05}_{-0.06}$ & 1220 &  & 1.46 & $^{+0.12}_{-0.08}$ & 1220 &  & 0.67 & $^{+0.06}_{-0.04}$ & 1220 &  & 0.32 & $^{+0.04}_{-0.02}$ & 1220 &  & 1.25 & $^{+0.04}_{-0.05}$ & 1220 &  & 0.59 & $^{+0.05}_{-0.05}$ & 1220 &  & 0.26 & $^{+0.06}_{-0.05}$ & 1220 &  & 0.16 & $^{+0.05}_{-0.05}$ & 1198 &  & 0.18 & $^{+0.03}_{-0.03}$ & 1198 &  & \nodata & \nodata & 0 \\
M5--M5.9\phn & 1.31 & $^{+0.04}_{-0.05}$ & 683 &  & 1.88 & $^{+0.10}_{-0.08}$ & 683 &  & 0.87 & $^{+0.04}_{-0.04}$ & 683 &  & 0.44 & $^{+0.03}_{-0.03}$ & 683 &  & 1.34 & $^{+0.05}_{-0.04}$ & 683 &  & 0.59 & $^{+0.06}_{-0.05}$ & 683 &  & 0.31 & $^{+0.06}_{-0.06}$ & 683 &  & 0.19 & $^{+0.05}_{-0.06}$ & 676 &  & 0.21 & $^{+0.03}_{-0.03}$ & 676 &  & 0.21 & $^{+0.16}_{-0.14}$ & 44 \\
M6--M6.9\phn & 1.33 & $^{+0.06}_{-0.07}$ & 399 &  & 2.13 & $^{+0.15}_{-0.12}$ & 388 &  & 0.98 & $^{+0.07}_{-0.05}$ & 387 &  & 0.51 & $^{+0.05}_{-0.03}$ & 387 &  & 1.40 & $^{+0.07}_{-0.05}$ & 391 &  & 0.60 & $^{+0.06}_{-0.05}$ & 401 &  & 0.33 & $^{+0.05}_{-0.05}$ & 402 &  & 0.20 & $^{+0.05}_{-0.04}$ & 388 &  & 0.22 & $^{+0.03}_{-0.03}$ & 388 &  & 0.25 & $^{+0.12}_{-0.07}$ & 40 \\
M7--M7.9\phn & 1.40 & $^{+0.09}_{-0.11}$ & 155 &  & 2.55 & $^{+0.23}_{-0.16}$ & 154 &  & 1.21 & $^{+0.13}_{-0.09}$ & 157 &  & 0.67 & $^{+0.09}_{-0.07}$ & 157 &  & 1.54 & $^{+0.08}_{-0.08}$ & 154 &  & 0.63 & $^{+0.04}_{-0.05}$ & 157 &  & 0.39 & $^{+0.04}_{-0.05}$ & 158 &  & 0.22 & $^{+0.04}_{-0.03}$ & 147 &  & 0.23 & $^{+0.02}_{-0.03}$ & 141 &  & 0.32 & $^{+0.10}_{-0.08}$ & 60 \\
M8--M8.9\phn & 1.53 & $^{+0.16}_{-0.13}$ & 90 &  & 2.69 & $^{+0.12}_{-0.20}$ & 108 &  & 1.38 & $^{+0.07}_{-0.10}$ & 119 &  & 0.81 & $^{+0.07}_{-0.09}$ & 120 &  & 1.66 & $^{+0.08}_{-0.08}$ & 108 &  & 0.68 & $^{+0.04}_{-0.05}$ & 111 &  & 0.43 & $^{+0.05}_{-0.04}$ & 112 &  & 0.26 & $^{+0.04}_{-0.05}$ & 99 &  & 0.23 & $^{+0.04}_{-0.04}$ & 97 &  & 0.35 & $^{+0.11}_{-0.11}$ & 52 \\
M9--M9.9\phn & 1.79 & $^{+0.16}_{-0.29}$ & 35 &  & 2.58 & $^{+0.11}_{-0.23}$ & 59 &  & 1.44 & $^{+0.05}_{-0.06}$ & 85 &  & 0.92 & $^{+0.05}_{-0.07}$ & 85 &  & 1.77 & $^{+0.09}_{-0.06}$ & 61 &  & 0.72 & $^{+0.08}_{-0.06}$ & 60 &  & 0.48 & $^{+0.06}_{-0.06}$ & 64 &  & 0.31 & $^{+0.07}_{-0.04}$ & 56 &  & 0.26 & $^{+0.05}_{-0.04}$ & 63 &  & 0.44 & $^{+0.18}_{-0.12}$ & 35 \\
L0--L0.9\phn & 1.85 & $^{+0.14}_{-0.29}$ & 16 &  & 2.35 & $^{+0.21}_{-0.21}$ & 202 &  & 1.47 & $^{+0.05}_{-0.05}$ & 344 &  & 0.95 & $^{+0.04}_{-0.07}$ & 347 &  & 1.82 & $^{+0.13}_{-0.13}$ & 307 &  & 0.76 & $^{+0.12}_{-0.13}$ & 277 &  & 0.48 & $^{+0.13}_{-0.13}$ & 281 &  & 0.32 & $^{+0.09}_{-0.09}$ & 263 &  & 0.27 & $^{+0.07}_{-0.05}$ & 311 &  & 0.51 & $^{+0.22}_{-0.13}$ & 21 \\
L1--L1.9\phn & 2.00 & $^{+0.24}_{-1.13}$ & 11 &  & 2.35 & $^{+0.13}_{-0.22}$ & 113 &  & 1.48 & $^{+0.09}_{-0.07}$ & 225 &  & 0.97 & $^{+0.04}_{-0.05}$ & 232 &  & 1.94 & $^{+0.11}_{-0.13}$ & 176 &  & 0.80 & $^{+0.10}_{-0.15}$ & 166 &  & 0.51 & $^{+0.11}_{-0.10}$ & 168 &  & 0.35 & $^{+0.11}_{-0.07}$ & 152 &  & 0.26 & $^{+0.06}_{-0.04}$ & 189 &  & 0.46 & $^{+0.15}_{-0.11}$ & 16 \\
L2--L2.9\phn & 2.30 & \nodata & 1 &  & 2.27 & $^{+0.08}_{-0.20}$ & 50 &  & 1.45 & $^{+0.08}_{-0.08}$ & 107 &  & 0.97 & $^{+0.04}_{-0.05}$ & 109 &  & 2.00 & $^{+0.13}_{-0.11}$ & 90 &  & 0.91 & $^{+0.10}_{-0.14}$ & 89 &  & 0.59 & $^{+0.12}_{-0.11}$ & 92 &  & 0.42 & $^{+0.12}_{-0.10}$ & 82 &  & 0.29 & $^{+0.06}_{-0.04}$ & 96 &  & 0.55 & $^{+0.21}_{-0.14}$ & 14 \\
L3--L3.9\phn & 2.70 & \nodata & 1 &  & 2.23 & $^{+0.14}_{-0.17}$ & 31 &  & 1.49 & $^{+0.16}_{-0.12}$ & 97 &  & 1.01 & $^{+0.05}_{-0.07}$ & 101 &  & 2.12 & $^{+0.09}_{-0.15}$ & 78 &  & 0.95 & $^{+0.13}_{-0.22}$ & 77 &  & 0.64 & $^{+0.13}_{-0.09}$ & 78 &  & 0.52 & $^{+0.12}_{-0.19}$ & 73 &  & 0.30 & $^{+0.06}_{-0.07}$ & 90 &  & 0.50 & $^{+0.32}_{-0.10}$ & 11 \\
L4--L4.9\phn & \nodata & \nodata & 0 &  & 2.11 & $^{+0.16}_{-0.15}$ & 17 &  & 1.63 & $^{+0.20}_{-0.17}$ & 68 &  & 1.02 & $^{+0.05}_{-0.08}$ & 84 &  & 2.15 & $^{+0.13}_{-0.14}$ & 54 &  & 1.06 & $^{+0.14}_{-0.21}$ & 51 &  & 0.64 & $^{+0.11}_{-0.20}$ & 53 &  & 0.61 & $^{+0.16}_{-0.12}$ & 50 &  & 0.32 & $^{+0.06}_{-0.08}$ & 71 &  & 0.64 & $^{+0.06}_{-0.29}$ & 8 \\
L5--L5.9\phn & \nodata & \nodata & 0 &  & 2.00 & $^{+0.16}_{-0.25}$ & 13 &  & 1.77 & $^{+0.15}_{-0.20}$ & 64 &  & 1.05 & $^{+0.07}_{-0.06}$ & 83 &  & 2.18 & $^{+0.15}_{-0.10}$ & 55 &  & 1.06 & $^{+0.16}_{-0.20}$ & 54 &  & 0.67 & $^{+0.11}_{-0.15}$ & 55 &  & 0.67 & $^{+0.20}_{-0.09}$ & 50 &  & 0.33 & $^{+0.11}_{-0.08}$ & 74 &  & 0.65 & $^{+0.26}_{-0.09}$ & 8 \\
L6--L6.9\phn & \nodata & \nodata & 0 &  & 1.91 & \nodata & 5 &  & 1.89 & $^{+0.25}_{-0.33}$ & 34 &  & 1.05 & $^{+0.05}_{-0.07}$ & 45 &  & 2.09 & $^{+0.19}_{-0.07}$ & 32 &  & 0.99 & $^{+0.22}_{-0.11}$ & 30 &  & 0.65 & $^{+0.27}_{-0.24}$ & 34 &  & 0.71 & $^{+0.14}_{-0.18}$ & 35 &  & 0.38 & $^{+0.10}_{-0.09}$ & 43 &  & 0.76 & \nodata & 5 \\
L7--L7.9\phn & \nodata & \nodata & 0 &  & \nodata & \nodata & 0 &  & 2.10 & $^{+0.22}_{-0.33}$ & 16 &  & 1.02 & $^{+0.12}_{-0.05}$ & 27 &  & 2.19 & $^{+0.23}_{-0.10}$ & 21 &  & 1.12 & $^{+0.28}_{-0.20}$ & 18 &  & 0.73 & $^{+0.10}_{-0.17}$ & 20 &  & 0.73 & $^{+0.16}_{-0.08}$ & 17 &  & 0.42 & $^{+0.12}_{-0.11}$ & 23 &  & 0.96 & \nodata & 2 \\
L8--L8.9\phn & \nodata & \nodata & 0 &  & 2.13 & \nodata & 1 &  & 2.45 & $^{+0.23}_{-0.55}$ & 17 &  & 1.05 & $^{+0.11}_{-0.04}$ & 28 &  & 2.14 & $^{+0.17}_{-0.12}$ & 22 &  & 1.13 & $^{+0.17}_{-0.13}$ & 23 &  & 0.65 & $^{+0.19}_{-0.08}$ & 28 &  & 0.85 & $^{+0.08}_{-0.13}$ & 26 &  & 0.53 & $^{+0.08}_{-0.09}$ & 27 &  & 1.12 & \nodata & 5 \\
L9--L9.9\phn & \nodata & \nodata & 0 &  & \nodata & \nodata & 0 &  & 2.45 & $^{+0.70}_{-0.41}$ & 8 &  & 1.16 & $^{+0.06}_{-0.15}$ & 40 &  & 2.11 & $^{+0.15}_{-0.11}$ & 28 &  & 1.07 & $^{+0.14}_{-0.13}$ & 27 &  & 0.58 & $^{+0.15}_{-0.10}$ & 27 &  & 0.78 & $^{+0.13}_{-0.11}$ & 22 &  & 0.57 & $^{+0.10}_{-0.13}$ & 32 &  & 0.88 & \nodata & 3 \\
T0--T0.9\phn & \nodata & \nodata & 0 &  & \nodata & \nodata & 0 &  & 2.38 & \nodata & 6 &  & 1.21 & $^{+0.20}_{-0.11}$ & 24 &  & 2.21 & $^{+0.18}_{-0.13}$ & 13 &  & 0.93 & $^{+0.12}_{-0.18}$ & 12 &  & 0.56 & $^{+0.04}_{-0.19}$ & 12 &  & 0.78 & $^{+0.15}_{-0.26}$ & 13 &  & 0.58 & $^{+0.13}_{-0.09}$ & 21 &  & 0.92 & \nodata & 1 \\
T1--T1.9\phn & \nodata & \nodata & 0 &  & \nodata & \nodata & 0 &  & 2.30 & \nodata & 3 &  & 1.39 & $^{+0.12}_{-0.26}$ & 25 &  & 2.19 & $^{+0.12}_{-0.13}$ & 14 &  & 0.86 & $^{+0.17}_{-0.08}$ & 12 &  & 0.33 & $^{+0.28}_{-0.18}$ & 12 &  & 0.63 & $^{+0.19}_{-0.20}$ & 13 &  & 0.72 & $^{+0.16}_{-0.10}$ & 25 &  & 1.41 & \nodata & 3 \\
T2--T2.9\phn & \nodata & \nodata & 0 &  & \nodata & \nodata & 0 &  & 3.53 & \nodata & 1 &  & 1.45 & $^{+0.14}_{-0.22}$ & 39 &  & 2.22 & $^{+0.07}_{-0.12}$ & 22 &  & 0.80 & $^{+0.21}_{-0.11}$ & 16 &  & 0.25 & $^{+0.29}_{-0.22}$ & 17 &  & 0.57 & $^{+0.21}_{-0.29}$ & 16 &  & 0.91 & $^{+0.14}_{-0.18}$ & 39 &  & 1.44 & \nodata & 5 \\
T3--T3.9\phn & \nodata & \nodata & 0 &  & \nodata & \nodata & 0 &  & 2.85 & \nodata & 1 &  & 1.55 & $^{+0.17}_{-0.23}$ & 23 &  & 2.24 & $^{+0.19}_{-0.07}$ & 11 &  & 0.59 & $^{+0.12}_{-0.29}$ & 7 &  & 0.17 & $^{+0.43}_{-0.13}$ & 7 &  & 0.56 & $^{+0.13}_{-0.26}$ & 8 &  & 1.07 & $^{+0.19}_{-0.37}$ & 19 &  & \nodata & \nodata & 0 \\
T4--T4.9\phn & \nodata & \nodata & 0 &  & \nodata & \nodata & 0 &  & \nodata & \nodata & 0 &  & 1.71 & $^{+0.08}_{-0.19}$ & 30 &  & 2.43 & $^{+0.11}_{-0.17}$ & 21 &  & 0.32 & $^{+0.09}_{-0.16}$ & 11 &  & 0.11 & $^{+0.11}_{-0.06}$ & 9 &  & 0.19 & $^{+0.07}_{-0.20}$ & 7 &  & 1.35 & $^{+0.25}_{-0.17}$ & 26 &  & 1.30 & \nodata & 1 \\
T5--T5.9\phn & \nodata & \nodata & 0 &  & \nodata & \nodata & 0 &  & 3.49 & \nodata & 1 &  & 1.74 & $^{+0.19}_{-0.11}$ & 32 &  & 2.49 & $^{+0.07}_{-0.20}$ & 30 &  & 0.18 & $^{+0.35}_{-0.10}$ & 10 &  & 0.03 & \nodata & 6 &  & 0.15 & \nodata & 3 &  & 1.85 & $^{+0.36}_{-0.32}$ & 40 &  & \nodata & \nodata & 0 \\
T6--T6.9\phn & \nodata & \nodata & 0 &  & \nodata & \nodata & 0 &  & \nodata & \nodata & 0 &  & 1.74 & $^{+0.15}_{-0.13}$ & 32 &  & 2.54 & $^{+0.09}_{-0.16}$ & 38 &  & 0.05 & $^{+0.18}_{-0.17}$ & 20 &  & 0.06 & $^{+0.17}_{-0.21}$ & 14 &  & 0.37 & $^{+0.24}_{-0.32}$ & 14 &  & 2.08 & $^{+0.30}_{-0.36}$ & 47 &  & 1.21 & \nodata & 2 \\
T7--T7.9\phn & \nodata & \nodata & 0 &  & \nodata & \nodata & 0 &  & \nodata & \nodata & 0 &  & 1.80 & $^{+0.05}_{-0.09}$ & 15 &  & 2.56 & $^{+0.19}_{-0.16}$ & 21 &  & 0.08 & $^{+0.12}_{-0.24}$ & 12 &  & 0.01 & $^{+0.21}_{-0.09}$ & 7 &  & 0.33 & \nodata & 5 &  & 2.26 & $^{+0.29}_{-0.34}$ & 27 &  & 1.35 & \nodata & 3 \\
T8--T8.9\phn & \nodata & \nodata & 0 &  & \nodata & \nodata & 0 &  & \nodata & \nodata & 0 &  & 1.76 & \nodata & 4 &  & 2.57 & $^{+0.27}_{-0.43}$ & 9 &  & 0.09 & \nodata & 1 &  & \nodata & \nodata & 0 &  & \nodata & \nodata & 0 &  & 2.79 & $^{+0.15}_{-0.44}$ & 10 &  & 1.40 & \nodata & 3 \\
T9--T9.9\phn & \nodata & \nodata & 0 &  & \nodata & \nodata & 0 &  & \nodata & \nodata & 0 &  & 1.37 & \nodata & 1 &  & 2.61 & \nodata & 2 &  & 0.24 & \nodata & 1 &  & \nodata & \nodata & 0 &  & \nodata & \nodata & 0 &  & 2.95 & \nodata & 3 &  & 1.82 & \nodata & 2 \\
\enddata
\tablecomments{For each spectral type and color, this table lists the median
  color and 68\% confidence limits followed by the number of objects ($N$) used
  to determine the median. Confidence intervals were calculated only when $N\ge7$.}
\end{deluxetable}
   % Table of median colors, step-by-step

\floattable
\begin{deluxetable}{lCCCcCCCcCCCcCCCcCCC}
\tablecaption{More Median Colors \label{tbl.meancols2}}
\tabletypesize{\tiny}
\tablewidth{0pt}
\setlength{\tabcolsep}{0.03in}
\tablehead{
  \colhead{} &
  \multicolumn{3}{c}{$\ips-\yps$} &
  \colhead{} &
  \multicolumn{3}{c}{$\ips-\jtwo$} &
  \colhead{} &
  \multicolumn{3}{c}{$\zps-\jtwo$} &
  \colhead{} &
  \multicolumn{3}{c}{\jkt} &
  \colhead{} &
  \multicolumn{3}{c}{$\yps-W1$} \\
  \cline{2-4}
  \cline{6-8}
  \cline{10-12}
  \cline{14-16}
  \cline{18-20}
  \colhead{SpT} &
  \colhead{Median} &
  \colhead{68\%} &
  \colhead{$N$} &
  \colhead{} &
  \colhead{Median} &
  \colhead{68\%} &
  \colhead{$N$} &
  \colhead{} &
  \colhead{Median} &
  \colhead{68\%} &
  \colhead{$N$} &
  \colhead{} &
  \colhead{Median} &
  \colhead{68\%} &
  \colhead{$N$} &
  \colhead{} &
  \colhead{Median} &
  \colhead{68\%} &
  \colhead{$N$} \\
  \colhead{} &
  \colhead{(mag)} &
  \colhead{(mag)} &
  \colhead{} &
  \colhead{} &
  \colhead{(mag)} &
  \colhead{(mag)} &
  \colhead{} &
  \colhead{} &
  \colhead{(mag)} &
  \colhead{(mag)} &
  \colhead{} &
  \colhead{} &
  \colhead{(mag)} &
  \colhead{(mag)} &
  \colhead{} &
  \colhead{} &
  \colhead{(mag)} &
  \colhead{(mag)} &
  \colhead{}
}
\startdata
M0--M0.9\phn & 0.48 & $^{+0.05}_{-0.05}$ & 991 &  & 1.61 & $^{+0.07}_{-0.08}$ & 991 &  & 1.29 & $^{+0.05}_{-0.05}$ & 991 &  & 0.84 & $^{+0.06}_{-0.07}$ & 991 &  & 2.05 & $^{+0.06}_{-0.06}$ & 969 \\
M1--M1.9\phn & 0.59 & $^{+0.05}_{-0.05}$ & 707 &  & 1.74 & $^{+0.07}_{-0.08}$ & 707 &  & 1.34 & $^{+0.05}_{-0.04}$ & 707 &  & 0.85 & $^{+0.06}_{-0.07}$ & 707 &  & 2.10 & $^{+0.06}_{-0.05}$ & 700 \\
M2--M2.9\phn & 0.69 & $^{+0.05}_{-0.04}$ & 1657 &  & 1.85 & $^{+0.07}_{-0.06}$ & 1657 &  & 1.39 & $^{+0.05}_{-0.05}$ & 1657 &  & 0.84 & $^{+0.07}_{-0.08}$ & 1657 &  & 2.12 & $^{+0.06}_{-0.05}$ & 1633 \\
M3--M3.9\phn & 0.82 & $^{+0.06}_{-0.06}$ & 2107 &  & 2.02 & $^{+0.09}_{-0.09}$ & 2106 &  & 1.46 & $^{+0.06}_{-0.05}$ & 2106 &  & 0.84 & $^{+0.06}_{-0.07}$ & 2106 &  & 2.17 & $^{+0.06}_{-0.06}$ & 2079 \\
M4--M4.9\phn & 0.99 & $^{+0.09}_{-0.06}$ & 1220 &  & 2.24 & $^{+0.13}_{-0.09}$ & 1220 &  & 1.57 & $^{+0.07}_{-0.06}$ & 1220 &  & 0.86 & $^{+0.06}_{-0.07}$ & 1220 &  & 2.26 & $^{+0.07}_{-0.07}$ & 1198 \\
M5--M5.9\phn & 1.31 & $^{+0.07}_{-0.07}$ & 683 &  & 2.66 & $^{+0.09}_{-0.11}$ & 683 &  & 1.78 & $^{+0.07}_{-0.07}$ & 683 &  & 0.91 & $^{+0.06}_{-0.07}$ & 683 &  & 2.44 & $^{+0.06}_{-0.07}$ & 676 \\
M6--M6.9\phn & 1.48 & $^{+0.13}_{-0.07}$ & 389 &  & 2.89 & $^{+0.17}_{-0.11}$ & 387 &  & 1.91 & $^{+0.11}_{-0.07}$ & 386 &  & 0.93 & $^{+0.07}_{-0.07}$ & 401 &  & 2.54 & $^{+0.10}_{-0.08}$ & 381 \\
M7--M7.9\phn & 1.87 & $^{+0.25}_{-0.13}$ & 158 &  & 3.41 & $^{+0.30}_{-0.21}$ & 153 &  & 2.21 & $^{+0.18}_{-0.14}$ & 152 &  & 1.02 & $^{+0.07}_{-0.07}$ & 157 &  & 2.78 & $^{+0.17}_{-0.17}$ & 145 \\
M8--M8.9\phn & 2.19 & $^{+0.14}_{-0.18}$ & 120 &  & 3.85 & $^{+0.19}_{-0.21}$ & 108 &  & 2.46 & $^{+0.14}_{-0.13}$ & 107 &  & 1.12 & $^{+0.06}_{-0.10}$ & 111 &  & 3.03 & $^{+0.15}_{-0.17}$ & 100 \\
M9--M9.9\phn & 2.35 & $^{+0.11}_{-0.10}$ & 85 &  & 4.16 & $^{+0.11}_{-0.17}$ & 61 &  & 2.71 & $^{+0.09}_{-0.15}$ & 61 &  & 1.19 & $^{+0.15}_{-0.07}$ & 60 &  & 3.33 & $^{+0.17}_{-0.15}$ & 68 \\
L0--L0.9\phn & 2.41 & $^{+0.07}_{-0.07}$ & 344 &  & 4.24 & $^{+0.15}_{-0.18}$ & 307 &  & 2.76 & $^{+0.15}_{-0.15}$ & 307 &  & 1.24 & $^{+0.13}_{-0.18}$ & 278 &  & 3.35 & $^{+0.26}_{-0.25}$ & 320 \\
L1--L1.9\phn & 2.45 & $^{+0.11}_{-0.10}$ & 225 &  & 4.38 & $^{+0.15}_{-0.10}$ & 175 &  & 2.90 & $^{+0.13}_{-0.12}$ & 176 &  & 1.32 & $^{+0.16}_{-0.19}$ & 165 &  & 3.59 & $^{+0.25}_{-0.26}$ & 201 \\
L2--L2.9\phn & 2.41 & $^{+0.12}_{-0.09}$ & 107 &  & 4.41 & $^{+0.18}_{-0.11}$ & 90 &  & 2.98 & $^{+0.11}_{-0.10}$ & 90 &  & 1.49 & $^{+0.19}_{-0.22}$ & 89 &  & 3.90 & $^{+0.36}_{-0.25}$ & 99 \\
L3--L3.9\phn & 2.51 & $^{+0.14}_{-0.18}$ & 97 &  & 4.63 & $^{+0.20}_{-0.28}$ & 78 &  & 3.11 & $^{+0.13}_{-0.18}$ & 78 &  & 1.58 & $^{+0.21}_{-0.24}$ & 77 &  & 4.26 & $^{+0.23}_{-0.57}$ & 94 \\
L4--L4.9\phn & 2.68 & $^{+0.20}_{-0.20}$ & 68 &  & 4.84 & $^{+0.26}_{-0.25}$ & 51 &  & 3.17 & $^{+0.15}_{-0.16}$ & 54 &  & 1.65 & $^{+0.29}_{-0.31}$ & 52 &  & 4.39 & $^{+0.31}_{-0.36}$ & 72 \\
L5--L5.9\phn & 2.83 & $^{+0.12}_{-0.23}$ & 64 &  & 5.02 & $^{+0.14}_{-0.25}$ & 51 &  & 3.25 & $^{+0.10}_{-0.16}$ & 55 &  & 1.75 & $^{+0.20}_{-0.26}$ & 53 &  & 4.59 & $^{+0.30}_{-0.30}$ & 74 \\
L6--L6.9\phn & 2.96 & $^{+0.25}_{-0.35}$ & 34 &  & 5.09 & $^{+0.24}_{-0.22}$ & 28 &  & 3.17 & $^{+0.15}_{-0.16}$ & 32 &  & 1.59 & $^{+0.38}_{-0.19}$ & 30 &  & 4.55 & $^{+0.23}_{-0.39}$ & 43 \\
L7--L7.9\phn & 3.07 & $^{+0.37}_{-0.23}$ & 16 &  & 5.15 & $^{+0.43}_{-0.24}$ & 12 &  & 3.25 & $^{+0.16}_{-0.11}$ & 20 &  & 1.82 & $^{+0.16}_{-0.17}$ & 18 &  & 4.68 & $^{+0.52}_{-0.52}$ & 23 \\
L8--L8.9\phn & 3.45 & $^{+0.32}_{-0.49}$ & 16 &  & 5.48 & $^{+0.32}_{-0.31}$ & 14 &  & 3.20 & $^{+0.23}_{-0.11}$ & 23 &  & 1.86 & $^{+0.15}_{-0.24}$ & 23 &  & 4.78 & $^{+0.37}_{-0.26}$ & 26 \\
L9--L9.9\phn & 3.60 & $^{+0.76}_{-0.37}$ & 8 &  & 5.63 & \nodata & 6 &  & 3.23 & $^{+0.27}_{-0.15}$ & 28 &  & 1.67 & $^{+0.27}_{-0.21}$ & 26 &  & 4.54 & $^{+0.19}_{-0.19}$ & 33 \\
T0--T0.9\phn & 3.52 & \nodata & 6 &  & 5.66 & \nodata & 5 &  & 3.42 & $^{+0.26}_{-0.18}$ & 13 &  & 1.48 & $^{+0.14}_{-0.35}$ & 11 &  & 4.29 & $^{+0.35}_{-0.36}$ & 21 \\
T1--T1.9\phn & 3.72 & \nodata & 3 &  & 5.85 & \nodata & 3 &  & 3.59 & $^{+0.12}_{-0.13}$ & 14 &  & 1.18 & $^{+0.38}_{-0.29}$ & 13 &  & 4.10 & $^{+0.43}_{-0.42}$ & 25 \\
T2--T2.9\phn & 5.17 & \nodata & 1 &  & 7.43 & \nodata & 1 &  & 3.72 & $^{+0.13}_{-0.26}$ & 21 &  & 1.06 & $^{+0.27}_{-0.18}$ & 16 &  & 3.83 & $^{+0.31}_{-0.36}$ & 41 \\
T3--T3.9\phn & 4.50 & \nodata & 1 &  & \nodata & \nodata & 0 &  & 3.82 & $^{+0.13}_{-0.18}$ & 10 &  & 0.82 & $^{+0.34}_{-0.30}$ & 7 &  & 3.65 & $^{+0.50}_{-0.57}$ & 21 \\
T4--T4.9\phn & \nodata & \nodata & 0 &  & \nodata & \nodata & 0 &  & 4.13 & $^{+0.10}_{-0.24}$ & 21 &  & 0.40 & $^{+0.04}_{-0.19}$ & 9 &  & 3.07 & $^{+0.44}_{-0.20}$ & 27 \\
T5--T5.9\phn & 5.23 & \nodata & 1 &  & 7.75 & \nodata & 1 &  & 4.26 & $^{+0.20}_{-0.27}$ & 27 &  & 0.17 & \nodata & 6 &  & 2.92 & $^{+0.23}_{-0.12}$ & 40 \\
T6--T6.9\phn & \nodata & \nodata & 0 &  & \nodata & \nodata & 0 &  & 4.29 & $^{+0.15}_{-0.18}$ & 27 &  & 0.16 & $^{+0.13}_{-0.18}$ & 15 &  & 3.02 & $^{+0.22}_{-0.22}$ & 49 \\
T7--T7.9\phn & \nodata & \nodata & 0 &  & \nodata & \nodata & 0 &  & 4.38 & $^{+0.15}_{-0.20}$ & 10 &  & 0.08 & $^{+0.22}_{-0.31}$ & 7 &  & 3.16 & $^{+0.39}_{-0.35}$ & 28 \\
T8--T8.9\phn & \nodata & \nodata & 0 &  & \nodata & \nodata & 0 &  & 4.34 & \nodata & 4 &  & \nodata & \nodata & 0 &  & 3.34 & $^{+0.60}_{-0.35}$ & 9 \\
T9--T9.9\phn & \nodata & \nodata & 0 &  & \nodata & \nodata & 0 &  & 4.07 & \nodata & 1 &  & \nodata & \nodata & 0 &  & 3.66 & \nodata & 2 \\
\enddata
\tablecomments{For each spectral type and color, this table lists the median
  color and 68\% confidence limits followed by the number of objects ($N$) used
  to determine the median. Confidence intervals were calculated only when $N\ge7$.}
\end{deluxetable}
   % Table of additional median colors

We construct empirical spectral energy distributions (SEDs) for field ultracool
dwarfs using the photometry in our catalog and parallaxes from the literature.
Excluding binaries, subdwarfs, and young objects, our catalog contains
\varnplxjmag~objects (spectral types M6--T9) with reported parallaxes and \jtwo\
photometry with no confusion or contamination flags and errors less than
0.2~mag.  We calculate absolute \jtwo\ magnitudes for these objects, and
determine the weighted mean and rms in bins of one spectral subtype.  We then
use the median colors relative to \jtwo\ from our full catalog
(Table~\ref{tbl.meancols}) to calculate absolute magnitudes for all other bands
from \gps\ to $W3$, adding the rms color for each band in quadrature with the
M$_{\jtwo}$ rms magnitude to determine errors.  We use our catalog colors rather
than directly calculating absolute magnitudes for each band because the colors
are derived from a much larger and carefully vetted sample.  We present our SEDs
for each spectral subtype between M6 and T9 in Table~\ref{tbl.seds}.

\floattable
\begin{deluxetable}{lCcCCcCCcCCcCCcCCcCCcCCcCCcCCcCCcCC}
\tablecaption{Spectral Energy Distributions for Field Ultracool Dwarfs \label{tbl.seds}}
\rotate
\tabletypesize{\tiny}
\tablewidth{0pt}
\setlength{\tabcolsep}{0.03in}
\tablehead{
  \colhead{} &
  \colhead{} &
  \colhead{} &
  \multicolumn{2}{c}{M$_{\gps}$} &
  \colhead{} &
  \multicolumn{2}{c}{M$_{\rps}$} &
  \colhead{} &
  \multicolumn{2}{c}{M$_{\ips}$} &
  \colhead{} &
  \multicolumn{2}{c}{M$_{\zps}$} &
  \colhead{} &
  \multicolumn{2}{c}{M$_{\yps}$} &
  \colhead{} &
  \multicolumn{2}{c}{M$_{\jtwo}$} &
  \colhead{} &
  \multicolumn{2}{c}{M$_{\htwo}$} &
  \colhead{} &
  \multicolumn{2}{c}{M$_{\ktwo}$} &
  \colhead{} &
  \multicolumn{2}{c}{M$_{W1}$} &
  \colhead{} &
  \multicolumn{2}{c}{M$_{W2}$} &
  \colhead{} &
  \multicolumn{2}{c}{M$_{W3}$} \\
  \cline{4-5}
  \cline{7-8}
  \cline{10-11}
  \cline{13-14}
  \cline{16-17}
  \cline{19-20}
  \cline{22-23}
  \cline{25-26}
  \cline{28-29}
  \cline{31-32}
  \cline{34-35}
  \colhead{SpT} &
  \colhead{$N$} &
  \colhead{} &
  \colhead{Mean} &
  \colhead{$\sigma$} &
  \colhead{} &
  \colhead{Mean} &
  \colhead{$\sigma$} &
  \colhead{} &
  \colhead{Mean} &
  \colhead{$\sigma$} &
  \colhead{} &
  \colhead{Mean} &
  \colhead{$\sigma$} &
  \colhead{} &
  \colhead{Mean} &
  \colhead{$\sigma$} &
  \colhead{} &
  \colhead{Mean} &
  \colhead{$\sigma$} &
  \colhead{} &
  \colhead{Mean} &
  \colhead{$\sigma$} &
  \colhead{} &
  \colhead{Mean} &
  \colhead{$\sigma$} &
  \colhead{} &
  \colhead{Mean} &
  \colhead{$\sigma$} &
  \colhead{} &
  \colhead{Mean} &
  \colhead{$\sigma$} &
  \colhead{} &
  \colhead{Mean} &
  \colhead{$\sigma$} \\
  \colhead{} &
  \colhead{} &
  \colhead{} &
  \colhead{(mag)} &
  \colhead{(mag)} &
  \colhead{} &
  \colhead{(mag)} &
  \colhead{(mag)} &
  \colhead{} &
  \colhead{(mag)} &
  \colhead{(mag)} &
  \colhead{} &
  \colhead{(mag)} &
  \colhead{(mag)} &
  \colhead{} &
  \colhead{(mag)} &
  \colhead{(mag)} &
  \colhead{} &
  \colhead{(mag)} &
  \colhead{(mag)} &
  \colhead{} &
  \colhead{(mag)} &
  \colhead{(mag)} &
  \colhead{} &
  \colhead{(mag)} &
  \colhead{(mag)} &
  \colhead{} &
  \colhead{(mag)} &
  \colhead{(mag)} &
  \colhead{} &
  \colhead{(mag)} &
  \colhead{(mag)} &
  \colhead{} &
  \colhead{(mag)} &
  \colhead{(mag)}
}
\startdata
M6--M6.9\phn & 30 &  & 16.71 & 0.44 &  & 15.37 & 0.43 &  & 13.25 & 0.34 &  & 12.27 & 0.32 &  & 11.76 & 0.30 &  & 10.36 & 0.30 &  & 9.76 & 0.30 &  & 9.43 & 0.31 &  & 9.22 & 0.30 &  & 9.01 & 0.30 &  & 8.72 & 0.34 \\
M7--M7.9\phn & 26 &  & 18.11 & 0.56 &  & 16.76 & 0.50 &  & 14.18 & 0.39 &  & 12.98 & 0.34 &  & 12.31 & 0.31 &  & 10.77 & 0.30 &  & 10.14 & 0.31 &  & 9.75 & 0.31 &  & 9.52 & 0.31 &  & 9.31 & 0.31 &  & 8.96 & 0.33 \\
M8--M8.9\phn & 21 &  & 19.19 & 0.59 &  & 17.74 & 0.51 &  & 15.00 & 0.48 &  & 13.61 & 0.45 &  & 12.81 & 0.43 &  & 11.15 & 0.42 &  & 10.47 & 0.43 &  & 10.03 & 0.43 &  & 9.78 & 0.43 &  & 9.55 & 0.44 &  & 9.18 & 0.47 \\
M9--M9.9\phn & 9 &  & 19.95 & 0.43 &  & 18.14 & 0.37 &  & 15.62 & 0.39 &  & 14.17 & 0.37 &  & 13.23 & 0.36 &  & 11.46 & 0.34 &  & 10.74 & 0.35 &  & 10.27 & 0.37 &  & 9.95 & 0.37 &  & 9.68 & 0.37 &  & 9.24 & 0.42 \\
L0--L0.9\phn & 17 &  & 20.33 & 0.40 &  & 18.37 & 0.31 &  & 16.00 & 0.26 &  & 14.52 & 0.25 &  & 13.58 & 0.23 &  & 11.76 & 0.18 &  & 11.00 & 0.23 &  & 10.52 & 0.24 &  & 10.22 & 0.26 &  & 9.94 & 0.26 &  & 9.35 & 0.32 \\
L1--L1.9\phn & 19 &  & 20.85 & 0.78 &  & 18.74 & 0.28 &  & 16.41 & 0.25 &  & 14.93 & 0.23 &  & 13.97 & 0.21 &  & 12.03 & 0.15 &  & 11.23 & 0.21 &  & 10.71 & 0.24 &  & 10.36 & 0.26 &  & 10.10 & 0.26 &  & 9.66 & 0.31 \\
L2--L2.9\phn & 12 &  & 21.24 & \nodata &  & 19.02 & 0.29 &  & 16.73 & 0.26 &  & 15.30 & 0.24 &  & 14.33 & 0.24 &  & 12.32 & 0.21 &  & 11.41 & 0.25 &  & 10.83 & 0.29 &  & 10.43 & 0.32 &  & 10.14 & 0.33 &  & 9.62 & 0.37 \\
L3--L3.9\phn & 8 &  & 22.51 & \nodata &  & 19.61 & 0.39 &  & 17.40 & 0.34 &  & 15.88 & 0.31 &  & 14.89 & 0.29 &  & 12.77 & 0.24 &  & 11.82 & 0.29 &  & 11.19 & 0.32 &  & 10.66 & 0.36 &  & 10.39 & 0.38 &  & 9.91 & 0.37 \\
L4--L4.9\phn & 8 &  & \nodata & \nodata &  & 20.60 & 0.56 &  & 18.35 & 0.38 &  & 16.68 & 0.33 &  & 15.66 & 0.32 &  & 13.51 & 0.28 &  & 12.45 & 0.33 &  & 11.85 & 0.40 &  & 11.18 & 0.43 &  & 10.84 & 0.46 &  & 10.05 & 0.44 \\
L5--L5.9\phn & 8 &  & \nodata & \nodata &  & 20.74 & 0.37 &  & 18.71 & 0.33 &  & 16.94 & 0.29 &  & 15.87 & 0.28 &  & 13.69 & 0.25 &  & 12.63 & 0.30 &  & 11.94 & 0.34 &  & 11.26 & 0.34 &  & 10.91 & 0.38 &  & 10.13 & 0.51 \\
L6--L6.9\phn & 9 &  & \nodata & \nodata &  & 21.21 & 0.78 &  & 19.27 & 0.65 &  & 17.35 & 0.62 &  & 16.27 & 0.61 &  & 14.18 & 0.60 &  & 13.19 & 0.62 &  & 12.58 & 0.68 &  & 11.85 & 0.71 &  & 11.49 & 0.75 &  & 10.53 & 0.87 \\
L7--L7.9\phn & 7 &  & \nodata & \nodata &  & \nodata & \nodata &  & 20.09 & 0.36 &  & 18.18 & 0.26 &  & 17.13 & 0.25 &  & 14.94 & 0.20 &  & 13.82 & 0.31 &  & 13.12 & 0.32 &  & 12.42 & 0.48 &  & 11.99 & 0.57 &  & 10.87 & 1.08 \\
L8--L8.9\phn & 10 &  & \nodata & \nodata &  & 22.88 & \nodata &  & 20.38 & 0.39 &  & 18.10 & 0.22 &  & 17.04 & 0.21 &  & 14.90 & 0.13 &  & 13.77 & 0.20 &  & 13.04 & 0.22 &  & 12.20 & 0.28 &  & 11.66 & 0.30 &  & 10.66 & 0.36 \\
L9--L9.9\phn & 3 &  & \nodata & \nodata &  & \nodata & \nodata &  & 20.09 & 0.79 &  & 17.69 & 0.73 &  & 16.57 & 0.72 &  & 14.46 & 0.71 &  & 13.39 & 0.73 &  & 12.79 & 0.74 &  & 11.94 & 0.74 &  & 11.45 & 0.77 &  & 10.48 & 0.75 \\
T0--T0.9\phn & 1 &  & \nodata & \nodata &  & \nodata & \nodata &  & 20.22 & \nodata &  & 17.98 & \nodata &  & 16.77 & \nodata &  & 14.56 & \nodata &  & 13.62 & \nodata &  & 13.08 & \nodata &  & 12.43 & \nodata &  & 11.94 & \nodata &  & 11.02 & \nodata \\
T1--T1.9\phn & 3 &  & \nodata & \nodata &  & \nodata & \nodata &  & 21.10 & 1.14 &  & 18.84 & 0.21 &  & 17.45 & 0.16 &  & 15.25 & 0.12 &  & 14.39 & 0.18 &  & 14.07 & 0.33 &  & 13.39 & 0.38 &  & 12.68 & 0.33 &  & 11.26 & 0.76 \\
T2--T2.9\phn & 5 &  & \nodata & \nodata &  & \nodata & \nodata &  & 21.97 & \nodata &  & 18.26 & 0.23 &  & 16.75 & 0.13 &  & 14.54 & 0.06 &  & 13.73 & 0.15 &  & 13.48 & 0.26 &  & 12.94 & 0.39 &  & 12.06 & 0.33 &  & 10.61 & 0.40 \\
T3--T3.9\phn & 2 &  & \nodata & \nodata &  & \nodata & \nodata &  & \nodata & \nodata &  & 18.08 & 0.25 &  & 16.50 & 0.22 &  & 14.26 & 0.16 &  & 13.67 & 0.24 &  & 13.44 & 0.32 &  & 12.84 & 0.61 &  & 11.72 & 0.45 &  & \nodata & \nodata \\
T4--T4.9\phn & 6 &  & \nodata & \nodata &  & \nodata & \nodata &  & \nodata & \nodata &  & 18.02 & 0.39 &  & 16.32 & 0.38 &  & 13.89 & 0.36 &  & 13.57 & 0.40 &  & 13.49 & 0.39 &  & 13.31 & 0.41 &  & 11.90 & 0.39 &  & 10.70 & \nodata \\
T5--T5.9\phn & 7 &  & \nodata & \nodata &  & \nodata & \nodata &  & 22.69 & \nodata &  & 19.20 & 0.22 &  & 17.43 & 0.18 &  & 14.94 & 0.12 &  & 14.76 & 0.24 &  & 14.77 & 0.23 &  & 14.52 & 0.25 &  & 12.69 & 0.24 &  & 11.37 & \nodata \\
T6--T6.9\phn & 11 &  & \nodata & \nodata &  & \nodata & \nodata &  & \nodata & \nodata &  & 19.82 & 0.32 &  & 18.06 & 0.32 &  & 15.53 & 0.27 &  & 15.48 & 0.37 &  & 15.37 & 0.41 &  & 15.08 & 0.35 &  & 13.02 & 0.42 &  & 11.61 & 0.37 \\
T7--T7.9\phn & 7 &  & \nodata & \nodata &  & \nodata & \nodata &  & \nodata & \nodata &  & 21.17 & 0.78 &  & 19.34 & 0.81 &  & 16.78 & 0.76 &  & 16.70 & 0.78 &  & 16.70 & 0.93 &  & 16.23 & 0.84 &  & 14.11 & 0.95 &  & 12.26 & 0.76 \\
T8--T8.9\phn & 4 &  & \nodata & \nodata &  & \nodata & \nodata &  & \nodata & \nodata &  & 21.52 & 0.52 &  & 19.75 & 0.59 &  & 17.18 & 0.51 &  & 17.09 & \nodata &  & \nodata & \nodata &  & 16.58 & 0.84 &  & 14.03 & 0.92 &  & 12.27 & 0.64 \\
T9--T9.9\phn & 1 &  & \nodata & \nodata &  & \nodata & \nodata &  & \nodata & \nodata &  & 21.82 & \nodata &  & 20.37 & \nodata &  & 17.75 & \nodata &  & 17.51 & \nodata &  & \nodata & \nodata &  & 16.70 & \nodata &  & 13.81 & \nodata &  & 11.97 & \nodata \\
\enddata
\tablecomments{SEDs constructed using parallaxes from the literature to
  determine absolute \jtwo~magnitudes, for which we report the weighted mean and
  rms for each spectral subtype bin.  The number of objects used to determine
  M$_{\jtwo}$ for each spectral subtype is listed in the second column ($N$).
  We use the median colors from our catalog to calculate absolute magnitudes in 
  other bands.  Binaries, subdwarfs, and young objects were excluded from these
  SEDs.}
\end{deluxetable}
   % Table of SEDs (absolute magnitudes)

\begin{figure*}
\begin{center}
  \begin{minipage}[t]{0.49\textwidth}
    \includegraphics[width=1.00\columnwidth, trim = 10mm 0 10mm 0]{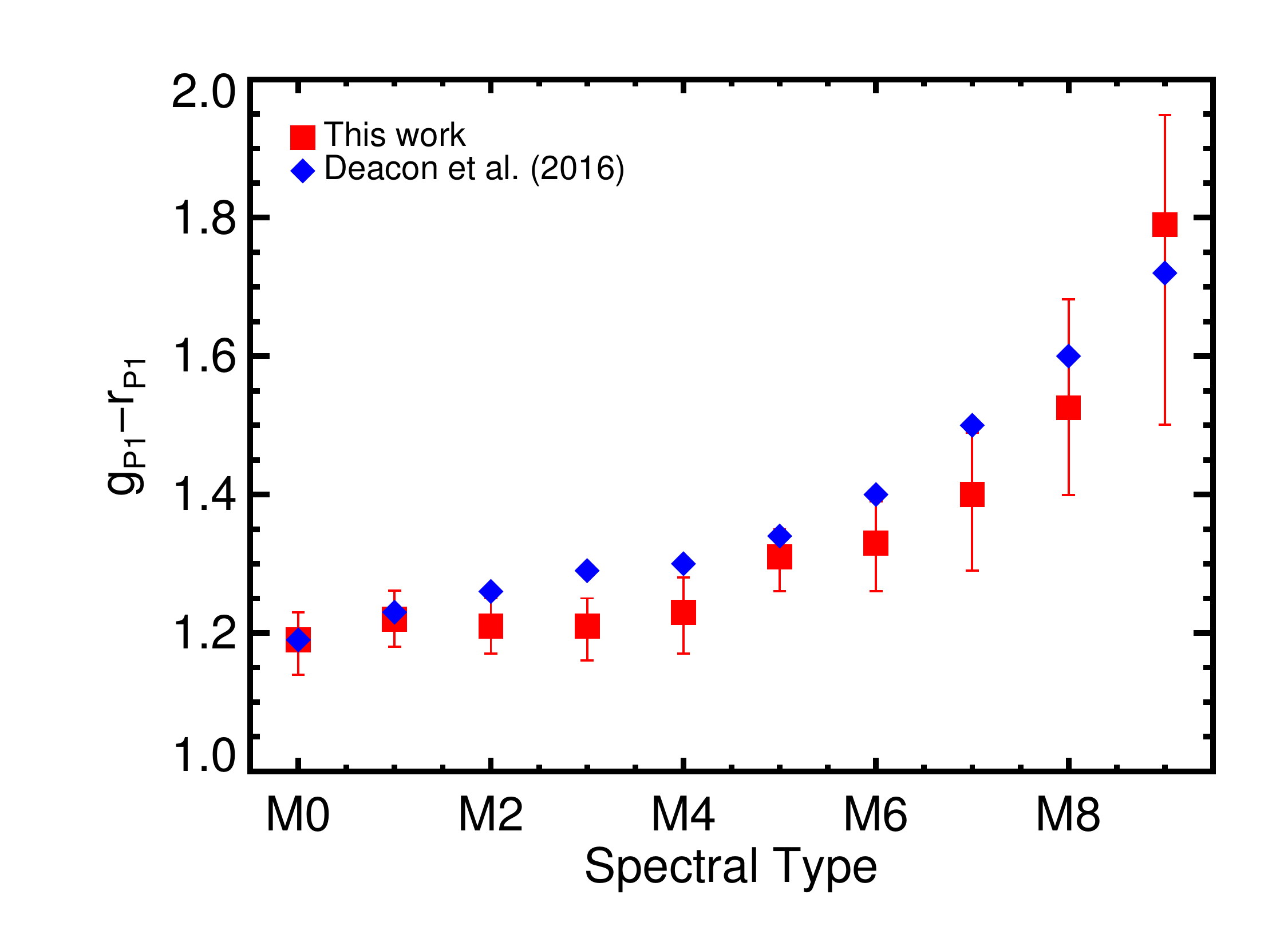}
  \end{minipage}
  \hfill
  \begin{minipage}[t]{0.49\textwidth}
    \includegraphics[width=1.00\columnwidth, trim = 10mm 0 10mm 0]{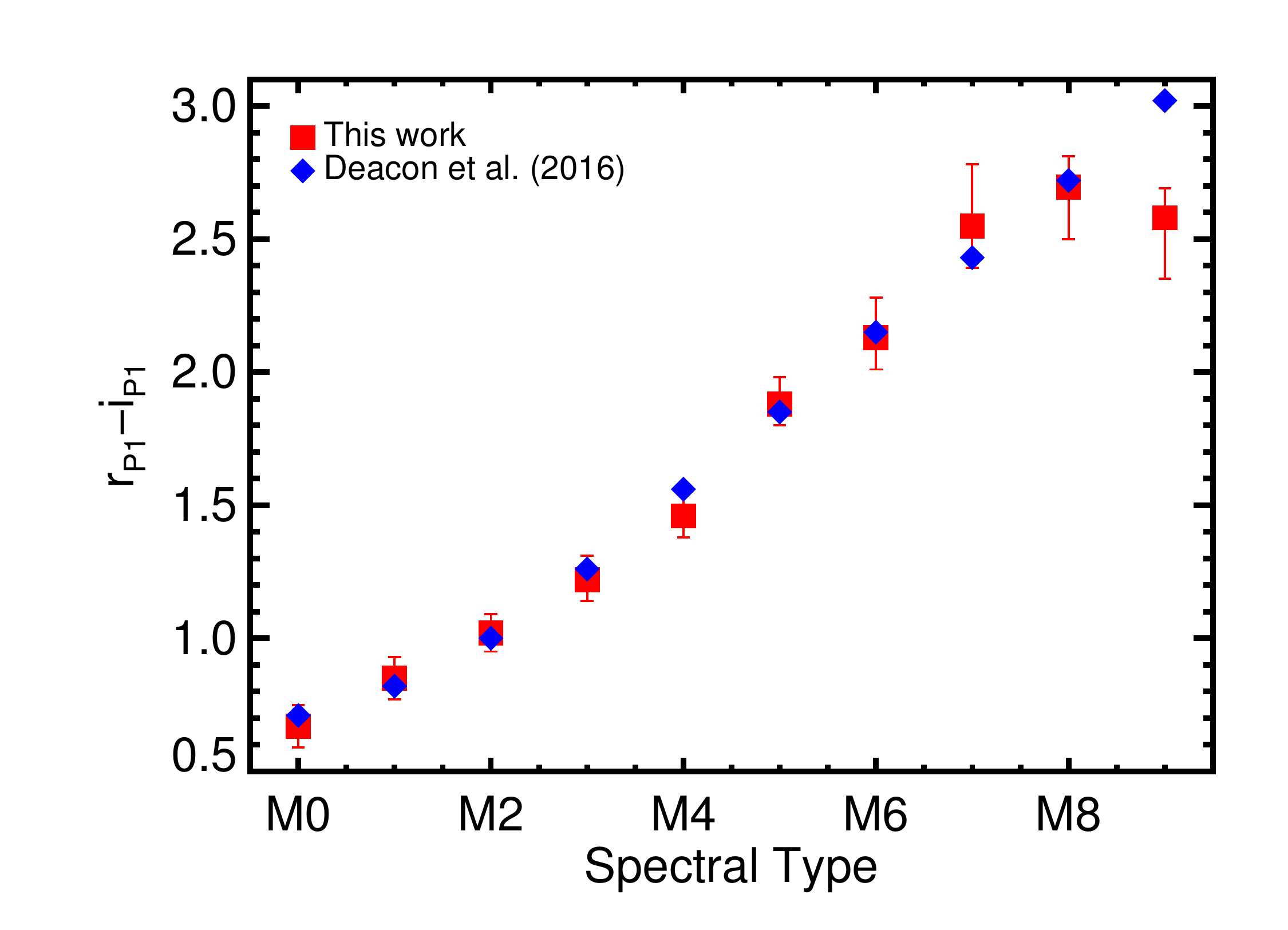}
  \end{minipage}
  \begin{minipage}[t]{0.49\textwidth}
    \includegraphics[width=1.00\columnwidth, trim = 10mm 0 10mm 0]{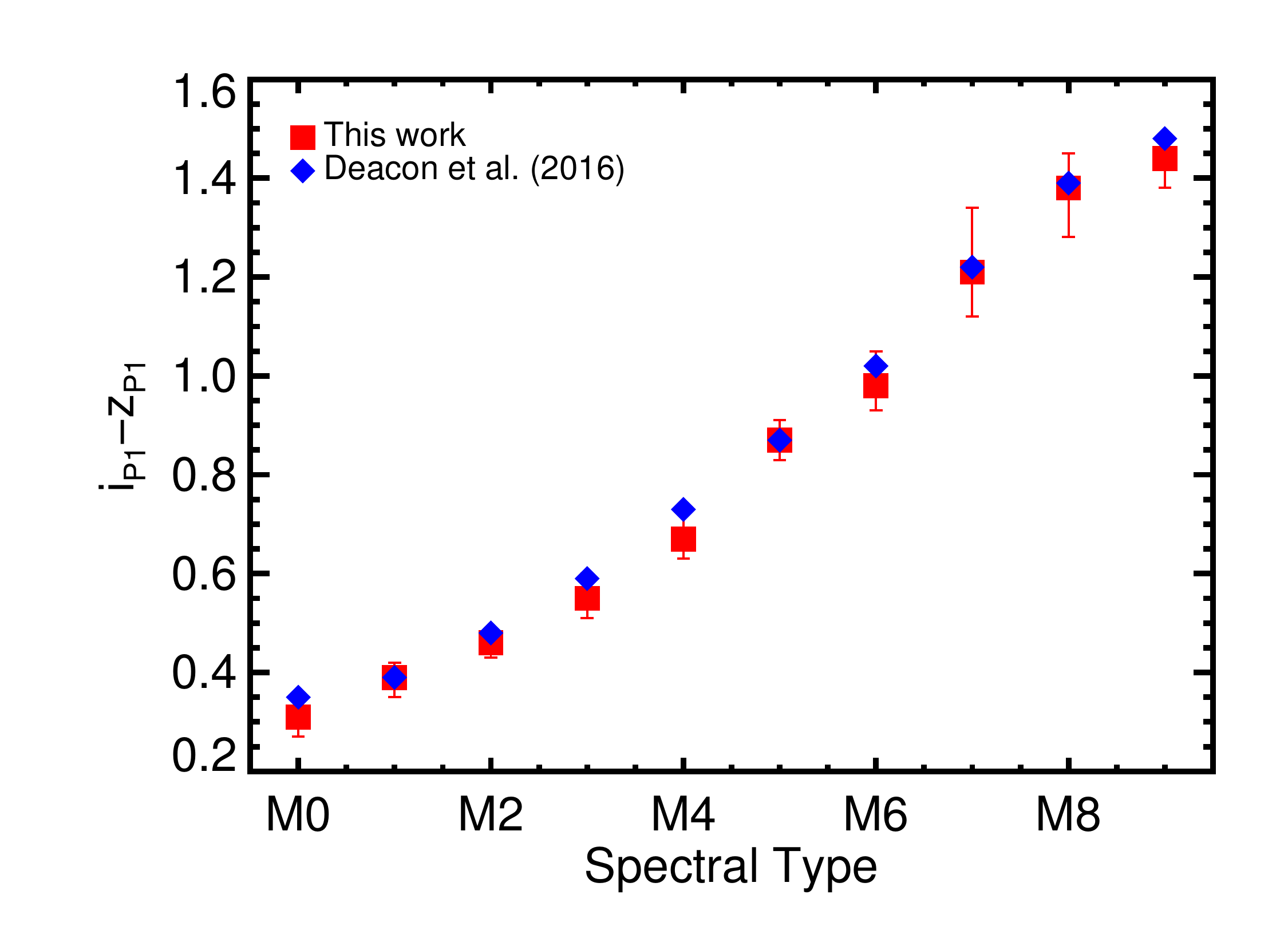}
  \end{minipage}
  \hfill
  \begin{minipage}[t]{0.49\textwidth}
    \includegraphics[width=1.00\columnwidth, trim = 10mm 0 10mm 0]{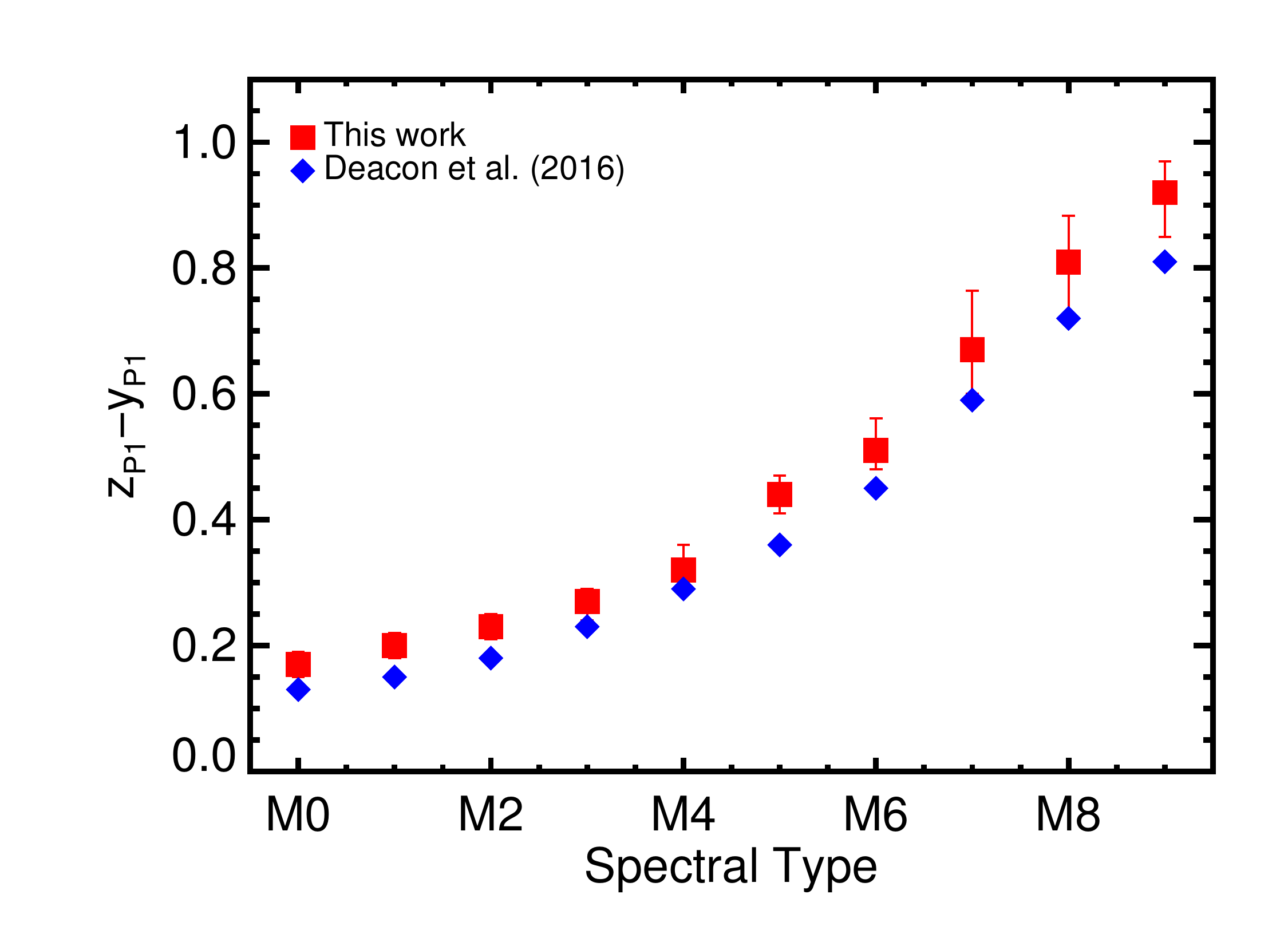}
  \end{minipage}
  \caption{The colors of M dwarfs (median and 68\% confidence intervals, red
    squares) as a function of spectral type, compared with the colors derived
    from the SED templates (blue diamonds) of \citet{Deacon:2016bh}.  Our colors
    are generally quite consistent with those of D16, with small variations
    resulting from D16's use of an earlier processing version of PS1 data (PV2)
    and differences in our input samples.  The blueward turn in \ri\ at spectral
    type M9 was not identified by D16 because of their use of \gi\ as a proxy
    for spectral type, which does not clearly distinguish between the late-M
    spectral subtypes.}
  \label{fig.colors.deacon}
\end{center}
\end{figure*}

Recently, \citet[hereinafter D16]{Deacon:2016bh} published empirical SEDs for
the \PS\ photometric system for spectral types B8V--M9V.  Our two sets of SEDs
have only spectral types M6--M9 in common.  For these late-M types, the SEDs are
consistent within our uncertainties, but we note that our absolute magnitudes
are mostly $\approx$$0.1-0.3$~mag fainter.  In Figure~\ref{fig.colors.deacon} we
compare our PS1 colors for M~dwarfs from Table~\ref{tbl.meancols} to the PS1
colors from D16.  The colors are generally quite consistent, although our \gr\
colors are $\approx$$0.5\sigma-1\sigma$ bluer and our \zy\ colors are
$\approx$~$1\sigma-2\sigma$ redder than those of D16.  The differences in 
colors are due to the fact that D16 used an earlier processing version (PV2) of
PS1 data, and likely also to differences in our input samples (DL16 used
$\approx$500~M~dwarfs, while our sample contains over 8000~M~dwarfs).  We also
find a significant blueward turn in the \ri\ colors of M9 dwarfs (that continues
into the L dwarfs) that D16 do not identify.  D16 converted spectral types into
absolute magnitudes using bolometric magnitudes and a series of color
transformations fitted with splines.  In particular, D16 used \gi\ as a proxy
for spectral type, and this relation does not clearly distinguish M9 dwarfs from
M6--M8 dwarfs, so the sudden turn for M9 in \ri\ could not be detected by their
method.

\section{Proper Motion}
\label{pm}
Table~\ref{tbl.pm} presents our proper motions for \varnpmtot ~M, L, and T
dwarfs based on PS1 astrometry, along with parallaxes (from the literature) or
photometric distances, tangential velocities, and proper motions from the
literature for comparison.  The full table is available for download in
electronic form from the online journal.  Like Table~\ref{tbl.phot},
Table~\ref{tbl.pm} is arranged in two parts: (1) the late-M, L, and T dwarfs
compiled from the literature, followed by (2) the M dwarfs from
\citet{West:2008eq}.  For reference, Table~\ref{tbl.pm} also repeats several
columns from Table~\ref{tbl.phot}, including spectral types, gravity
classifications, and flags for binaries and young objects.

\floattable
\begin{splitdeluxetable*}{lccccccccCCCBCCCCCCCCCCCBCCCCCCcCCCCc}
\centering
\tablecaption{Positions and Proper Motions of M, L, and T Dwarfs from the \PS\ 3$\pi$ Survey \label{tbl.pm}}
\tablewidth{0pt}
\tabletypesize{\tiny}
\tablehead{   % column headings
  \colhead{} &
  \multicolumn{3}{c}{Spectral Type\tablenotemark{a,b}} &
  \colhead{} &
  \multicolumn{2}{c}{Gravity\tablenotemark{b}} &
  \colhead{} &
  \colhead{} &
  \multicolumn{3}{c}{PS1 Mean Position\tablenotemark{d}} &
  \multicolumn{11}{c}{PS1 Proper Motion} &
  \multicolumn{4}{c}{} &
  \multicolumn{2}{c}{PS1 \vtan} &
  \colhead{} &
  \multicolumn{4}{c}{Literature Proper Motion} &
  \colhead{} \\
%   \cline{2-4}
%   \cline{6-7}
%   \cline{10-12}
%   \cline{13-23}
%   \cline{28-29}
%   \cline{31-34}
  \colhead{Discovery Name} &
  \colhead{Opt} &
  \colhead{NIR} &
  \colhead{Adopted} &
  \colhead{} &
  \colhead{Opt} &
  \colhead{NIR} &
  \colhead{Binary} &
  \colhead{Young\tablenotemark{c}} &
  \colhead{$\alpha_{\rm J2000}$} &
  \colhead{$\delta_{\rm J2000}$} &
  \colhead{Epoch} &
  \colhead{\mua} &
  \colhead{err$_{\mua}$} &
  \colhead{\mud} &
  \colhead{err$_{\mud}$} &
  \colhead{$\mu$} &
  \colhead{err$_\mu$} &
  \colhead{${\rm PA}_\mu$} &
  \colhead{err$_{{\rm PA}_\mu}$} &
  \colhead{\rchi} &
  \colhead{$N_{\rm ep}$} &
  \colhead{$\Delta t$} &
  \colhead{$\pi$} &
  \colhead{err$_{\pi}$} &
  \colhead{dist\tablenotemark{e}} &
  \colhead{err$_{\rm dist}$} &
  \colhead{\vtan} &
  \colhead{\stan} &
  \colhead{} &
  \colhead{\mua} &
  \colhead{err$_{\mua}$} &
  \colhead{\mud} &
  \colhead{err$_{\mud}$} &
  \colhead{References\tablenotemark{f}} \\
  \multicolumn{9}{c}{} &
  \colhead{(deg)} &
  \colhead{(deg)} &
  \colhead{(MJD)} &
  \colhead{(\my)} &
  \colhead{(\my)} &
  \colhead{(\my)} &
  \colhead{(\my)} &
  \colhead{(\my)} &
  \colhead{(\my)} &
  \colhead{(deg)} &
  \colhead{(deg)} &
  \colhead{} &
  \colhead{} &
  \colhead{(yr)} &
  \colhead{(mas)} &
  \colhead{(mas)} &
  \colhead{(pc)} &
  \colhead{(pc)} &
  \colhead{(\kms)} &
  \colhead{(\kms)} &
  \colhead{} &
  \colhead{(\my)} &
  \colhead{(\my)} &
  \colhead{(\my)} &
  \colhead{(\my)} &
  \colhead{($\pi$; $\mu_{\rm lit}$)}
}
\startdata
SDSS J000013.54+255418.6 & T5 & T4.5 & T4.5 &  & \nodata & \nodata & \nodata & \nodata & 0.056400 & 25.905389 & 55326.38 & -18.4 & 5.5 & 123.1 & 3.3 & 124.5 & 3.4 & 351.5 & 2.5 & 1.4 & 28 & 15.2 & 70.80 & 1.90 & 14.12 & 0.38 & 8.3 & 0.3 &  & -19.1 & 1.5 & 126.7 & 1.3 & 43; 43 \\
SDSS J000112.18+153535.5 & \nodata & L3.7 \intg & L3.7 \intg &  & \nodata & \intg & \nodata & Y & 0.301175 & 15.592648 & 55750.95 & 137.3 & 2.2 & -181.2 & 2.8 & 227.3 & 2.6 & 142.8 & 0.6 & 1.6 & 34 & 14.1 & \nodata & \nodata & 24.34 & 2.93 & 26.2 & 3.2 &  & 135.2 & 10.7 & -169.6 & 13.7 & --; 54 \\
WISEA J000131.93$-$084126.9 & \nodata & L1 pec (blue) & L1 pec (blue) &  & \nodata & \nodata & \nodata & \nodata & 0.383130 & -8.690953 & 55632.76 & 339.9 & 2.6 & -304.9 & 3.3 & 456.6 & 2.9 & 131.9 & 0.4 & 0.7 & 46 & 15.8 & \nodata & \nodata & \nodata & \nodata & \nodata & \nodata &  & 331.0 & 14.0 & -299.0 & 14.0 & --; 90 \\
SDSS J000250.98+245413.8 & \nodata & L5.5 & L5.5 &  & \nodata & \nodata & \nodata & \nodata & 0.712466 & 24.903795 & 55309.02 & 22.4 & 17.4 & -45.6 & 7.2 & 50.8 & 9.9 & 153.8 & 17.6 & 1.1 & 28 & 15.2 & \nodata & \nodata & 53.33 & 6.54 & 12.8 & 3.0 &  & \nodata & \nodata & \nodata & \nodata & --; -- \\
2MASSI J0003422$-$282241 & M7.5 & M7: \fldg & M7.5 &  & \nodata & \fldg & \nodata & \nodata & 0.927389 & -28.378609 & 56120.45 & 285.8 & 1.5 & -142.6 & 1.3 & 319.4 & 1.5 & 116.5 & 0.2 & 1.5 & 67 & 16.1 & 25.70 & 0.93 & 38.91 & 1.41 & 58.9 & 2.2 &  & 280.8 & 0.9 & -141.5 & 0.8 & 131; 131 \\
2MASS J00044144$-$2058298 & M8 & \nodata & M8 &  & \nodata & \nodata & \nodata & \nodata & 1.175450 & -20.974698 & 55862.58 & 751.9 & 1.9 & 91.9 & 2.5 & 757.5 & 1.9 & 83.0 & 0.2 & 2.0 & 74 & 15.1 & \nodata & \nodata & 18.94 & 3.14 & 68.0 & 11.3 &  & 825.5 & 75.7 & -9.2 & 75.5 & --; 70 \\
2MASS J00054844$-$2157196 & M9 & \nodata & M9 &  & \nodata & \nodata & \nodata & \nodata & 1.454361 & -21.955894 & 55708.99 & 709.4 & 2.2 & -122.9 & 2.5 & 720.0 & 2.2 & 99.8 & 0.2 & 1.7 & 66 & 14.9 & \nodata & \nodata & 25.43 & 4.22 & 86.8 & 14.4 &  & 703.0 & 24.0 & -119.0 & 4.0 & --; 76 \\
ULAS J000613.24+154020.7 & \nodata & L9 & L9 &  & \nodata & \nodata & \nodata & \nodata & 1.555343 & 15.672398 & 56032.02 & 84.7 & 40.6 & -38.5 & 20.4 & 93.0 & 35.6 & 114.4 & 21.4 & 0.9 & 16 & 3.9 & \nodata & \nodata & 43.24 & 5.35 & 19.1 & 7.7 &  & 92.9 & 23.3 & -73.8 & 23.3 & --; 74 \\
SDSS J000614.06+160454.5 & L0 & \nodata & L0 &  & \nodata & \nodata & \nodata & \nodata & 1.558610 & 16.081685 & 55734.32 & 4.4 & 4.1 & -41.8 & 3.9 & 42.0 & 3.9 & 174.0 & 5.6 & 0.7 & 40 & 17.2 & \nodata & \nodata & 103.52 & 17.46 & 20.6 & 4.0 &  & -5.0 & 19.0 & -40.0 & 19.0 & --; 139 \\
\enddata
\tablenotetext{a}{Spectral types taken from the literature
  (Section~\ref{catalog.spt}).  When both optical and near-IR types are
  available, we adopt the optical type for M and L~dwarfs and the near-IR type
  for T~dwarfs.  Most spectral types have an uncertainty of $\pm0.5$~subtypes;
  ``:'' = $\pm1$~subtype; ``::'' = $\pm2$ or more~subtypes. ``sd'' = subdwarf;
  ``esd'' = extreme subdwarf \citep{Gizis:1997cj}.}
\tablenotetext{b}{$\beta$, $\gamma$, and $\delta$ indicate classes of
  increasingly low gravity based on optical
  \citep{Kirkpatrick:2005cv,Cruz:2009gs} or near-infrared \citep{Gagne:2015dc}
  spectra.  \textsc{fld-g} indicates near-infrared spectral signatures of
  field-age gravity, \textsc{int-g} indicates intermediate gravity, and
  \textsc{vl-g} indicates very low gravity \citep{Allers:2013hk}.}
\tablenotetext{c}{Young objects identified by low-gravity classifications or
  other spectroscopic evidence for youth, membership in star-forming regions or
  young moving groups, or companionship to a young star
  (Section~\ref{catalog.young}).}
\tablenotetext{d}{The mean PS1 position is the position calculated
  for the weighted mean epoch, in which the epochs are weighted by the rms of the
  R.A. and Decl. astrometric uncertainties.}
\tablenotetext{e}{Distances were calculated from trigonometric parallaxes when
  available.  If no parallax was available, we calculated photometric distances
  using \ips~magnitudes for spectral types M0--M5 and $W2$~magnitudes (when
  available) for spectral types M6 and later.}
\tablenotetext{f}{References for spectral type, gravity, and binarity are given
  in Table~\ref{tbl.phot}.}
\tablenotetext{g}{Although classified as \fldg, the spectrum shows hints of
  intermediate gravity \citep[as described in][]{Aller:2016kg}.}
\tablenotetext{h}{The R.A. and Decl. proper motion components listed in Table 2
  of \citet{Lodieu:2012il} appear to be reversed.  We quote the corrected order
  here.}
\tablerefs{(1) \citet{Aberasturi:2014cc}, (2) \citet{Albert:2011bc}, (3) \citet{Aller:2016kg}, (4) \citet{Andrei:2011jm}, (5) \citet{Artigau:2011hr}, (6) \citet{BardalezGagliuffi:2014fl}, (7) \citet{Baron:2015fn}, (8) \citet{Bartlett:2007tk}, (9) \citet{Beamin:2013fx}, (10) \citet{Bejar:2008ev}, (11) \citet{Best:2015em}, (12) \citet{Best:2017ih}, (13) \citet{Bihain:2013gw}, (14) \citet{Bouvier:2008kf}, (15) \citet{Bouy:2015gl}, (16) \citet{Burgasser:2003ij}, (17) \citet{Burgasser:2004hg}, (18) \citet{Burgasser:2007ki}, (19) \citet{Burgasser:2008cj}, (20) \citet{Burgasser:2008ke}, (21) \citet{Burgasser:2012ga}, (22) \citet{Burgasser:2015fd}, (23) \citet{Burgasser:2016cx}, (24) \citet{Burningham:2013gt}, (25) \citet{Caballero:2007dk}, (26) \citet{Cardoso:2015eu}, (27) \citet{Casewell:2008gn}, (28) \citet{Castro:2013bb}, (29) \citet{Castro:2016dr}, (30) \citet{Costa:2005fd}, (31) \citet{Costa:2006gd}, (32) \citet{Dahn:2002fu}, (33) \citet{Dahn:2008bo}, (34) \citet{Deacon:2005jf}, (35) \citet{Deacon:2007jl}, (36) \citet{Deacon:2009bb}, (37) \citet{Deacon:2011gz}, (38) \citet{Deacon:2012gf}, (39) \citet{Deacon:2014ey}, (40) \citet{Deacon:2017kd}, (41) \citet{Dieterich:2014ic}, (42) \citet{Dittmann:2014cr}, (43) \citet{Dupuy:2012bp}, (44) \citet{Dupuy:2013ks}, (45) \citet{Dupuy:2015gl}, (46) \citet{Faherty:2009kg}, (47) \citet{Faherty:2010gt}, (48) \citet{Faherty:2012cy}, (49) \citet{Faherty:2016fx}, (50) \citet{Finch:2016hw}, (51) \citet{Folkes:2012ep}, (52) \citet{Forbrich:2016ek}, (53) \citet{Gagne:2014gp}, (54) \citet{Gagne:2015ij}, (55) \citet{Gatewood:2009cd}, (56) \citet{Gauza:2015fw}, (57) \citet{Gawronski:2016kl}, (58) \citet{Gelino:2014ie}, (59) \citet{Girard:2011jo}, (60) \citet{Gizis:2011jv}, (61) \citet{Gizis:2013ik}, (62) \citet{Gizis:2015ey}, (63) \citet{Hambly:2001be}, (64) \citet{Harrington:1993jc}, (65) \citet{Henry:2006jp}, (66) \citet{Hog:2000wk}, (67) \citet{Jameson:2008ha}, (68) \citet{Kellogg:2016fo}, (69) \citet{Kendall:2004kb}, (70) \citet{Kendall:2007fd}, (71) \citet{Kirkpatrick:2010dc}, (72) \citet{Kirkpatrick:2011ey}, (73) \citet{Kirkpatrick:2014kv}, (74) \citet{Lawrence:2012wh}, (75) \citet{Leggett:2012gg}, (76) \citet{Lepine:2002gc}, (77) \citet{Lepine:2005jx}, (78) \citet{Lepine:2009hi}, (79) \citet{Lepine:2011gl}, (80) \citet{Liu:2011hc}, (81) \citet{Liu:2016co}, (82) \citet{Lodieu:2005kd}, (83) \citet{Lodieu:2006di}, (84) \citet{Lodieu:2012go}, (85) \citet{Lodieu:2012il}, (86) \citet{Lodieu:2013cj}, (87) \citet{Lodieu:2013eo}, (88) \citet{Lodieu:2014jo}, (89) \citet{Luhman:2012ir}, (90) \citet{Luhman:2014hj}, (91) \citet{Luyten:1979vs}, (92) \citet{Manjavacas:2013cg}, (93) \citet{Marocco:2010cj}, (94) \citet{Marocco:2013kv}, (95) \citet{Marsh:2013hk}, (96) \citet{McCaughrean:2004ey}, (97) \citet{Monet:1992bs}, (98) \citet{Monet:2003bw}, (99) \citet{Muzic:2012hc}, (100) \citet{PhanBao:2008kz}, (101) \citet{PhanBao:2011cb}, (102) \citet{Pokorny:2004fd}, (103) \citet{Qi:2015kr}, (104) \citet{Radigan:2008jd}, (105) \citet{Reid:2003kx}, (106) \citet{Riedel:2014ce}, (107) \citet{Roeser:2010cr}, (108) \citet{Sahlmann:2014hu}, (109) \citet{Sahlmann:2015ku}, (110) \citet{Sahlmann:2016iy}, (111) \citet{Salim:2003gv}, (112) \citet{Schilbach:2009ja}, (113) \citet{Schmidt:2007ig}, (114) \citet{Schmidt:2010ex}, (115) \citet{Schneider:2016iq}, (116) \citet{Scholz:2009fp}, (117) \citet{Scholz:2014hk}, (118) \citet{Schweitzer:1999wb}, (119) \citet{Seifahrt:2010jd}, (120) \citet{Sheppard:2009ed}, (121) \citet{Shkolnik:2012cs}, (122) \citet{Smart:2013km}, (123) \citet{Smith:2014df}, (124) \citet{Smith:2014gw}, (125) \citet{Stern:2007ge}, (126) \citet{Thompson:2013kv}, (127) \citet{Tinney:1995fp}, (128) \citet{Tinney:1996vd}, (129) \citet{Tinney:2003eg}, (130) \citet{vanAltena:1995vg}, (131) \citet{vanLeeuwen:2007dc}, (132) \citet{Vrba:2004ee}, (133) \citet{Wang:2014fe}, (134) \citet{Weinberger:2016gy}, (135) \citet{West:2008eq}, (136) \citet{Wright:2013bo}, (137) \citet{Zacharias:2005tx}, (138) \citet{Zhang:2009kw}, (139) \citet{Zhang:2010bq}.}
\tablecomments{This table is available in its entirety in machine-readable form
  in the online journal.  A portion is shown here for guidance regarding its
  form and content. The full table contains \varnobjtot~rows.  Columns 2--8 are
  the same as in Table~\ref{tbl.phot}, and are repeated here for reference.}
\end{splitdeluxetable*}

\subsection{Method}
\label{pm.method}

\subsubsection{PS1 database proper motions}
\label{pm.method.pv3}
The procedure used to calculate the proper motions and errors for the objects in
the PS1 database will be described in detail in \citet{Magnier:2017vq}.
Briefly, all detections for an object in all filters were fit simultaneously for
proper motion and parallax using iteratively-reweighted least-squares regression
with outlier clipping.  The detections retained in the fit were then
bootstrap-resampled to determine the errors on the proper motion and parallax.
For objects having a 2MASS or \textit{Gaia}~DR1 counterpart within 1'' of the
mean PS1 position, the PS1 proper motion calculation includes these positions.
(The parallaxes will be presented in E. A. Magnier et al., 2018, in prep).  For
this paper we also convert the PS1-measured \mua\ and \mud\ to a combined proper
motion $\mu$ and position angle PA, and we calculate errors for these in a Monte
Carlo fashion.  To establish a minimum quality for our PS1 proper motions, if
either \mua\ or \mud\ for an object has an error greater than 100~\my\ and
$\mu/\sigma_\mu<3$, we do not report a proper motion for the object.  This
rejects large but very uncertain proper motions that may erroneously identify
fast-moving objects (e.g., $500\pm300$~\my), but preserves high-precision
measurements of very small proper motions that have a formal S/N~$<3$ (e.g.,
$5\pm3$~\my).

\subsubsection{PS1 mean positions and epoch}
\label{pm.method.pos}
Table~\ref{tbl.pm} includes a mean PS1 position and epoch for each object.  To
determine these, the PS1 astrometric pipeline calculates a weighted mean epoch
$t_0$, in which the epochs are weighted by the rms of the R.A. and
Decl. astrometric uncertainties.  The mean position~$(\alpha_0,\delta_0)$ is
calculated (in simple terms) by fitting the positions~$(\alpha,\delta)$ at each
epoch~$t$ to
\begin{eqnarray}
\alpha(t) & = & \alpha_0 + \pi p_\alpha(t) + \mu_\alpha (t-t_0)/{\rm cos}(\delta)
  \nonumber \\
\delta(t) & = & \delta_0 + \pi p_\delta(t) + \mu_\delta(t-t_0)
\end{eqnarray}
where $\pi$ is the parallax, and $p_\alpha$ and $p_\delta$ are the parallax
factors in R.A. and Decl., respectively.  This is also, by construction, the
position at which the covariance with the proper motion is minimized, and is
therefore the best position given the set of observations.  In cases where the
pipeline is unable to fit a proper motion to the data, the mean position
coordinates are simply the weighted means of the individual epochs.  For objects
with an associated 2MASS or \textit{Gaia}~DR1 detection, these are included in
determining the mean epoch and position.

\subsubsection{Our proper motion calculation}
\label{pm.method.recalc}
The PS1 database builds astronomical objects from individual detections at
different epochs by grouping detections within 1'' of each other.  In cases
where two distinct objects are detected within 1'' of each other at a single
epoch, detections at other epochs are associated with the closer of the two
objects \citep[for details, see][]{Magnier:2017vq}.  This procedure is
successful for stationary and slow-moving point sources, but for an object that
moved $\gtrsim$1'' over the four-year timespan of the PS1 survey, the detections
may not all be associated in the PS1 database.  In our ultracool catalog, we
found that objects with proper motions $\gtrsim200$~\my\ were often split into
two or more ``partial objects'', identifiable by their proximity on the sky,
similar photometry, and astrometric consistency with proper motions from the
literature.  These partial objects often had significantly different proper
motion measurements but offered no \textit{a priori} way to determine which (if
any) of the measurements was correct.  This motivated us to recalculate the
proper motions for fast-moving objects, using an automated procedure to identify
all the detections for an object along its path of motion.  While our primary
goal was to improve the proper motions of fast-moving objects, we performed the
recalculation for all objects of spectral type M6 and later in our catalog.

To determine where to search for detections for each object, we calculated the
distance between the PS1 and 2MASS positions and generated a search box for the
PS1 data based on the implied proper motion of the distance divided by a
baseline of 12~years.  We added 4'' in quadrature to 5~years times the implied
proper motion and used the resulting value for the length of a square search
box.  We searched the PS1 and 2MASS databases for all detections within this box
centered on the PS1 position.  In cases where the 2MASS and PS1 positions were
more than 2'' apart, we also searched around the 2MASS position within a 1''
box.  Once we identified candidate detections, we determined the path on the sky
from the 2MASS position to the PS1 position, and if the 2MASS--PS1 distance was
more than 3'', we excluded any points that were more than 1'' away from the
path.  For objects in our catalog with no associated 2MASS detection, we used
the proper motion from the PS1 database (if available) or the best available
proper motion from the literature to predict a 2MASS position by projecting
backward 10~years from the PS1 position.  For objects with no associated 2MASS
detection and no PS1 or literature proper motion, we simply searched in a 4''
box around the PS1 position.  Finally, we searched the \textit{Gaia}~DR1
database for detections within 1'' of the PS1 position, and included the
astrometry from a \textit{Gaia} match if one was found.

We then calculated proper motions using the method summarized in
Section~\ref{pm.method.pv3} for objects having at least seven detections (to
ensure robust astrometric fits).  We applied the same quality standards that we
used for the PS1 database proper motions, rejecting those for which \mua\ or
\mud\ had an error greater than 100~\my\ and $\mu/\sigma_\mu<3$.  Our
recalculations produced proper motions for 2376 M6--T9 dwarfs.  These included
63 objects that do not have a proper motion in the PS1 database, most of which
are moving faster than 200~\my, demonstrating the success of our method.  The
largest proper motion for objects in our catalog reported in the PS1 database is
1561~\my, but our recalculated measurements include 16 objects with greater
proper motions, up to a maximum of 3507~\my\ (SSSPM~J1444$-$2019).  The
recalculation was unable to converge on a proper motion fit to the identified
detections for 154 objects.  41 of the objects for which the recalculation
failed do have PS1 database proper motions, but these are mostly poor
measurements ($\sigma_{\mu}\gtrsim40$~\my) for faint objects ($y\gtrsim20$~mag)
with relatively few detections ($N\lesssim10$).

\begin{figure}
\begin{center}
  \includegraphics[width=0.76\columnwidth, trim = 0mm 0 0 0]{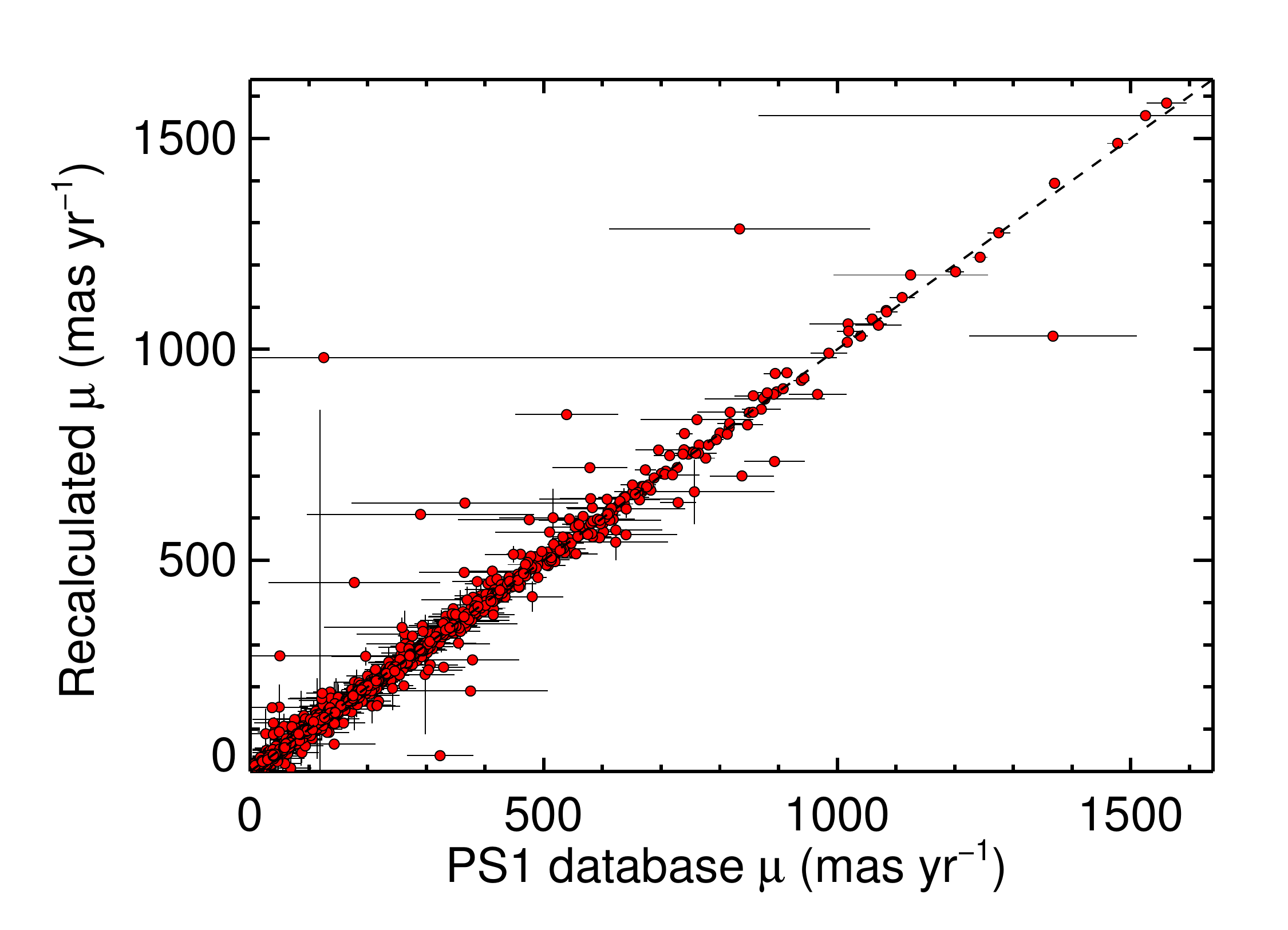}
  \includegraphics[width=0.76\columnwidth, trim = 0mm 0 0 0]{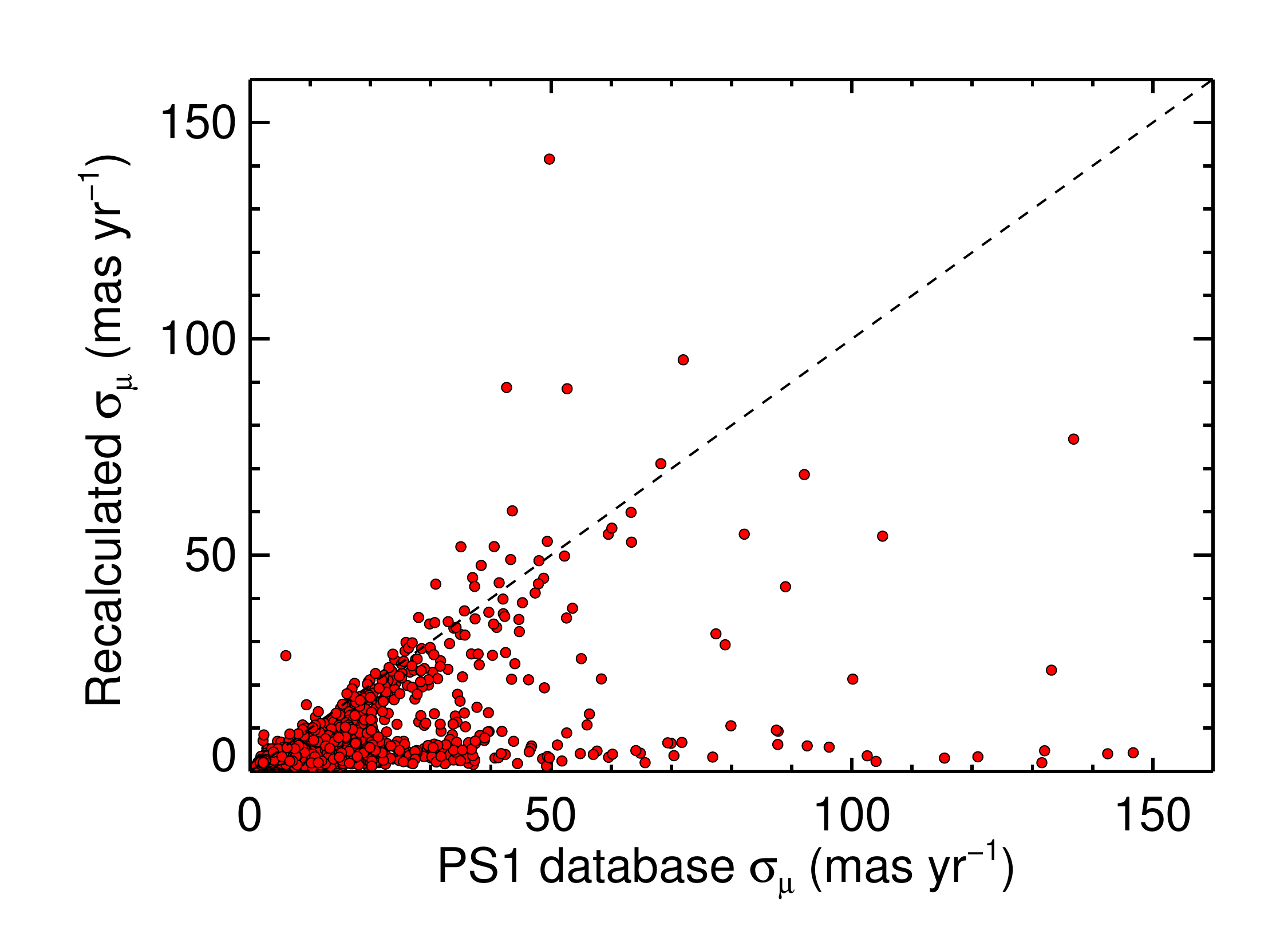}
  \caption{Comparison of the recalculated and PS1 database proper motions
    (\textit{top}) and errors (\textit{bottom}) for the M6--T9 dwarfs in our PS1
    ultracool catalog. The dashed lines indicate equal values.  The recalculated
    proper motions are strongly consistent with the database values, and in the
    cases of large discrepancy the recalculated proper motions are consistent
    with literature values (except for six cases of clear contamination by a
    nearby object).  The errors for the recalculated proper motions are lower
    for 87\% of the objects, with a median reduction of 35\%.}
  \label{fig.pm.pv3.recalc}
\end{center}
\end{figure}

Figure~\ref{fig.pm.pv3.recalc} compares our recalculated proper motions and
errors to the values in the PS1 database.  The recalculated proper motions are
strongly consistent with the database measurements, and the errors for the
recalculated proper motions are lower for 87\% of the objects, with a median
improvement in precision of 35\%.  We examined all objects for which the
recalculated \mua\ or \mud\ differed by more than three times the error on the
corresponding value in the PS1 database, and identified only six cases in which
the recalculation produced a result clearly inconsistent with the separation
between the PS1 and 2MASS positions or with a literature measurement.  In each
of these six cases, a nearby object appears to have significantly contaminated
the proper motion calculation.  We rejected the recalculated values for these
six objects.  In all other cases of significant discrepancy, the recalculated
proper motion was consistent with a value in the literature (except for one
object with no literature value), confirming that our recalculation improved the
accuracy of these proper motions.

For our catalog (Table~\ref{tbl.pm}), we adopt the recalculated proper motions
by default.  We also present mean positions and epochs from our recalculations,
which incorporate the 2MASS and \textit{Gaia}~DR1 positions used in the
calculations.  We use the PS1 database proper motions and positions only for the
six contaminated recalculations described above and in 41 cases where our
recalculation was unable to fit a proper motion.  We also recalculated chip
photometry for each object along with the proper motion, but we did not apply
the rigorous outlier-clipping procedure used for the PS1 database photometry, so
by default we use the database photometry in our catalog.  We use our
recalculated chip photometry in 55 cases where the database reports photometry
of insufficient quality (or none at all) and our recalculated photometry passes
the quality standards described in Section~\ref{phot.ps1.catalog}.

\subsection{Characteristics}
\label{pm.char}
For the remainder of Section~\ref{pm}, we restrict our discussion to objects
with spectral types M6 and later in order to focus our kinematic analysis on
ultracool dwarfs near the Sun.  Early-M dwarfs are visible at distances well
beyond the solar neighborhood ($\gtrsim$200~pc) where large-scale galactic
motions dominate the kinematics, entailing a discussion that is beyond the scope
of this paper.  We report PS1 proper motions for a total of \varnpmucool~M6--T9
dwarfs, including the largest sets of uniformly-calculated proper motions for
confirmed L~dwarfs (\varnpml~objects) and T~dwarfs (\varnpmt~objects) to date.
We caution that the proper motions in our catalog do not comprise a
clearly-defined sample and reflect biases inherited from the programs that
discovered the objects (Section~\ref{catalog.completeness}), but our proper
motions nevertheless serve as a large and illustrative sample of the local
ultracool population.

\begin{figure}   % update with ps1_pm_figures.pro
\begin{center}
  \includegraphics[width=0.76\columnwidth, trim = 10mm 0 0 0]{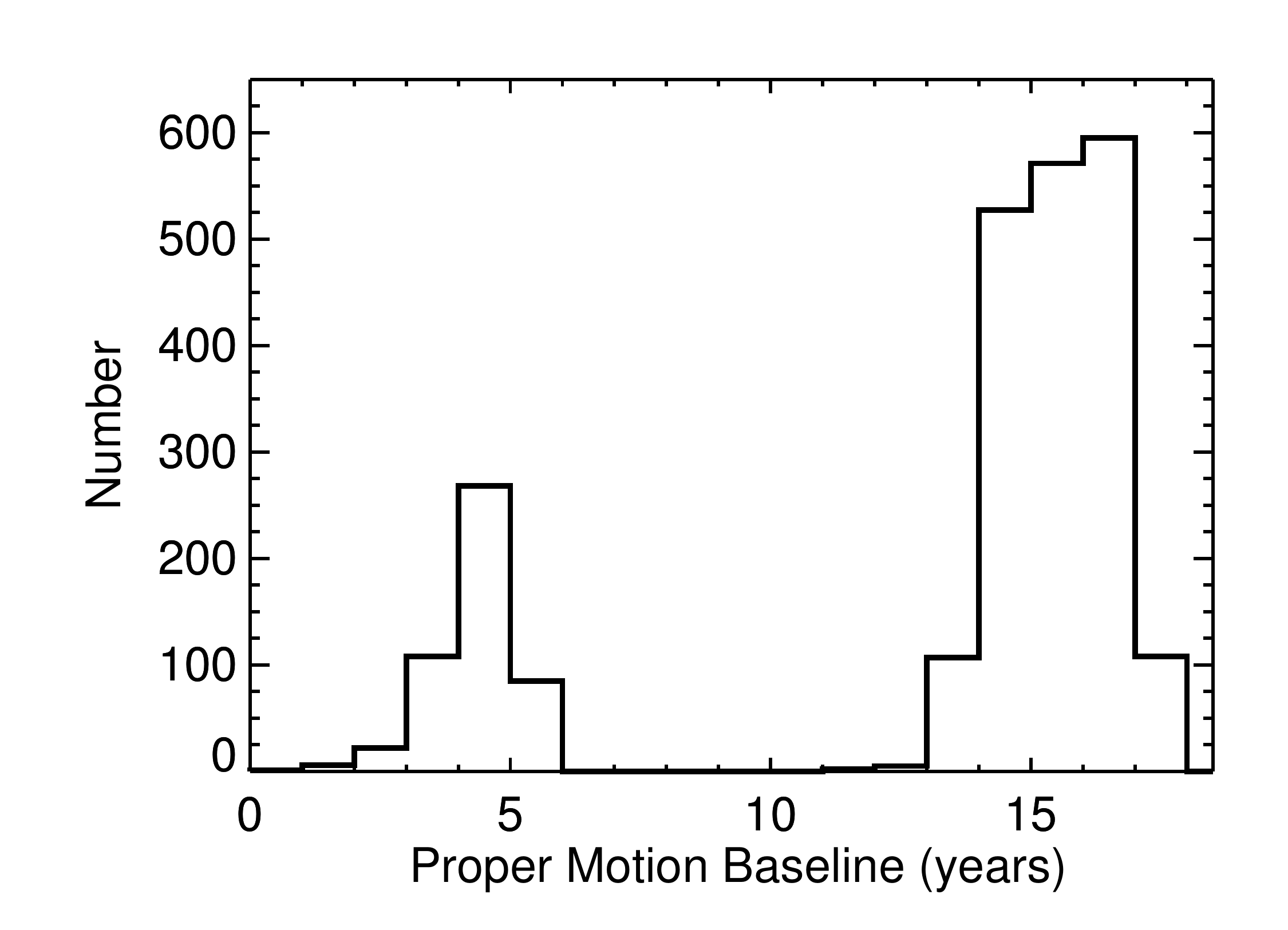}
  \includegraphics[width=0.76\columnwidth, trim = 10mm 0 0 0]{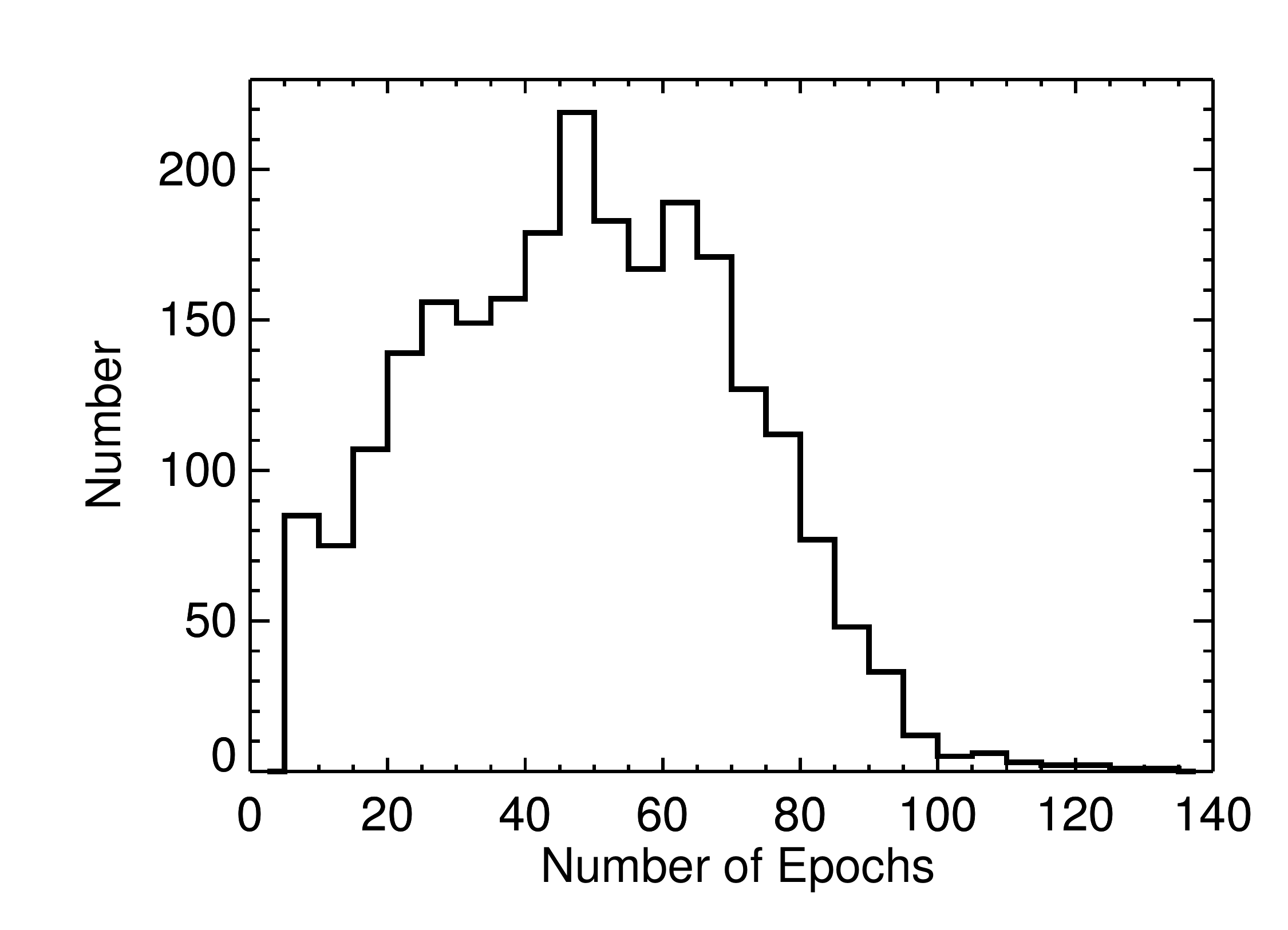}
  \caption{Distributions of the time baselines (\textit{top}) and number of
    epochs (\textit{bottom}) used to calculate the proper motions of the M6--T9
    dwarfs in our PS1 ultracool catalog.  For most objects (baselines
    $\approx$13-17~years) a 2MASS position was used in the motion fit.
    Baselines less than 10~years indicate that only PS1 astrometry was used.
    When available (for about one-quarter of the objects), a \textit{Gaia}~DR1
    position was also included. We required a minimum of seven epochs in order
    to calculate a proper motion fit.}
  \label{fig.pm.baselines}
\end{center}
\end{figure}

Figure~\ref{fig.pm.baselines} shows the time baselines and number of epochs used
for the ultracool proper motions in our catalog.  Most of our proper motions
were calculated using a 2MASS position and have time baselines spanning
13--17~years.  Proper motions using only PS1 astrometry have time baselines
spanning $1-5$~years (data taken 2009--2014, including during PS1
commissioning).  For about one-quarter of our sample (593 objects), a
\textit{Gaia}~DR1 position was included in our proper motion.

\begin{figure}
\begin{center}
  \includegraphics[width=0.76\columnwidth, trim = 10mm 0 0 0]{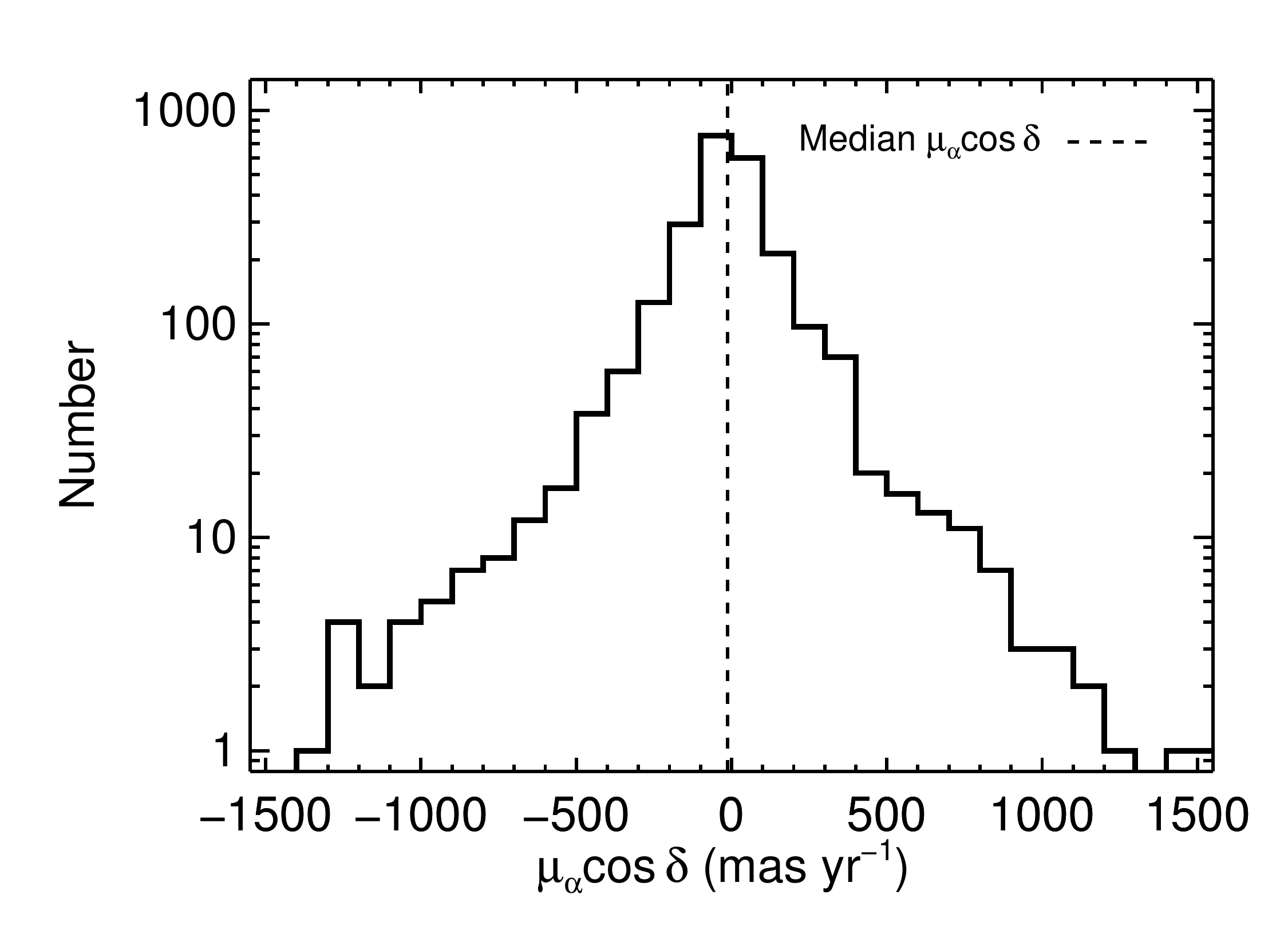}
  \includegraphics[width=0.76\columnwidth, trim = 10mm 0 0 0]{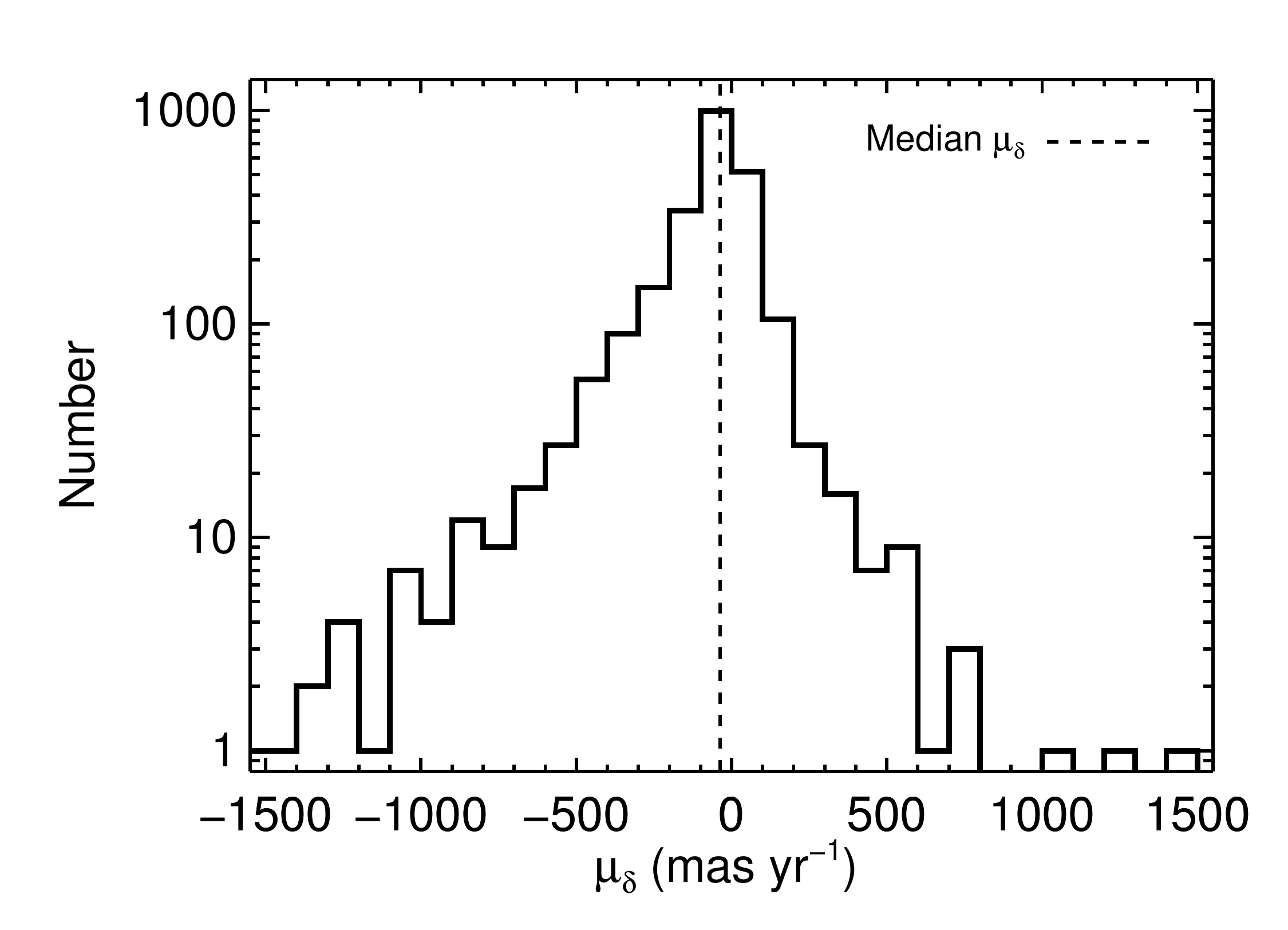}
  \caption{Distributions of the R.A. (\textit{top}) and Decl. (\textit{bottom})
    proper motion components of the M6--T9 dwarfs in our PS1 ultracool
    catalog. The vertical dashed lines indicate the median values.  The
    distributions are similar to those found in previous ultracool surveys (S10,
    BDKP, BASS, and LaTE-MoVeRS).}
  \label{fig.pm.distrib.radec}
\end{center}
\end{figure}

\begin{figure}
\begin{center}
  \includegraphics[width=0.76\columnwidth, trim = 10mm 0 0 0]{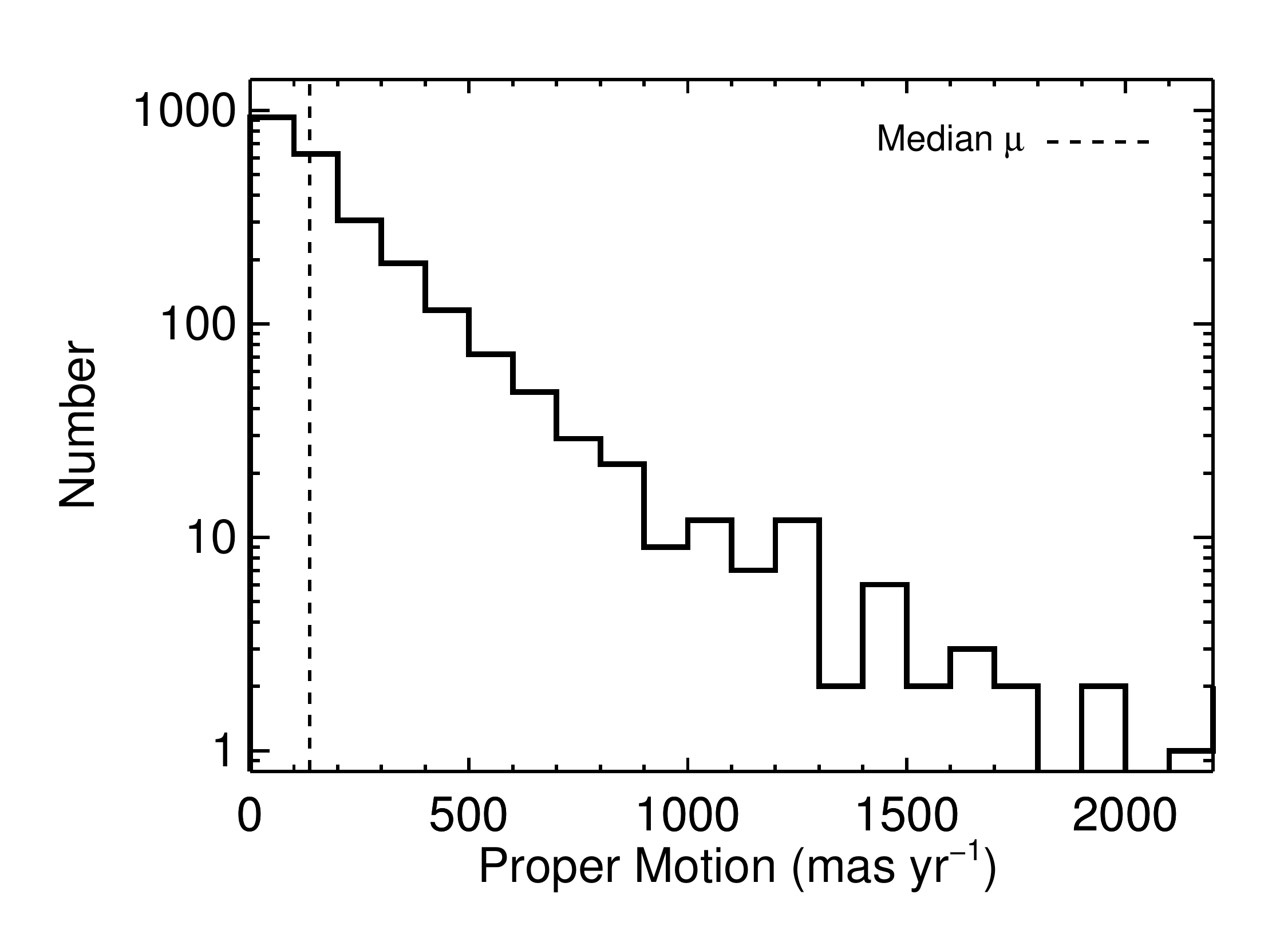}
  \includegraphics[width=0.76\columnwidth, trim = 0 0 0 0]{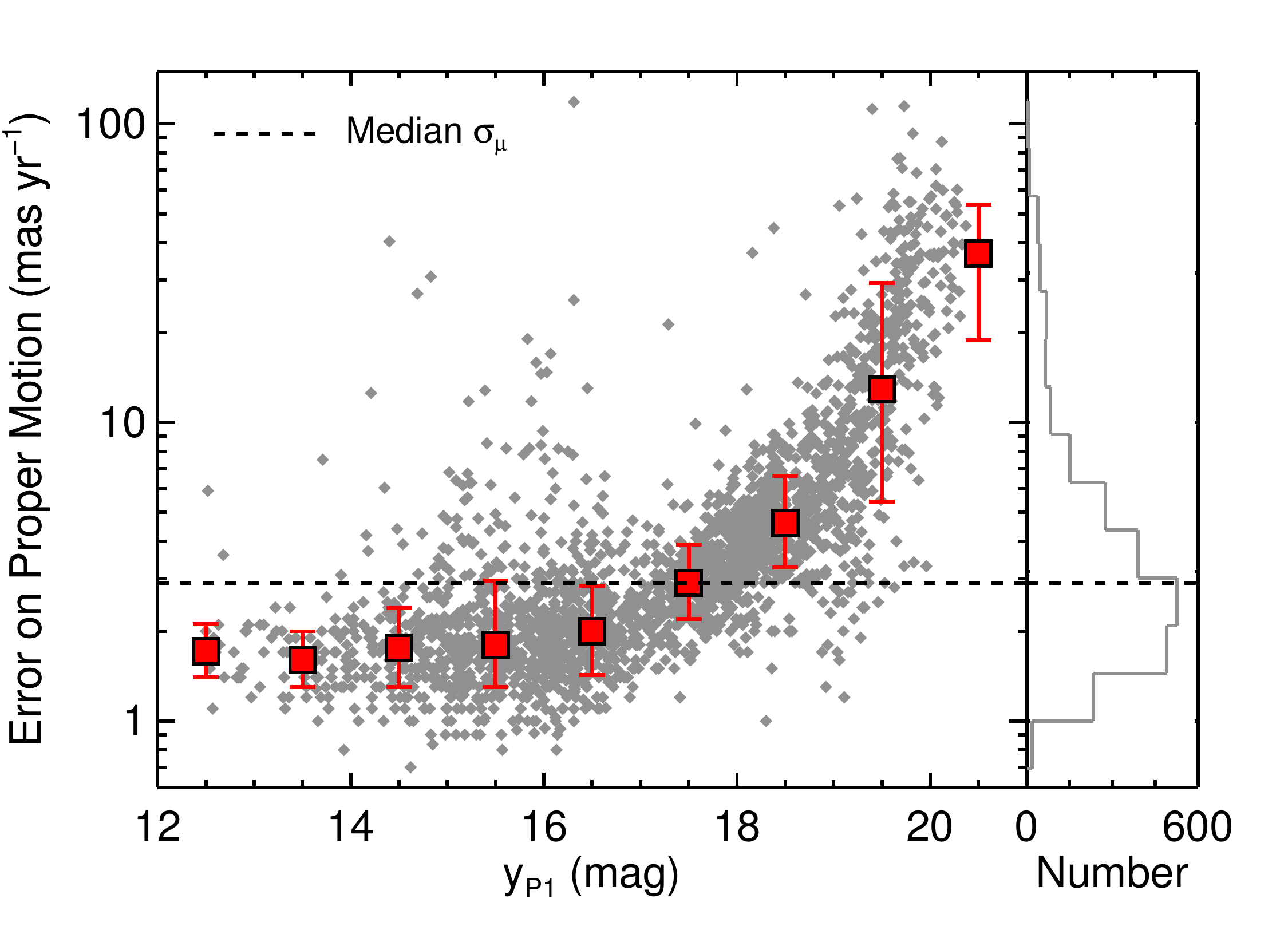}
  \caption{Distribution of the total proper motions (\textit{top}) and proper
    motion errors as a function of \yps\ (\textit{bottom}) for the M6--T9 dwarfs
    in our PS1 ultracool catalog.  The dashed lines indicate the median values,
    including a median error of \varnpmerr~\my.  The large red squares indicate
    median errors for bins of one magnitude in \yps, with error bars showing
    68\% confidence limits.  Our errors are $\approx$9~times smaller than those
    in S10, $\approx$5~times smaller than those in BDKP, and $\approx$3~times
    smaller than those in BASS and LaTE-MoVeRS.}
  \label{fig.pm.distrib.total}
\end{center}
\end{figure}

Figures \ref{fig.pm.distrib.radec} and~\ref{fig.pm.distrib.total} show the
proper motion distributions for our PS1 ultracool catalog.  We find median
proper motion components of $\mua=-13.4$~\my\ and $\mud=-37.1$~\my.  Our
distributions are similar to those found in previous catalogs of ultracool
proper motions, including the L~dwarf catalog of \citet[hereinafter
S10]{Schmidt:2010ex}, the Brown Dwarf Kinematics Project
\citep[BDKP;][]{Faherty:2009kg,Faherty:2012cy}, the BANYAN All-Sky Survey (BASS)
Input Sample \citep{Gagne:2015ij}, and the Late-Type Extension to MoVeRS
\citep[LaTE-MoVeRS;][]{Theissen:2017df}.  Our median proper motion error of
\varnpmerr~\my\ (Figure~\ref{fig.pm.distrib.total}) is a factor of $\approx$9
smaller than that of S10, $\approx$5 smaller than BDKP, and $\approx$3 smaller
than BASS and LaTE-MoVeRS.  The high precision of the PS1 measurements is a
consequence of the astrometric precision of PS1 and the large number of epochs
($N>20$ for 90\% of proper motions), as well as the long time baseline for
objects with a 2MASS position.

Figure~\ref{fig.pm.compare} compares our PS1 proper motions to those of S10,
BDKP, BASS, and LaTE-MoVeRS, each of which shares more than 350 objects in
common with our catalog, as well as other literature sources.  We note that the
overlaps of our PS1 catalog and the other catalogs are predominantly L~dwarfs,
with very few T~dwarfs in common.  We also include a comparison to Motion
Verified Red Stars \citep[MoVeRS;][]{Theissen:2016gn}. MoVeRS contains mostly
earlier-type M~dwarfs and hotter stars and has only 132 ultracool dwarfs in
common with our PS1 proper motion catalog (mostly late-M dwarfs), so it does not
provide as robust a comparison for our full spectral type range as do the other
catalogs.  Our proper motions are consistent with all of these large catalogs,
within $2\sigma$ for $\approx$95\% of objects in common from BASS, MoVeRS, and
LaTE-MoVeRS, and for $\approx$90\% of objects in common from BDKP and S10.  In
addition, we see no systematic offset between our measurements and those from
any of the comparison catalogs.  We do see a slight offset from the aggregate of
other published proper motions (many sources), indicating that proper motions in
the literature tend be slightly larger than those from PS1 and the other large
catalogs listed here.  Nevertheless, 90\% of our values are consistent at
$3\sigma$ or less with these diverse literature sources.

\begin{figure*}
\begin{center}
  \begin{minipage}[t]{0.66\textwidth}
    \includegraphics[width=1.00\columnwidth, trim = 10mm 0 0 0]{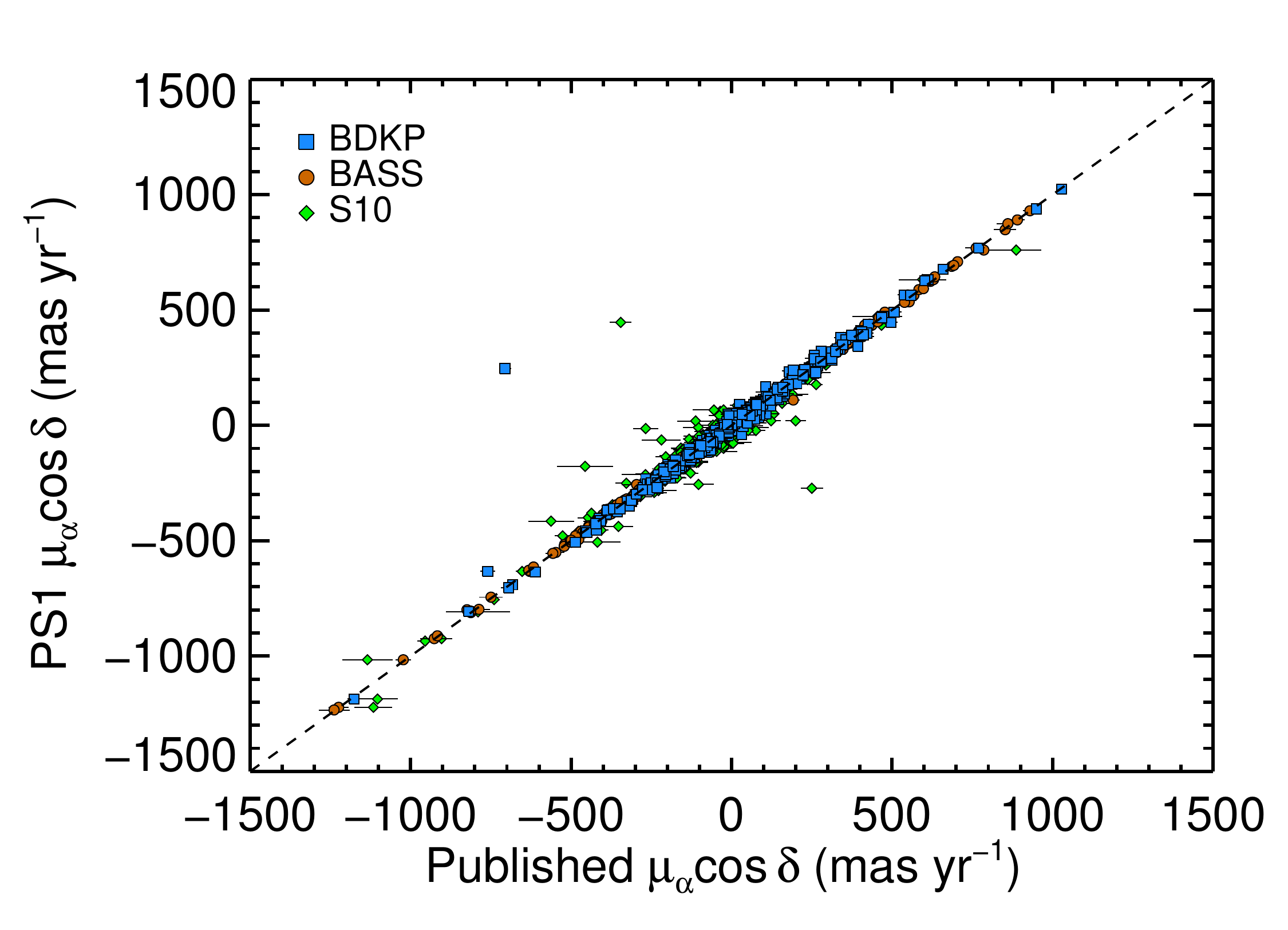}
  \end{minipage}
  \hfill
  \begin{minipage}[t]{0.32\textwidth}
    \includegraphics[width=1.00\columnwidth, trim = 10mm 0 0 0]{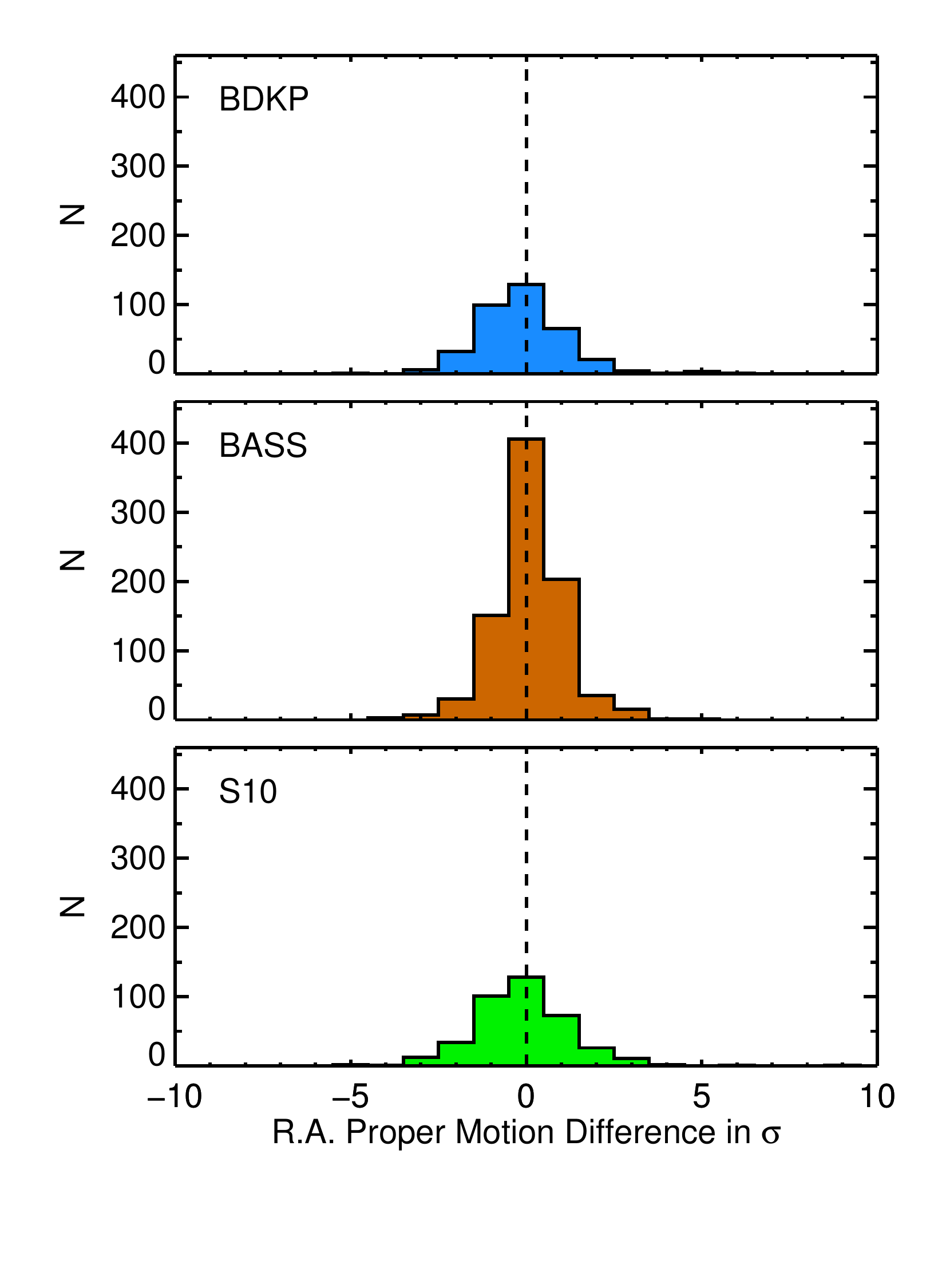}
  \end{minipage}
  \begin{minipage}[t]{0.66\textwidth}
    \includegraphics[width=1.00\columnwidth, trim = 10mm 0 0 0]{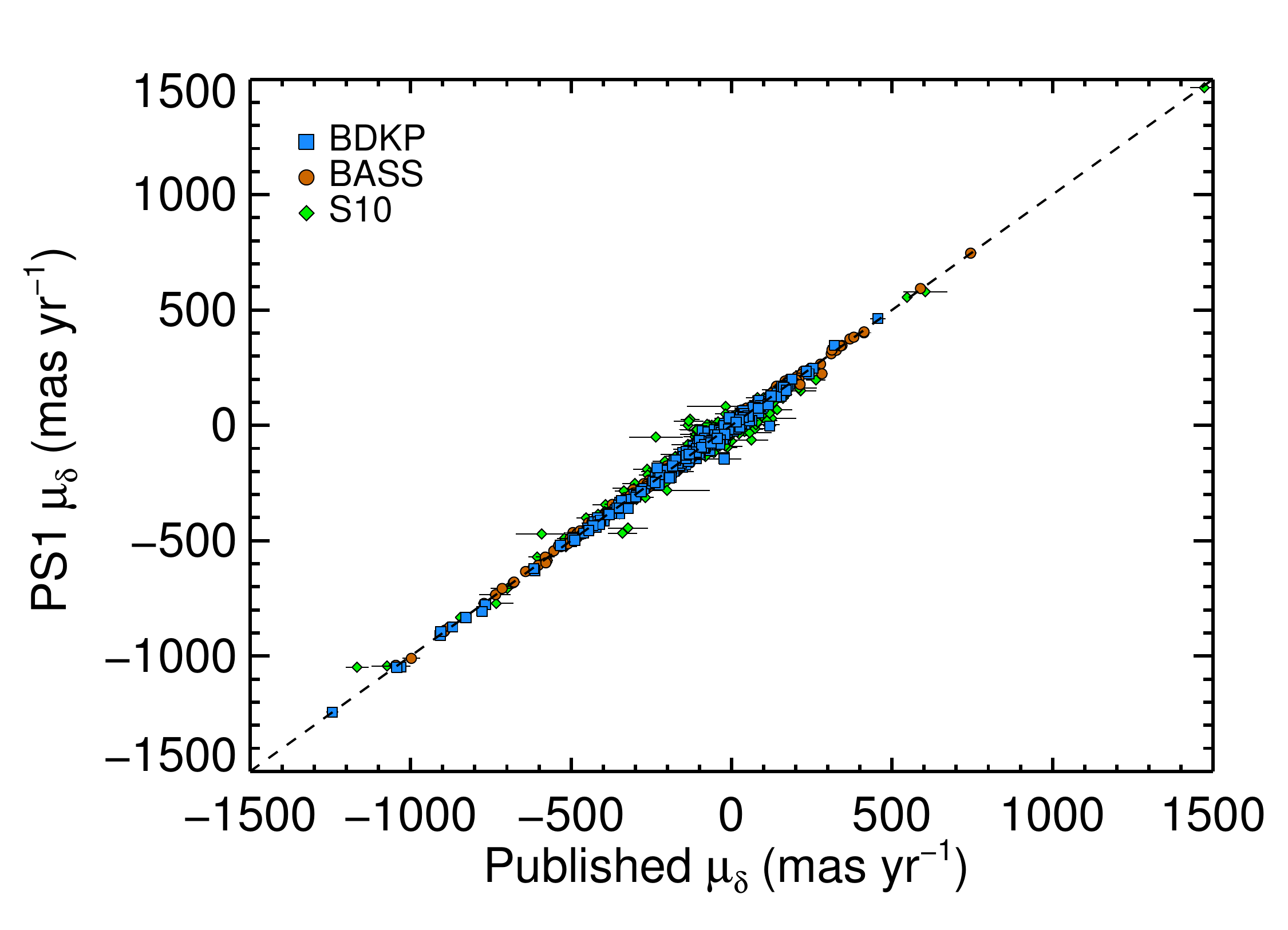}
  \end{minipage}
  \hfill
  \begin{minipage}[t]{0.32\textwidth}
    \includegraphics[width=1.00\columnwidth, trim = 10mm 0 0 0]{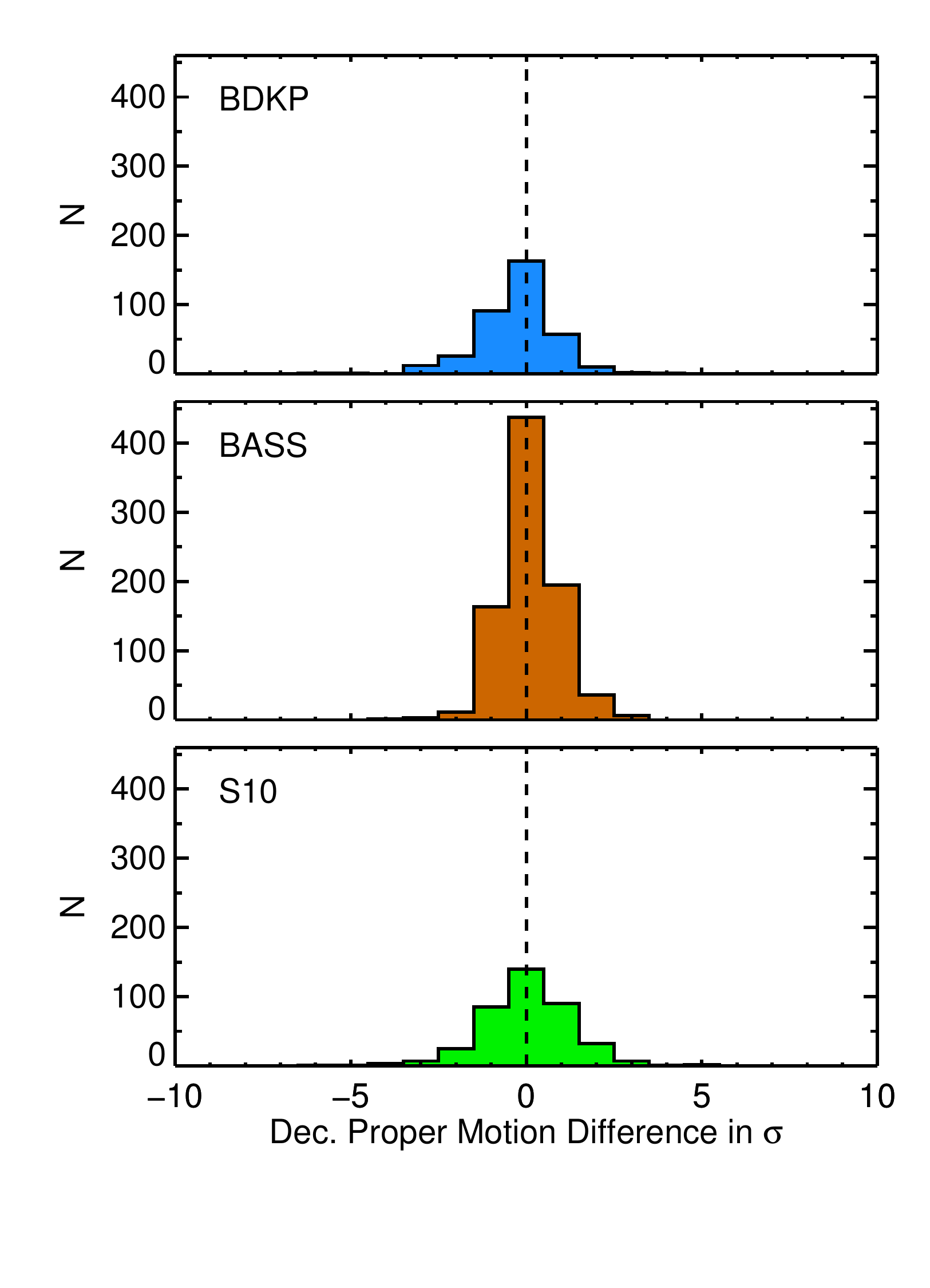}
  \end{minipage}
  \caption{\textit{Left}: Comparison of our PS1 proper motions in
    R.A.~(\textit{top}) and Decl.~(\textit{bottom}) with other proper motions
    from the literature, highlighting the large BDKP, BASS, S10, and (on the
    following page) MoVeRS and LaTE-MoVeRS catalogs. \textit{Right}: Histograms
    showing differences in R.A. and Decl.~proper motions (computed as
    $\mathrm{PS1~value}-\mathrm{literature~value}$).  $\approx$95\% of are
    within $2\sigma$ of the BASS, MoVeRS, and LaTE-MoVeRS measurements.
    $\approx$90\% of our proper motions are within $2\sigma$ of the BDKP and S10
    measurements, and within $3\sigma$ of other literature values.  The
    histograms show no systematic offset between our PS1 proper motions and
    those of the S10, BDKP, BASS, MoVeRS, and LaTE-MoVeRS catalogs, but suggest
    that other literature sources tend to have slightly larger proper motions.}
\figurenum{fig.pm.compare.1}
\end{center}
\end{figure*}

\begin{figure*}
\begin{center}
  \begin{minipage}[t]{0.66\textwidth}
    \includegraphics[width=1.00\columnwidth, trim = 10mm 0 0 0]{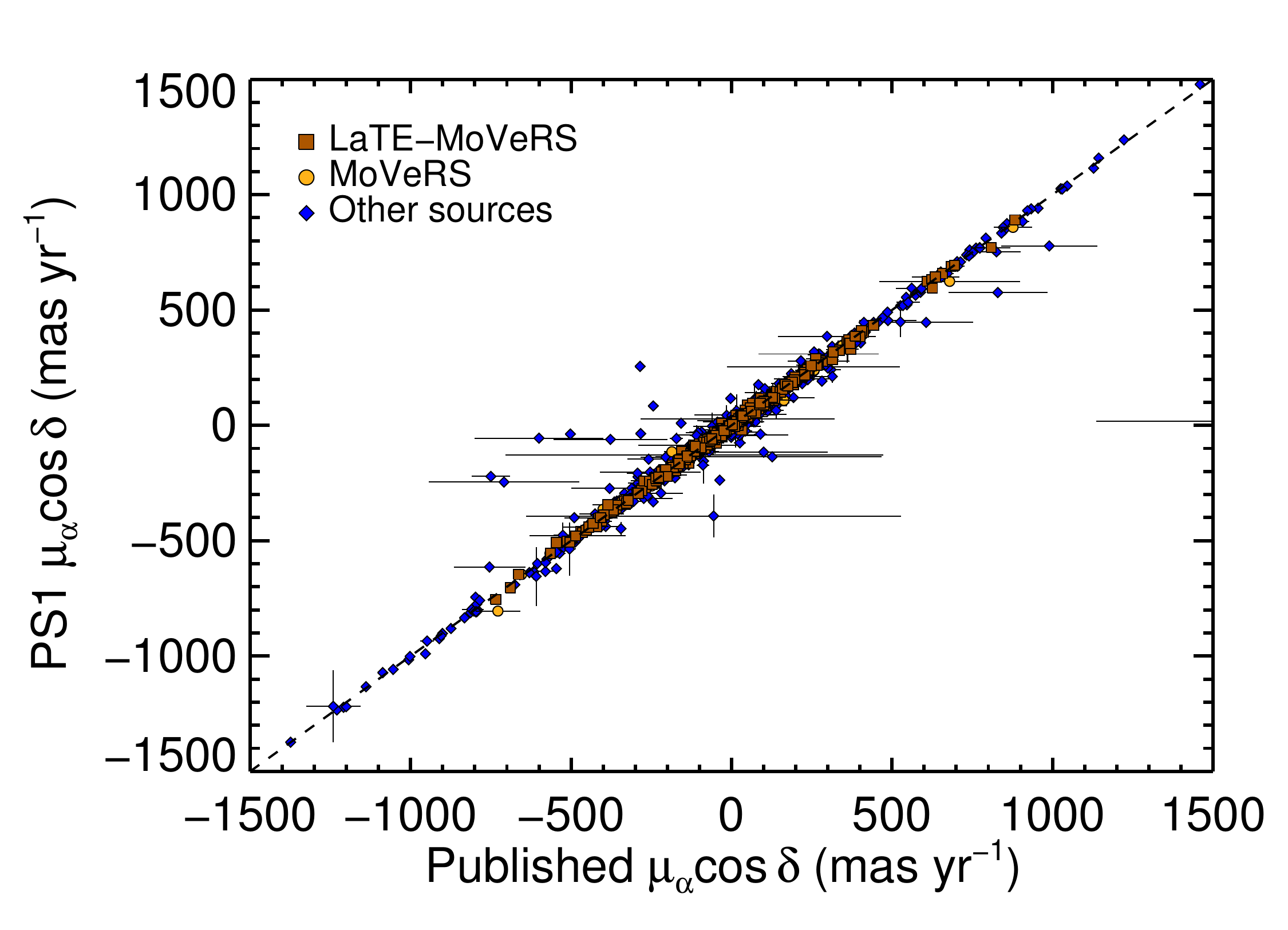}
  \end{minipage}
  \hfill
  \begin{minipage}[t]{0.32\textwidth}
    \includegraphics[width=1.00\columnwidth, trim = 10mm 0 0 0]{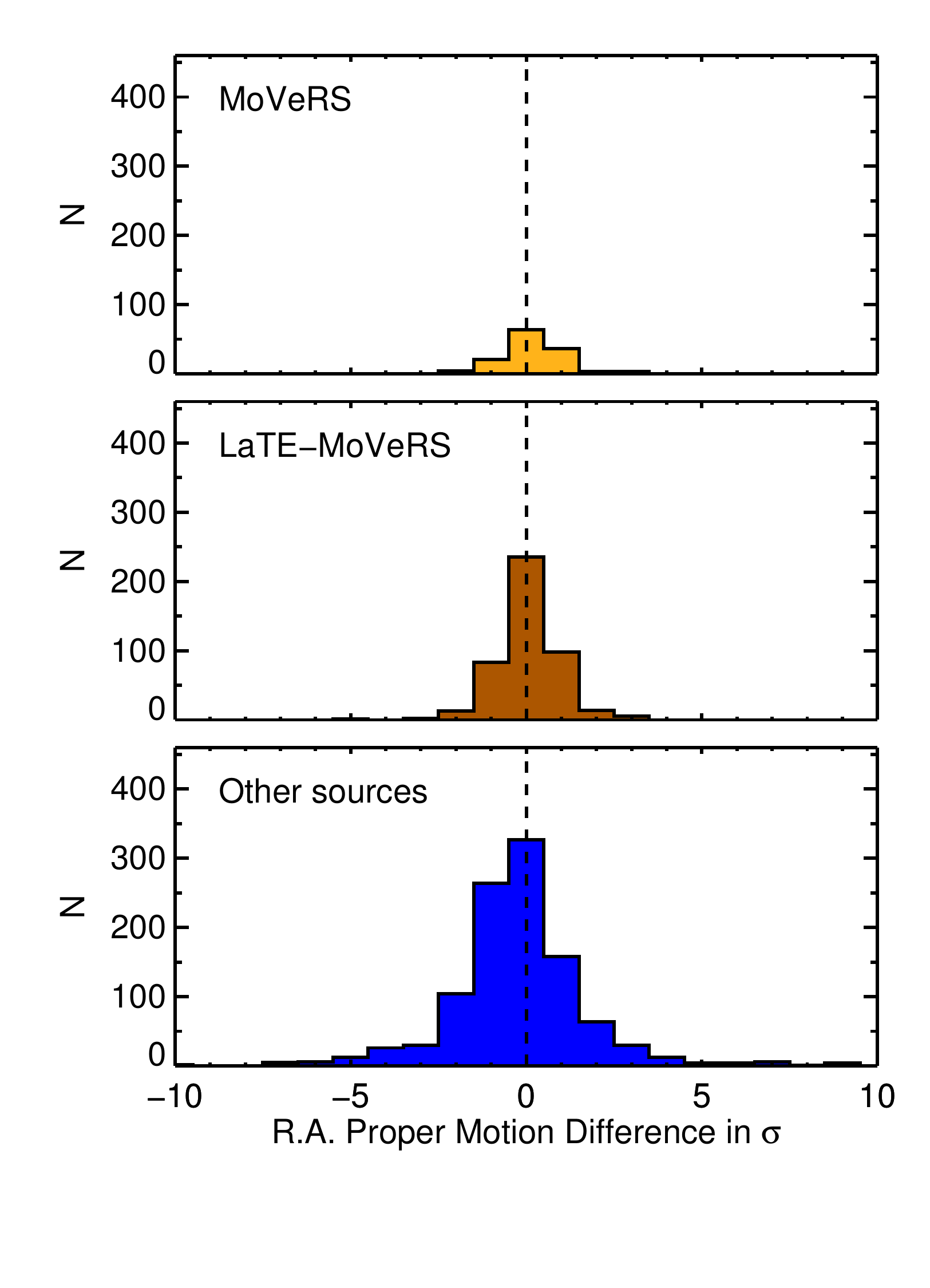}
  \end{minipage}
  \begin{minipage}[t]{0.66\textwidth}
    \includegraphics[width=1.00\columnwidth, trim = 10mm 0 0 0]{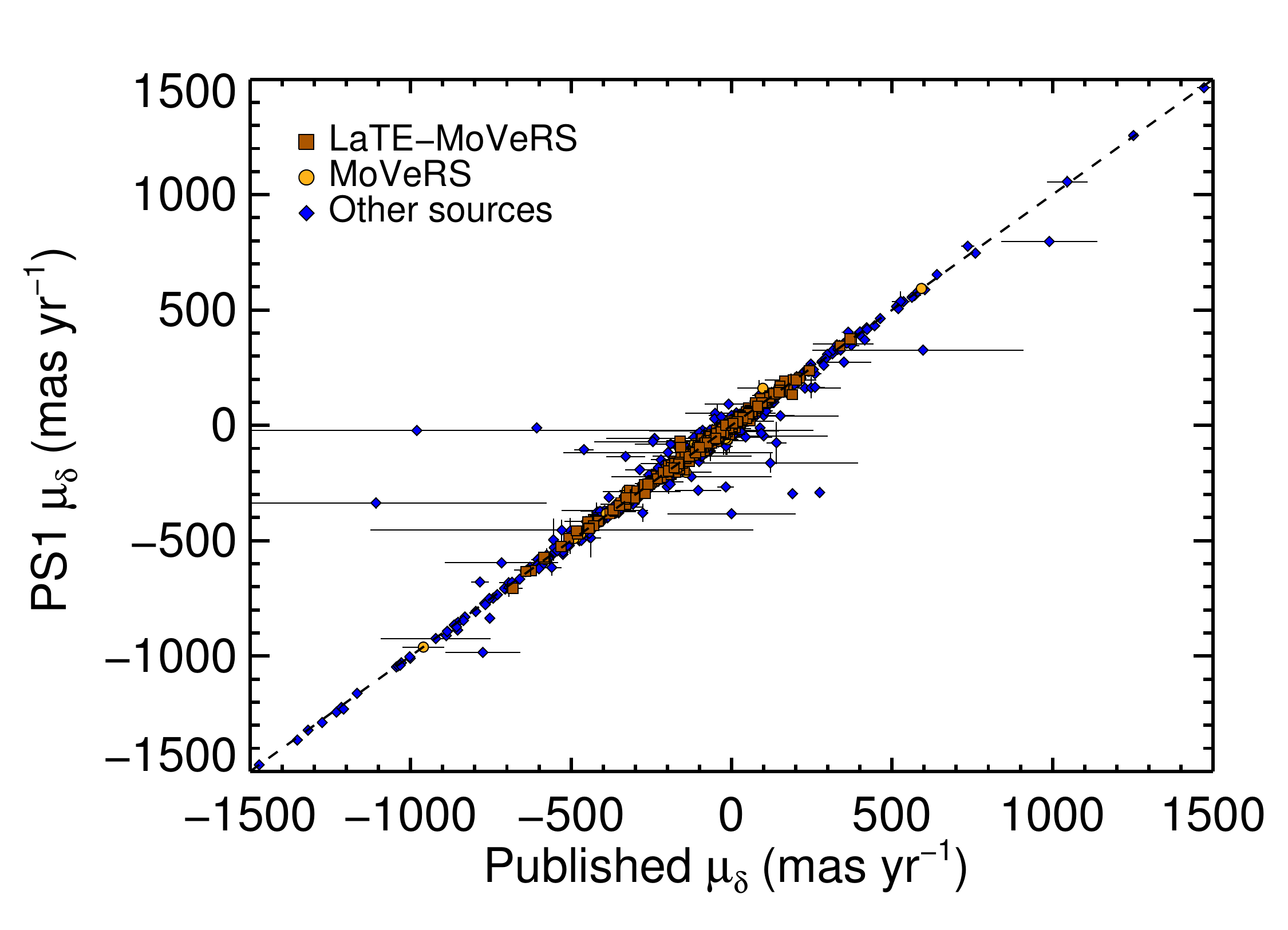}
  \end{minipage}
  \hfill
  \begin{minipage}[t]{0.32\textwidth}
    \includegraphics[width=1.00\columnwidth, trim = 10mm 0 0 0]{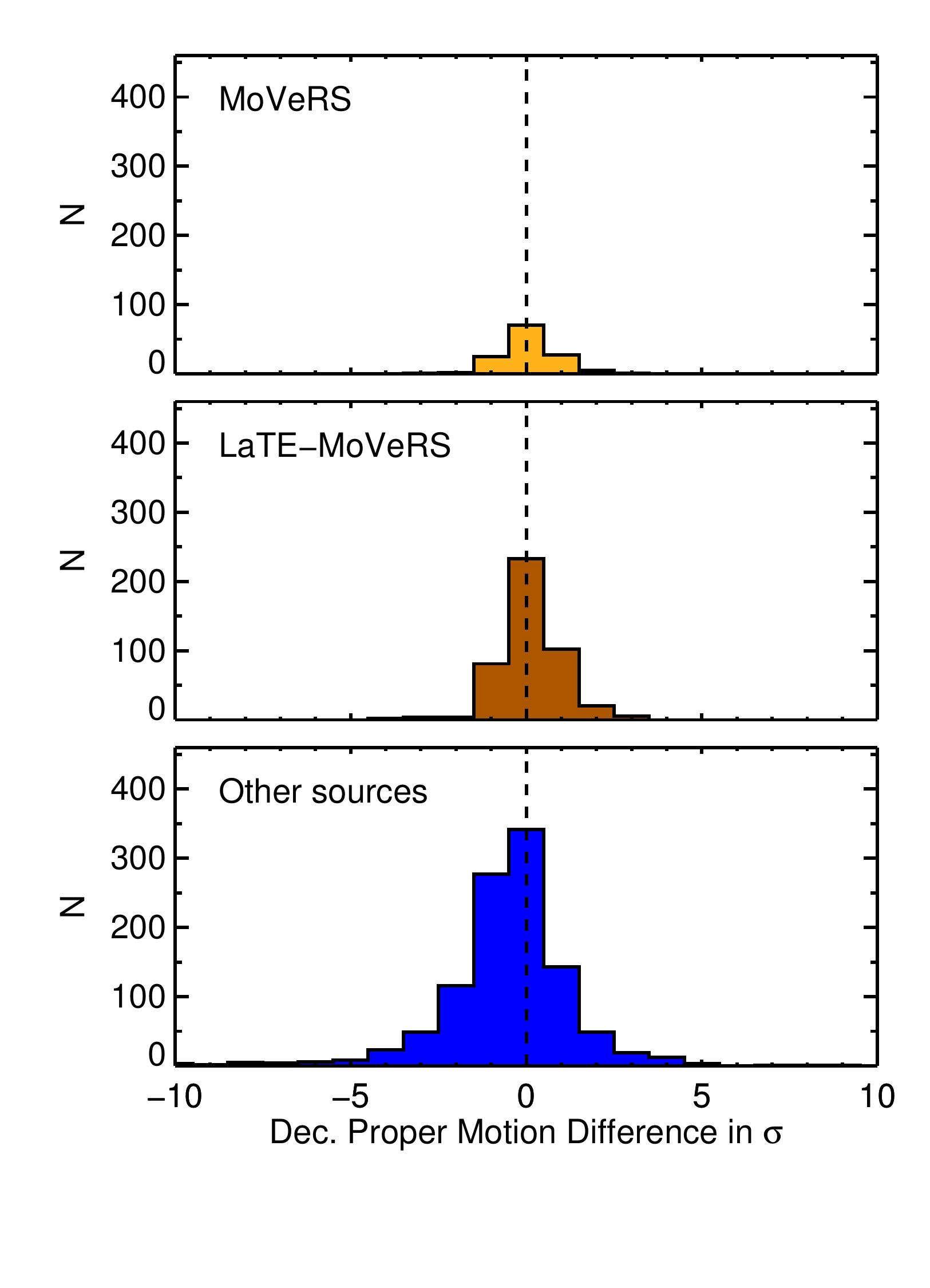}
  \end{minipage}
  \caption{continued.}
  \label{fig.pm.compare}
\end{center}
\end{figure*}

\vspace{30pt}

\subsection{Kinematics}
\label{pm.kin}
Our catalog includes distances and tangential velocities (\vtan) for each object
with a proper motion (Table~\ref{tbl.pm}).  We calculated distances from
parallaxes in the literature (also in Table~\ref{tbl.pm}) when possible.  When
no parallax was available, we used \ips\ photometry and the SED templates from
D16 for M0--M5 dwarfs.  D16 do not quote uncertainties for their SED templates,
so based on the apparent scatter in their color transformations we adopt an
uncertainty of 0.2~mag, and we add this in quadrature with the photometry errors
to determine distance errors.  For M6 and later-type dwarfs lacking parallaxes,
we use $W2$ photometry with the spectral type-absolute magnitude polynomial and
rms from \citet{Dupuy:2012bp} to calculate photometric distances and errors.

\begin{figure}
\begin{center}
    \includegraphics[width=1.00\columnwidth]{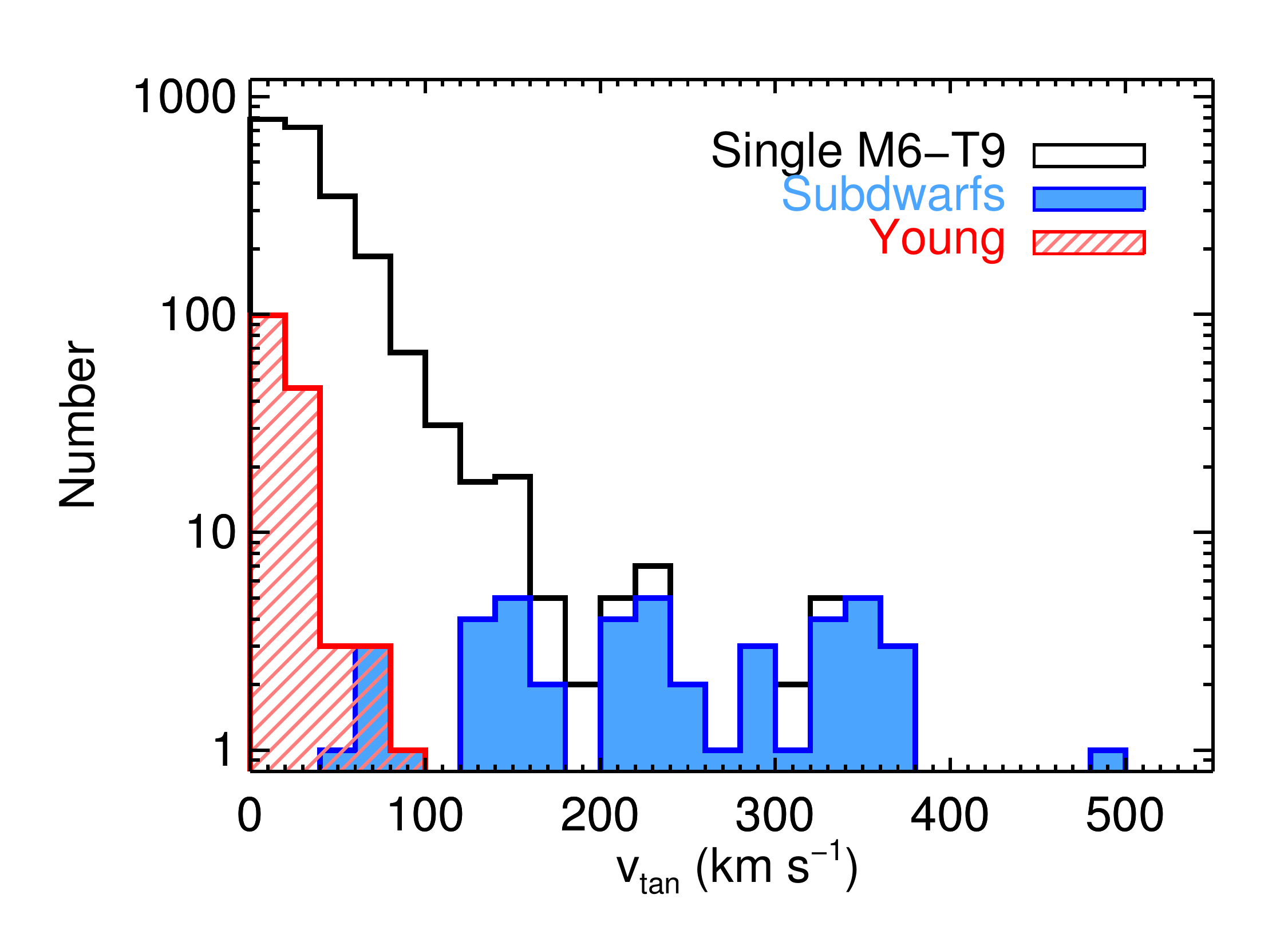}
    \caption{Distribution of tangential velocities for the single M6--T9 dwarfs
      in our catalog (black outline).  Over 90\% of objects have
      $\vtan<75$~\kms\ indicating membership in the thin disk population
      \citep{Dupuy:2012bp}.  The solid blue histogram highlights the subdwarfs
      in our catalog, which have the high \vtan\ values typical of members of
      the older thick disk and halo populations.  The hatched red histogram
      shows young objects in our catalog, which typically have smaller \vtan\
      values than field-age objects.}
  \label{fig.pm.vtan}
\end{center}
\end{figure}

Figure~\ref{fig.pm.vtan} shows the distribution of tangential velocities among
our single M6--T9 dwarfs, highlighting the young objects
(Section~\ref{catalog.young}) and subdwarfs.  Excluding the young objects and
subdwarfs, we find a median of $\vtan=29$~\kms\ and a dispersion of
$\stan=29$~\kms, consistent with the $\vtan=26$~\kms\ and $\stan=25$~\kms\ found
by \citet[hereinafter F09]{Faherty:2009kg} for their 20~pc volume-limited
ultracool sample and the $\vtan=28$~\kms\ and $\stan=25$~\kms\ found by S10 for
their L~dwarf sample.  90\% of our PS1 sample has $\vtan<75$~\kms, indicating a
very high probability of membership in the thin disk population
\citep{Dupuy:2012bp}. 98\% of our sample has $\vtan<200$~\kms, indicating that
most of the remaining objects are likely to be in the thick disk population
\citep{Dhital:2010ir,Dupuy:2012bp}, consistent with the kinematics of the
LaTE-MoVeRS sample \citep{Theissen:2017df}.

Figure~\ref{fig.pm.vtan} demonstrates that the young objects and subdwarfs are
members of distinct kinematic populations, corroborating \citet{Faherty:2012cy}.
Almost all young objects have $\vtan<60$~\kms, with a median of 16~\kms\ and
dispersion 13~\kms, slower than the rest of the thin-disk population.  On the
other hand, all but one subdwarf have $\vtan>60$~\kms\ and subdwarfs comprise
85\% of the objects with $\vtan>200$~\kms, extending to much higher velocities
than other objects.  The high \vtan\ of the subdwarfs implies they are likely to
be members of the older thick disk or halo ($\vtan>200$~\kms) populations, as
expected for low-metallicity objects.

F09 and S10 used \jkt\ colors, tangential velocities, and velocity dispersions
to identify young populations of late-M and L~dwarfs with kinematics distinct
from the field population.  F09 found that red \jkt\ outliers in their sample
have lower \vtan\ and \stan, while blue outliers have higher \vtan\ and \stan.
S10 found evidence that this correlation between \jkt\ and velocity dispersion
extends throughout the field L~dwarf population and is not limited to outliers.
Both studies link the reddest objects to young, low-velocity thin disk
populations and the bluest objects to older, high-velocity thick disk and halo
populations.

\begin{figure}
\begin{center}
  \includegraphics[width=1.00\columnwidth]{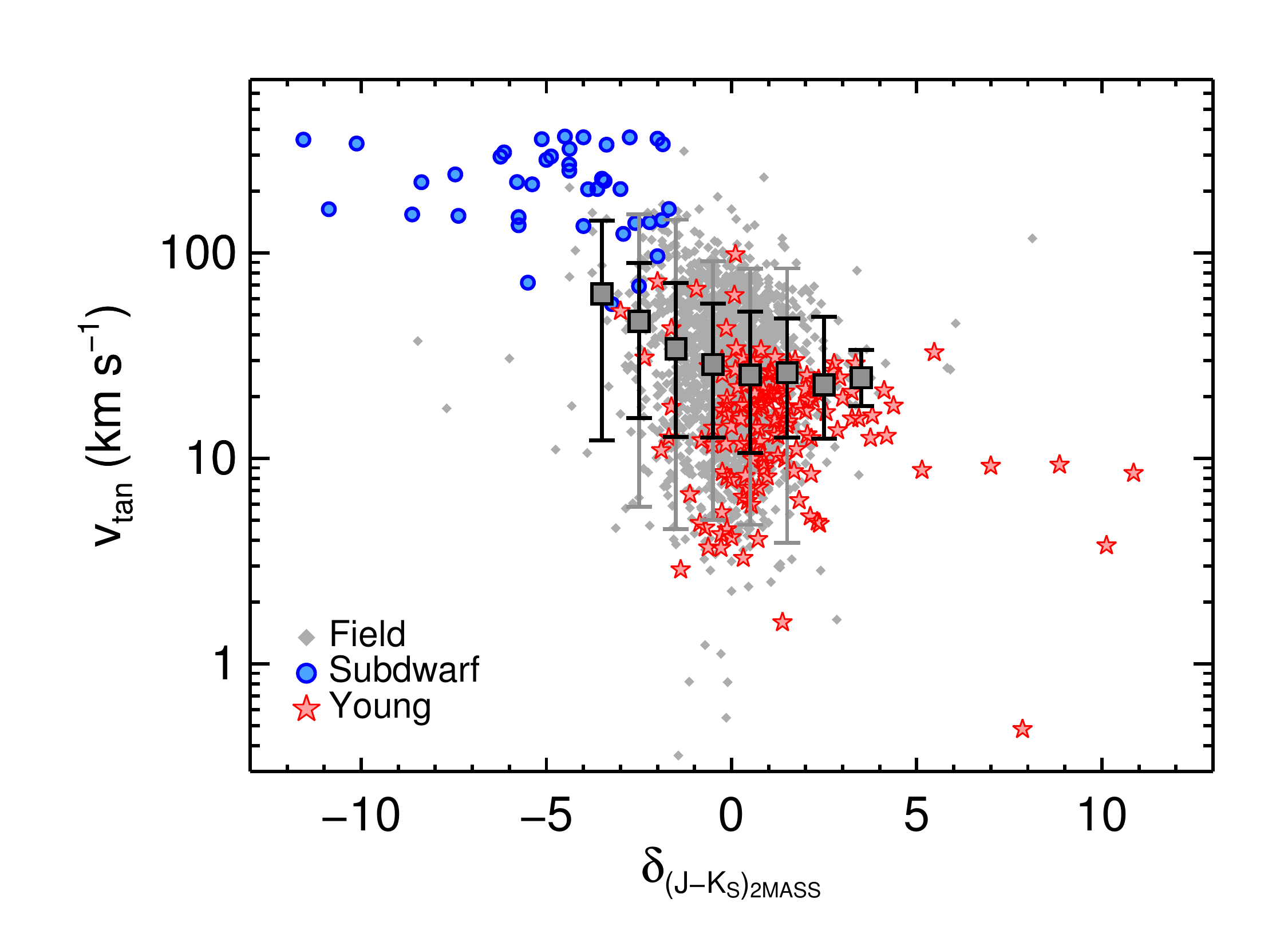}
  \caption{Tangential velocity as a function of $\delta_{(J-K_S)_{\rm 2MASS}}$
    (the number of standard deviations by which an object's \jkt\ color differs
    from the median \jkt\ for its spectral type) for single M6--T9 dwarfs.  The
    large gray squares indicate median \vtan\ values for $1\sigma$ bins of
    $\delta_{J-K_S}$, with dark and light error bars marking 68\% and 95\%
    confidence intervals, respectively.  The medians do not include subdwarfs or
    young objects, but we overplot these with blue circles and red stars,
    respectively, for comparison.  The blue end of the field population tends
    toward higher \vtan, but on the red end there is no correlation with \vtan.}
  \label{fig.pm.vtan.colors}
\end{center}
\end{figure}

To examine the relationship between \vtan\ and color differences in our catalog,
we use the $\delta_{J-K_S}=[(J-K_S)-(J-K_S)_{\rm med}]/\sigma_{J-K_S}$ defined
by S10, where $(J-K_S)_{\rm med}$ and $\sigma_{J-K_S}$ are the median and rms
$J-K_S$ colors, respectively, for each spectral type.  $\delta_{J-K_S}$
therefore gives us a spectral type-independent measurement of the extent to
which an object's $J-K_S$ color differs from the median, where a negative
$\delta_{J-K_S}$ value means the object is bluer than the median color.  In
Figure~\ref{fig.pm.vtan.colors}, we compare our \vtan\ to $\delta_{J-K_S}$ for
single M6--T9 dwarfs.  We also calculated median \vtan\ values for bins of
$1\sigma$ in $\delta_{J-K_S}$, excluding subdwarfs and young objects from the
medians in order to assess the color dependence of \vtan\ for the generic
ultracool field population.  We show these median \vtan\ values in
Figure~\ref{fig.pm.vtan.colors}, and we overplot the subdwarfs and young objects
for comparison.  The blue outliers are mostly subdwarfs with $\vtan>100$~\kms\
while the red outliers are primarily young objects with $\vtan<10$~\kms,
supporting the link between color outliers and age found by F09 and S10.
Excluding the subdwarfs and young objects, the field population in our catalog
shows a trend toward higher \vtan\ for bluer-than-median objects that is
consistent with the \stan\ trend identified by S10, but we see no correlation
between \vtan\ and $J-K_S$ for redder objects.

\begin{figure}
\begin{center}
  \includegraphics[width=1.00\columnwidth]{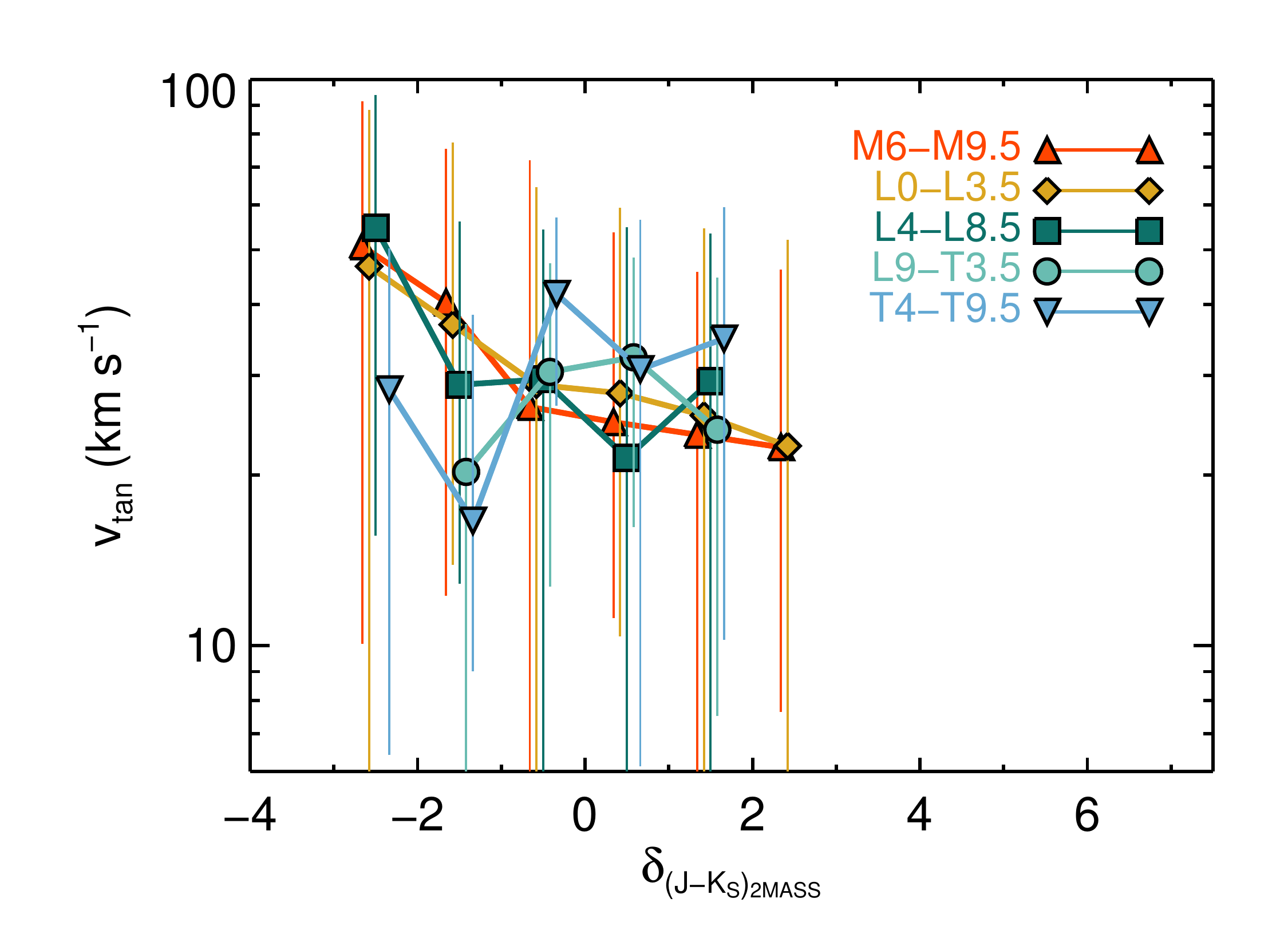}
  \caption{Median tangential velocities for $1\sigma$ bins of $\delta_{J-K_S}$
    (same as in Figure~\ref{fig.pm.vtan.colors}), split into five spectral type
    ranges (see legend).  The $\delta_{J-K_S}$ bins are the same for each
    spectral type range; the median symbols have been offset slightly to improve
    visibility. The colored lines indicate the 68\% confidence intervals for
    each $\delta_{J-K_S}$ bin.  The trend toward higher tangential velocities
    for bluer objects holds true for the late-M and L~dwarfs, but not for
    T~dwarfs.}
  \label{fig.pm.vtan.colors.sptbins}
\end{center}
\end{figure}

We explored whether the trend of higher \vtan\ for bluer-than-median objects
held true across the full spectral type range of our sample, because color
variations in different spectral types will have different physical causes
(e.g., different types of atmospheric clouds).  We calculated the same median
\vtan\ values for narrower ranges of spectral type (M6--M9.5, L0--L3.5,
L4--L8.5, L9--T3.5, and T4--T9).  Figure~\ref{fig.pm.vtan.colors.sptbins} shows
these medians, which make clear the that the trend toward higher \vtan\ for
bluer field objects is a late-M and L~dwarf phenomenon but does not apply to
T~dwarfs.

\begin{figure}
\begin{center}
  \includegraphics[width=1.00\columnwidth]{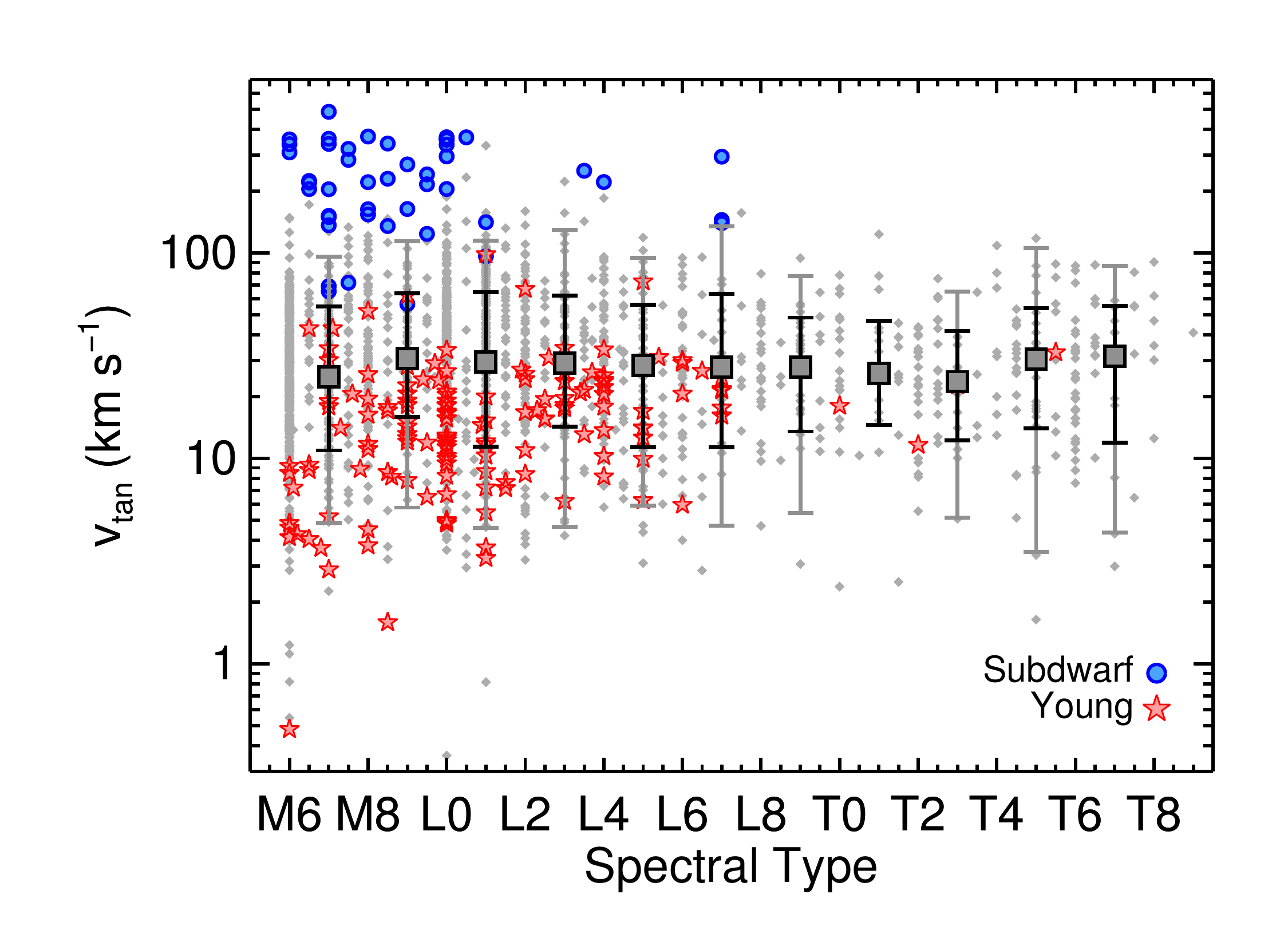}
  \caption{Tangential velocity as a function of spectral type for the single
    M6--T9 dwarfs in our catalog, using the same format as in
    Figure~\ref{fig.pm.vtan.colors}.  Large gray symbols indicate median
    tangential velocities for bins of two spectral subtypes, excluding subdwarfs
    and young objects.  We find no correlation between spectral type and
    tangential velocity in our catalog.}
  \label{fig.pm.vtan.spt}
\end{center}
\end{figure}

We also explored the direct relationship between spectral type and tangential
velocity, as any correlation would imply an age trend in ultracool spectral
types.  Figure~\ref{fig.pm.vtan.spt} shows \vtan\ as a function of spectral
type.  We include median \vtan\ values for bins of two spectral subtypes,
excluding subdwarfs and young objects from the medians but overplotting them for
reference.  We see no evidence for a dependence of \vtan\ on spectral type for
ultracool dwarfs.

\section{A Binary Discovered in PS1}
\label{binary}
During outlier inspection we discovered a new visual binary.
2MASS~J09033514$-$0637336 (hereinafter 2MASS~J0903$-$0637), first identified and
assigned a spectral type of M7 by \citet{Cruz:2003fi}, is clearly resolved in
individual \rps, \ips, \zps, and \yps\ warp images, even though it appears in
the PS1 database as a single object (see Section~\ref{catalog.binaries} for a
discussion of similar objects).  Using the \ips\ warp images, we measure a
difference in flux of $0.10\pm0.03$~mag, separation of

$1.\!\!''14\pm0.\!\!''02$, and position angle of
$239.\!\!^\circ3\pm1.\!\!^\circ0$ for 2MASS~J0903$-$0637.  We detect no change
in separation or position angle over the four-year span of the PS1 survey.
2MASS~J0903$-$0637 has a proper motion from PS1 of $63.9\pm3.4$~\my; if one of
the two components were actually a stationary background object, we would see
the other component move by $\approx$0.$\!\!''25$ in four years, a change in
position significantly larger than our measurement uncertainties. We therefore
conclude that the components are gravitationally bound.  Using the \ips\
photometry for 2MASS~J0903$-$0637 and the absolute magnitude for M7 dwarfs from
D16, and correcting for a binary with the measured 0.10~mag flux ratio, we
calculate a photometric distance of $60.9\pm5.7$~pc.  This places the projected
binary separation at $69.5\pm6.6$~AU, unusually wide for ultracool field
binaries \citep[e.g.,][]{Duchene:2013il}.  As expected given the small angular
separation, the pair is unresolved in 2MASS and \WISE\ images, and therefore was
not previously identified as a binary.  This was a serendipitous discovery, and
we have not undertaken a comprehensive search for more binaries that are newly
resolved by PS1 for this catalog.

\section{Summary}
\label{summary}
We present a catalog of \varnobjtot\ M, L, and T dwarfs with photometry and
proper motions from the \PS\ 3$\pi$~Survey.  This catalog contains all known L
and T dwarfs as of 2015 December having well-measured photometry in at least one
of the five PS1 bands (\grizy), including \varnldwarf~L~dwarfs and
\varntdwarf~T~dwarfs.  The catalog also contains \varnmdwarfmega~late-M dwarfs
chosen to represent the diversity of the nearby population, including
low-gravity objects, high proper motion objects, young moving group members,
known or suspected binaries, and wide companions to more massive stars, along
with a large sample of \varnmdwarfwest~field M dwarfs identified by SDSS.  We
cross-matched our catalog with 2MASS, AllWISE, and \textit{Gaia}~DR1 to obtain
photometry spanning 0.55 to 12~\um.  We carefully vetted the detections in PS1,
2MASS, and AllWISE to ensure their association with previously identified M, L,
and T dwarfs.

We use two types of photometry from PS1 in our catalog: chip photometry (highest
accuracy) for most objects and warp photometry (greater depth) for faint,
slow-moving objects.  We identified a number of false detections (i.e.,
measurements of background noise) at the warp limiting magnitude for objects in
our catalog, and we develop a method for screening warp detections to ensure
they are real.

We use the photometry along with parallaxes from the literature to create
empirical SEDs for field ultracool dwarfs covering \gps\ to $W3$~bands.  We
determine typical colors of M0--T9 dwarfs, and we present numerous
color-spectral type and color-color diagrams, along with median colors for each
spectral subtype.  We separate binaries, young objects, and subdwarfs from the
rest of the field population, and we compare the colors of the different groups.
Our catalog includes \varnrbandldwarf~L~dwarfs detected in \rps, the largest
sample of L~dwarfs detected in this optical band.  \rps\ L~dwarf colors show
striking features, including a sharp blueward turn at the M/L transition due to
decreasing TiO absorption in \rps, and a handful of young objects with colors
bluer than the median for their spectral type.

We calculate proper motions for our catalog using multiple-epoch astrometry from
PS1 along with 2MASS and \textit{Gaia} when available.  Our method allows us to
link the epochs of fast-moving objects that are split into more than one
``object'' in PS1, improving the precision of our proper motions compared to the
PS1 database for 87\% of the M6 and later dwarfs in our catalog and producing
measurements for 63 objects lacking proper motions in PS1.  Our catalog contains
proper motions for \varnpmtot~objects with a median precision of \varnpmerr~\my\
(a factor of $\approx$3$-$10 improvement over previous large catalogs), tied to
the \textit{Gaia}~DR1 reference frame.  The catalog includes proper motions for
a total of \varnpmucool~M6--T9 dwarfs, including \varnnewpmucool~objects with no
previously published values and \varnbetterpmucool~measurements that improve
upon previous literature values.  Our catalog incorporates the largest set of
homogeneous proper motions for L and T dwarfs published to date.

We assess the kinematics of the late-M, L, and T dwarfs in our sample and find
evidence that bluer late-M and L~dwarfs with field ages (i.e., not subdwarfs)
have higher tangential velocities, consistent with the trend towards higher
\stan\ for bluer L dwarfs found by S10.  More work is needed using well-defined
(i.e., volume-limited) samples with accurate distances to precisely characterize
the relationship between colors and kinematics for nearby ultracool dwarfs.

Tables in this paper are available in electronic form from the online journal
and at \url{http://www.ifa.hawaii.edu/users/wbest/Will_Best/PS1_MLT_Dwarfs.html}.

\vspace{20pt} We thank the anonymous referee for detailed comments that improved
the paper in several significant ways, and Gabriel Bihain for helpful comments
on the catalog.  The \PS\ Surveys (PS1) have been made possible through
contributions of the Institute for Astronomy, the University of Hawaii, the
Pan-STARRS Project Office, the Max-Planck Society and its participating
institutes, the Max Planck Institute for Astronomy, Heidelberg and the Max
Planck Institute for Extraterrestrial Physics, Garching, The Johns Hopkins
University, Durham University, the University of Edinburgh, Queen's University
Belfast, the Harvard-Smithsonian Center for Astrophysics, the Las Cumbres
Observatory Global Telescope Network Incorporated, the National Central
University of Taiwan, the Space Telescope Science Institute, the National
Aeronautics and Space Administration under Grant No. NNX08AR22G issued through
the Planetary Science Division of the NASA Science Mission Directorate, the
National Science Foundation under Grant No. AST-1238877, the University of
Maryland, Eotvos Lorand University (ELTE), and the Los Alamos National
Laboratory.  This publication makes use of data products from the Two Micron All
Sky Survey, which is a joint project of the University of Massachusetts and the
Infrared Processing and Analysis Center/California Institute of Technology,
funded by the National Aeronautics and Space Administration and the National
Science Foundation.  This publication makes use of data products from the
Wide-field Infrared Survey Explorer, which is a joint project of the University
of California, Los Angeles, and the Jet Propulsion Laboratory/California
Institute of Technology, and NEOWISE, which is a project of the Jet Propulsion
Laboratory/California Institute of Technology. \WISE\ and NEOWISE are funded by
the National Aeronautics and Space Administration.  This work has made use of
data from the European Space Agency (ESA) mission {\it Gaia}
(\url{http://www.cosmos.esa.int/gaia}), processed by the {\it Gaia} Data
Processing and Analysis Consortium (DPAC,
\url{http://www.cosmos.esa.int/web/gaia/dpac/consortium}). Funding for the DPAC
has been provided by national institutions, in particular the institutions
participating in the {\it Gaia} Multilateral Agreement.  This research has made
use of NASA's Astrophysical Data System, the UKIDSS data products, and the
Database of Ultracool Parallaxes, maintained by Trent Dupuy at
\url{https://www.cfa.harvard.edu/~tdupuy/plx}.  This research has made extensive
use of the SIMBAD and Vizier databases operated at CDS, Strasbourg, France.
This work was greatly facilitated by many features of the TOPCAT software
written by Mark Taylor (\url{http://www.starlink.ac.uk/topcat/}).  WMJB received
support from NSF grant AST09-09222.  WMBJ, MCL, and EAM received support from
NSF grant AST-1313455.

% \facility{\PS}

\bibliography{\string~/Astro/LaTeX/willastro}

\end{document}